\documentclass[]{aastex62}

\usepackage{amsmath}
\usepackage{soul}
\usepackage[normalem]{ulem}
\usepackage{bm}

\usepackage{xcolor} 
\definecolor{darkpink}{rgb}{0.8,0.47,0.47}

\begin{document}

\title{GRB Redshift Classifier to Follow-up High-Redshift GRBs Using Supervised Machine Learning}

\author[0000-0003-4442-8546]{Maria Giovanna Dainotti}
\affiliation{Division of Science, National Astronomical Observatory of Japan, 2-21-1 Osawa, Mitaka, Tokyo 181-8588, Japan}
\affiliation{Department of Astronomical Sciences, The Graduate University for Advanced Studies, SOKENDAI, Shonankokusaimura, Hayama, Miura District, Kanagawa 240-0115, Japan}
\affiliation{Space Science Institute, 4765 Walnut Street, Suite B, Boulder, CO 80301, USA}
\affiliation{Nevada Center for Astrophysics, University of Nevada, 4505 Maryland Parkway, Las Vegas, NV 89154, USA}

\author[0000-0003-4709-0915]{Shubham Bhardwaj}
\affiliation{Division of Science, National Astronomical Observatory of Japan, 2-21-1 Osawa, Mitaka, Tokyo 181-8588, Japan}
\affiliation{Department of Astronomical Sciences, The Graduate University for Advanced Studies, SOKENDAI, Shonankokusaimura, Hayama, Miura District, Kanagawa 240-0115, Japan}

\author{Christopher Cook}
\author[0000-0001-7368-7104]{Joshua Ange}
\affiliation{Department of Physics, Southern Methodist University, Dallas, TX 75275, USA}

\author[0000-0002-9007-7962]{Nishan Lamichhane}
\affiliation{Department of Physics, Amrit Campus, Tribhuvan University, Kathmandu 44600, Nepal}

\author[0000-0002-0657-4342]{Malgorzata Bogdan}
\affiliation{Department of Mathematics, University of Wroclaw, 50-384, Poland}
\affiliation{Department of Statistics, Lund University, SE-221 00 Lund, Sweden}

\author[0000-0001-9150-7832mm]{Monnie McGee}
\affiliation{Department of Statistics and Data Science, Southern Methodist University, Dallas, TX 75205, USA}

\author[0000-0003-3732-0860]{Pavel Nadolsky}
\affiliation{Department of Physics, Southern Methodist University, Dallas, TX 75275, USA}

\author[0009-0007-0050-9762]{Milind Sarkar}
\affiliation{Department of Physical Sciences, Indian Institute of Science Education and Research (IISER), Mohali,
Sector – 81, Knowledge City, SAS Nagar, PO Manauli, Pin – 140306, Punjab, India}

\author[0000-0003-3358-0665]{Agnieszka Pollo}
\affiliation{National Centre for Nuclear Research, 02-093 Warsaw, Poland}
\affiliation{Astronomical Observatory of Jagiellonian University, Orla 171, 30-244 Krakow, Poland}

\author[0000-0002-7025-284X]{Shigehiro Nagataki}
\affiliation{Astrophysical Big Bang Laboratory (ABBL), Cluster of Pioneering Research, RIKEN, 2-1, Wako, Hirosawa, Saitama 351-0198, Japan} 
\affiliation{Interdisciplinary Theoretical and Mathematical Sciences Program (iTHEMS), RIKEN, 2-1, Wako, Hirosawa, Saitama 351-0198, Japan}
\affiliation{Astrophysical Big Bang Group (ABBG), Okinawa Institute of Science and Technology Graduate University (OIST), 1919-1 Tancha, Onna-son, Kunigami-gun, Okinawa, 904-0495, Japan}

%



\begin{abstract}
Gamma-ray bursts (GRBs) are intense, short-lived bursts of gamma-ray radiation observed up to a high redshift ($z \sim 10$) due to their luminosities. Thus, they can serve as cosmological tools to probe the early Universe. However, we need a large sample of high$-z$ GRBs, currently limited due to the difficulty in securing time at the large aperture Telescopes. Thus, it is painstaking to determine quickly whether a GRB is high$-z$ or low$-z$, which hampers the possibility of performing rapid follow-up observations. Previous efforts to distinguish between high$-$ and low$-z$ GRBs using GRB properties and machine learning (ML) have resulted in limited sensitivity. In this study, we aim to improve this classification by employing an ensemble ML method on 251 GRBs with measured redshifts and plateaus observed by the Neil Gehrels Swift Observatory. Incorporating the plateau phase with the prompt emission, we have employed an ensemble of classification methods to enhance the sensitivity unprecedentedly. Additionally, we investigate the effectiveness of various classification methods using different redshift thresholds, $z_{threshold}$=$z_t$ at $z_{t}=$ 2.0, 2.5, 3.0, and 3.5. We achieve a sensitivity of 87\% and 89\% with a balanced sampling for both $z_{t}=3.0$ and $z_{t}=3.5$, respectively, representing a 9\% and 11\% increase in the sensitivity over Random Forest used alone. Overall, the best results are at $z_{t} = 3.5$, where the difference between the sensitivity of the training set and the test set is the smallest. This enhancement of the proposed method paves the way for new and intriguing follow-up observations of high$-z$ GRBs.
\end{abstract}


\section{Introduction} 
\label{sec:intro}
Gamma-ray bursts (GRBs) are the most energetic astronomical events in the Universe after the Big Bang. 
They emit massive amounts of radiation across the electromagnetic spectrum, encompassing $\gamma$-rays, X-rays, and optical and radio wavelengths. 
Traditionally, they are classified into two main classes: long GRBs (LGRBs) and short GRBs (SGRBs), based on $T_{90}$, the duration over which the GRB emits 90\% of its total observed fluence (energy emitted in $\gamma$-rays) \citep{1993ApJ...413L.101K} from 5 to 95\% of the total background-subtracted emission. 
LGRBs have $T_{90} > 2$ seconds, while SGRBs have $T_{90} < 2$ seconds. 
Additional classifications over the years have been proposed, dividing GRBs into Type I (binary merger events) and Type II (collapsars) \citep{Zhang2004IJMPA..19.2385Z, Zhang2006, Zhang2009, zhang2014long}.
GRB progenitors are challenging to identify, but LGRBs are typically linked to massive stars that collapse into black holes in the final stages of their lifespan \citep{woosley1993ApJ...405..273W, woosley1993, paczynski1998ApJ...494L..45P, Woosley2006ARA&A}, while SGRBs are linked to the coalescence of two compact objects (black holes or neutron stars) \citep{woosley1993ApJ...405..273W, woosley1993, paczynski1998ApJ...494L..45P, Woosley2006ARA&A}.
GRBs have two distinct phases: the prompt emission and the afterglow. The first is the primary phase, characterized by intense radiation observed from high-energy $\gamma$-rays to hard and soft X-rays and occasionally in the optical spectrum \citep{Vestrand2005Natur,Beskin2010ApJ,2012MNRAS.421.1874G,2014Sci...343...38V,2014IJMPD..2330002Z, zhang_2018}. 
The afterglow follows the prompt, with a prolonged multi-wavelength emission observed in X-rays, optical, and sometimes radio frequencies \citep{costa1997,vanParadijs1997,Piro1998,2015PhR...561....1K, zhang_2018}. 
Often, a significant feature observed in the GRB light curve (LC) is the plateau phase characterized by a relatively constant flux that precedes the subsequent decay of the afterglow \citep{OBrien2006,Zhang2006,Nousek2006,Sakamoto2007,Liang2007,Dainotti2008,Zaninoni2013,Rowlinson2014}.
According to the sample size and the fitting functions investigated to determine the plateau, the plateau phases are detected in approximately 42\% of X-ray afterglows \citep{Evans2009, Li2018b, Srinivasaragavan2020, 2021ApJS..255...13D}, 30\% of optical afterglows \citep{Vestrand2005Natur,Kann2006,Zeh2006,panaitescu2008taxonomy,panaitescu2011optical,Oates2012,dainotti2020b,Dainotti2022}, and 6.6\% of radio afterglows \citep{2022ApJ...925...15L}.

GRBs are have been observed at $z \sim 9.4$ \citep{2011ApJ...736....7C} and in principle can be detected up to $z \sim 20$ \citep{lamb_reichart2000, 2002luml.conf..157L}. Thus, they can be used as powerful tools to understand the evolution of the early Universe, including the epoch of reionization and the formation of population III stars. Indeed, the availability of redshifts is essential for many science cases, especially population studies.
Multi-wavelength afterglow studies of high$-z$ GRBs can also cast light on the metal and dust content of early faint galaxies \citep{frail2006,cusumano2007,cusumano2006}.
There is, however, a significant caveat: GRB afterglows dim extremely rapidly,
and performing high$-z$ measurements is challenging because even observations of optically bright GRBs are hindered by reduced telescope time and the limited number of follow-up programs for GRBs. Hence, the bottleneck is to craft a system to identify whether a newly detected GRB event is observed at high$-z$ to allow follow-up observations in other wavelengths \citep{Dainotti2024b, Dainotti_2024}.
Past studies have attempted to classify GRB redshift; for example, \cite{morgan2012} used a training set of 135 Swift GRBs with known redshifts with only one ML method: the Random Forest \citep{hastie2009elements,breiman2001randomforest}. We use a sample of 251 GRBs (86\% larger than used by \cite{morgan2012}) and obtain an 11\% improvement in the sensitivity using a balanced dataset.

The Neil Gehrels Swift Observatory \citep{2004ApJ...611.1005G}, hereafter \textit{Swift}, with its localization capabilities, has opened a new window to the high$-z$ Universe during its two decades of operation. 
The \textit{Swift} satellite is equipped with three primary instruments: Burst Alert Telescope (BAT, 15 -- 150 keV, \cite{2005SSRv..120..165B}) allows the detection of prompt emission, X-ray Telescope (XRT, 0.2 -- 10 keV, \cite{Barthelmy2005}) enables rapid follow-up of the afterglow and the Ultraviolet/Optical Telescope (UVOT, 1700 -- 6500 \AA, \cite{Roming2005}). Together, these instruments synergistically detect, localize, and gather data on GRBs and their afterglows, spanning a range of wavelengths, including $\gamma$-rays, X-rays, and ultraviolet/optical spectra.

Despite the \textit{Swift} advantages in accurately pinpointing GRB locations, only 23\% (431 out of 1859) of the GRBs detected by \textit{Swift} in general, even if triggered by other satellites or instruments, up to August 7, 2024, have spectroscopic redshift measurements. These measurements are essential for utilizing GRBs for population studies and as cosmological distance ladders.
A rapid follow-up is required to infer useful information from such high$-z$ bursts before the optical emission fades significantly to be detectable.
Furthermore, the decision to point a ground-based telescope at the new GRB must be made within hours of initial discovery based mainly on the information \textit{Swift} can provide. 
Thus, in this paper, we aim to infer which GRBs are at high$-z$ within hours, given the properties associated with the prompt and the plateau emission phase.
Studies have been conducted in the past to identify high$-z$ GRBs, but results were not sufficiently accurate \citep{2007AJ....133.2216G, 2007A&A...464L..25C, 2007MNRAS.380L..45S, 2008AIPC.1000...80V, 2008AIPC.1000..166U, 2009AIPC.1133..437U, 2009MNRAS.396.1499K, 2010MNRAS.401.1369K}. 
Additional studies have employed supervised machine learning (ML) techniques to identify high$-z$ GRBs \citep{morgan2012, ukwatta2016machine}, 
each utilizing only a single algorithm for training the model. Equally importantly, their feature set did not incorporate the plateau emission phase. 
However, our analysis includes X-ray plateaus and an ensemble method with multiple models within SuperLearner. The aim is to craft a reliable method that would allow tackling follow-up programs in optical and radio since we have already shown in \cite{dainotti2020b, Dainotti2022} that the plateau duration in X-rays is indeed shorter than the ones in optical roughly 40 hours for the longer duration plateau.
We investigate to which extent and precision several classification methods can distinguish between high$-$ and low$-z$ events using different redshift thresholds, $z_{threshold} = z_{t}$ at $z_{t} = 2.0, 2.5, 3.0,$ and $3.5$. These thresholds have been chosen arbitrarily to test our method with different sample sizes and redshift ranges. 
Our results show that we can start from predicting the high$-z$ class, with a sensitivity of 58\%, when we use raw data with no outlier removal or imputation methods and without balancing the sample. Adding more ML methods (removing outliers, imputing missing variables, and balancing the dataset) has led to 89\% sensitivity for the classification, an improvement of 53\% compared to our initial analysis, paving the way for new, exciting follow-up observations. To the best of our knowledge, with all these methods, the current classification is the only one that includes plateau emission in the literature. The paper is structured as follows. 
In Section \ref{sec:sample}, we outline the data sample and the features utilized in our study. Section \ref{sec:methodology} details our methodology, including data cleaning, outlier removal, data imputation, feature selection, the algorithms employed, data upsampling, and the implementation of the SuperLearner. In Section \ref{sec:results}, we present the results of the SuperLearner classification. We discuss our findings in Section \ref{sec:discussion} and compare our results to prior work in Section \ref{sec:comparison}. Finally, Section \ref{sec:broader impact} and Section \ref{sec:conclusion} summarize the broader impact of our work and conclusions, respectively.

\section{Data Sample}
\label{sec:sample}
In this study, we employed a sample of 251 GRBs observed in $\gamma-$rays and X-rays, utilizing the data from BAT and XRT aboard the \textit{Swift} satellite. The data contains measurements of various GRB observed properties, including the prompt, the afterglow phase, and the plateau phase, which marks the unique aspect of our data set. We accessed this data through the NASA \textit{Swift} GRB Search Tool \footnote{\url{https://swift.gsfc.nasa.gov/archive/grb_table/}} and the Third \textit{Swift}-BAT GRB Catalog \citep{2016ApJ...829....7L}.
We have derived the plateau emission by performing our fitting of the LCs with the \cite{Willingale2007} model, and we have derived 251 GRBs from January 2005 until August 2023, thus extending the previous sample in \cite{Dainotti_2024}, where the sample gathered data from January 2005 to August 2019.
Our data set includes the following 10 GRB explanatory variables, the features of the GRBs for which we can determine the classification, with the redshift being the response variable and the rest being predictors:

\begin{enumerate}
    \item $z$ - redshift of the GRB.
    \item $T_{90}$ - the duration in which 90\% of GRB's total $\gamma-$ray observed fluence is emitted.
    \item $F_{a}$ - the flux at the end of the plateau emission.
    \item  $T_{a}$ - the end time of the plateau emission.
    \item $\alpha$ - the temporal power-law decay index following the end of the plateau emission.
    \item $\beta$ - the spectral index obtained during the plateau emission phase by assuming a power-law distribution of the spectral energy.
    \item $\Gamma$ - the spectral index determined through the time-averaged spectral fit derived from the \textit{Swift} XRT photon counting mode data.
    \item NH - the neutral hydrogen column density along the line of sight.
    \item Fluence - the energy fluence of the prompt emission over $T_{90}$ in erg cm$^{-2}$.
    \item PeakFlux - the peak photon flux during the prompt emission in a number of photons cm$^{-2}$ sec$^{-1}$.
    \item PhotonIndex - the prompt photon index obtained by modeling the BAT Telescope's photon energy distribution with a power-law.
\end{enumerate}

\section{Methodology}
\label{sec:methodology}
\begin{figure*}[htbp]
\centering
    \includegraphics[width=1.0\textwidth,height=0.87\textheight]{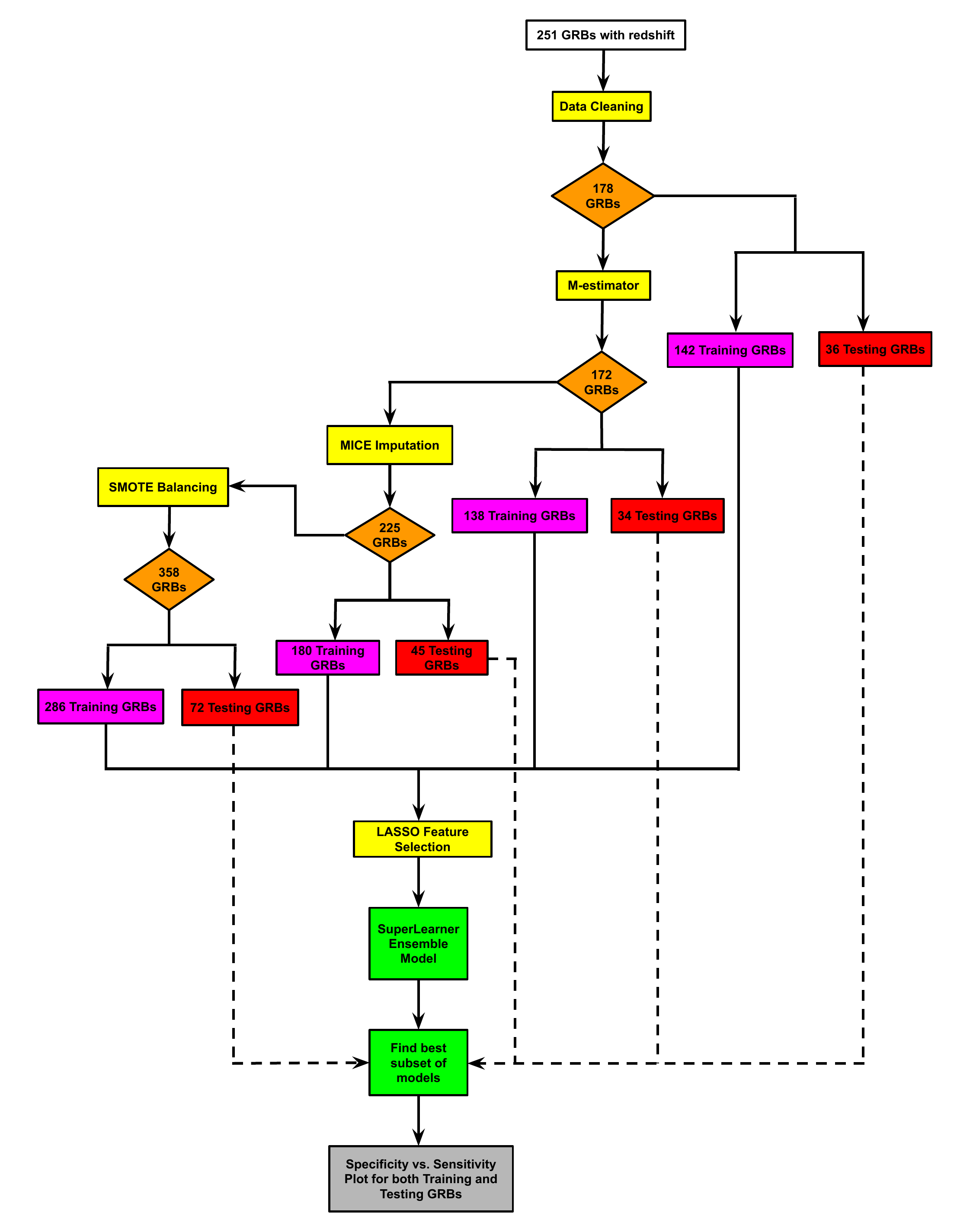}
\caption{Flowchart detailing each pipeline step, from the initial data to the SuperLearner ensemble model. Yellow boxes indicate the sequence of steps followed in the pipeline. Orange boxes display the total number of GRBs after each step. Magenta boxes show the number of GRBs in the training, and red boxes show the number in the test sets at each process stage. Green boxes highlight the steps involved in the model construction. The gray box represents the outcome of the constructed model; both applied on the test and training sets.}
\label{Fig:flowchart}
\end{figure*}

The analysis procedure consists of several parts: the data cleaning, the outlier removal, the data imputation, balancing the data between low$-$ and high$-z$, feature selection, selecting the algorithms included in the SuperLearner ensemble method such as the parametric, semi-parametric, and fully non-parametric models, choosing the metrics for assessing the goodness of our methods. All these steps come subsequently. Namely, in all the procedures, we always start with data cleaning and then perform all the rest of the steps on this data. The use of the M-estimator is performed only once on the cleaned data. In addition, the balanced sampling procedure is always applied after Multivariate Imputation by Chained Equations  \citep[MICE,][]{schafer2002missing, van2011mice}. Figure \ref{Fig:flowchart} presents the flowchart summarizing our methodology.

\subsection{Data Cleaning}
\label{sec:DataCleaning}
Due to the wide range of most of the variables, we perform the so-called variable transformation, which in this case is the change of some of the variables from a linear to a log scale. We change the observed features $T_{90}$, Fluence, PeakFlux, NH, $F_a$, and $T_a$ into log base-10.  Additionally, we clean our data following \cite{Dainotti_2024} by assigning Not-Available (NA) to the values where $\log(\text{NH}) < 20$, $\text{PhotonIndex} < 0$, $\text{PeakFlux} = 0$, $\alpha > 3$, $\beta > 3$, and $\Gamma > 3$, including their corresponding error bars if available. These thresholds are used because such characteristics are non-physical or unusual for most GRBs (for example, the $\alpha$, $\beta$, and $\gamma > 3$ belong to the tail of their respective distributions). We later impute these NA values in our analysis using the MICE technique, as described in Section \ref{sec:mice}. 
In addition, we introduce an extra variable, $\log(z$ + 1), as it provides a more natural parametrization of the cosmological variable $z$.

\subsection{Outlier Removal}
\label{sec:Mestimator}
The dataset contains outliers that could, in principle, affect the prediction of our ML models because they cover a different parameter space in some of the variables; therefore, we applied an M-estimator, a generalized maximum likelihood method described in \cite{DEMENEZES2021107254} and the references therein, to locate potential outliers and increase the robustness of our predictions.
The M-estimator is a robust alternative to the ordinary least squares (OLS) method for fitting data using the function specified in Eq. \ref{Eq:1}. The OLS method minimizes the squared residuals (L2 norm regression), which can give excessive weight to outliers and distort the regression equation and subsequent inferences. Conversely, the M-estimator reduces the sum of a chosen function of the residuals, thus mitigating the impact of outliers. For this purpose, we use the Huber function \citep{10.1214/aoms/1177703732, huber2009robust}, following \cite{Dainotti_2024}. 
The Huber method is not the only one that could be applied. There are other methods, such as the Tukey bi-square method, which also offer robustness against outliers. However, the Tukey bi-square method may be less efficient for datasets with fewer outliers and can potentially discard meaningful extreme observations \citep{huber2009robust, maronna2006robust}. In contrast, the Huber function provides a good trade-off between robustness and efficiency \citep{huber2009robust, maronna2006robust}, making it a more versatile choice for our case. 
To this end, we employed the methodology described in \cite{Dainotti_2024} to identify the most predictive model. This approach utilizes a generalized linear model \citep[GLM,][]{68aee965-a8a0-3e72-9f89-8d89ae91a62b}), 
which incorporates squared terms of one or multiple data features to capture potential nonlinear correlations between the response variable ($z$) and the predictors. The selected model is represented as follows:

\begin{align}
\log(1+z) = & \; (\log(T_{a}) + \log(F_{a}))^{2} + \log(\rm{PeakFlux}) + \log(\rm{NH}) + \rm{PhotonIndex} + \log(T_{90}) + \alpha \notag  \\
& + (\log(\rm{PeakFlux}))^{2} + (\log(\rm{NH}))^{2} + \rm{PhotonIndex}^{2} + (\log(F_{a}))^{2} + (\log(T_{a}))^{2} + (\log(T_{90}))^{2} + \alpha^{2} \notag \\
\label{Eq:1}
\end{align}

\begin{figure*}[htbp]
    \centering
    \gridline{\fig{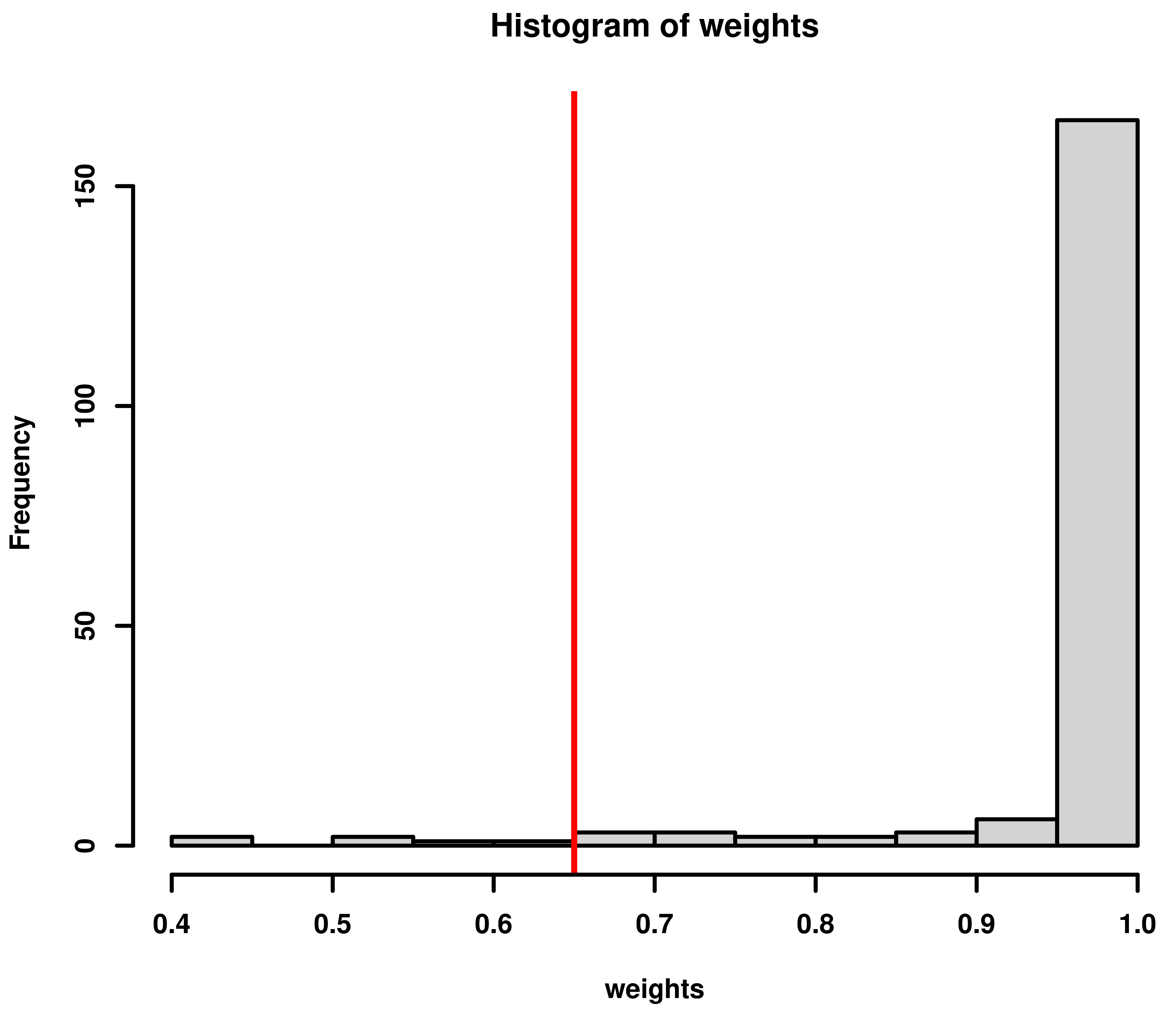}{0.82\textwidth}{}
              }
\caption
{Distribution of weights assigned by the M-estimator to each GRB on the raw data. The red vertical line shows the cutoff line of 0.65 for the outliers. GRBs below this cutoff line are considered outliers.}
\label{Fig:M-estimator weights}
\end{figure*}

We used the MASS package in the open-source statistical software \textsc{R} \citep{R} to implement the M-estimator.
The findings (see Section \ref{sec:results}) demonstrate the M-estimator's efficacy in capturing the dataset's main trends while mitigating the influence of outliers.
Figure \ref{Fig:M-estimator weights} shows the distribution of weights assigned by the M-estimator to each GRB on the raw data. The cutoff employed to remove the outliers from the dataset is set to 0.65 (red vertical line in Figure \ref{Fig:M-estimator weights}). Observations with a weight assigned by the M-estimator of less than 0.65 are classified as outliers.
We used this weight cutoff because we wish to remove $<$ 5\% data points from our sample to retain enough data for reliable model estimation. In fact, with this choice, we remove 3.4\% of data points. 
We also computed the variance inflation factor (VIF), which measures the extent to which predictors are multicollinear. We found that the VIF values were between 1.1 and 1.8, indicating a low to moderate level of collinearity among the predictors \citep{Brown2009}.
This range of VIF is generally considered acceptable, as a VIF above ten typically indicates severe collinearity, which can inflate standard errors and produce unreliable regression coefficients \citep{Brown2009}. Given that our VIF values are well below this threshold,  we conclude that our M-estimator model is sufficiently robust, allowing for effective outlier removal.

\begin{figure*}[htbp]
\centering
    \includegraphics[width = 1\textwidth]{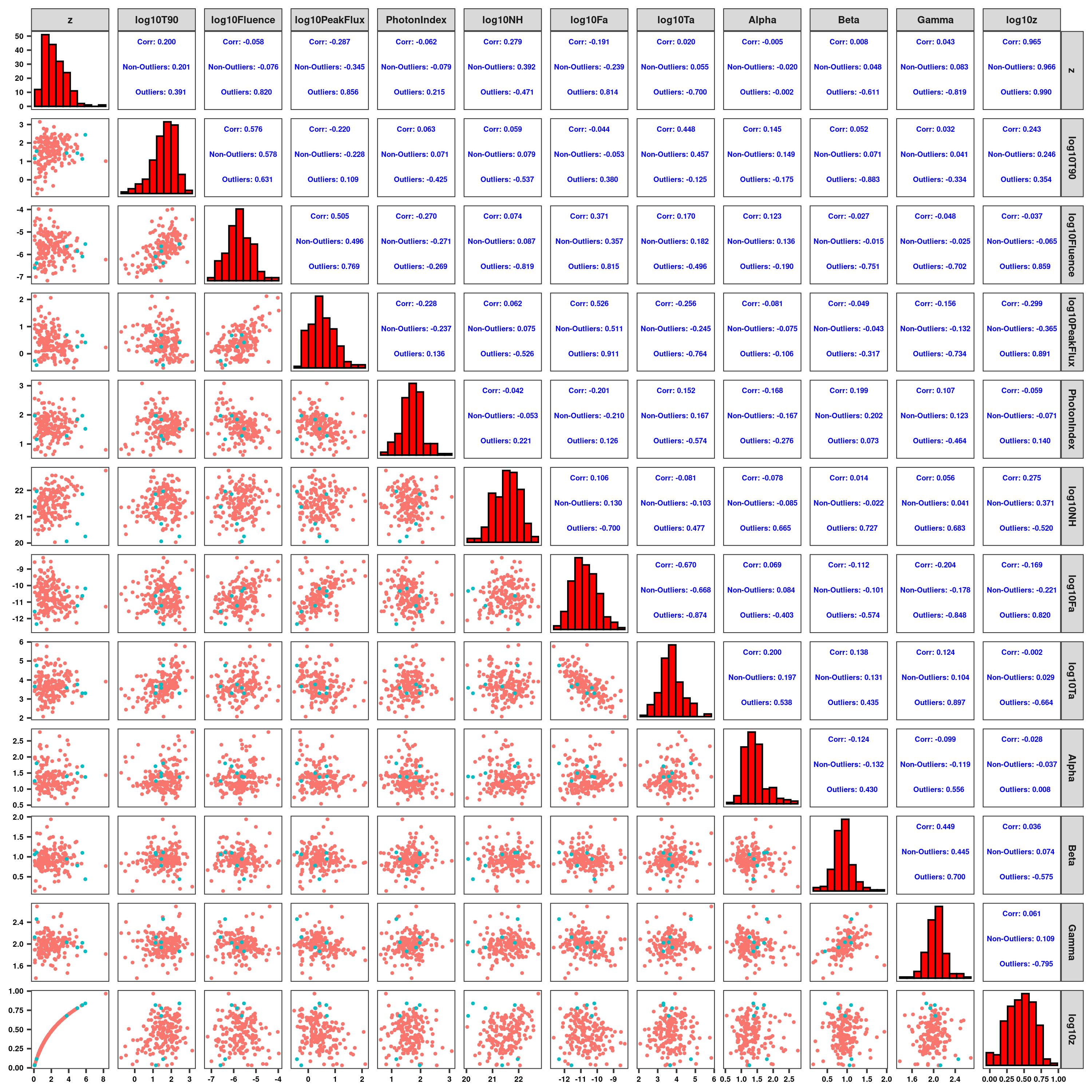}
\caption{Scatter matrix plot for the raw data showing the outliers determined by the M-estimator in cyan and the rest of the data in red.}
\label{Fig:M-estimator Scatter matrix plot-raw}
\end{figure*}

By applying this weight cutoff of 0.65 for the outliers removal with the M-estimator, we systematically identified and excluded from the raw data six GRBs as outliers in the GRB features: GRB 050826A, GRB 051109B, GRB 130408A, GRB 130606A, GRB 160327A, and GRB 220521A. After segregating the dataset into non-outlier and outlier groups, a categorical label is assigned to each observation, as shown in Figure \ref{Fig:M-estimator Scatter matrix plot-raw}. This scatter matrix plot also displays the correlation of features in the upper right half, which comprehensively explores variable relationships within the dataset, emphasizing the impact of outliers on the overall distribution.
After removing the outliers via the M-estimator, we also discard the data points where $\Delta x/x > 1$ for any given feature $x$, with errorbar $\Delta x$.

\subsection{Data Imputation}
\label{sec:mice}
\begin{figure*}[htbp]
\centering
\includegraphics[width = 1\textwidth]{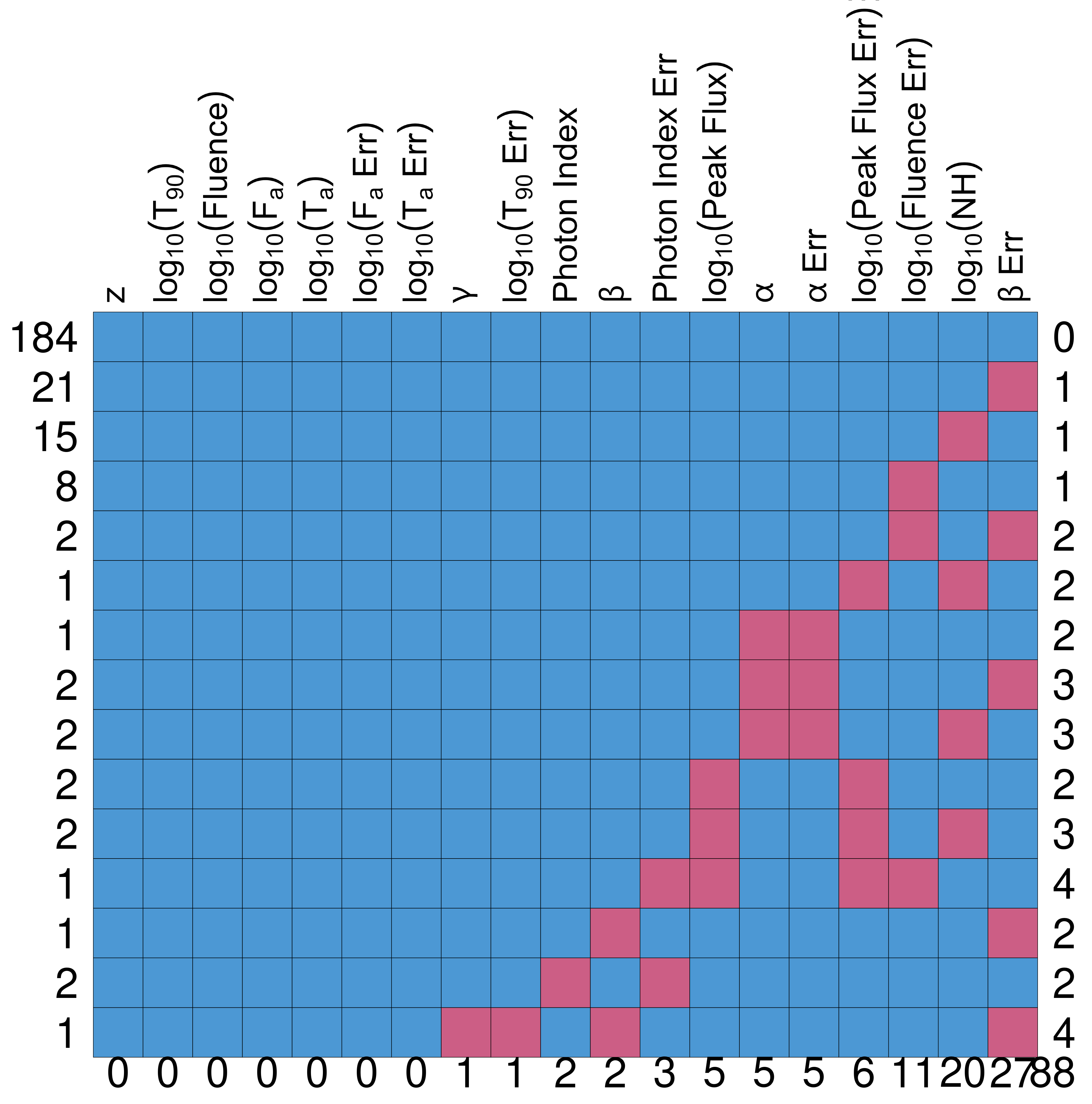}
\caption{The distribution of missing data in our sample. Red boxes highlight GRBs with missing data points, while blue boxes indicate GRBs with complete data for a given variable, as noted on the top axis. The bottom axis enumerates the number of missing variables according to the GRB number shown on the left axis. The left axis represents the count of observations with missing data for specific features. For instance, there are 184 GRBs with complete data, 21 GRBs missing only $\beta_{\rm{err}}$ values, 15 GRBs missing $\log(\rm{NH})$ values, and so on. The right axis indicates the number of features with missing data for each row.}
\label{fig:mice}
\end{figure*}

MICE \citep{schafer2002missing, van2011mice} is an iterative approach for imputing missing values in multivariate datasets that leverages the information from other variables within the dataset. This method operates under the assumption that data is missing at random (MAR) \citep{rubin1976inference}. 
MAR means that missing observations are missing in such a way that the probability of their missingness is unrelated to the value of the observed variable. 
This is a reasonable assumption for GRB measurements because many missing variables are due to the orbital period gaps and the Satellite's later follow-up repointing. 
Using various methods, MICE can generate imputed values in \textsc{R}. In this study, we utilize the predictive mean-matching (PMM) method, specifically the \textit{midastouch} approach introduced by \cite{little2019statistical}. This method starts by filling in missing values of a feature with its mean and then refines these values by training a model on the complete data available. Each prediction is assigned a probability based on its proximity to the imputed value for the missing variable. The missing entry is then credited by randomly selecting from the observed values of the respective predictor, weighted according to the previously defined probability. This procedure is repeated $N$ times, and the final imputed value for each missing entry is determined by averaging the predictions from each iteration.
Our data contains missing observations (as seen in Figure \ref{fig:mice}); therefore, we employ MICE to impute these NA variables. This methodology has already been successfully applied to active galactic nuclei (AGN) \citep{gibson2022} and GRB data \citep{10.1093/mnras/stad2593, Dainotti_2024}. We imputed the missing values $N=20$ times per iteration to enhance deterministic results, following the approach in \citep{10.1093/mnras/stad2593}. The final imputed value is obtained by averaging the results from each iteration. After MICE imputation, we also remove the data points if the $\Delta x/x > 1$ for each given feature $x$, with errorbar $\Delta x$.

\begin{figure}[htbp]
\centering
    \includegraphics[width = 1\textwidth]{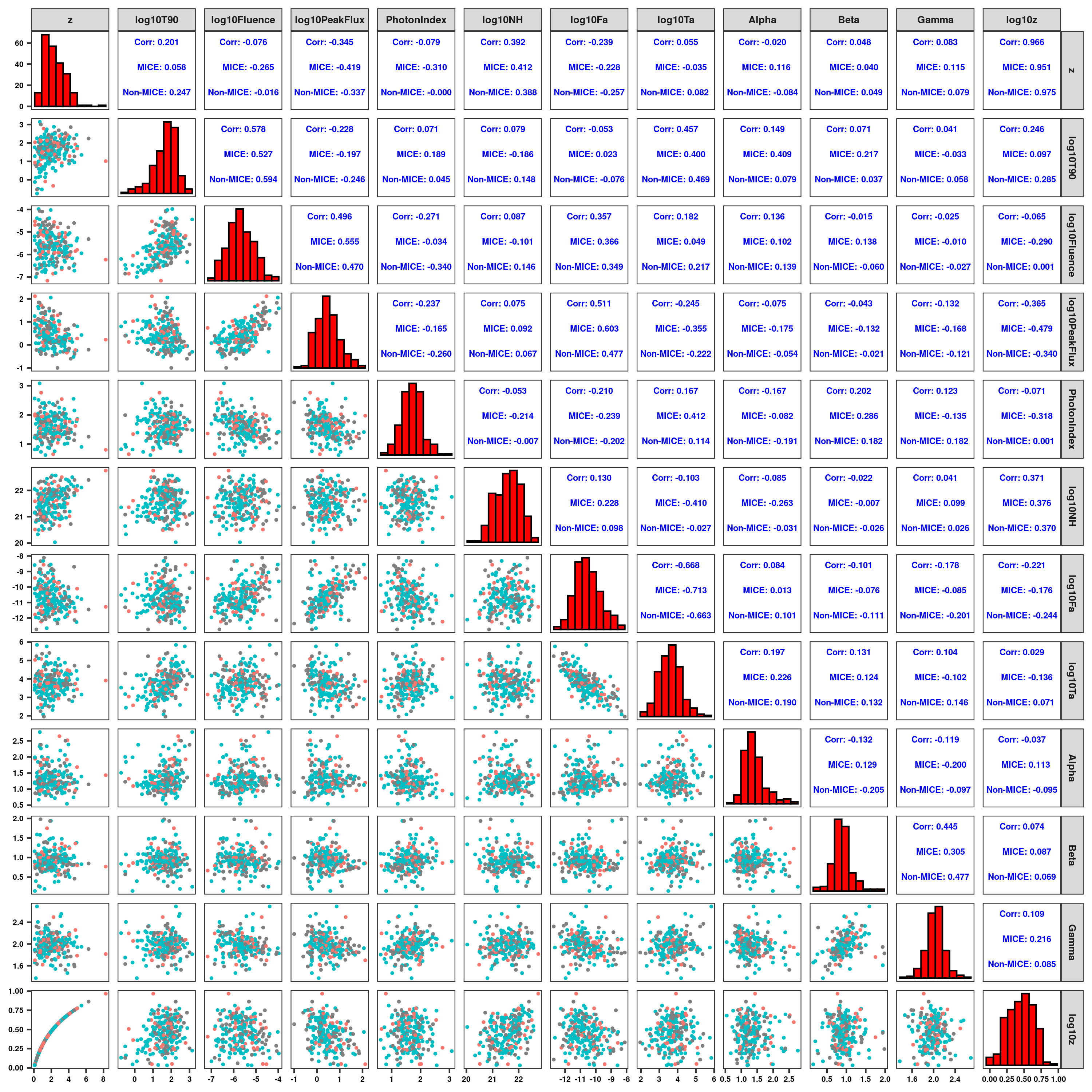}
\caption{Scatter matrix plot for the MICE-imputed data showing the MICE-imputed data in red and the original data in cyan.}
\label{Fig:MICE scatter plot}
\end{figure}

\begin{figure*}[htbp]
    \centering
    \gridline{\fig{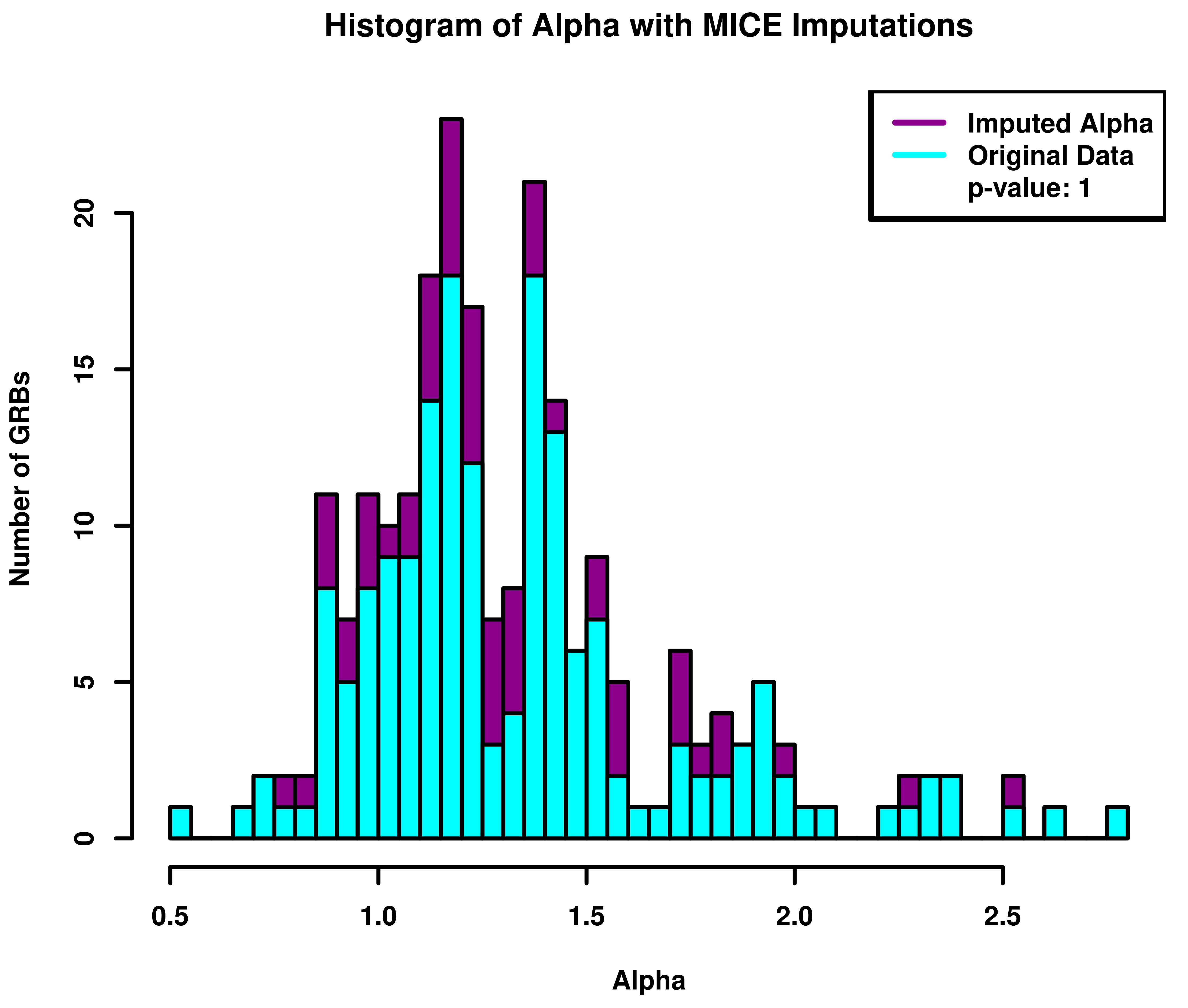}{0.30\textwidth}{(a)}
              \fig{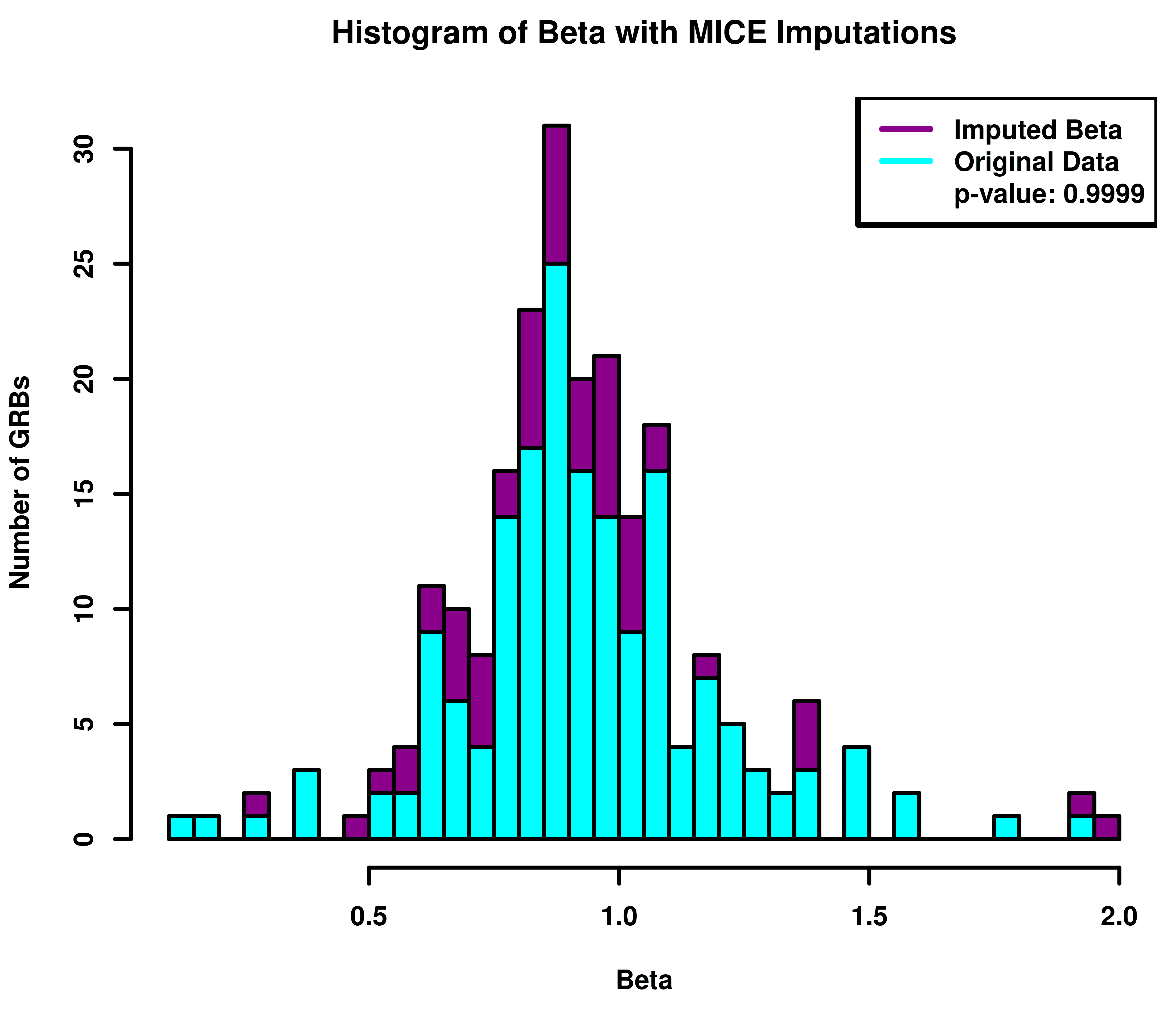}{0.30\textwidth}{(b)}
              \fig{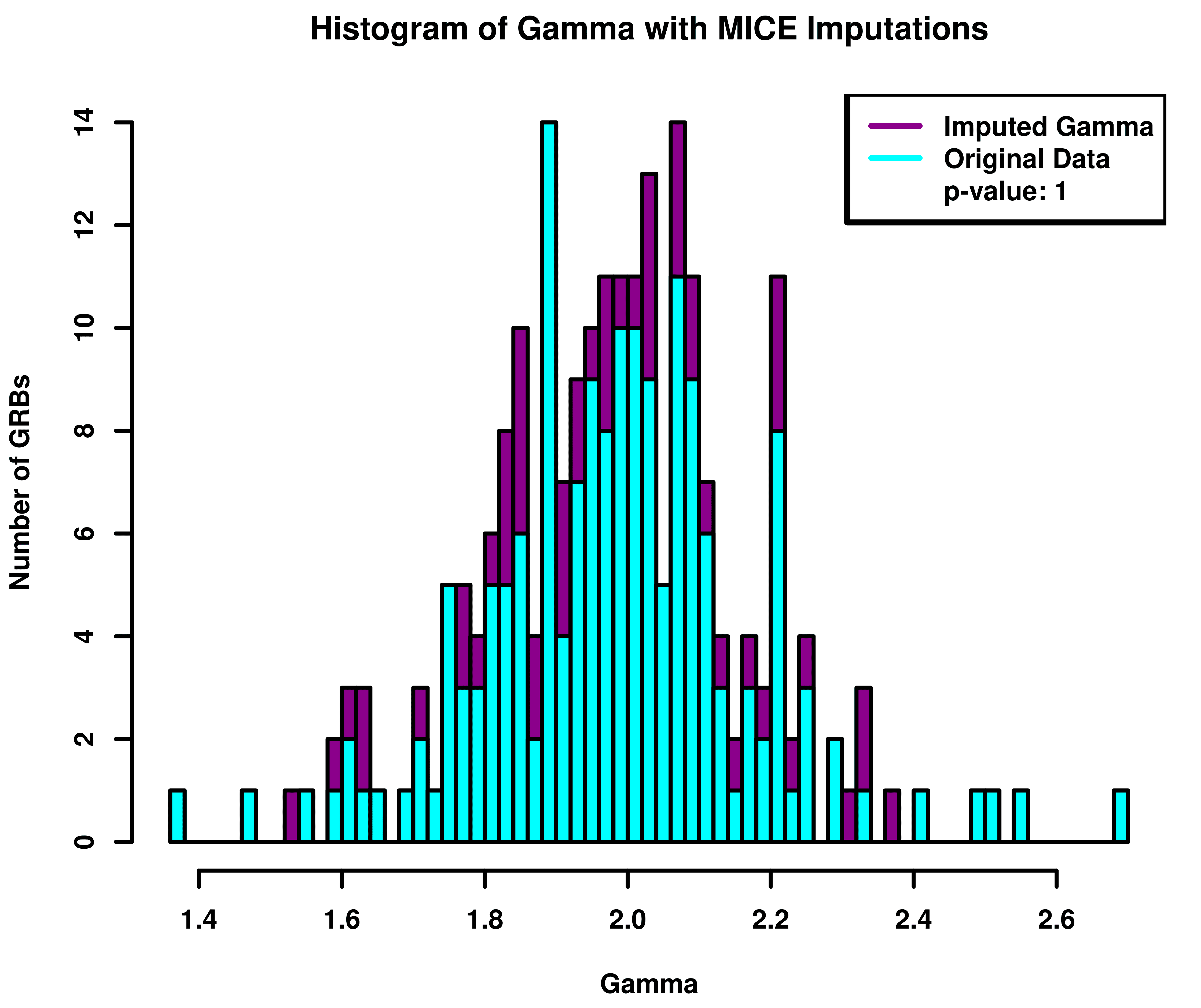}{0.30\textwidth}{(c)}
              }
    \vspace{-10pt}
    \gridline{\fig{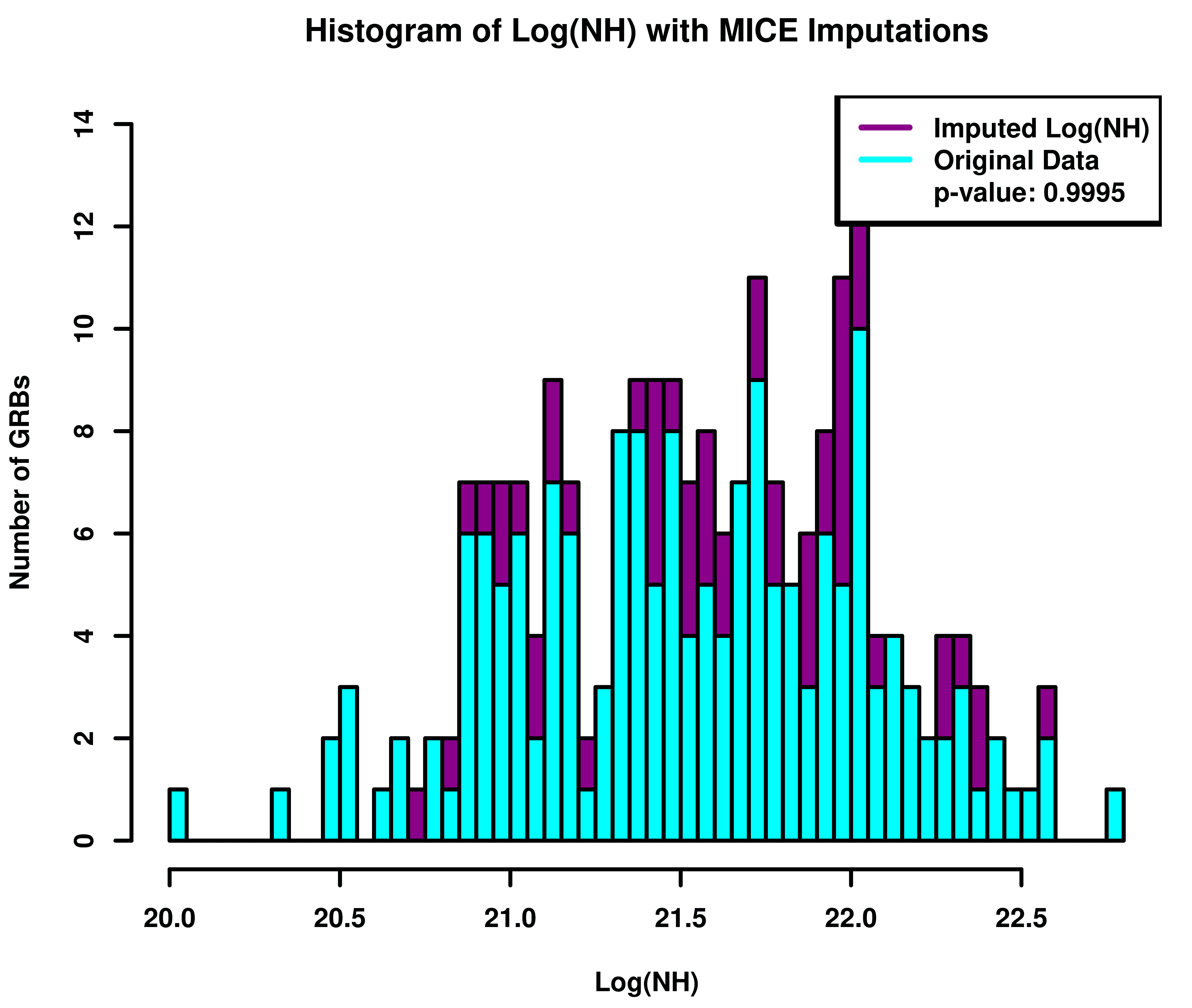}{0.30\textwidth}{(d)}
              \fig{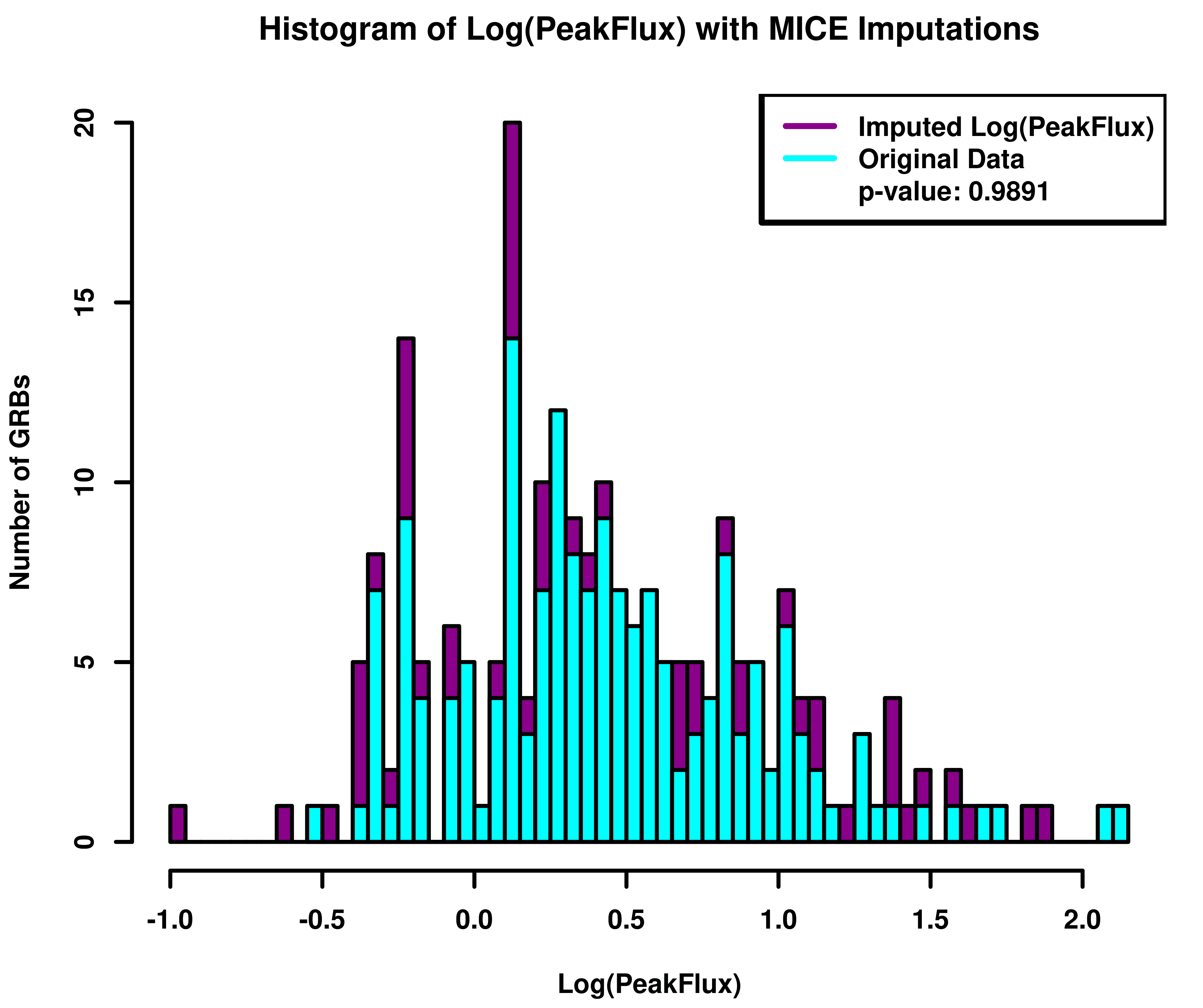}{0.30\textwidth}{(e)}
              \fig{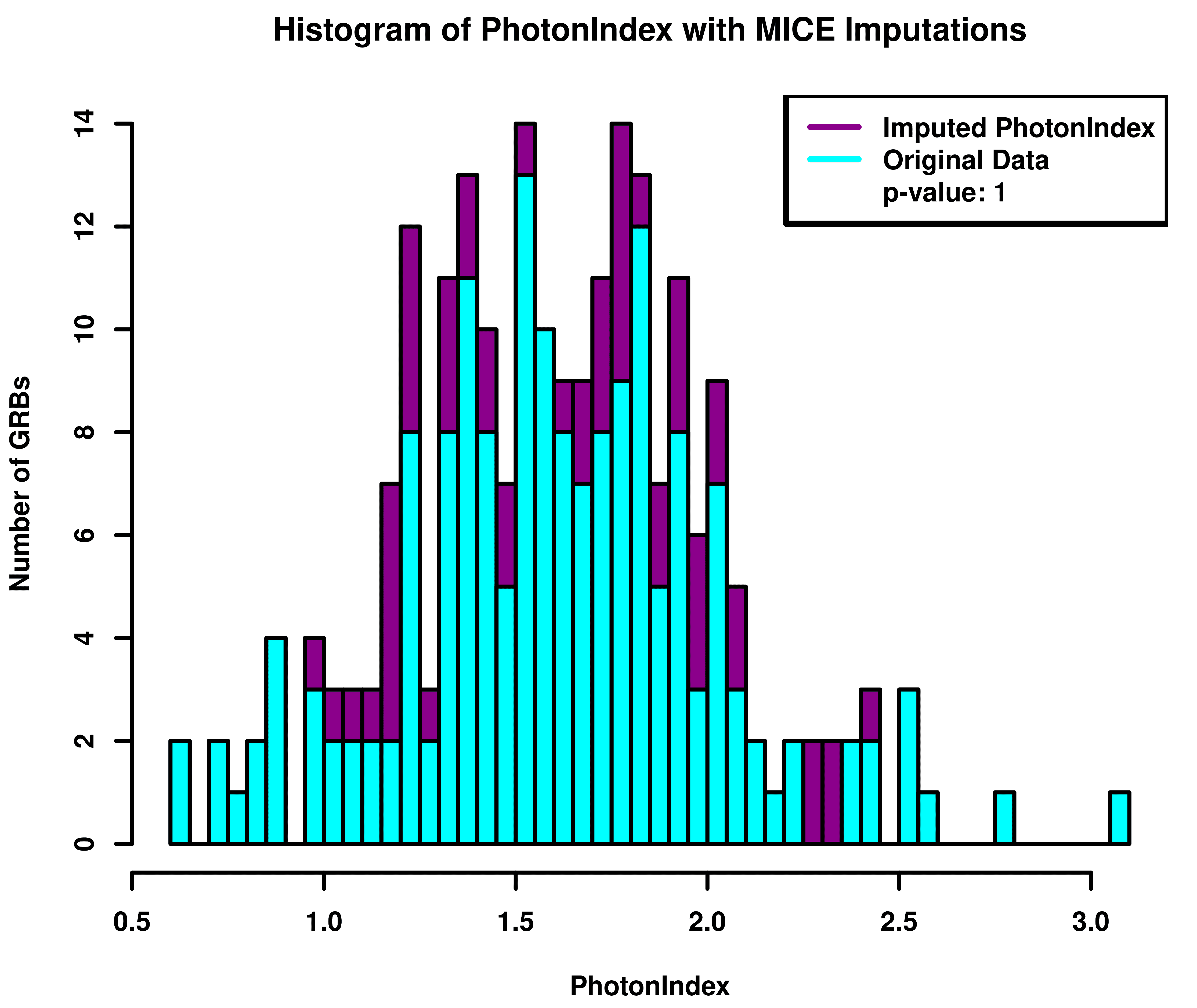}{0.30\textwidth}{(f)}
              }
    \vspace{-10pt}
    \gridline{\fig{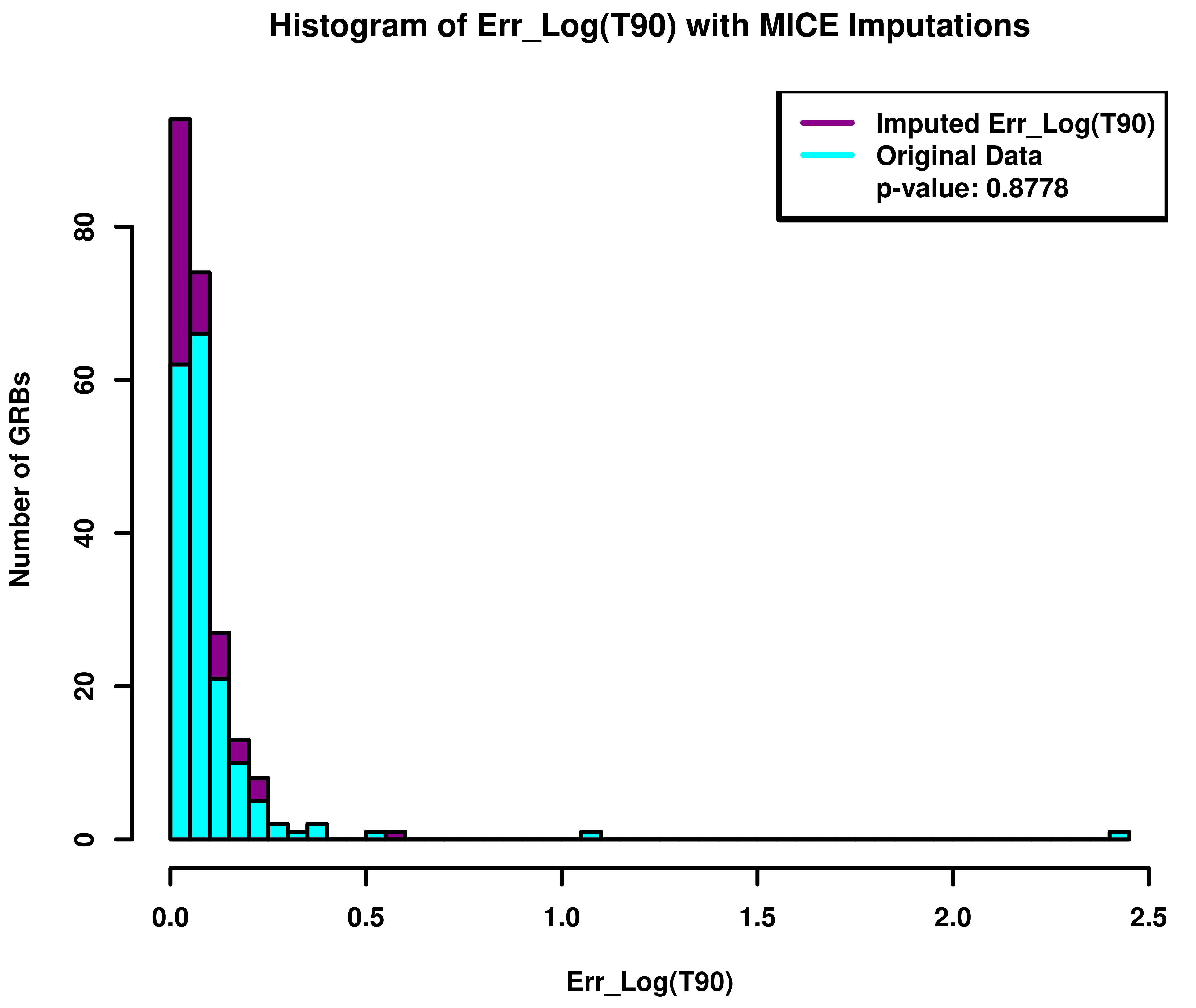}{0.30\textwidth}{(g)}
              \fig{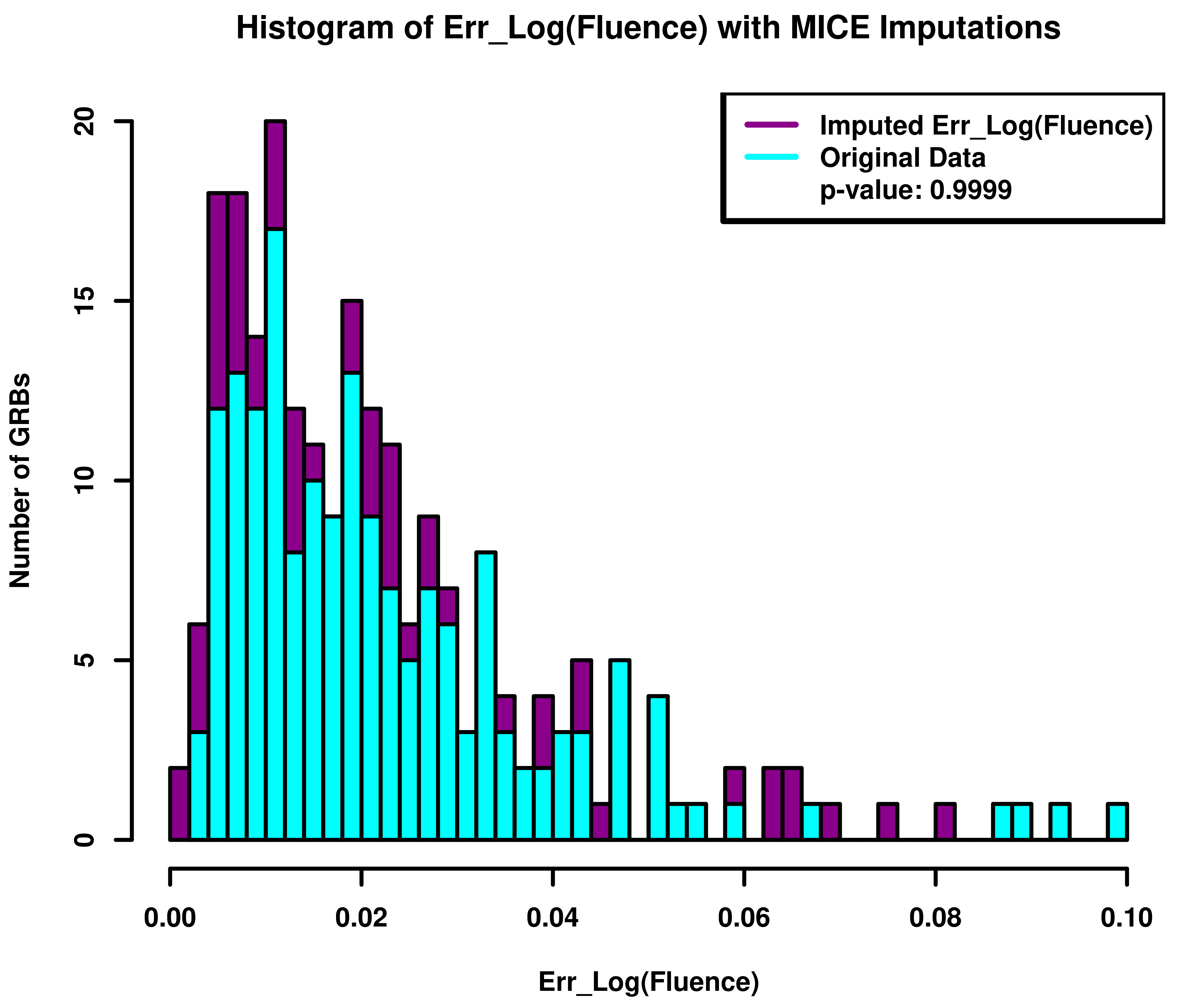}{0.30\textwidth}{(h)}
              \fig{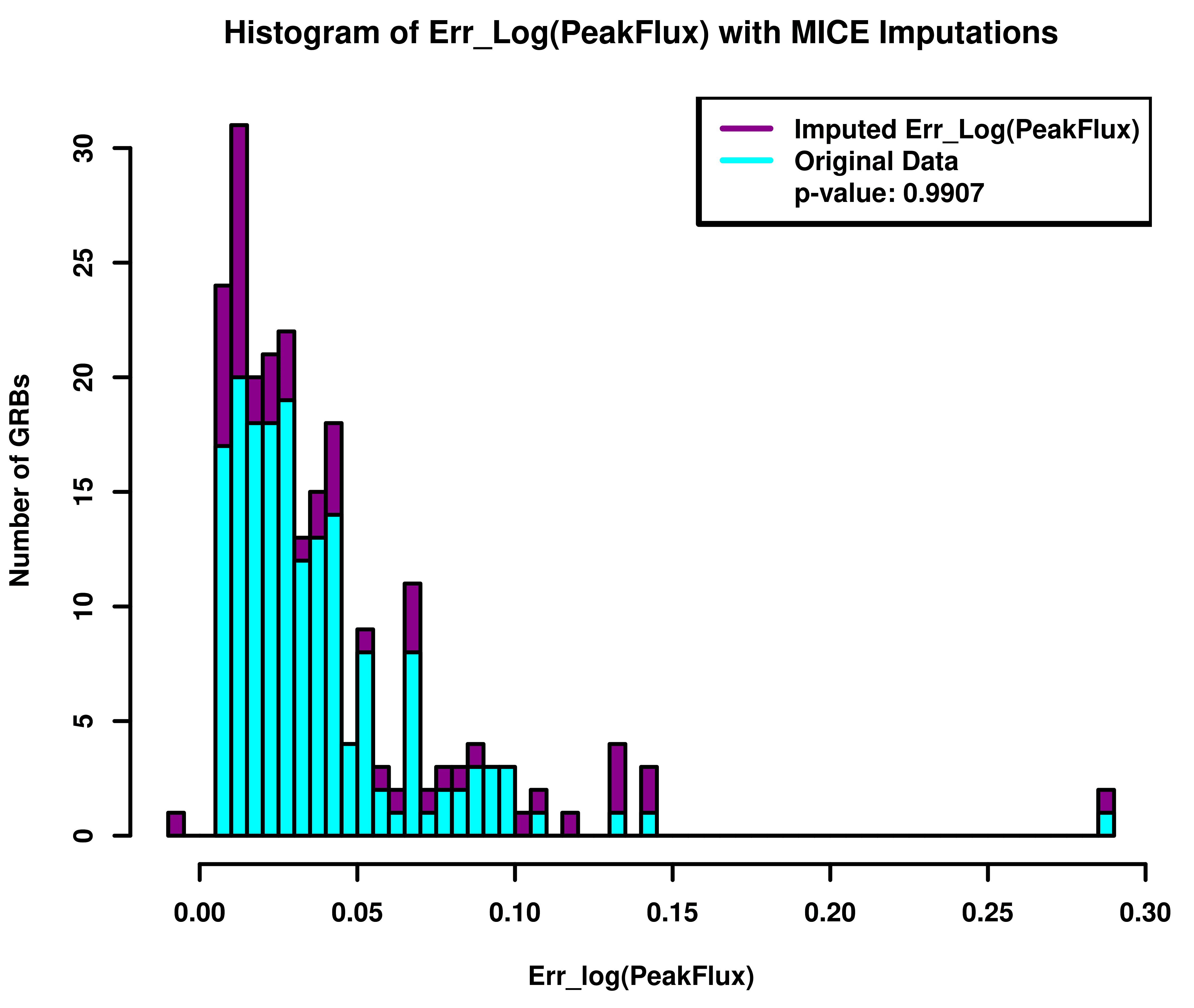}{0.30\textwidth}{(i)}
              }
    \vspace{-10pt}
    \gridline{\fig{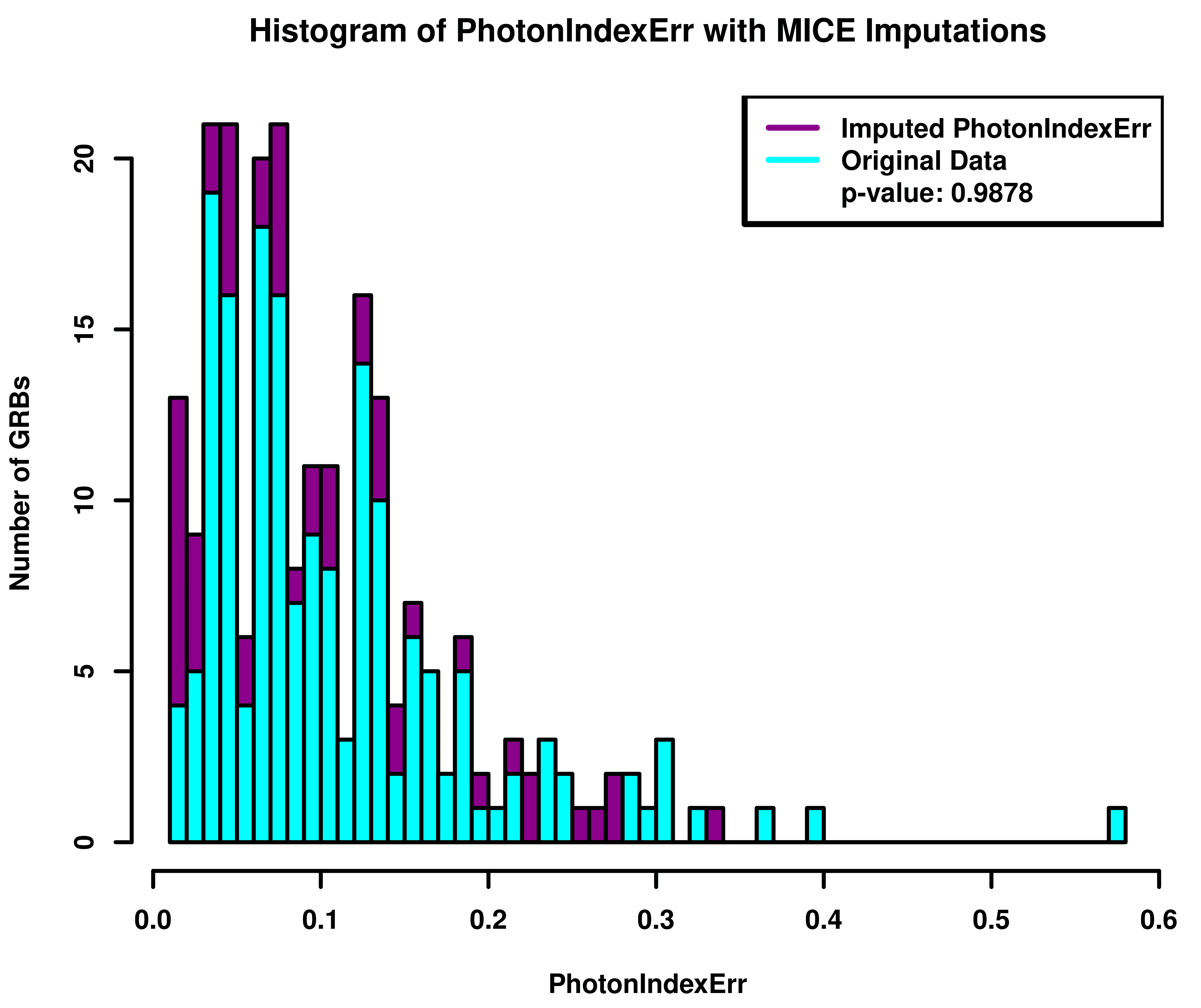}{0.30\textwidth}{(j)}
              \fig{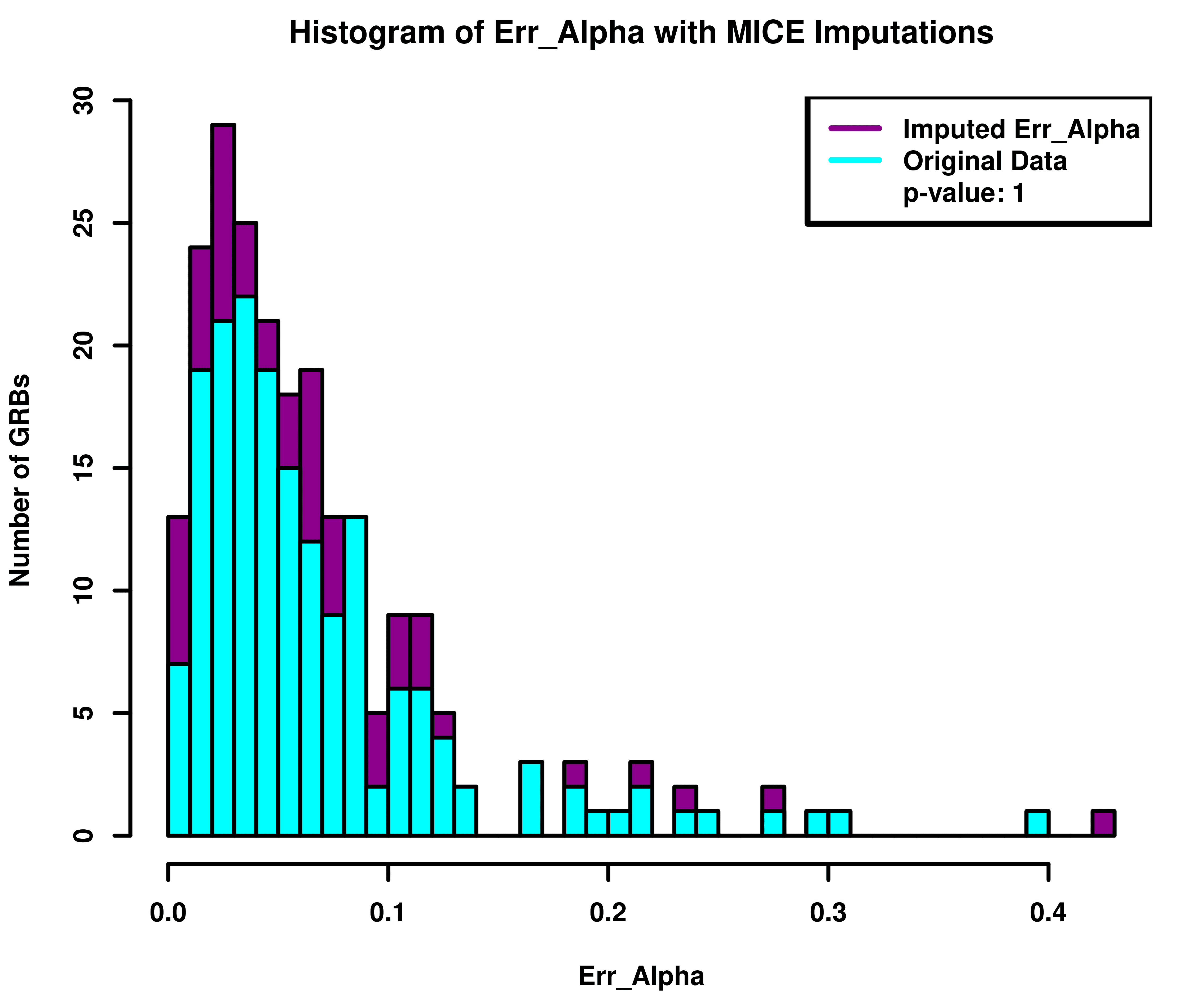}{0.30\textwidth}{(k)}
              \fig{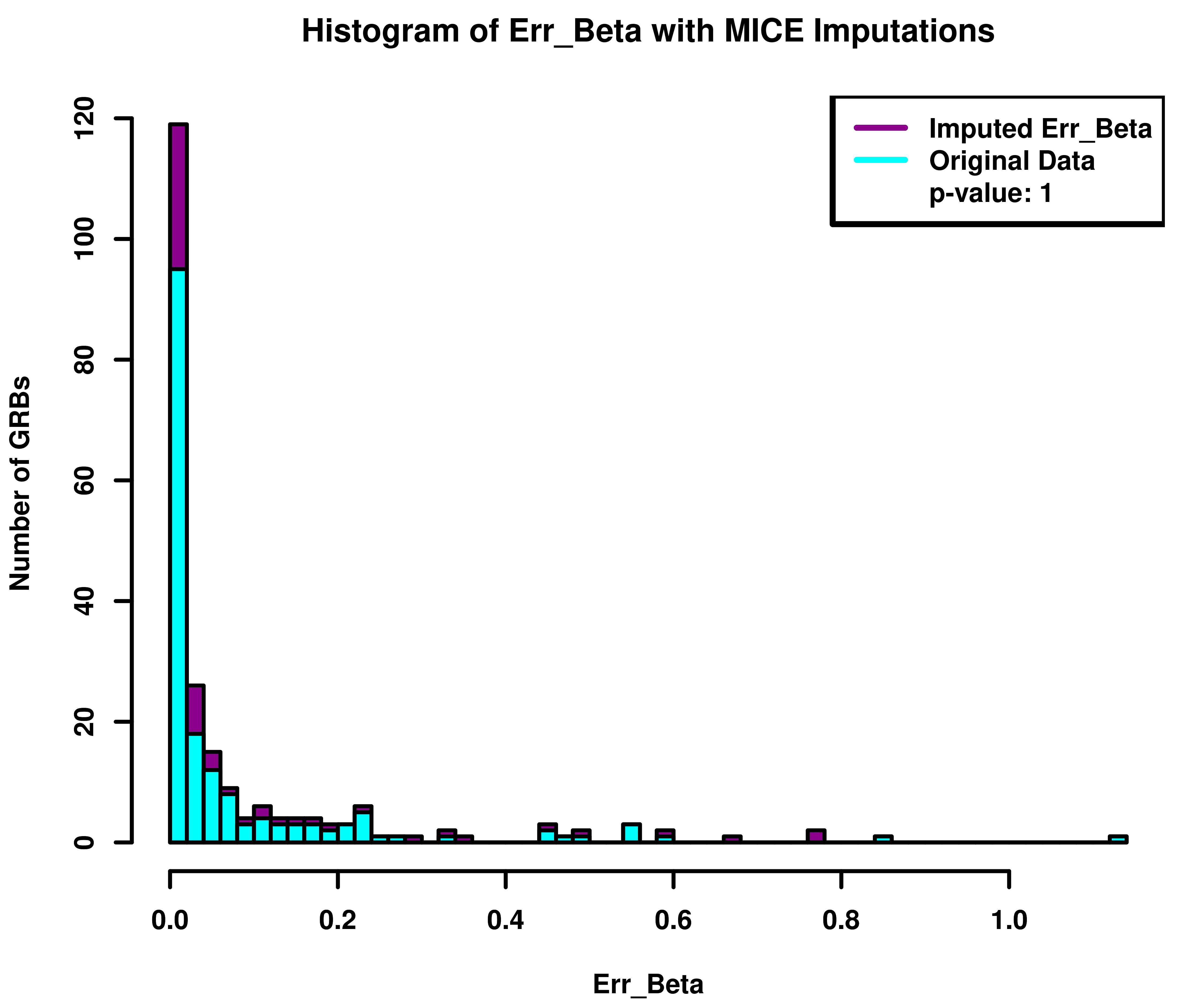}{0.30\textwidth}{(l)}
              }
    \vspace{-2pt}
    \caption{Distributions of $\alpha$, $\beta$, $\Gamma$, $\log({\rm{NH})}$, $\log({\rm{PeakFlux})}$, PhotonIndex, Err $\log({T_{90})}$, Err $\log({\rm{Fluence})}$, Err $\log({\rm{PeakFlux})}$, PhotonIndexErr, Err $\alpha$, and Err $\beta$ are shown in panels (a) to (l) with MICE imputed data in magenta and the original data in blue. The plots also show the $p$-value from the KS test.}
    \label{fig:combined_hist}
\end{figure*}

After dividing the dataset into non-MICE imputed and MICE-imputed groups, a categorical label is assigned to each observation, as illustrated in Figure \ref{Fig:MICE scatter plot}. This scatter matrix plot also includes correlations among the features in the upper right half, providing a comprehensive exploration of variable relationships within the dataset. It highlights the positive impact of using MICE imputation on the overall distribution.
Figure \ref{fig:combined_hist} shows histograms of the imputed missing values for $\alpha$, $\beta$, $\Gamma$, $\log({\rm{NH})}$, $\log({\rm{PeakFlux})}$, PhotonIndex, Err $\log({T_{90})}$, Err $\log({\rm{Fluence})}$, Err $\log({\rm{PeakFlux})}$, PhotonIndexErr, Err $\alpha$, and Err $\beta$ against the observed values. To ensure the MICE-imputed data is reliable, these imputed values must originate from the same parent population as the observed feature values, so we performed the Kolmogorov-Smirnov (KS) test with a two-sided alternative hypothesis \citep{doi:10.1080/01621459.1968.11009335} on the imputed variables of $\alpha$, $\beta$, $\Gamma$, $\log(\rm{NH})$, $\log(\rm{PeakFlux})$, PhotonIndex, Err $\log(T_{90})$, Err $\log(\rm{Fluence})$, Err $\log(\rm{PeakFlux})$, PhotonIndexErr, Err $\alpha$, and Err $\beta$, which determines whether the MICE-imputed data is consistent with the parent population. The resulting $p$-value of each MICE-imputed variable varies from 0.88 for Err $\log(T_{90})$, being this the lowest value, to 0.99 for $\beta$, $\log(\rm{NH})$, $\log(\rm{PeakFlux})$, $\log(\rm{Fluence})$, Err $\log(\rm{PeakFlux})$, PhotonIndexErr, and 1 for the rest of the variables. In all instances, the $p$-values are very high (close to 1), indicating no significant difference between the imputed and original data distribution. 
We also conducted the Anderson-Darling (AD) test \citep{AD-1, AD} on these MICE-imputed variables. The $p$-values from the AD test varies from 0.68 for Err $\log(\rm{Fluence})$, being the smallest value, to 0.75 for $\log(\rm{PeakFlux})$, to 0.79 for Err $\log(\rm{PeakFlux})$, to 0.83 for PhotonIndexErr, to 0.89 for Err $\log(T_{90})$, to 0.99 for $\Gamma$, PhotonIndex, $\beta$, $\log(\rm{NH})$, Err $\alpha$, and Err $\beta$, to 1 for $\alpha$. All these values are much $>$ 5\%, again indicating that the imputed data and original data come from the same distribution.
Thus, the imputation method does not introduce biases, supporting the robustness of our analysis.

\subsection{Data Balancing}
\label{sec:databalancing}
An imbalanced dataset, where the dataset is highly skewed, can significantly affect the performance of ML algorithms and subsequently impact the classifier's effectiveness. To address this issue, we employed a data balancing technique to equalize the minority (high$-z$ GRBs in our case) and majority (low$-z$ GRBs in our case) classes. Specifically, we used the Synthetic Minority Over-sampling Technique \citep[SMOTE,][]{2011arXiv1106.1813C}. In this method, the minority class is augmented by generating synthetic samples using the existing data rather than simply over-sampling with replacement. We know the possible limitations of such a synthetic sample, as it does not capture the possible outliers events that often exist in actual observations. However, this analysis gives an idea of how, in the future, with more observations, we can balance the sample. In a forthcoming paper, we plan to consider also this effect. 
A successful technique inspires this approach in handwritten character recognition \citep{Bunke2007}. Here, the synthetic samples are generated in a feature space rather than a data space. Oversampling of the minority class is done by taking each minority class sample and creating synthetic samples along the line segments that connect any or all of the $k$ nearest minority class neighbors. Depending on the required amount of over-sampling, neighbors are selected randomly from the $k$ nearest neighbors. After balancing the data using SMOTE, we again exclude the data points where the $\Delta x/x > 1$ for any given feature $x$, with errorbar $\Delta x$.

\subsection{Feature Selection}
\begin{figure*}[htbp]
    \centering
    \gridline{\fig{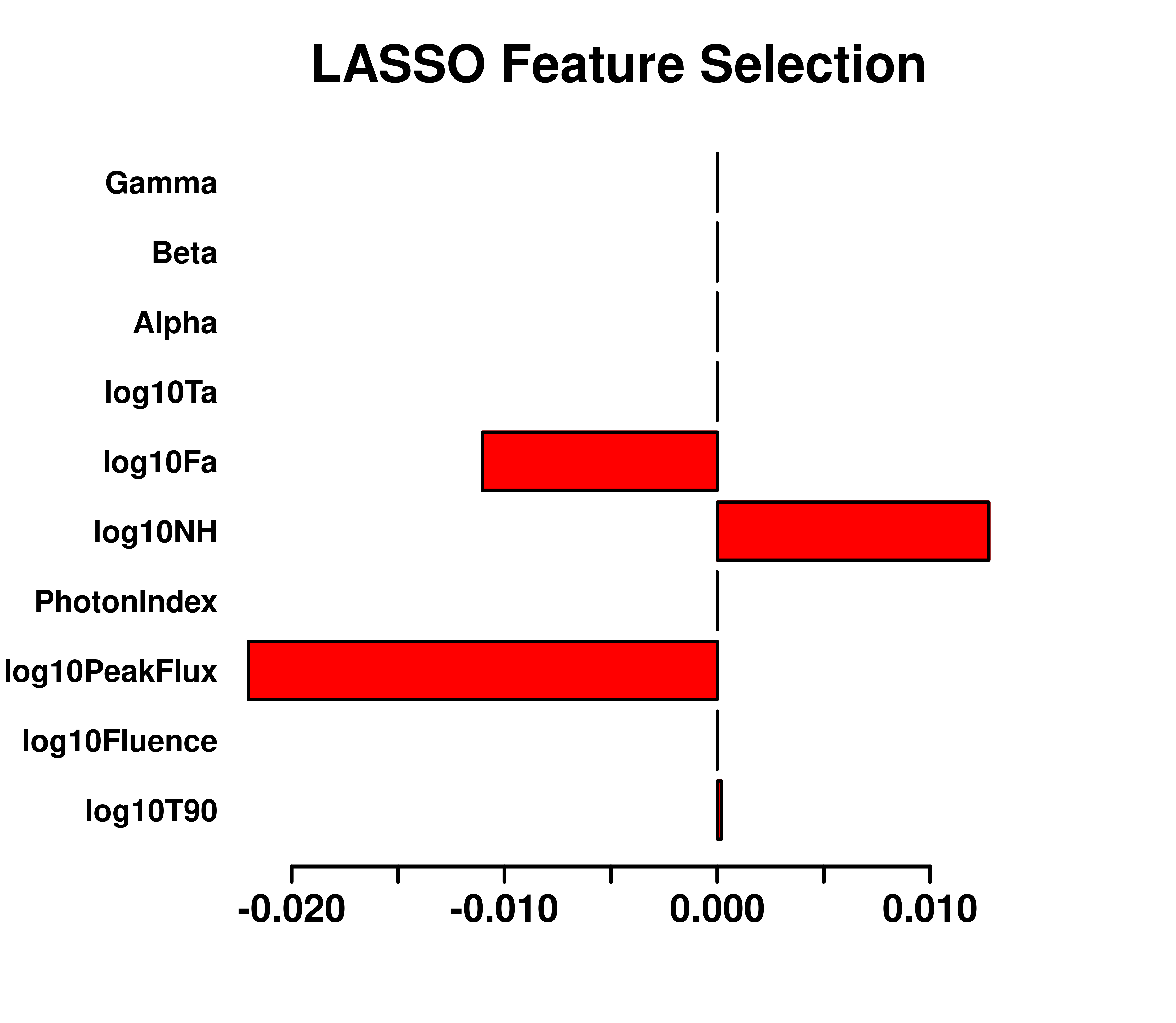}{0.40\textwidth}{}
              \fig{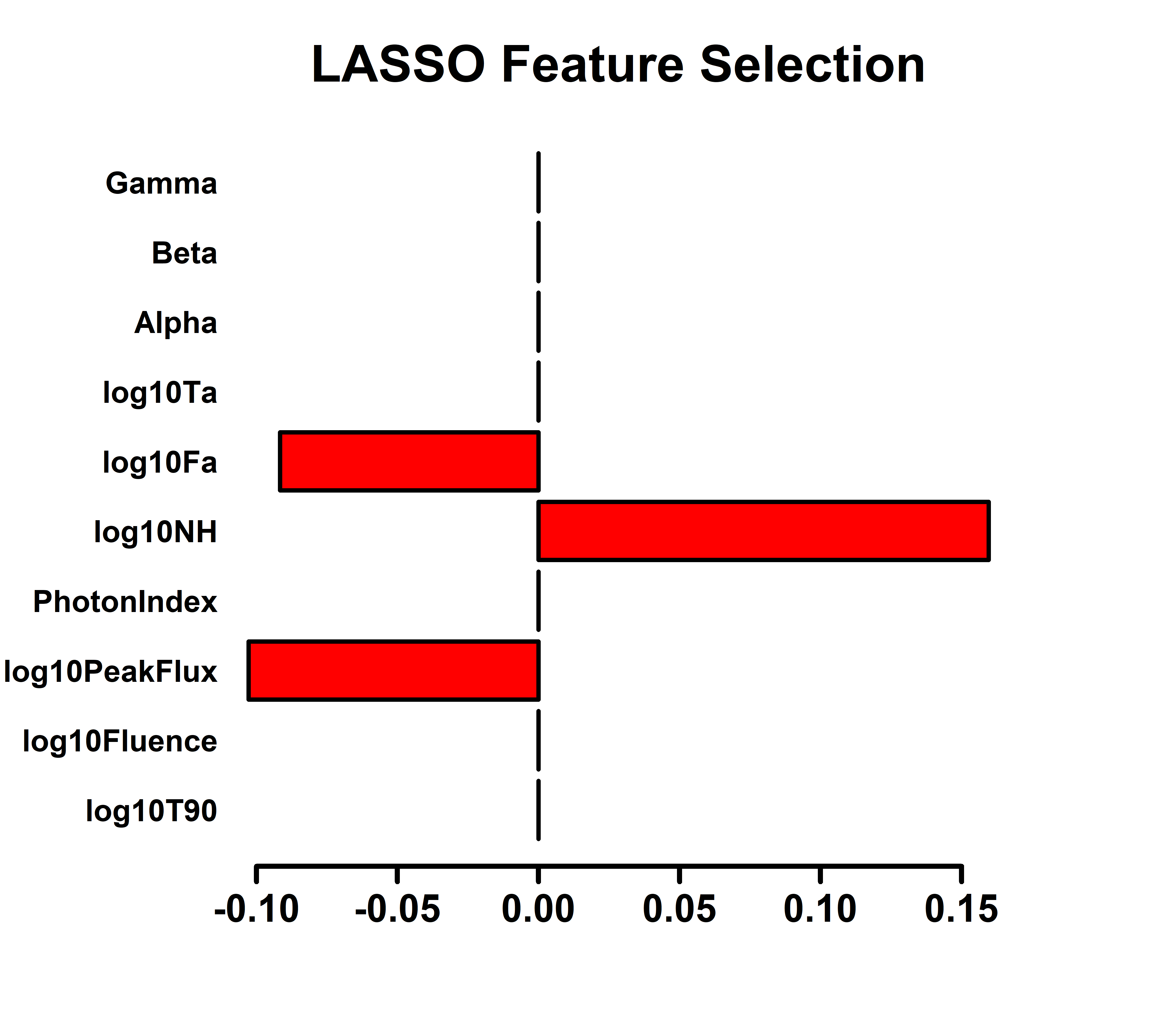}{0.40\textwidth}{}
              }
    \vspace{-25pt}
    \gridline{\fig{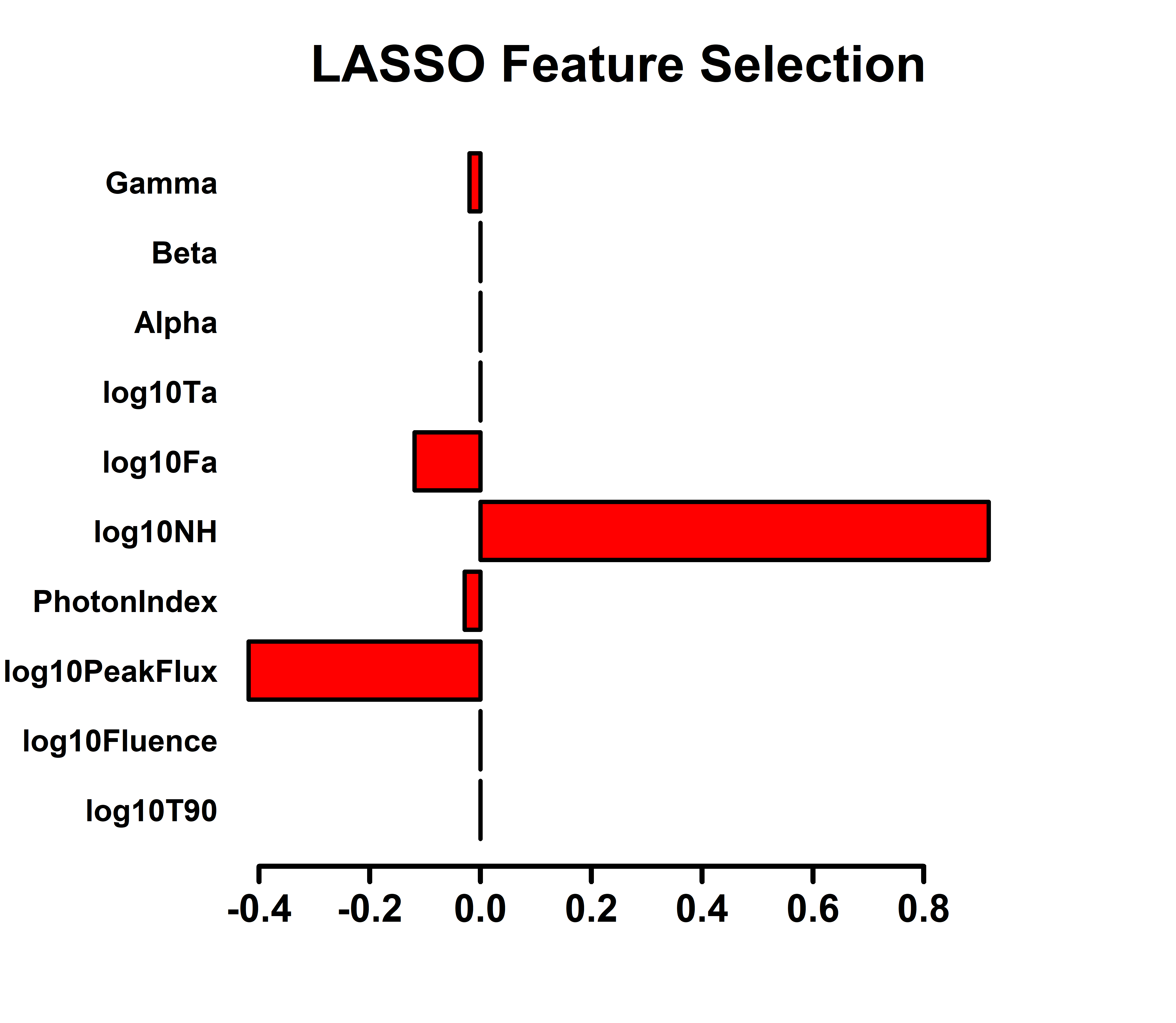}{0.40\textwidth}{}
              \fig{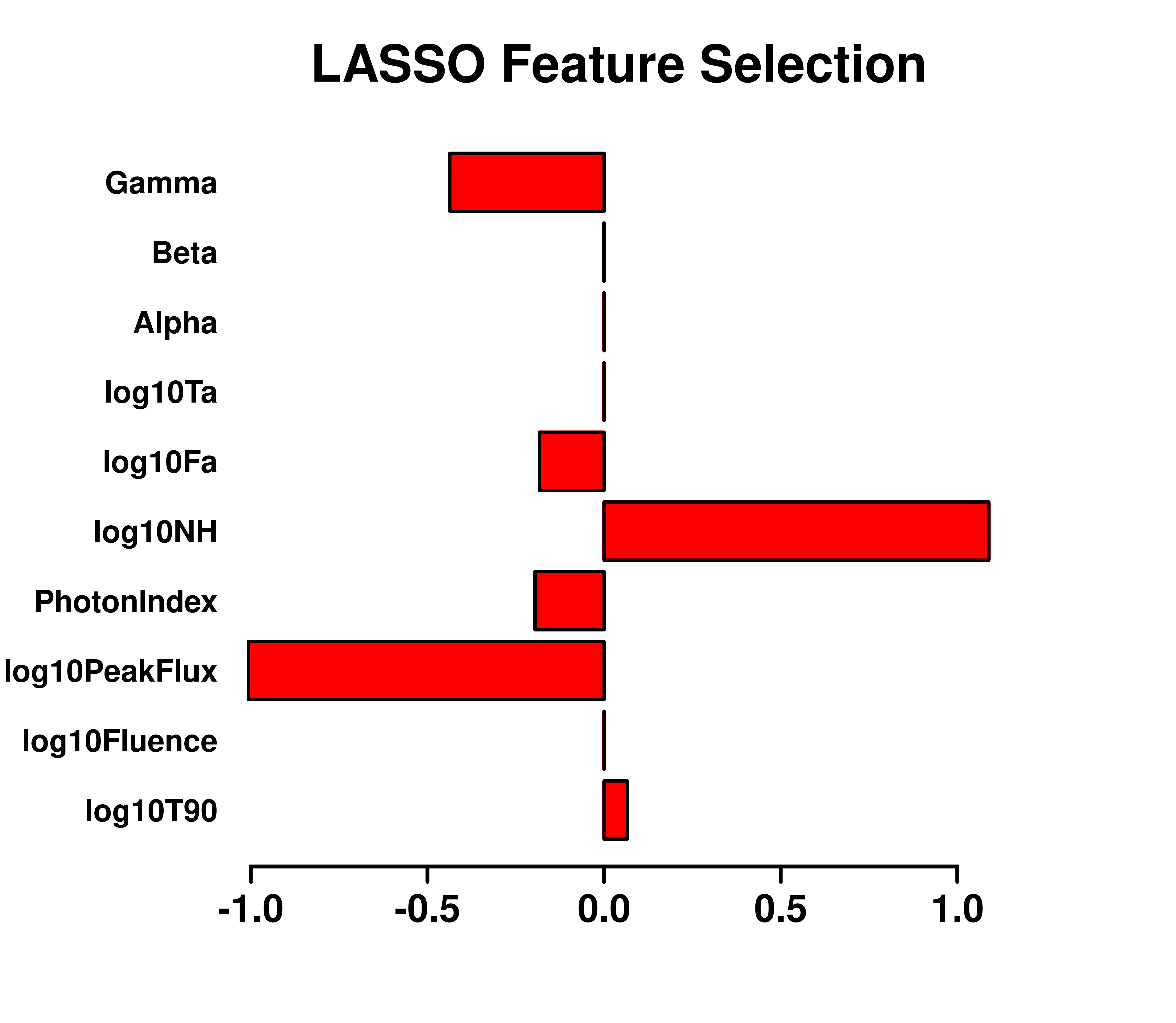}{0.40\textwidth}{}
              }
\caption{Weight assigned to each feature by LASSO for $z_{t}$ = 3.5 with the raw dataset without the M-estimator (top left), the raw dataset with the M-estimator (top right), MICE-imputed dataset (bottom left), and SMOTE balanced dataset (bottom right). Features with weights $>$ 2\% are selected as the best features for the classification.}
\label{Fig:LASSO}
\end{figure*}

After the data cleaning and outlier removal, the data was split into two subsets: an 80\% training set used for model development and a 20\% test set reserved exclusively for performance evaluation. The test set remained untouched during the model development process to ensure unbiased assessment. 
We note that, although in other previous works, cross-validation has been used, generally, 20\% of the sample is not kept separated from the rest of the sample, see \cite{morgan2012, ukwatta2016machine}.
Feature selection is a critical step as it enhances model interpretability and reduces the dimensionality of the data, thereby enabling algorithms to operate more efficiently. This process also mitigates the risk of overfitting. Identifying relevant features is also essential for improving the overall accuracy of the SuperLearner algorithm.
Thus, the Least Absolute Shrinkage and Selecting Operator (LASSO) approach is used in the next step to select the best features for classifying high$-$ and low$-z$ GRBs. LASSO minimizes the residual sum of squares provided that the sum of the coefficients' absolute values is less than a constant \citep{TibshiraniLasso}. Furthermore, LASSO chooses a subset of all the predictors in the data that better predict the response variable here, $z$. 
To enhance result stability, we performed multiple iterations of LASSO in sets of 100, 200, 300, 500, and 1000 loops, yielding the averaged weights for each variable as the outcome. We observed consistent outcomes of features selected across these different loop counts.
Variables with weights exceeding 2\% were identified as the most significant features and selected for use in subsequent classifier model development. 
We found that setting the LASSO feature selection threshold to 2\% ensures that at least 2-3 properties of the plateau emission are always pinpointed. 
If we reduce the threshold to 1\%, more features are included, but it complicates the algorithm parameterization without improving prediction accuracy due to our limited data points. The extra variables selected at a 1\% threshold often correlate with the variables already present, making the 2\% threshold a good compromise between balancing the number of features and the data points. An example of the LASSO feature selection is shown in Figure \ref{Fig:LASSO} for a $z_{t}=3.5$ with the raw dataset without the M-estimator, the raw dataset with the M-estimator, MICE-imputed dataset, and SMOTE balanced dataset. 

\subsection{Algorithms}
\label{sec:algorithms}
We employ a combination of parametric, semi-parametric, and non-parametric methods within the ML framework. Parametric models are efficient and straightforward to train but are constrained by their predefined functional forms. On the other hand, non-parametric models do not assume a specific functional form, offering greater predictive power but risking overfitting and requiring large sample sizes and extensive computational time. Semi-parametric models integrate the strengths of both parametric and non-parametric approaches, balancing simplicity and flexibility. Below, we categorize the algorithms used in this study into parametric, non-parametric, and semi-parametric groups.

\subsubsection{The parametric models}
\begin{itemize}
    \item Linear Model (lm): lm is the simplest classification or regression algorithm, which enacts OLS for classification or regression. This method assumes a linear relationship between features and the response variable and estimates the coefficients for a linear boundary that minimizes the sum of squared differences between them \citep{fox_2015}.
    \item Generalized Linear Model (glm): glm is a generalization of the lm where the relationship between features and the response is specified through a link function. It allows for various probability distributions, and its parameters are estimated through an iterative maximum likelihood estimation (MLE), meaning glm can handle distributions like binomial, Poisson, and gamma \citep{nelder1972generalized}. We use the glm algorithm from the SuperLearner package based on the work of \cite{fox_2015}.
    \item Lasso and Elastic-Net Regularized Generalized Linear Model (glmnet): glmnet is an extremely efficient algorithm that fits the entire regularization path for various models, including LASSO and elastic-net regularization. These models encompass linear regression, logistic and multinomial regression, Poisson regression, Cox models, multiple-response Gaussian models, and grouped multinomial regression \citep{10.1111/j.1467-9868.2005.00503.x,friedman2010regularization, JSSv039i05}. Glmnet uses a regularization parameter and fits glm to a penalized maximum likelihood, thereby removing the less important features. Thus, glmnet can handle high-dimensional data and focus on highly correlated features.
    \item Speedglm and speedlm: They are faster implementations of glm and lm, respectively, designed to handle medium to large-scale datasets efficiently \citep{speedglmspeedlm}. These packages significantly reduce computation time by leveraging optimized Basic Linear Algebra Subprograms (BLAS) for matrix operations, enhancing performance compared to standard implementations. Speedglm and speedlm maintain the accuracy and reliability of glm and lm while providing a practical solution for large datasets that fit into \textsc{R}'s memory, making them valuable tools for statistical analysis in high-dimensional data contexts.
    \item Bayesglm: A Bayesian inference of glm used to estimate coefficients, enhancing the stability and reliability of the estimates, especially in the context of small sample sizes or when independent variables in a model are correlated, namely multi-collinearity. It determines the most likely estimate of the response variable given the particular set of predictors and the prior distribution on the set of regression parameters. The Bayesglm method is more numerically and computationally stable than the normal glm model. It employs a Student-t prior distribution for the regression coefficients. Then, given the observed data, the likelihood function for these parameters is calculated. The likelihood function and priors are combined to produce the posterior distributions from which we obtain the Maximum A Posteriori MAP estimates of the desired parameters \citep{birnbaum1962foundations, hastie1987generalized, hastie1990generalized, friedman2010regularization}. This approach provides more robust parameter estimates, reducing the risk of overfitting and improving the predictive performance.
    \item Linear Discriminant Analysis (lda): The lda algorithm computes the linear combination of features to maximize the separation between classes, assuming that underlying data for each class is normally distributed and characterized by certain parameters, specifically the mean and covariance matrix \citep{mclachlan2004, hastie2009elements}. Thus, it can handle high-dimensional and multicollinear data.
    \item Quadratic Discriminant Analysis (qda): qda is an adaption of lda for handling data with significantly varying variance-covariance matrices between classes \citep{hastie2009elements}, thus allowing for varying distributions. This implies that qda can model nonlinear decision boundaries where lda cannot do so.
\end{itemize}

\subsubsection{The Semi-parametric models}
\begin{itemize}
    \item BigLASSO: A computationally efficient implementation of the LASSO algorithm in \textsc{R}. BigLASSO extends the capabilities of traditional LASSO and elastic-net linear and logistic regression models. This leads it to accommodate ultra-high-dimensional, multi-gigabyte data sets that exceed memory capacity. 
    It achieves efficient model fitting and regularization by utilizing memory-mapped files, which store the extensive data on disk and load it into memory only as needed during the fitting process. This approach introduces flexibility in model fitting and significantly enhances computational efficiency for big data applications \citep{zeng2017biglasso}.
    \item Enhanced Adaptive Regression Through Hinge (earth): It is an implementation of the Multivariate Adaptive Regression Splines (MARS) method \citep{FriedmanMARS}. This technique extends linear models by incorporating the ability to capture nonlinear relationships between predictors and the response variable.
    Unlike traditional regression, which relies on polynomial terms, in this algorithm, MARS begins with a constant term and sequentially adds functions that can be constants, hinge functions, or products of hinge functions. These hinge functions enable the model to adjust to changes in slope, effectively capturing nonlinear interactions. Each new term added to the model maximizes the reduction of the sum-of-squares residual error. The advantage of earth is that MARS automatically selects important variables and interactions through its forward and backward stepwise process, enhancing model accuracy and interpretability.
\end{itemize}

\subsubsection{The non-parametric models}
\begin{itemize}
\item Random Forest: The Random Forest \citep{breiman2001randomforest} algorithm operates by constructing multiple decision trees. During the tree-building process, the data is divided based on the values of a randomly selected subset of features from the training dataset \citep{ho1995random}. This division continues until the tree reaches a predetermined depth. At each node, the response variable values are averaged, and this average is assigned as the prediction for any GRB that falls within that node. Finally, the algorithm  provides the prediction by averaging the outcomes from all the independent decision trees.
    \item Conditional Random Forest (cforest): The cforest algorithm is an expansion on the Random Forest model, which generates multiple decision trees from a subset of features in the training data. Cforest constructs decision trees using conditional inference  \citep{doi:10.1198/106186006X133933}, which splits decision trees on significance tests that measure node purity. Thus, more successful features are prioritized and can fit more complex models \citep{wager2017estimation}.
    \item Ranger: It is an efficient implementation of the Random Forest algorithm designed for high-dimensional data. It supports ensembles of classification, regression, and survival trees. Compared to Random Forest, ranger offers a faster execution and the capability to utilize highly randomized trees (ERT). Following a top-down approach, ERTs are constructed similarly to the decision trees in Random Forest. However, in ranger, the cuts in ERTs are generated randomly. The cut that most reduces prediction error is chosen \citep{geurts06extremetrees}. ERTs can reduce variance further than Random Forest due to the increased randomness in the node-splitting procedure.
    \item caret.rpart: ``caret.rpart" refers to the application of the ``rpart" (Recursive Partitioning and Regression Trees) algorithm within the ``caret" (Classification and Regression Training) package in \textsc{R}. The ``caret" package is a comprehensive toolkit that makes the process of building predictive models easier, including data preprocessing, model training, tuning, and validation \citep{caret-rpart}. Within this package, ``rpart" is used to create decision trees for classification and regression tasks \citep{caret-rpart}.
    The algorithm operates on the principles of recursive partitioning, where the data is iteratively split into subsets of similar values (homogeneous subsets). The tree-like model is generated through a process that begins with identifying a single variable that best splits the data into two groups. This division is then recursively applied to each resulting subgroup until a minimum subgroup size is reached or further improvement is no longer possible.
    \item Xgboost: The xgboost algorithm, or extreme gradient boosting algorithm, is a scalable, distributed gradient-boosted decision tree (GBDT) algorithm. It provides parallel tree boosting and is a widely used ML technique for regression, classification, and ranking problems. It employs a more regularized model formalization to prevent overfitting and provides improved performance compared to a single decision tree \citep{friedman2000additive, friedman2001elements, chen2016xgboost}.
    The advantage of Xgboost is that the trees are dependent on each other and this depends helps to remember and remove the weakest learners, differently from Random Forest where the trees do not remember the weak and strong learners.
    
    \item Kernel-based k-Nearest Neighbors (kernelKnn): The algorithm is an extension of the traditional k-Nearest Neighbors (Knn). The standard Knn algorithm identifies the k nearest data points (neighbors) to a given query point based on some distance metric, typically Euclidean distance. However, in the kernelKnn, a kernel function assigns weights to the neighbors based on their distance from the query point. This is used to construct a This is used to construct a mapping from the initial feature space to a higher dimensional space \citep{kernelkNN}. Common kernel functions include Gaussian, polynomial, and sigmoid kernels.
    \item Support Vector Machine (svm): A supervised ML algorithm used for classification, regression, and outlier detection tasks \citep{vapnik1995nature, cortes1995support, hastie2009elements}. 
    The svm works by finding the hyperplane in the multidimensional feature space that maximizes the margin between different classes (in our case, high$-z$ and low$-z$ GRBs). It incorporates a regularization parameter (C) to control the trade-off between maximizing the margin, the distance between the separating decision boundary and the nearest data points from each class, and minimizing the classification error.
    \item Kernlab Support Vector Machine (ksvm): It is an extension of svm that maps the input data to a higher-dimensional space using the kernel function such that more complex relationships of the data can be handled without assuming a specific parametric form \citep{JSSv011i09}.
\end{itemize}

\subsection{SuperLearner}
\label{sec:SL}
SuperLearner, as described in \cite{van2007super}, is an ML algorithm that works with ensemble learning. SuperLearner combines multiple algorithms into a single model, thereby leveraging the predictive power of each constituent model such that the overall performance of the model is enhanced compared to what could be achieved by any individual component.
SuperLearner uses a default 10-fold Cross-Validation (10fCV) to estimate the predictive performance of the trained model. 10fCV involves partitioning the dataset into ten distinct subsets. The model is iteratively trained on nine of these subsets while using the remaining subset for testing. This process is repeated so that each subset serves as the testing set exactly once. The prediction results from each iteration are averaged to obtain the mean prediction, and the standard deviation of these results is used to estimate the prediction error.
Within SuperLearner, the base learner algorithms, as outlined in Section \ref{sec:algorithms}, are trained using 10fCV. SuperLearner assigns coefficients to each algorithm based on their Root Mean Squared Error (RMSE). These coefficients reflect the relative importance of each algorithm in the overall prediction model. Each coefficient is always equal to or greater than 0, and their total sum is always equal to 1. Consequently, the highest-weighted algorithms can be gathered into an ensemble for an overall lower RMSE than any individual algorithm \citep{polley2010super}. 

\begin{figure*}[htbp]
    \centering
    \gridline{\fig{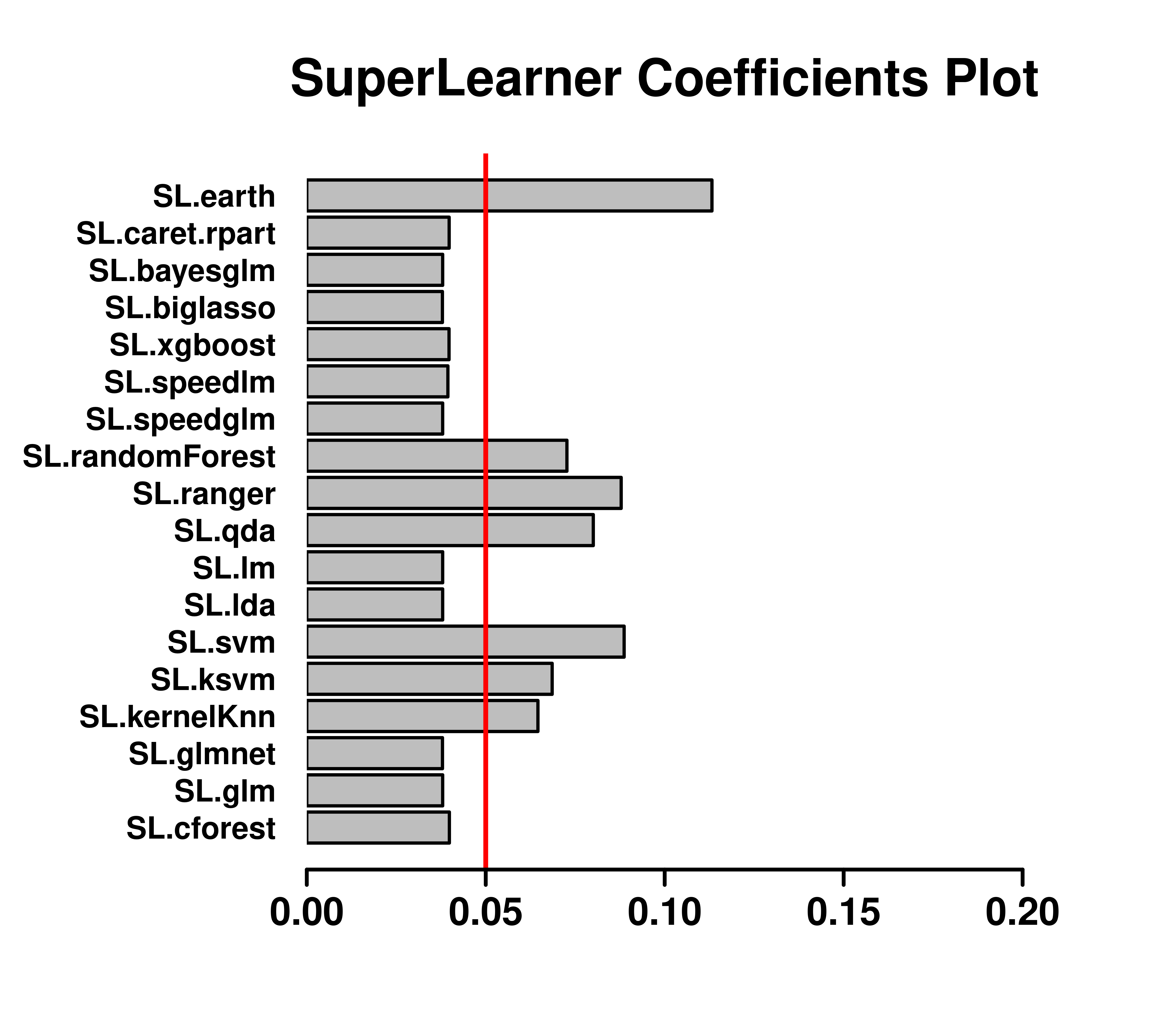}{0.46\textwidth}{}
              \fig{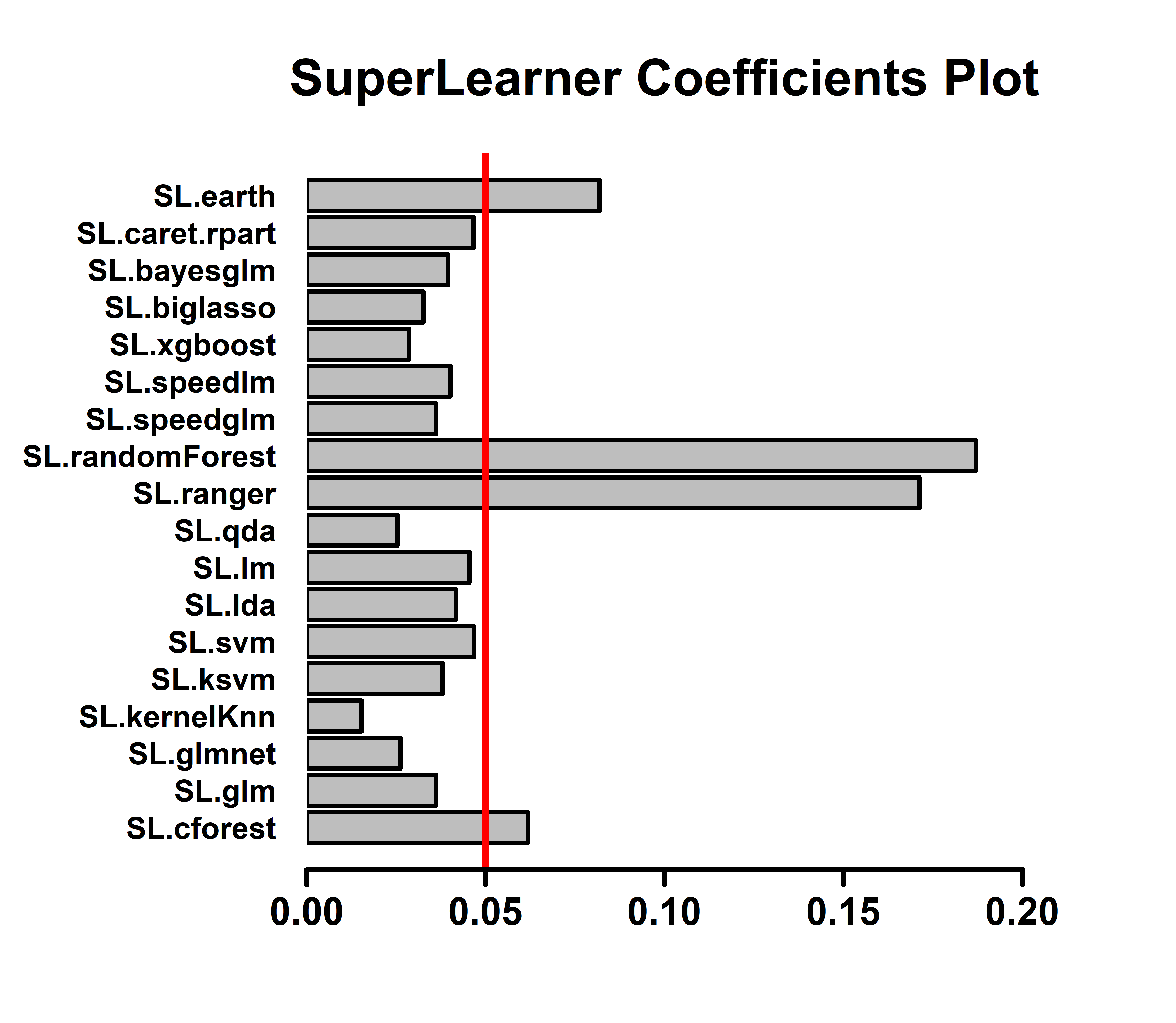}{0.46\textwidth}{}
              }
    \vspace{-25pt}
    \gridline{\fig{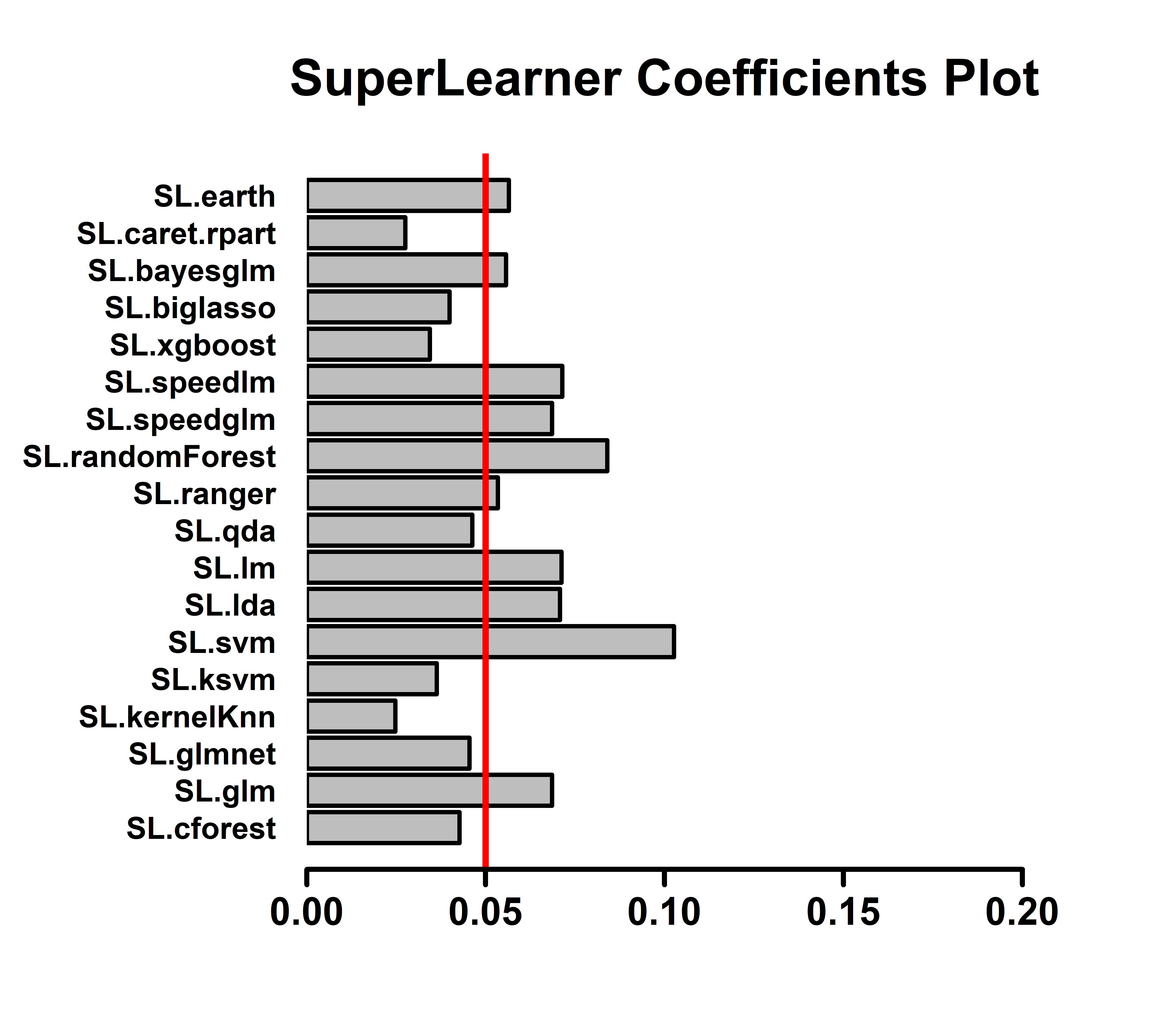}{0.46\textwidth}{}
              \fig{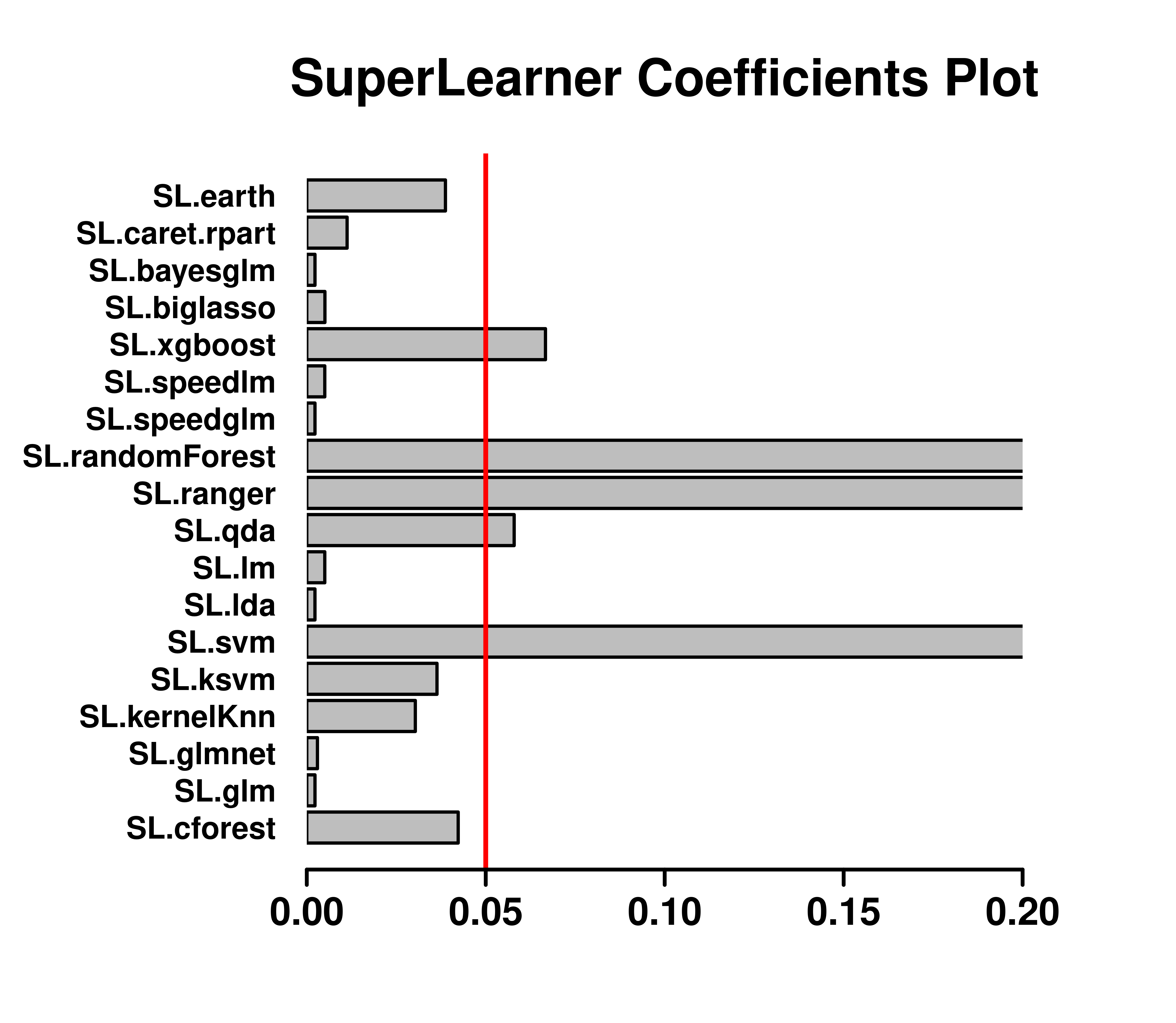}{0.46\textwidth}{}
              }
\caption{Weight assigned to each algorithm by SuperLearner for $z_{t}$ = 3.5 with the raw dataset without the M-estimator (top left), the raw dataset with the M-estimator (top right), MICE-imputed dataset (bottom left), and SMOTE balanced dataset (bottom right). The red vertical line shows the weight cutoff line of 0.05 for algorithm selection. Algorithms above this threshold are selected as the best algorithms.}
\label{Fig:SL weights}
\end{figure*}

We implemented the SuperLearner algorithm in a nested loop, where the 10fCV procedure was repeated $n$ times to increase the overall determinism of the process. We tested with $n$ = 100, 200, 300, 500, and 1000. Our findings indicated that the Area Under the Curve (AUC) of the training remains stable across all runs (see Section \ref{sec:discussion}). For our study, we fixed $n = 100$ because the AUC, explained in Section \ref{sec:results}) since we do not need to run with more loops to obtain compatible results at the 1\% value. This choice also aligns with the number of iterations used in prior studies with SuperLearner \citep{dainotti2021predicting, narendra2022predicting,Dainotti2024b,Dainotti_2024}. In our analysis, models with weights below 0.05 (5\%) are discarded after performing 100 iterations of 10fCV to ensure the stability of the most effective models and to enhance the overall performance of the classifier. A cutoff of 0.05 eliminates less important models while maintaining a reliable prediction power. 
We tried adding models with lower weights, but these did not enhance the prediction on the metrics, and thus, we stopped at this threshold. On the other hand, a threshold greater than 0.05 eliminates certain models that can enhance the performance. Figure \ref{Fig:SL weights} illustrates an example of the weights assigned by SuperLearner for a $z_{t}$ of 3.5 using the raw dataset without the M-estimator, the raw dataset with the M-estimator, MICE-imputed dataset, and SMOTE balanced dataset. The red vertical line indicates the threshold of 0.05. Algorithms with weights exceeding this threshold are selected as the best-performing models.

We check the model's stability by running the SuperLearner algorithm with multiple iterations and analyzing the consistency of the results. This ensures that the conclusions drawn from the model are stable and reliable. Indeed, having a small sample size prevents us from using more models than the one already used, but in the future, we can add more algorithms with a larger sample.

\subsection{Metrics}
\label{sec:Metrics}
\begin{figure*}[htbp]
\centering
    \fig{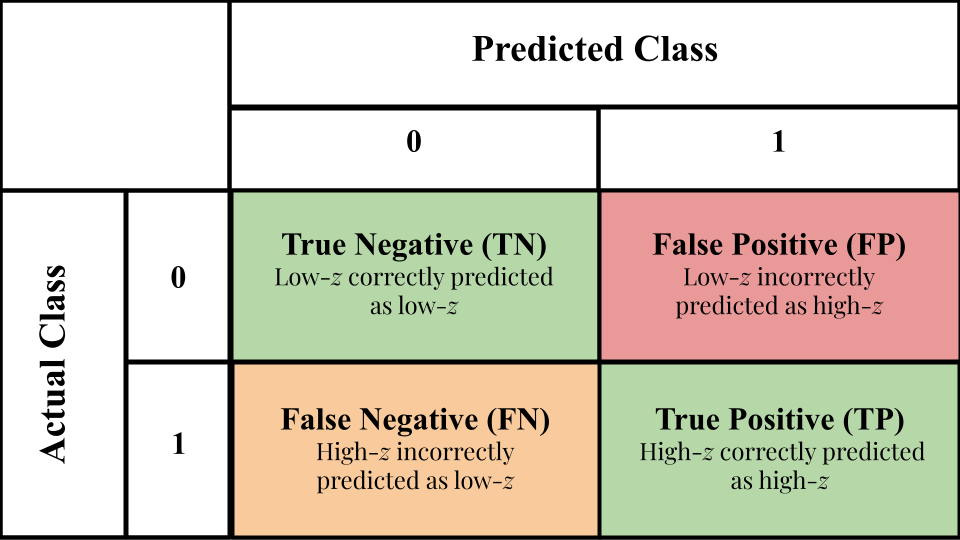}{0.55\textwidth}{}
\caption{Confusion matrix in relation to our data where the class 0 refers to low-z and 1 to high-z GRBs.}
\label{Fig:LegConfusionMatrix}
\end{figure*}

This section describes the terminology and classification metrics used to evaluate our results and the classifier's performance.
In this work, we assigned GRB with high$-z$ as class ``1" and GRB with low$-z$ as class ``0", respectively.
In relation to these two classes, four definitions are summarized in the confusion matrix plot; see Figure \ref{Fig:LegConfusionMatrix}.

\begin{itemize}
    \item True Positive (TP): Denotes the percentage for which the model correctly predicts the positive class (high$-z$ GRB in our case) and is accurately categorized as positive.
    \item True Negative (TN): Denotes the percentage of data that the model correctly predicts as true negative (low$-z$ GRB in our case) and is accurately categorized as negative.
    \item False Positive (FP): Denotes the percentage of the data incorrectly predicted as the positive class, but they are indeed the negative class.
    \item False Negative (FN): Denotes the percentage of the data points incorrectly predicted as the negative class, but they belong to the positive class.
    \item Sensitivity or Recall: It is a metric that evaluates the effectiveness of an ML model in identifying true positive instances \citep{altman1994diagnostic}. It is also called Recall or True Positive Rate (TPR):
    \begin{equation}
        \rm{Sensitivity = \frac{TP}{TP + FN}} 
    \end{equation}
    \item Specificity: It is a metric that evaluates the effectiveness of an ML model in identifying true negative instances \citep{altman1994diagnostic}. The specificity is called True Negative Rate (TNR). Specificity and sensitivity have an inverse relationship, which means that the specificity decreases with the increase in sensitivity and vice versa \citep{parikh2008understanding}:
    \begin{equation}
        \rm{Specificity = \frac{TN}{TN + FP}} 
    \end{equation}
    \item Accuracy: It refers to the proportion of correct assessments in our case, GRBs whose redshifts are correctly identified both as low and high-z GRBs): 
    \begin{equation}
        \rm{Accuracy = \frac{TP + TN}{TP + TN + FP + FN}}
    \end{equation}
    \item We also use precision and recall (sensitivity), as accuracy can be misleading for unbalanced classifications. Precision: It is a metric that assesses the number of true positives (i.e., the number of GRBs correctly classified as high$-z$) to the total number of positive classifications (i.e., the total number of GRBs classified as high$-z$):
    \begin{equation}
        \rm{Precision = \frac{TP}{TP + FP}}
    \end{equation}
\end{itemize}

\section{Results}
\label{sec:results}
We present our results with the following primary goal: to offer a binary classifier that the community can freely use in a web app to perform follow-up observations. We used four different $z_{t}$ (2.0, 2.5, 3.0, and 3.5) to understand how these different $z_{t}$ impact the classifier performance. 
These thresholds were chosen to examine datasets of different lengths, favoring cases with more or less data on high$-$ and low$-z$ events. Adjusting the $z_t$ value from 2.0 and 3.5 does not produce mutually exclusive and exhaustive classes, as a GRB with $z \leq 2$ will be classified as low$-z$ in all $z_t$ samples. However, a GRB with $z \leq 2.5$ will be considered low$-z$ for $z_t = 2.5$, 3.0, and 3.5, but not for $z_t = 2$, and so on.
Thus, the counts of high$-$ and low$-z$ GRBs in each class defined by the $z_{t}$ will vary depending on the chosen $z_{t}$. Table \ref{tab:GRBdata} shows the number of GRBs in each class. 
This allows us to assess for which $z_{t}$ our classifier performs the best in terms of sensitivity for our dataset. 
Indeed, one can also try additional redshifts; however, the analysis of these new redshifts is beyond the scope of the current paper. 
We used four datasets in Superlearner: the raw dataset without and with the M-estimator, the MICE-imputed, and the SMOTE balanced datasets. 

We plotted the Receiver Operating Characteristic (ROC) curve to visualize the classifier's performance across all classification thresholds. The ROC space here is defined as the specificity on the x-axis vs. the sensitivity on the y-axis. Each point on the ROC curve represents a different threshold for the classification decision. The area beneath this ROC curve, known as AUC, evaluates the ability of each model, shown in the picture label, and the combined ability of SuperLearner to discriminate between high$-$ and low$-z$ GRBs. A higher AUC indicates a better-performing model, with a maximum value of 1 indicating the perfect discrimination between the two classes. Below, we discuss the four  $z_{t}$ results for the four different datasets.

\begin{figure*}[htbp]
    \centering
    \gridline{\fig{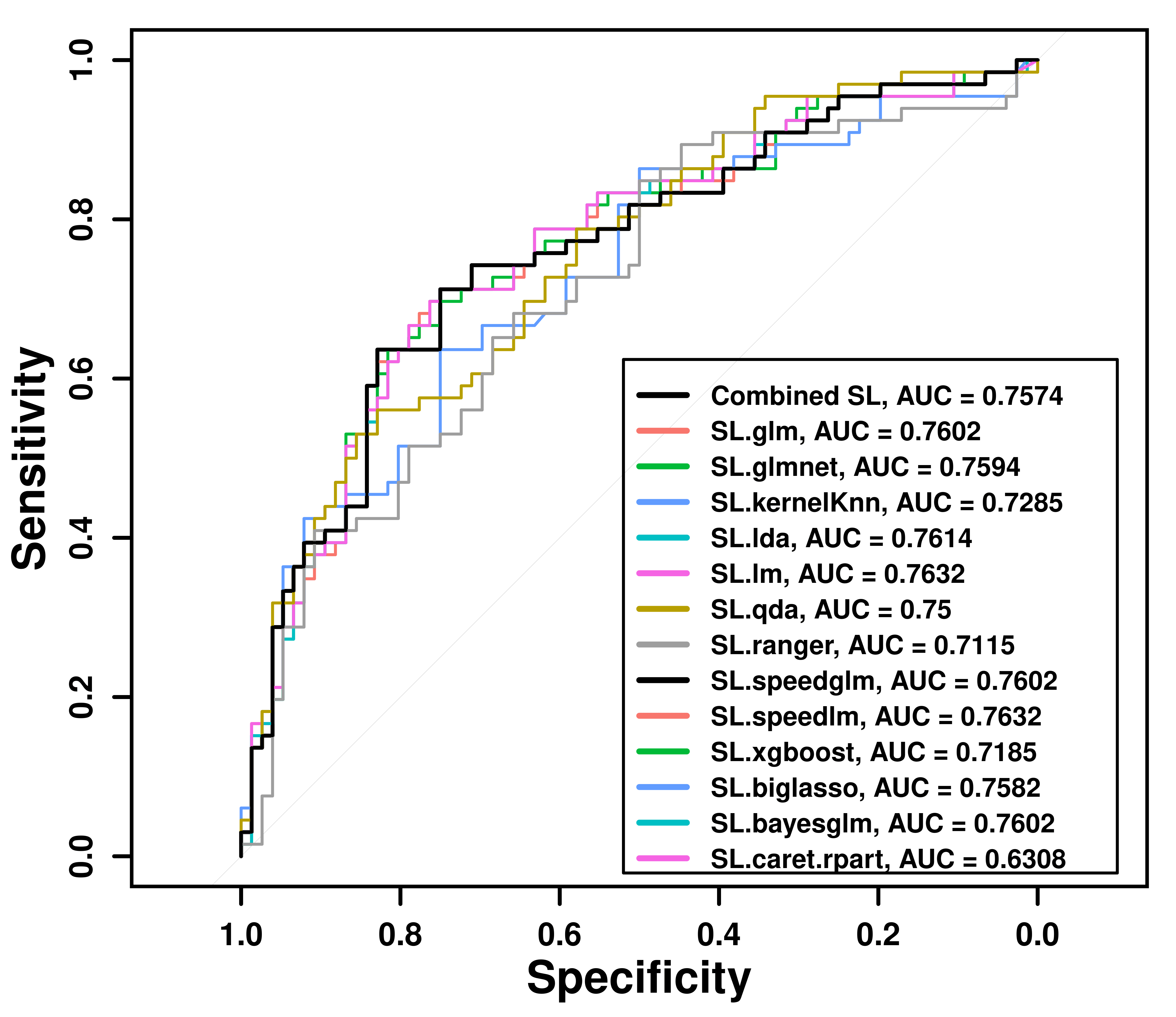}{0.33\textwidth}{}
              \fig{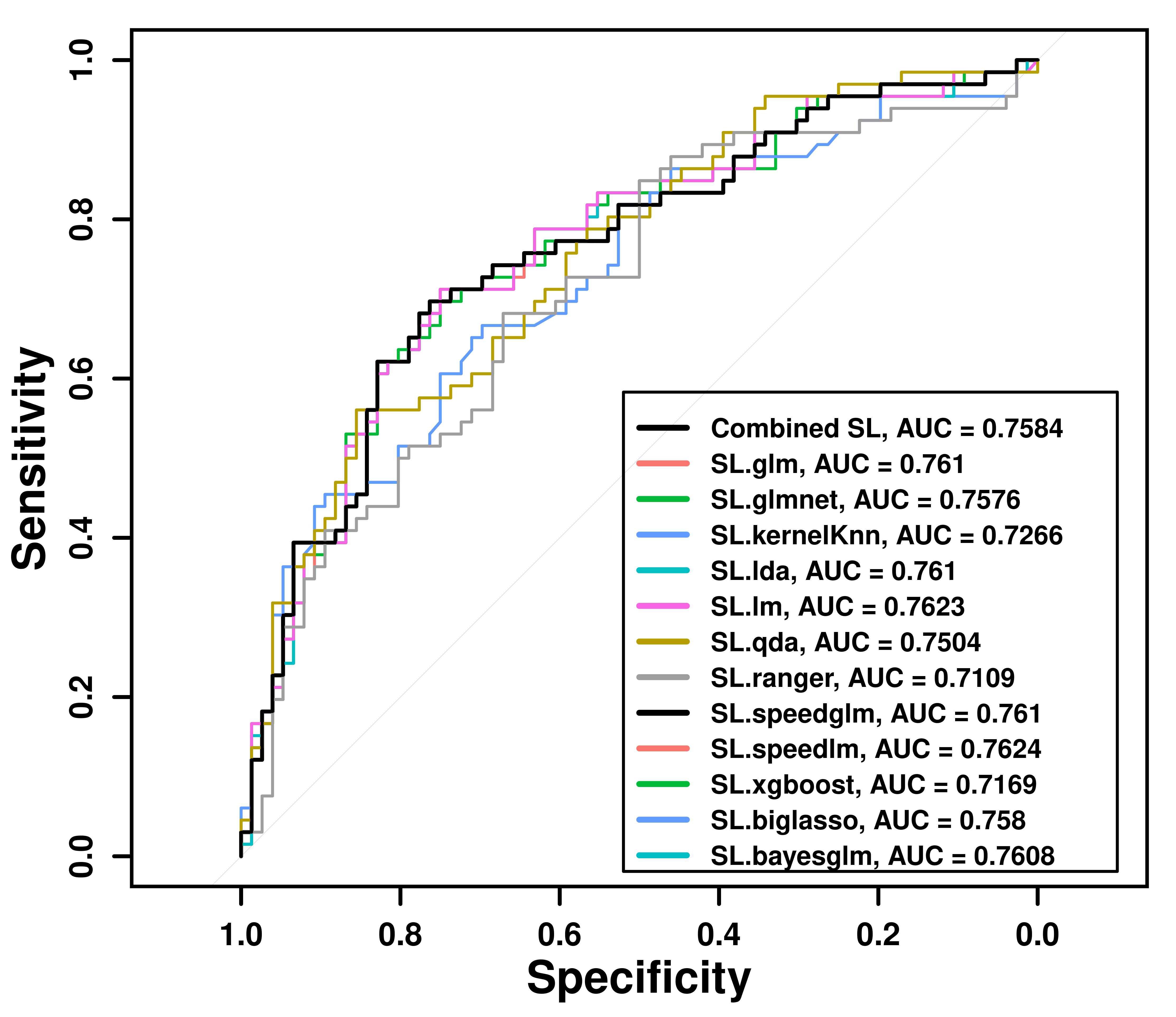}{0.33\textwidth}{}
              }
    \vspace{-30pt}
    \gridline{\fig{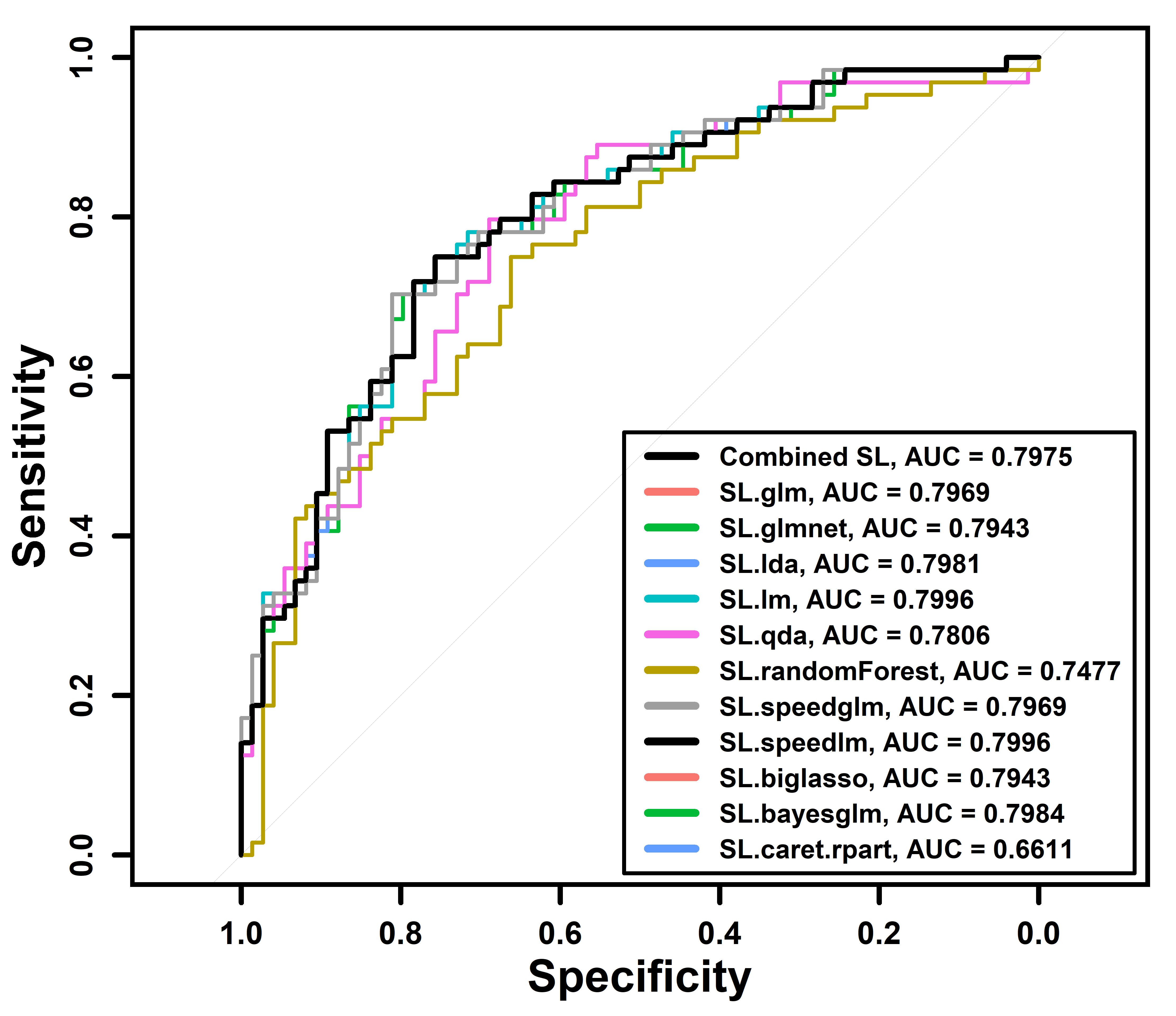}{0.33\textwidth}{}
              \fig{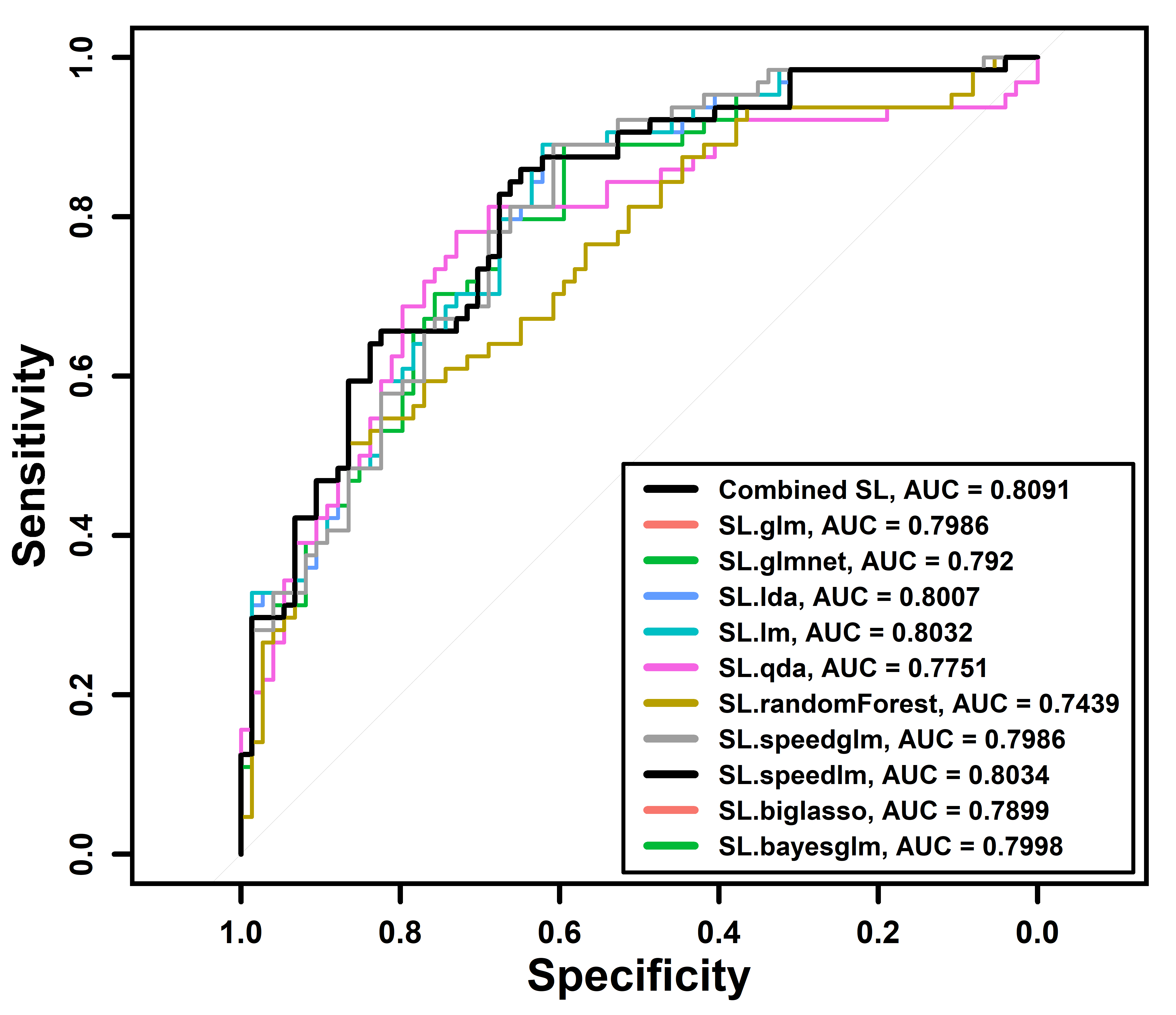}{0.33\textwidth}{}
              }
    \vspace{-30pt}
    \gridline{\fig{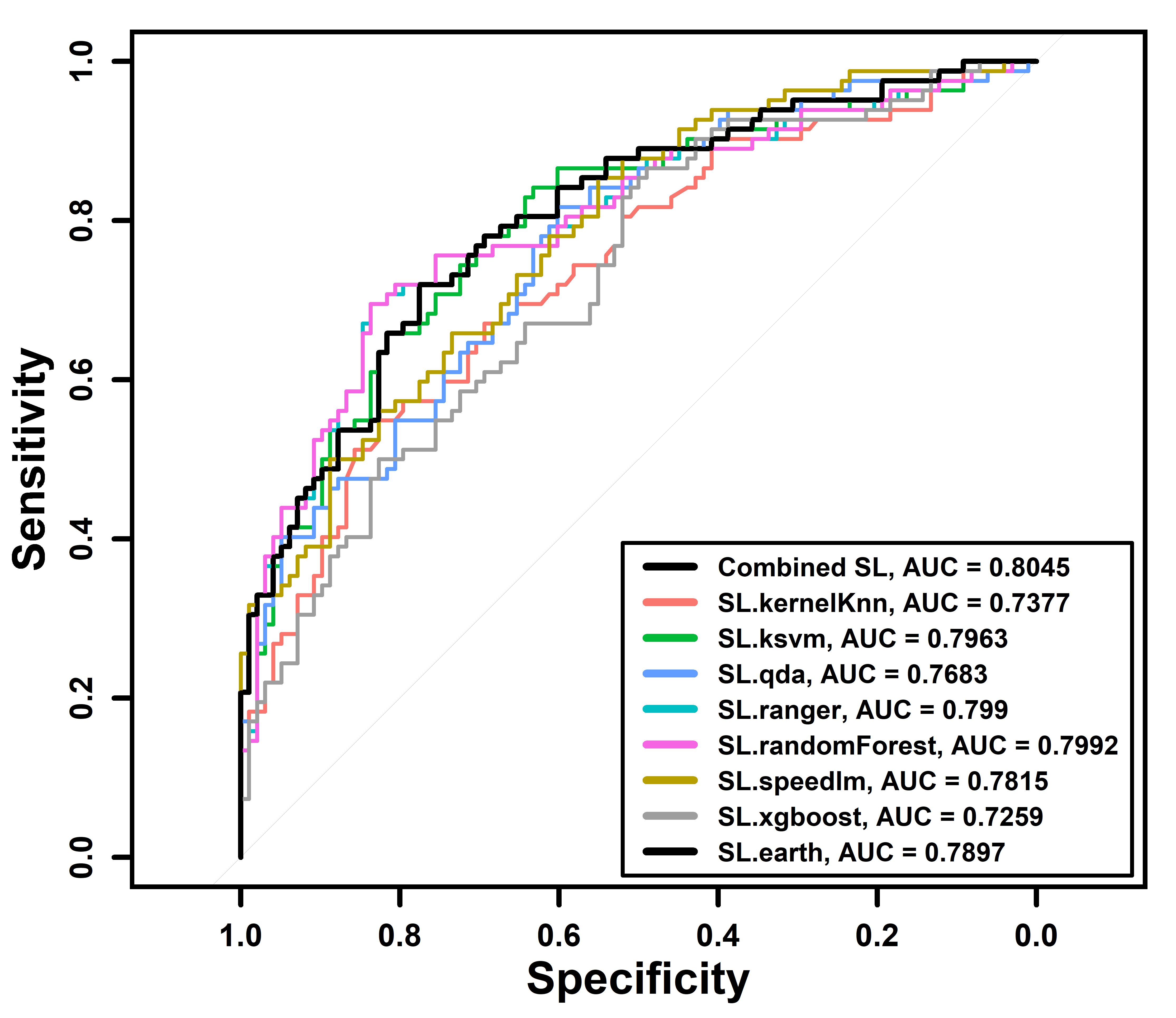}{0.33\textwidth}{}
              \fig{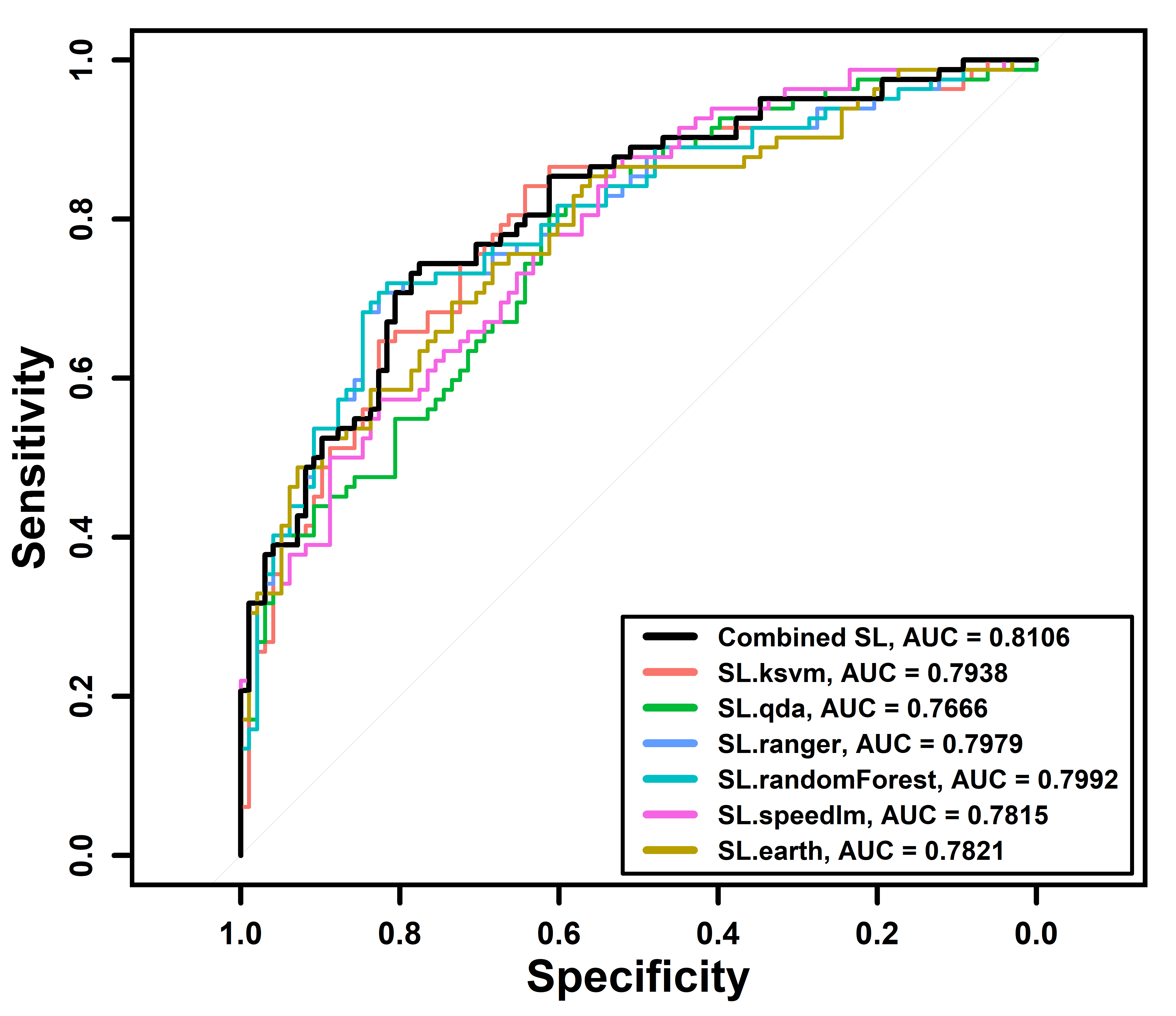}{0.33\textwidth}{}
              }
    \vspace{-30pt}
    \gridline{\fig{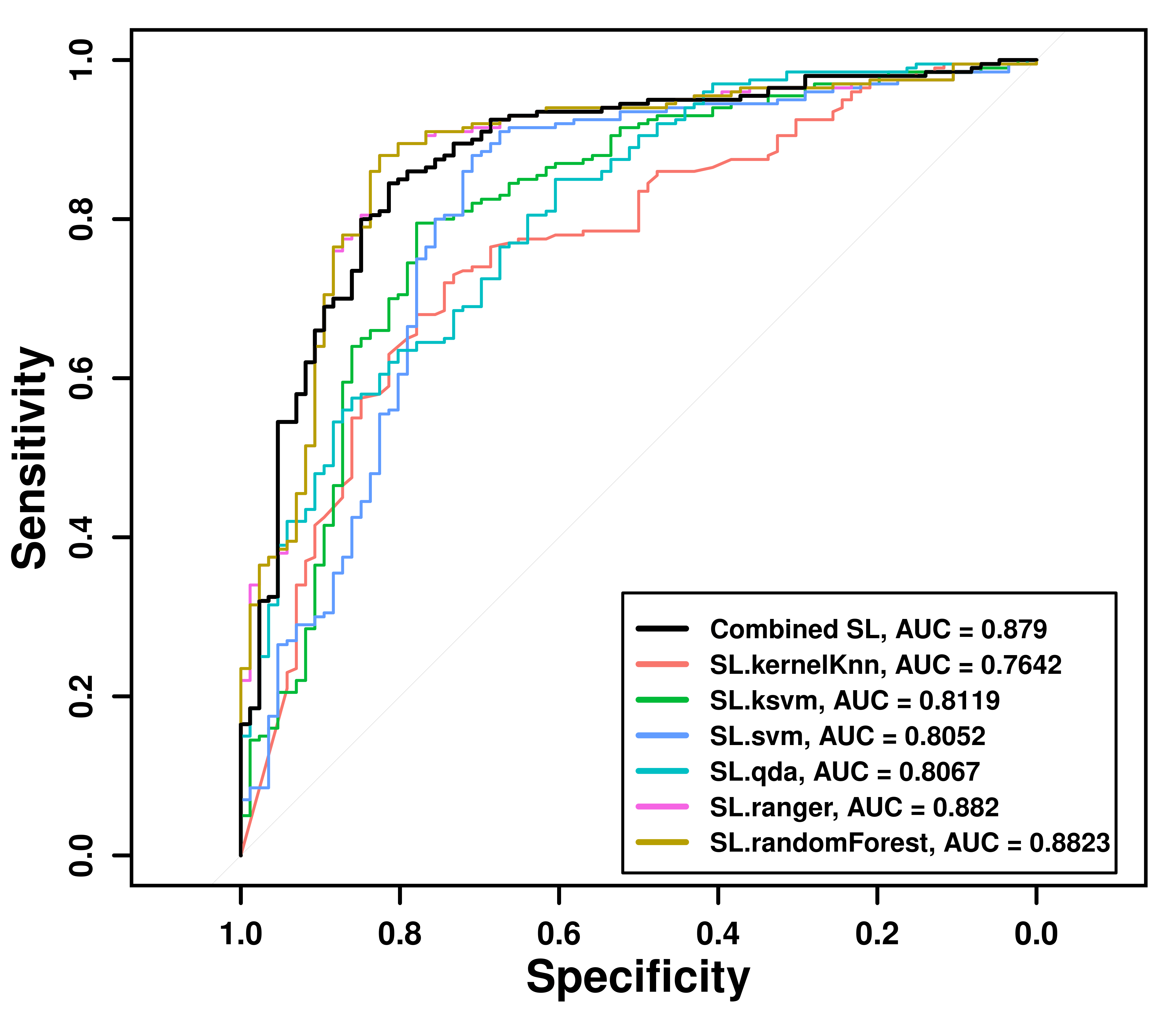}{0.33\textwidth}{}
              \fig{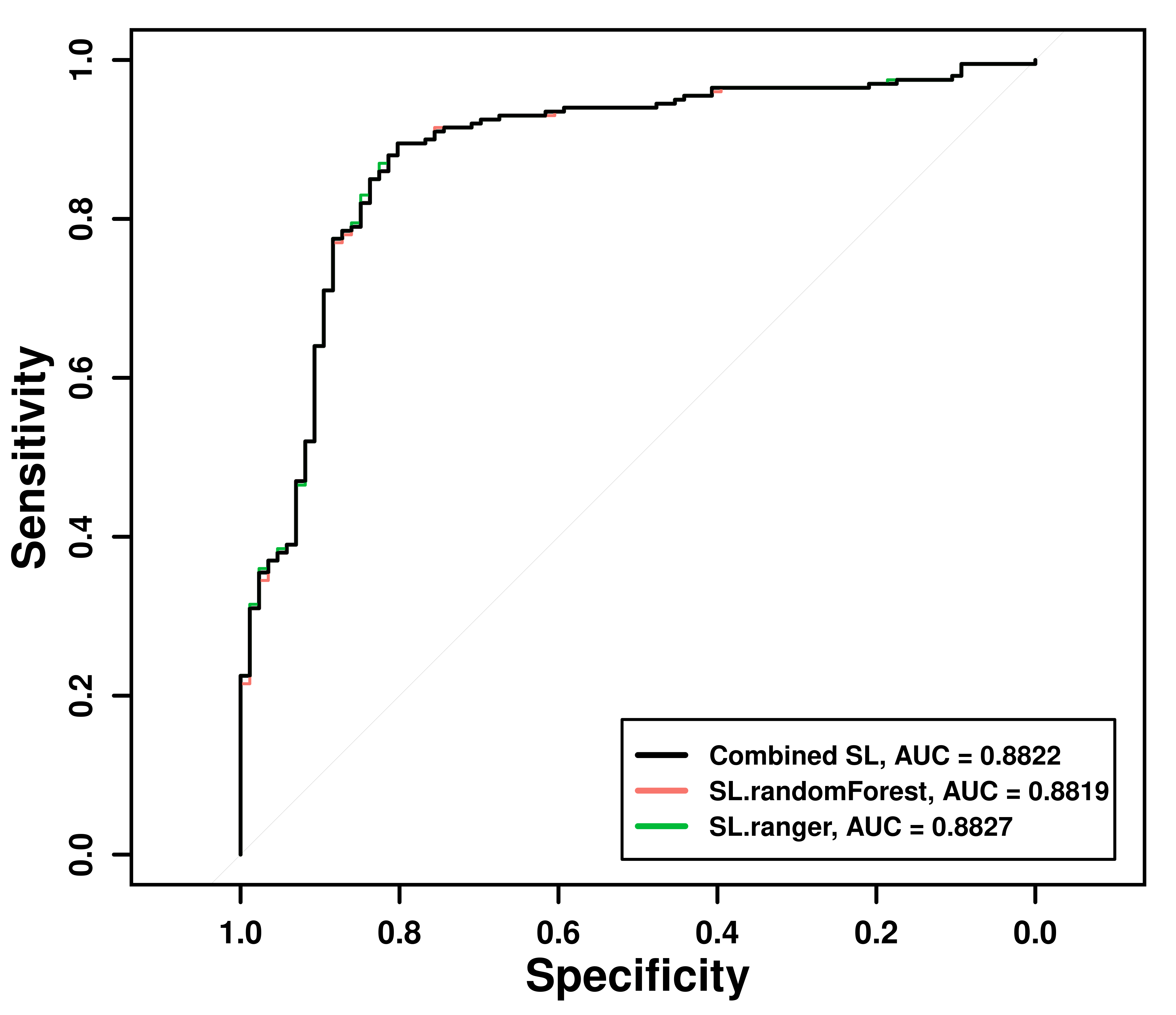}{0.33\textwidth}{}
              }
    \vspace{-25pt}
    \caption{
    For $z_{t}=2.0$. Top row: raw without M-estimator. Second row: raw with M-estimator. Third row: MICE-imputed. Bottom row: SMOTE balanced. First column: ROC-AUC curve with best algorithms chosen by SuperLearner with weights $>$ 0.05 in each dataset. Top right: ROC-AUC curve without `caret.rpart'. Second right: ROC-AUC curve without `caret.rpart'. Third right: ROC-AUC curve without `kernelKnn' and `xgboost'. Bottom right: ROC-AUC curve without `kernelKnn', `ksvm', `svm', and `qda'.}
    \label{fig:combined_raw_z2.0}
\end{figure*}

\begin{figure*}[htbp]
\vspace{-30pt}
\gridline{\fig{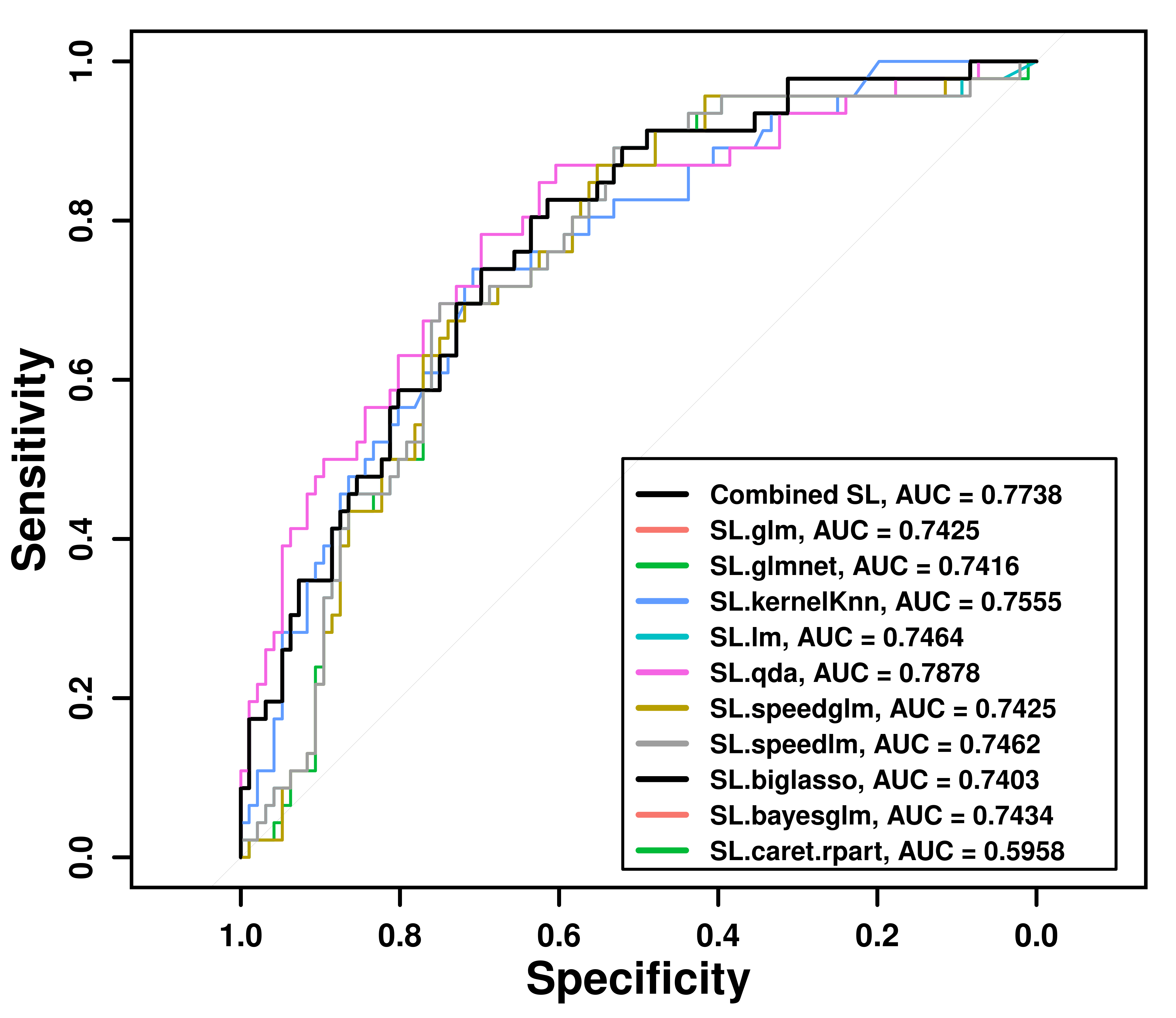}{0.33\textwidth}{}
          \fig{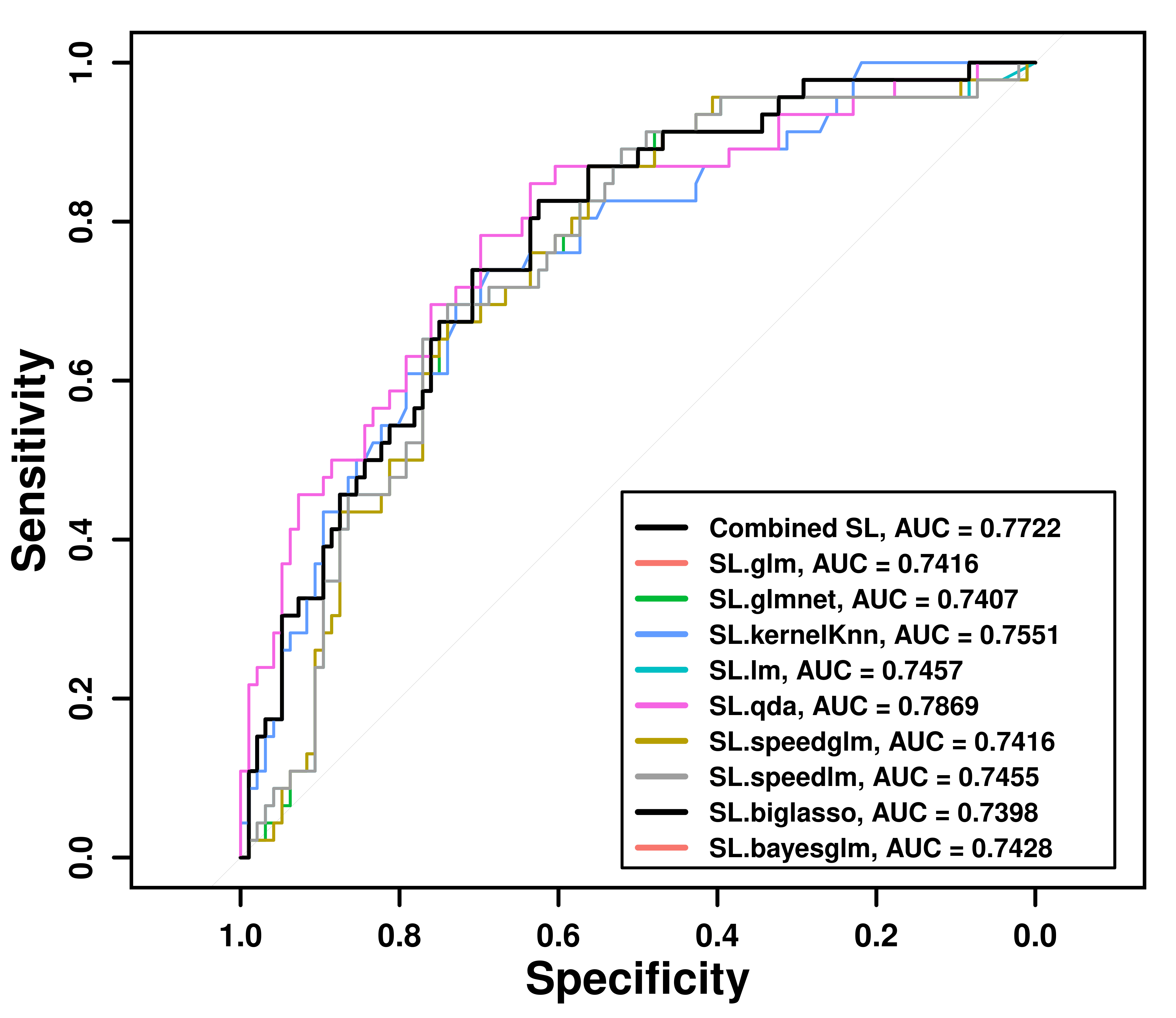}{0.33\textwidth}{}
         }
\vspace{-30pt}
\gridline{\fig{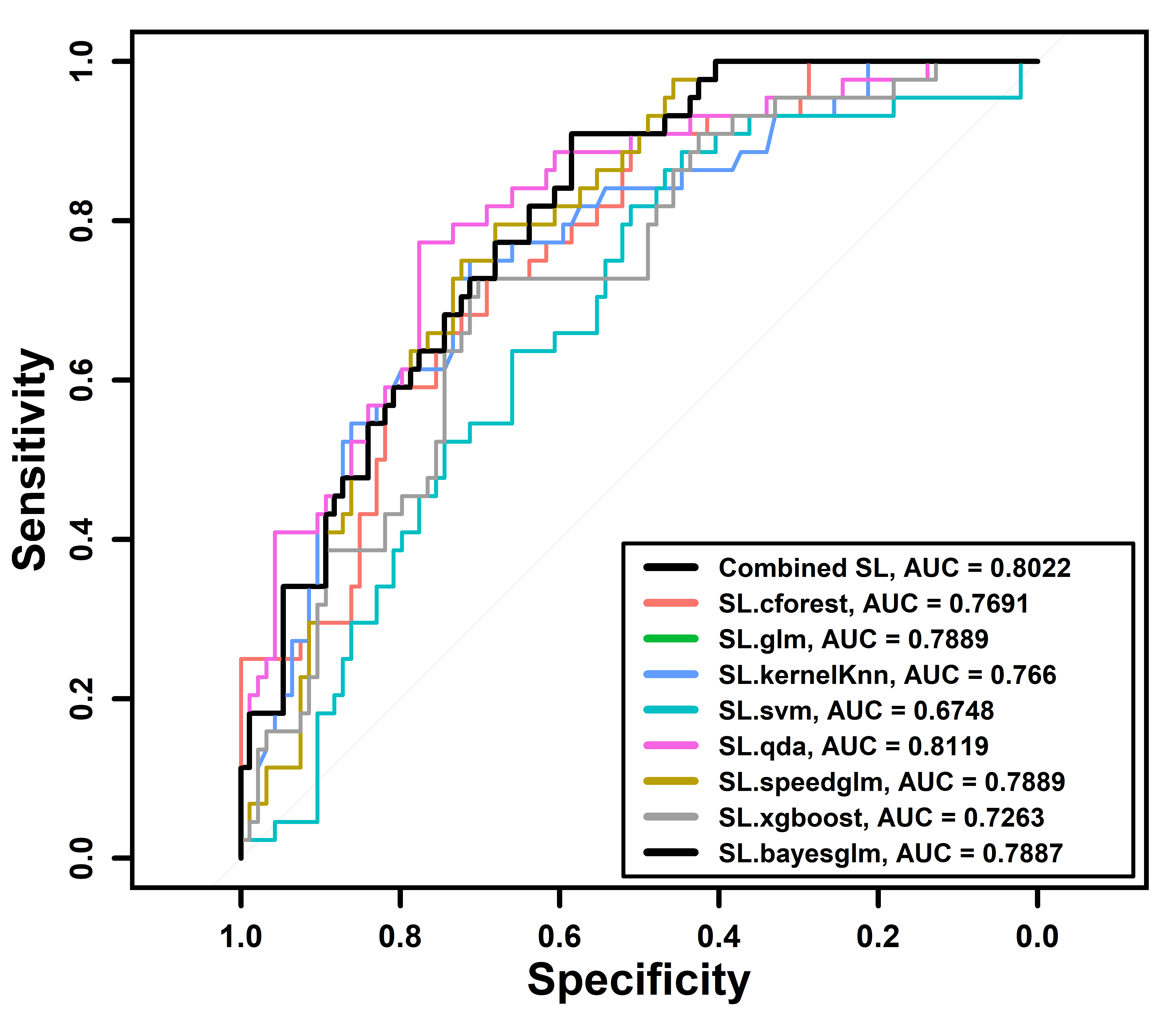}{0.33\textwidth}{}
          \fig{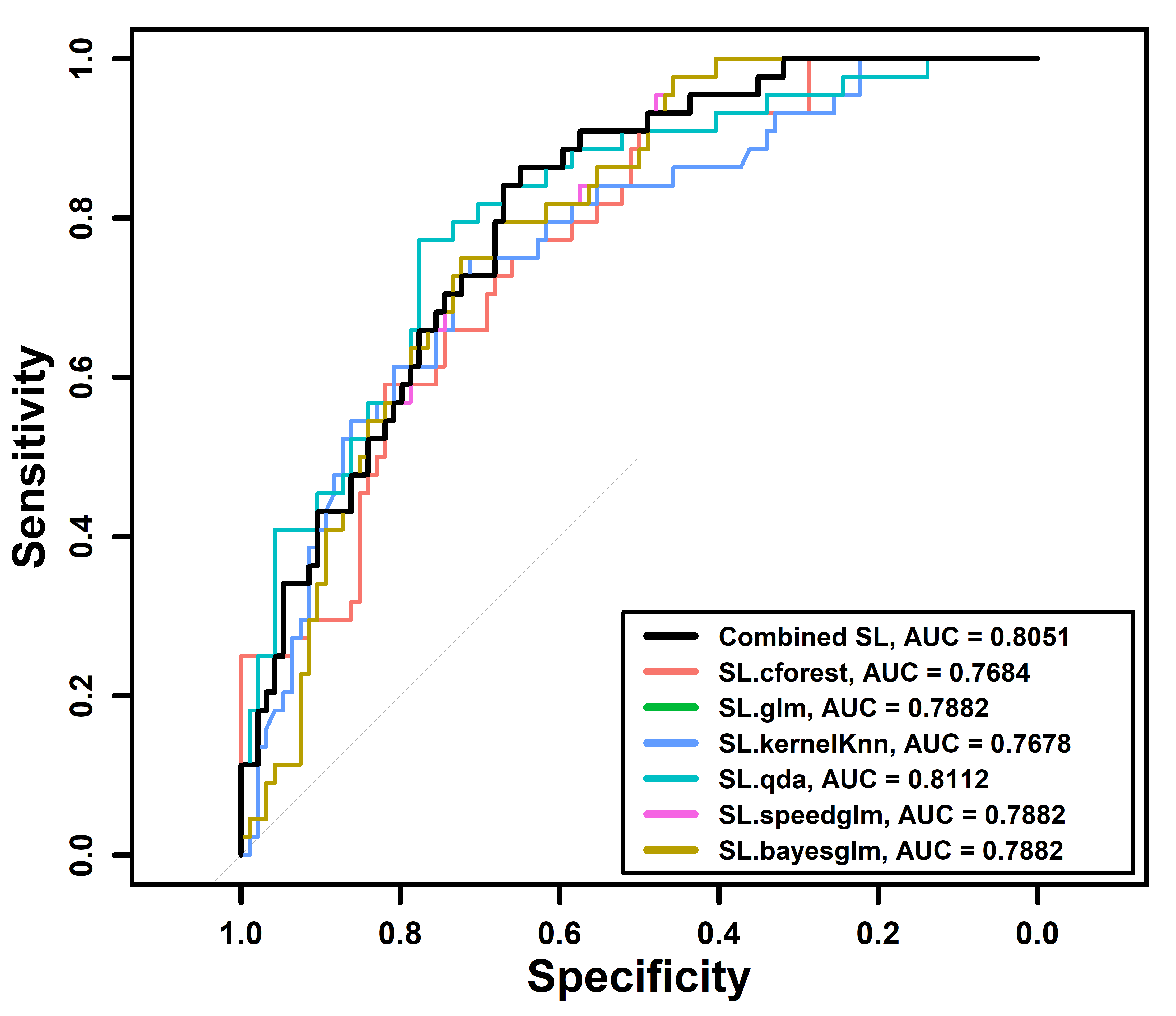}{0.33\textwidth}{}
         }
\vspace{-30pt}

\gridline{\fig{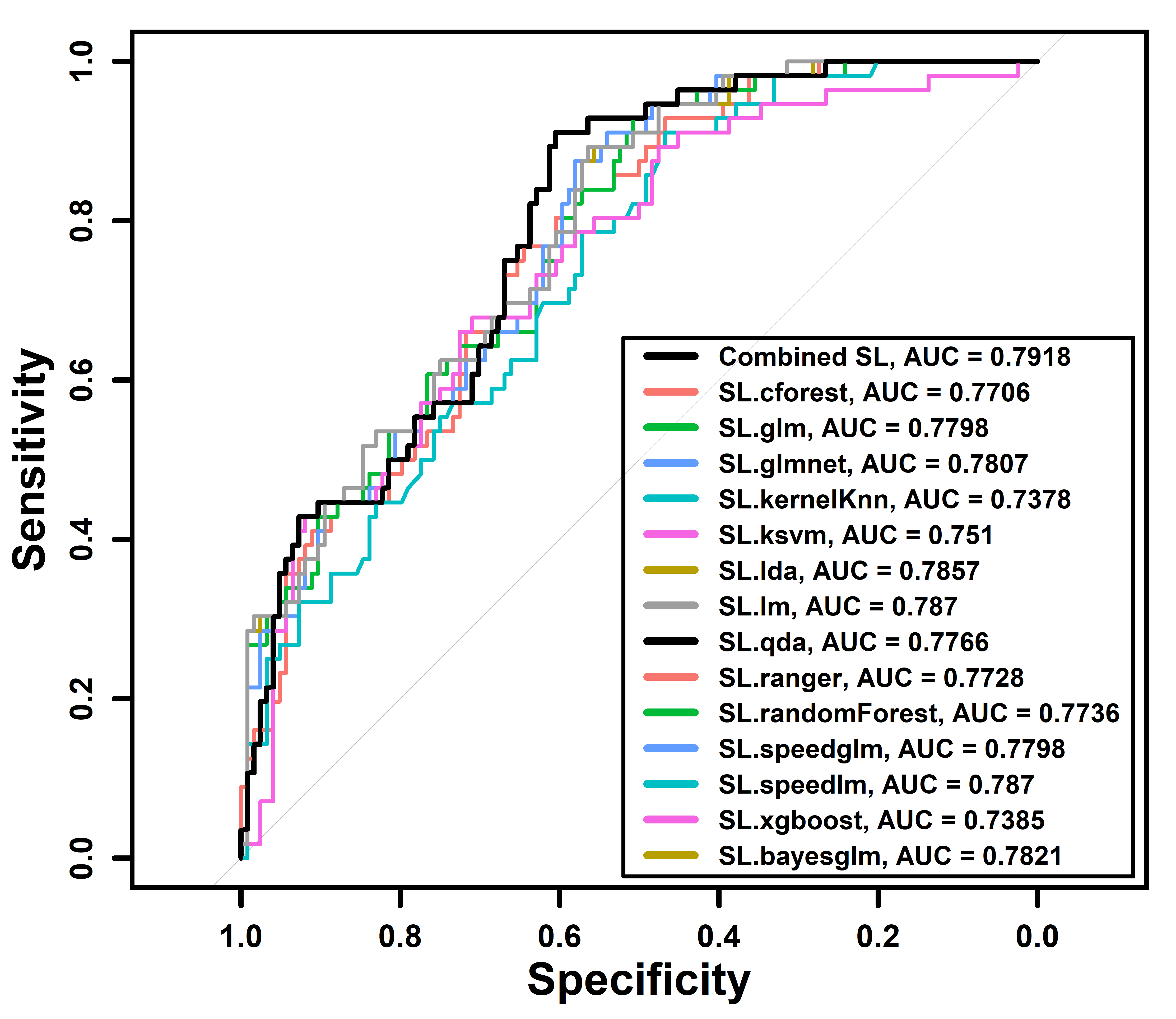}{0.33\textwidth}{}
          \fig{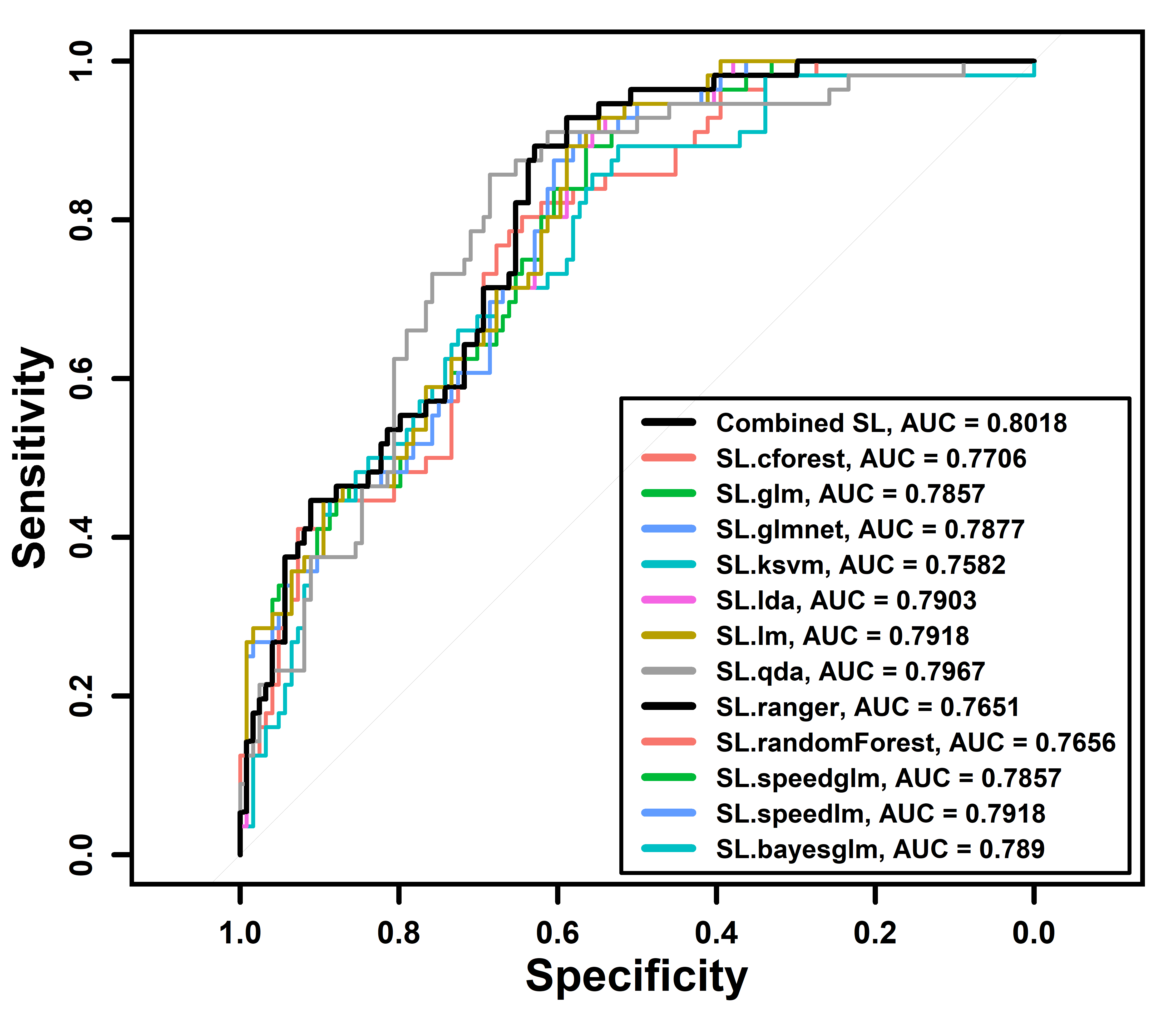}{0.33\textwidth}{}
         }
\vspace{-30pt}
\gridline{\fig{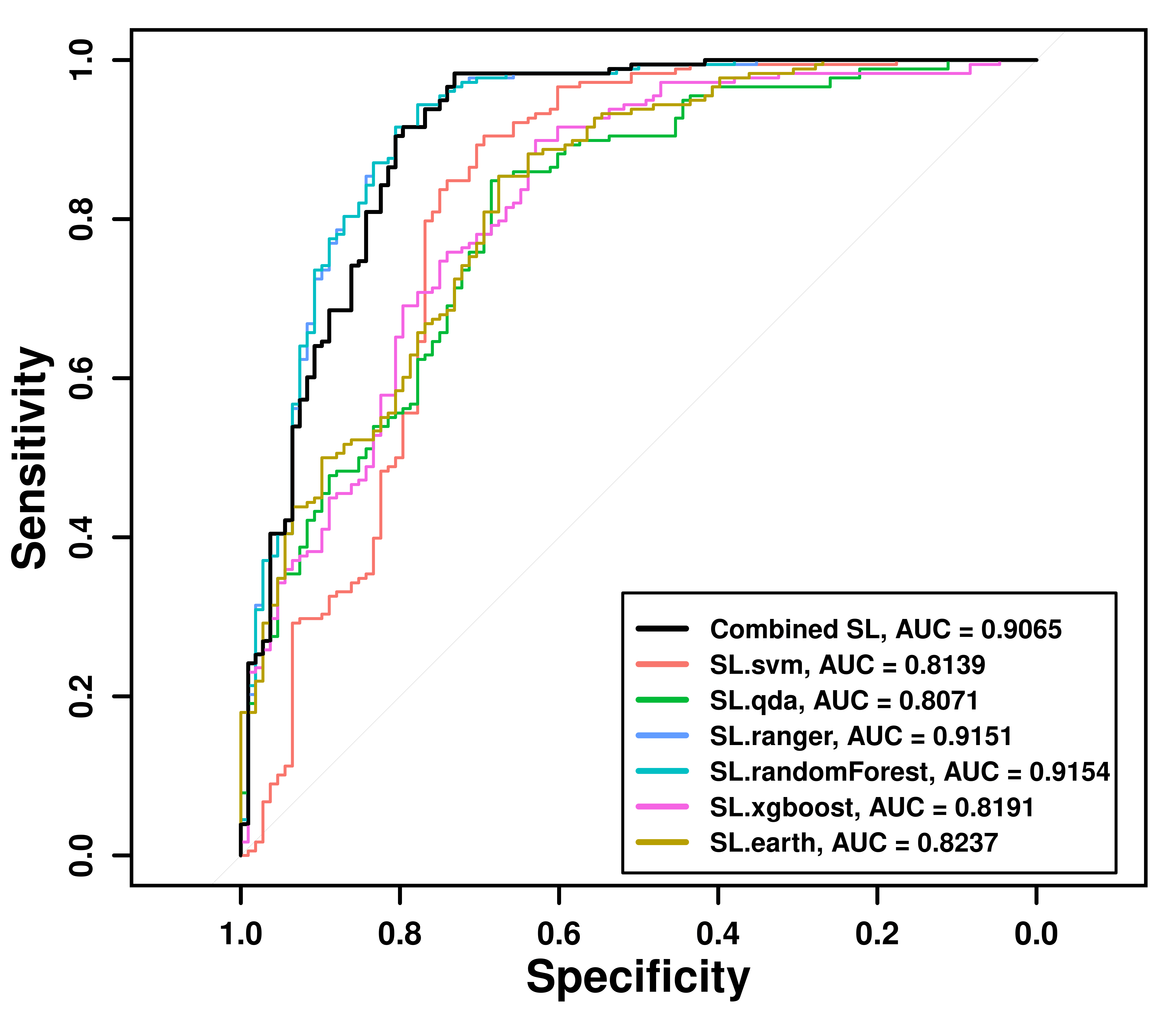}{0.35\textwidth}{}
          \fig{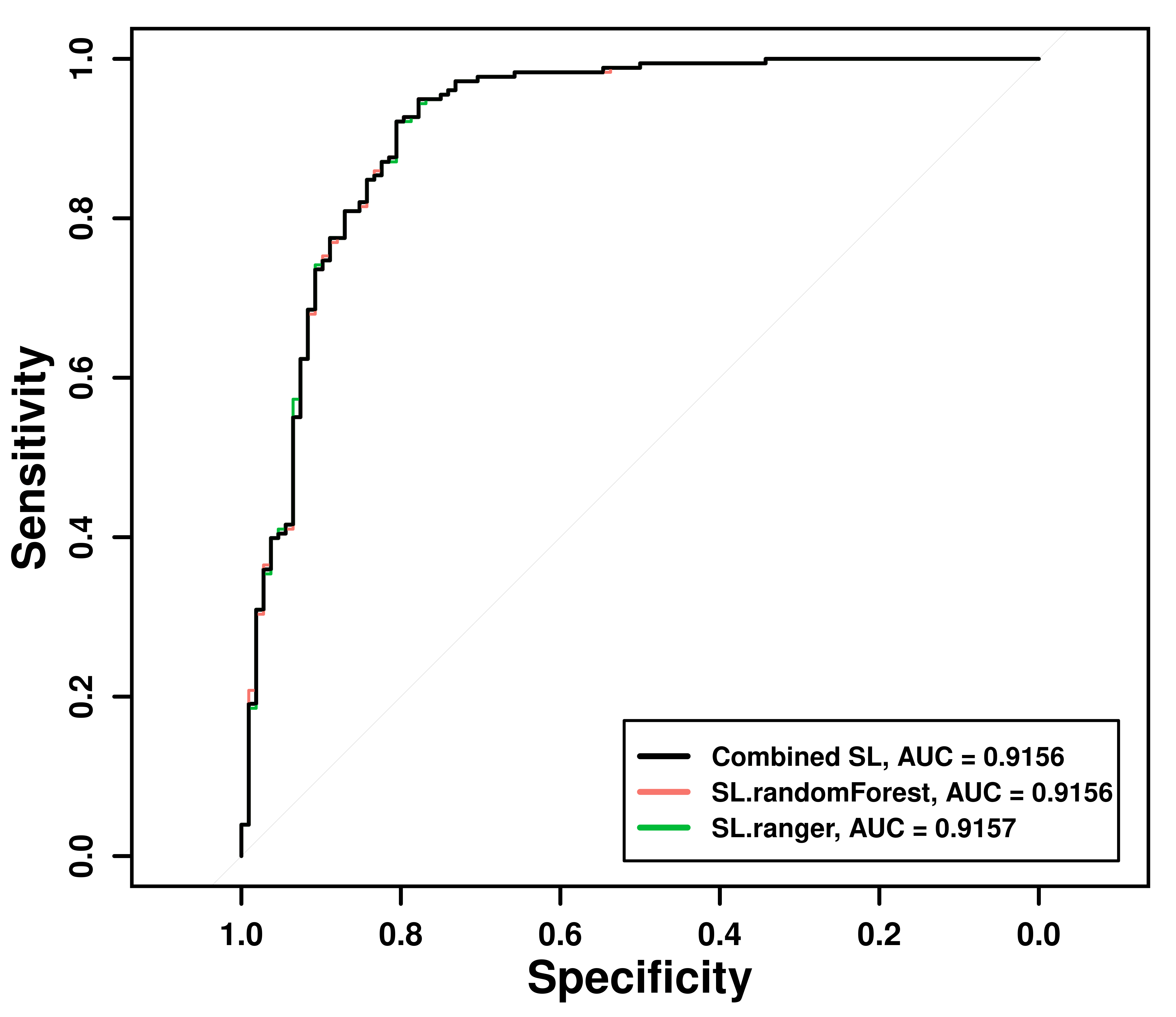}{0.35\textwidth}{}
         }

\vspace{-15pt}
\caption{
For $z_{t}=2.5$. Top row: raw without M-estimator. Second row: raw with M-estimator. Third row: MICE-imputed. Bottom row: SMOTE balanced. First column: ROC-AUC curve with best algorithms chosen by SuperLearner with weights $>$ 0.05 in each dataset. Top right: ROC-AUC curve without `caret.rpart'. Second right: ROC-AUC curve without `svm' and 'xgboost'. Third right: ROC-AUC curve without `kernelKnn' and `xgboost'. Bottom right: ROC-AUC curve without `svm', `qda', `xgboost', and `earth'.
}
\label{fig:combined_raw_z2.5}
\end{figure*}

\begin{figure*}[htbp]
\gridline{\fig{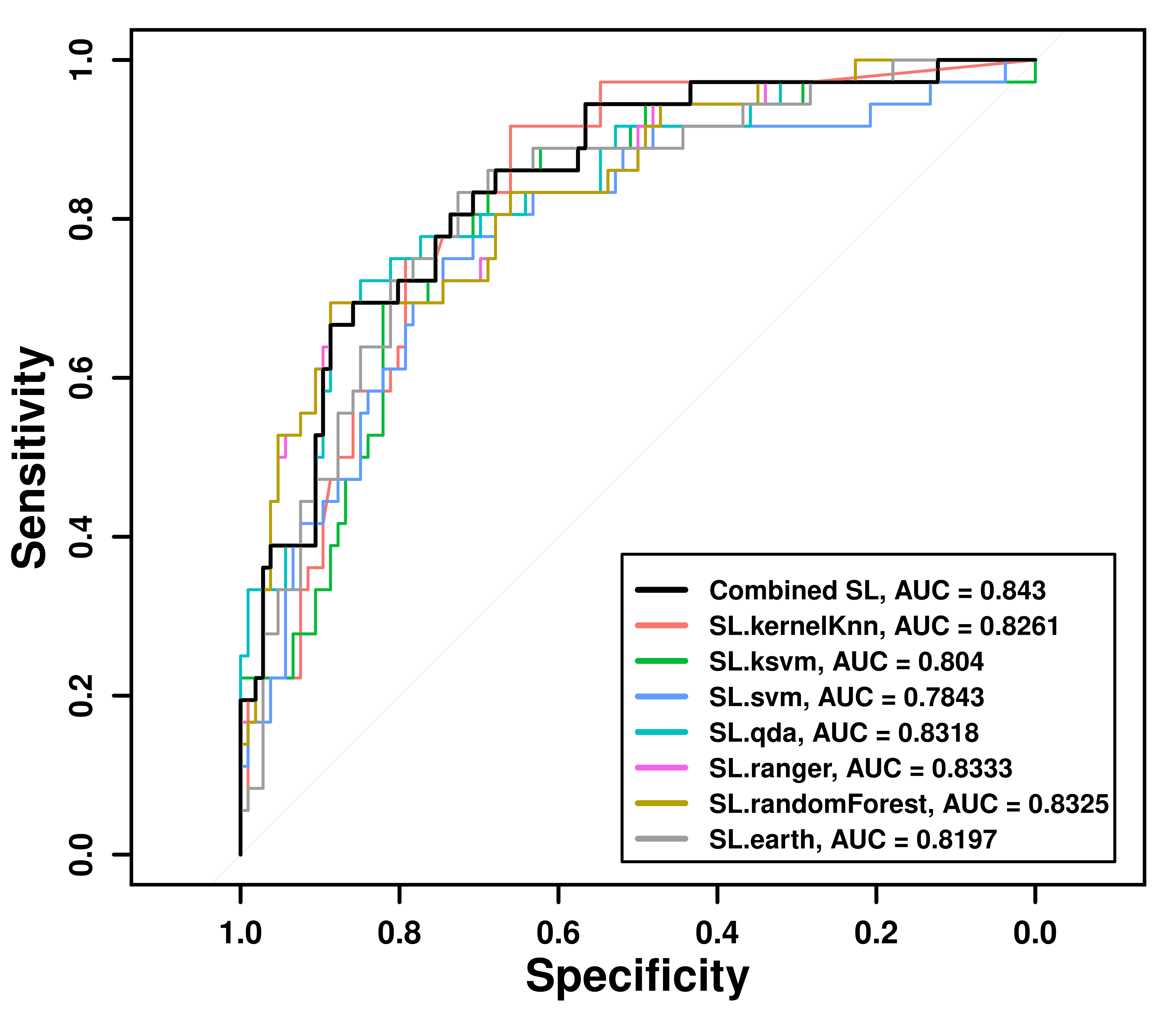}{0.33\textwidth}{}
          \fig{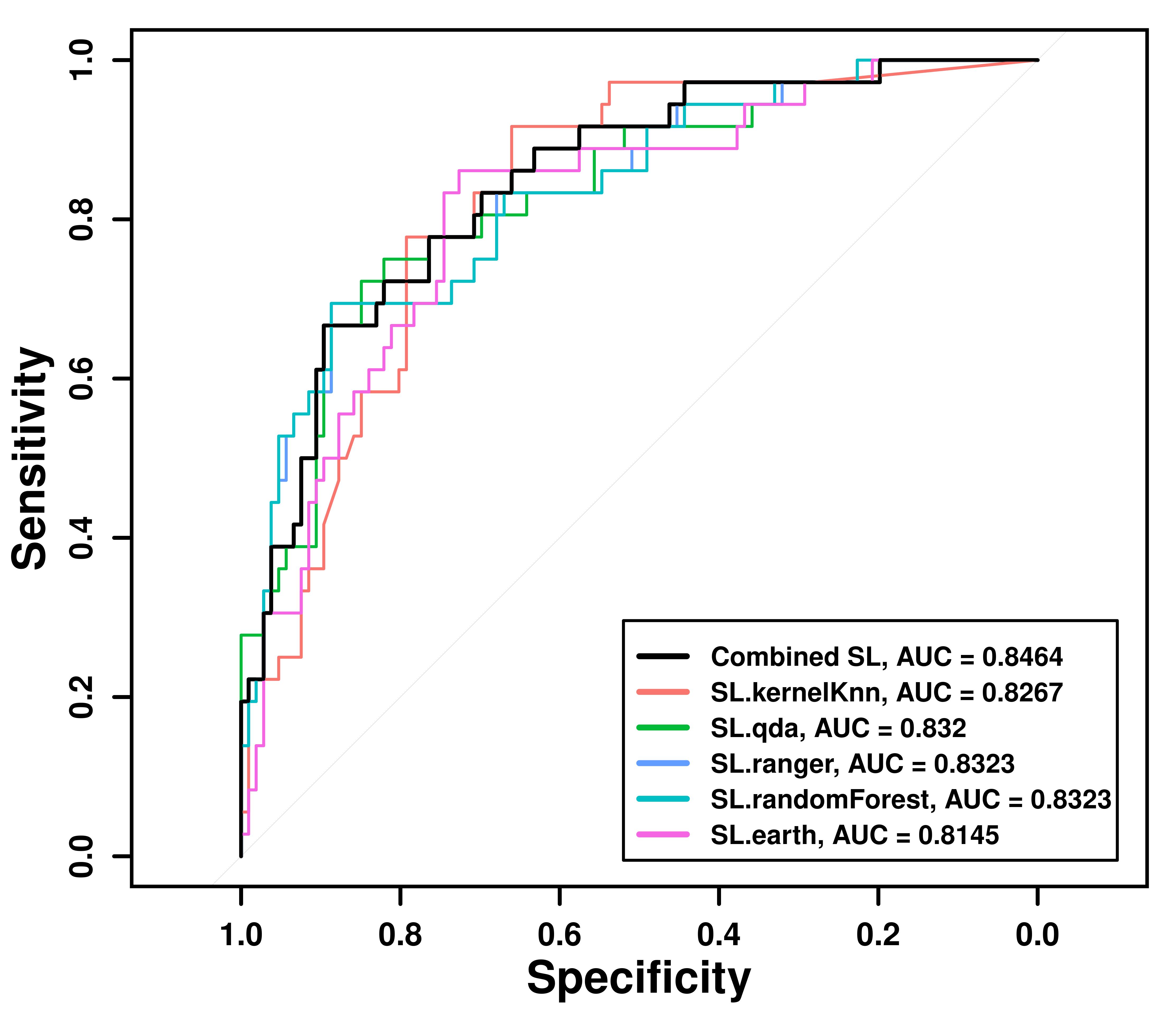}{0.33\textwidth}{}
         }
\vspace{-30pt}
\gridline{\fig{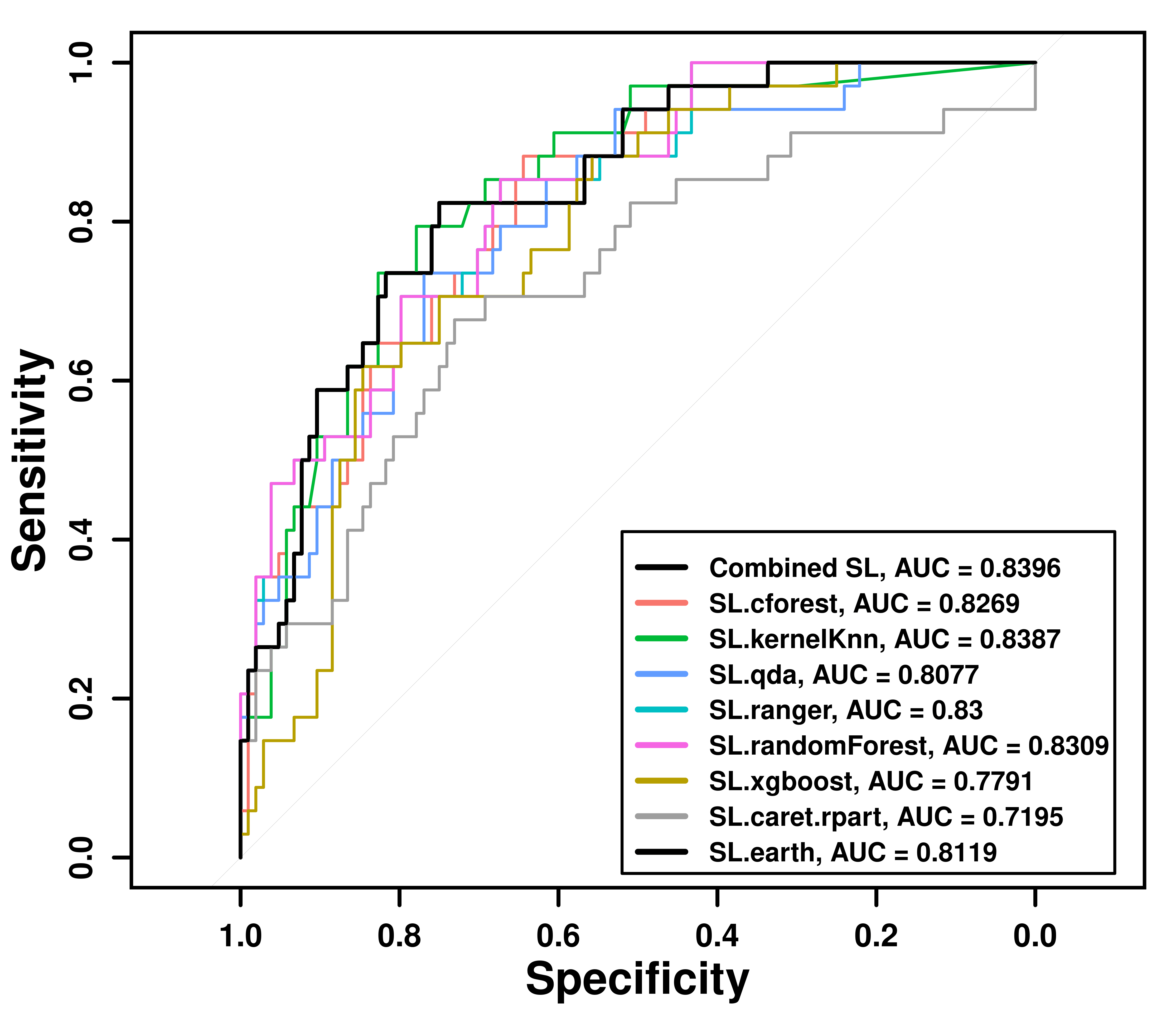}{0.33\textwidth}{}
          \fig{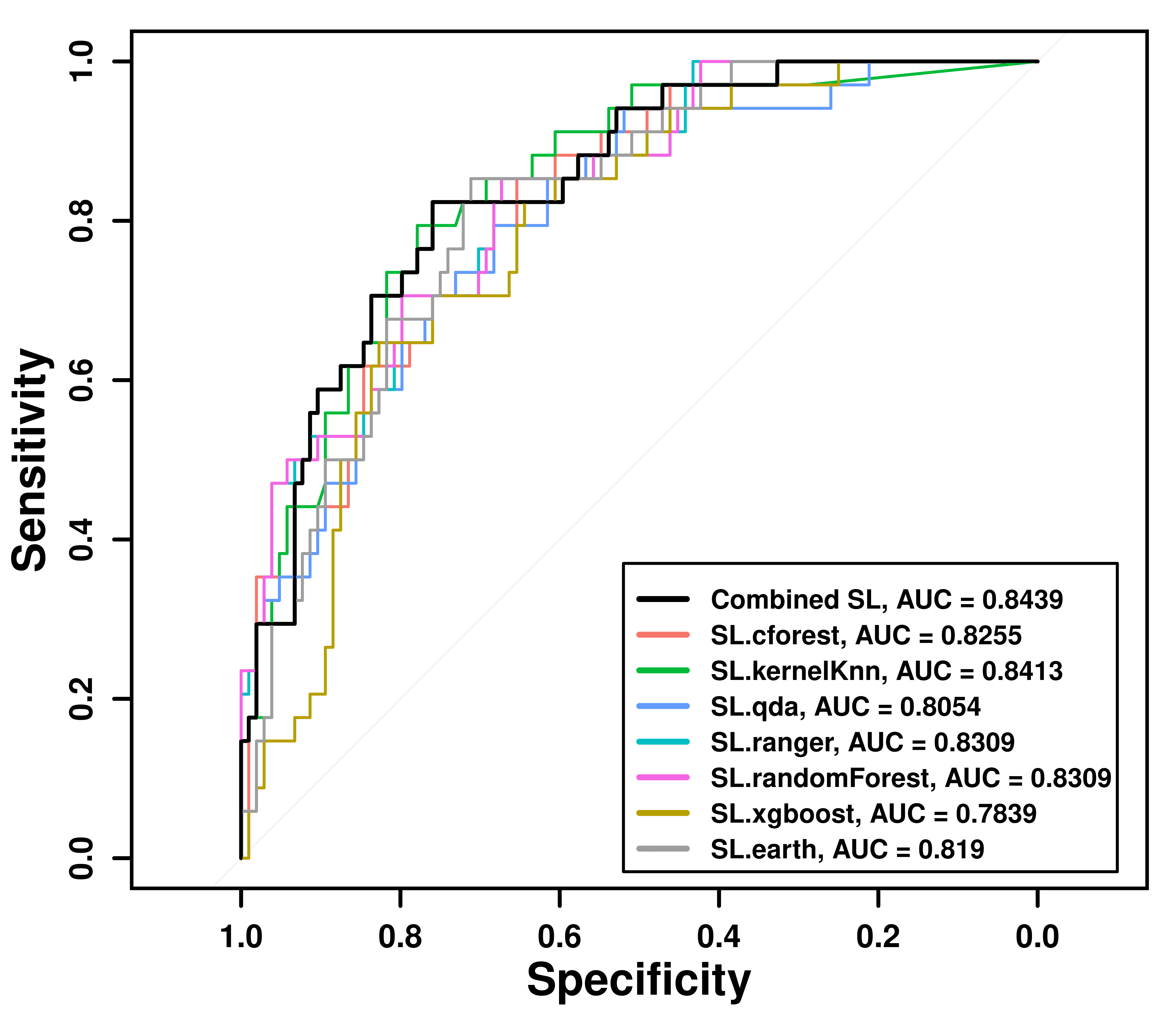}{0.33\textwidth}{}
         }

\vspace{-30pt}
\gridline{\fig{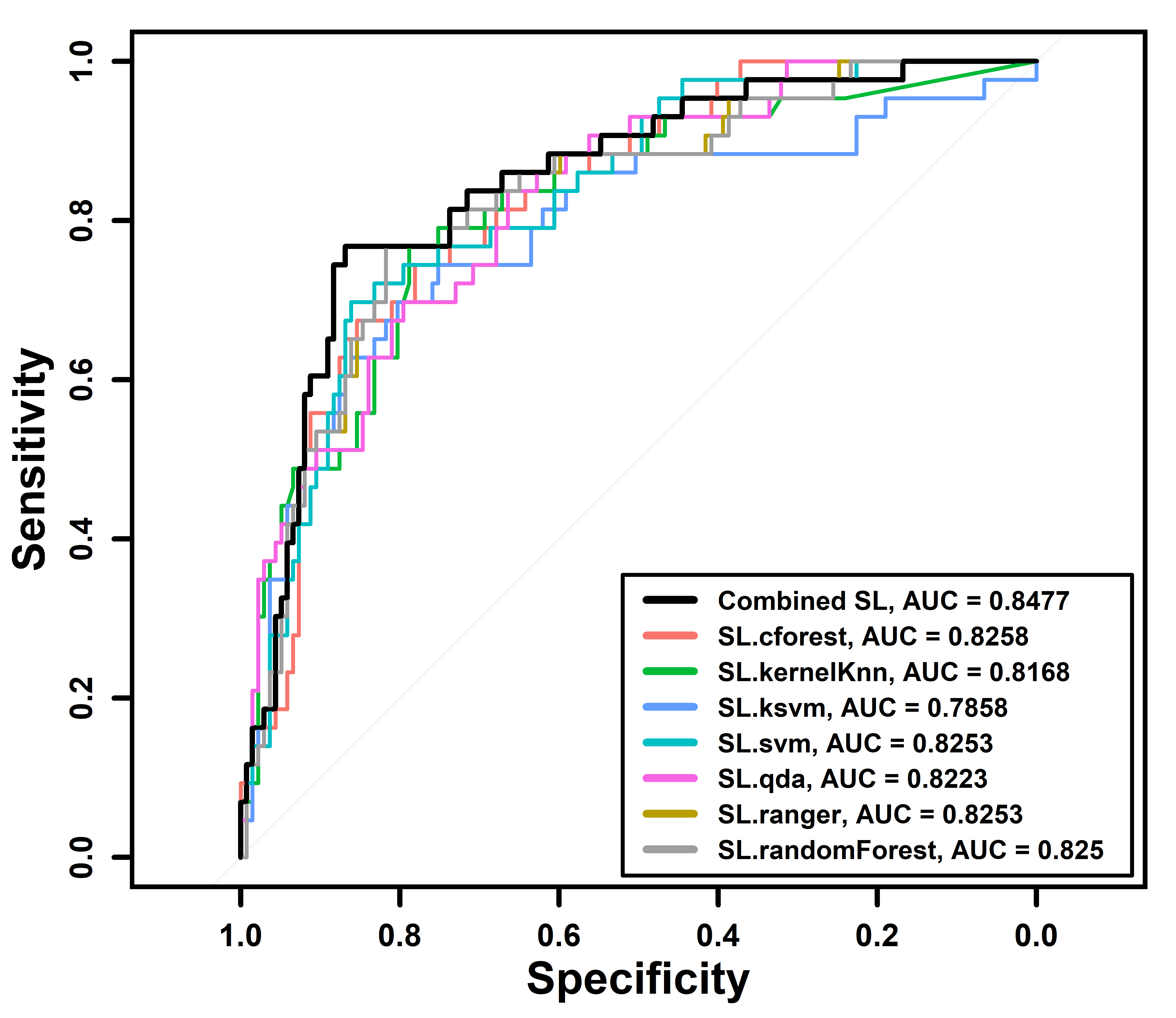}{0.33\textwidth}{}
          \fig{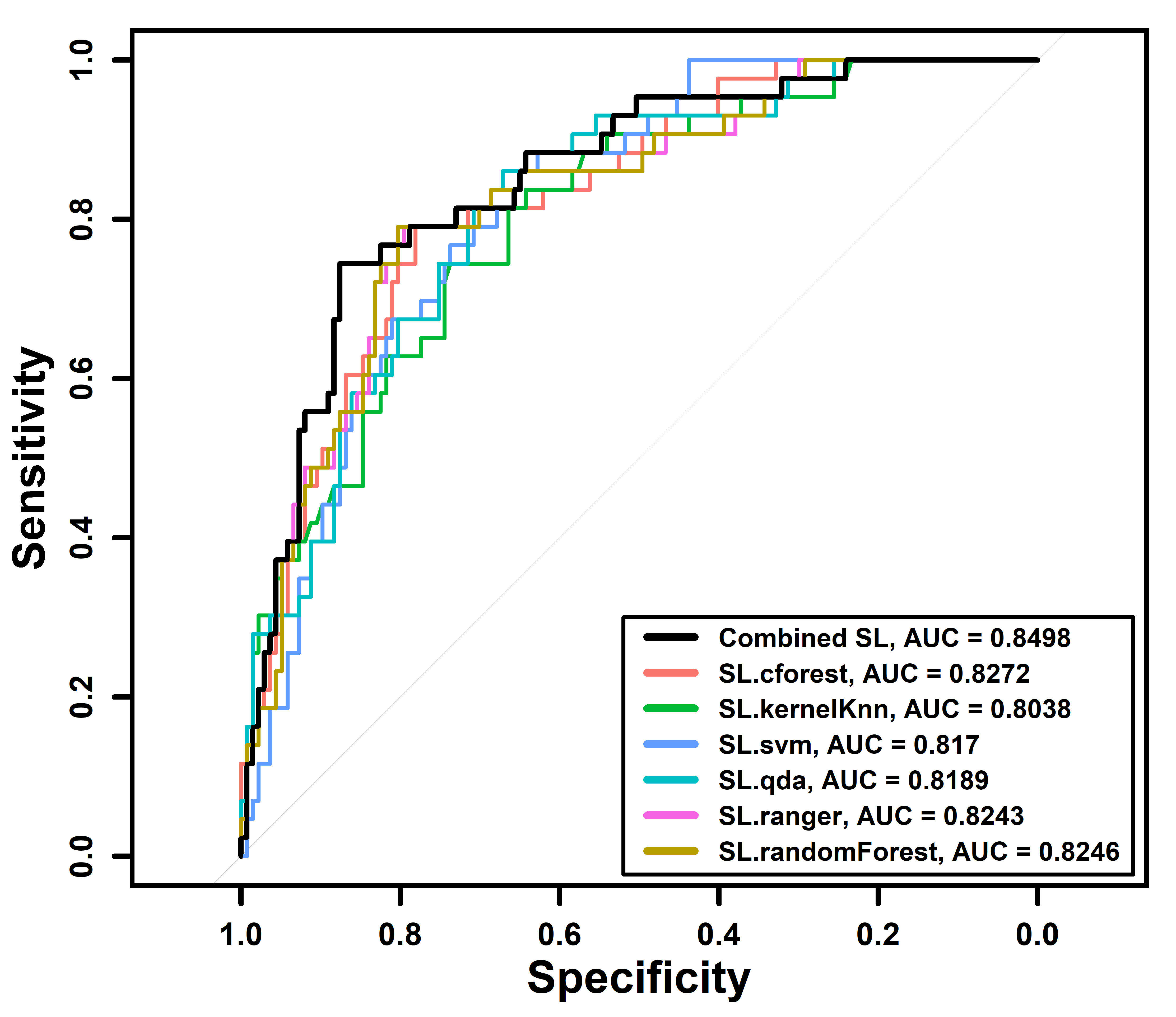}{0.33\textwidth}{}
         }
\vspace{-30pt}
\gridline{\fig{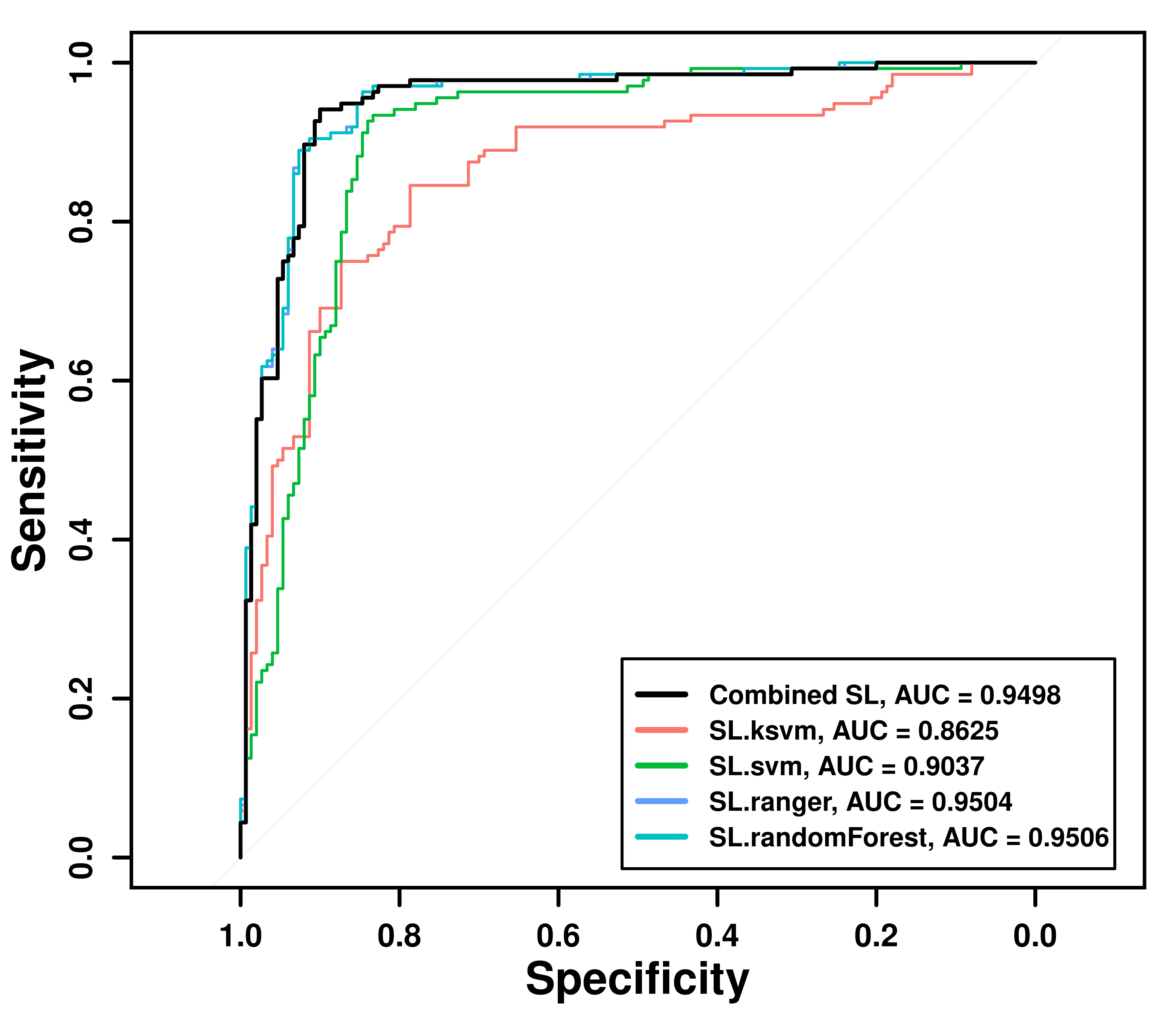}{0.33\textwidth}{}
          \fig{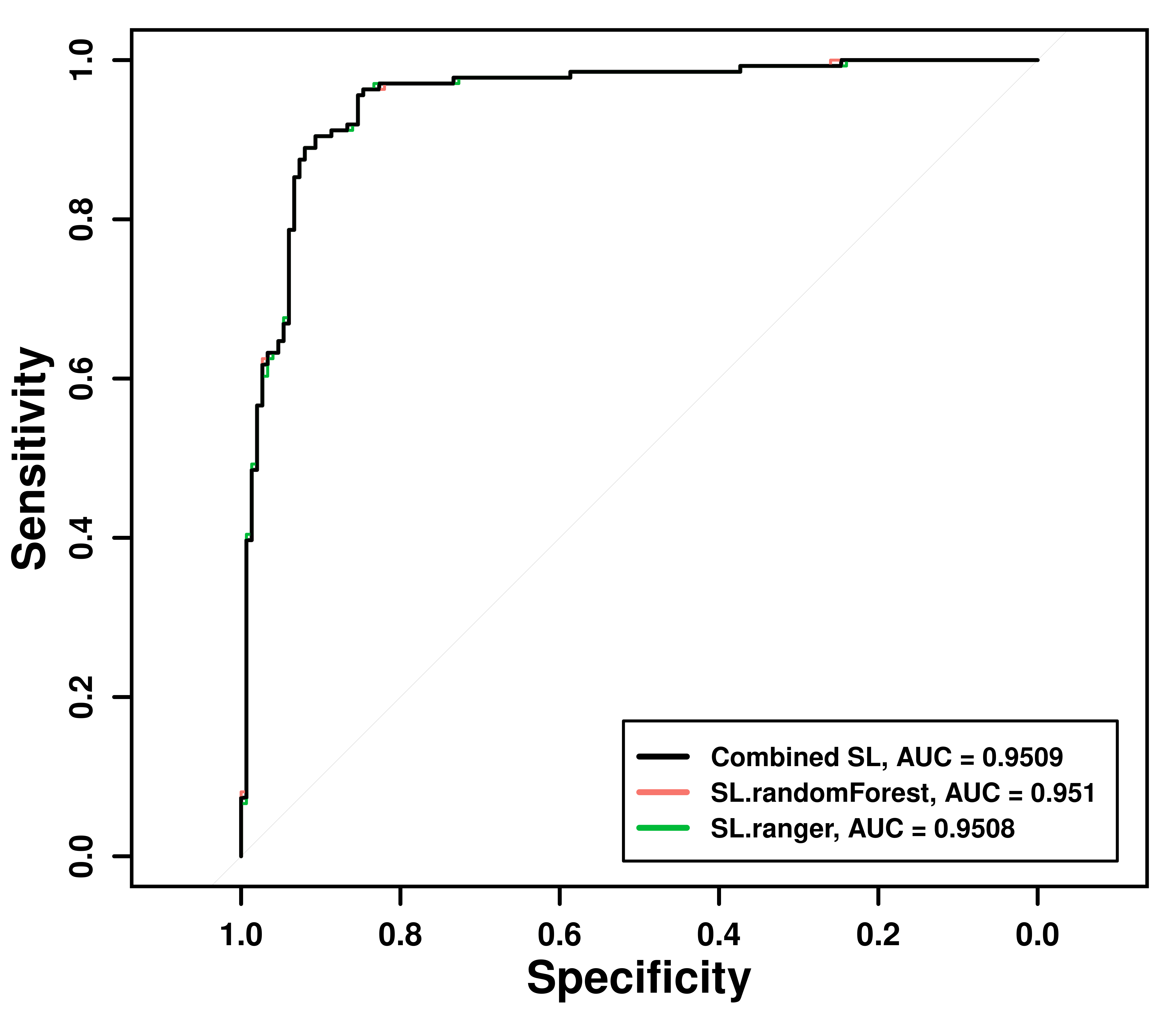}{0.33\textwidth}{}
         }

\vspace{-10pt}
\caption{
For $z_{t}=3.0$. Top row: raw without M-estimator. Second row: raw with M-estimator. Third row: MICE-imputed. Bottom row: SMOTE balanced. First column: ROC-AUC curve with best algorithms chosen by SuperLearner with weights $>$ 0.05 in each dataset. Top right: ROC-AUC curve without `ksvm' and 'svm'. Second right: ROC-AUC curve without `caret.rpart'. Third right: ROC-AUC curve without `ksvm'. Bottom right: ROC-AUC curve without `ksvm' and `svm'.
}
\label{fig:combined_raw_z3.0}
\end{figure*}

\begin{figure*}[htbp]
\gridline{\fig{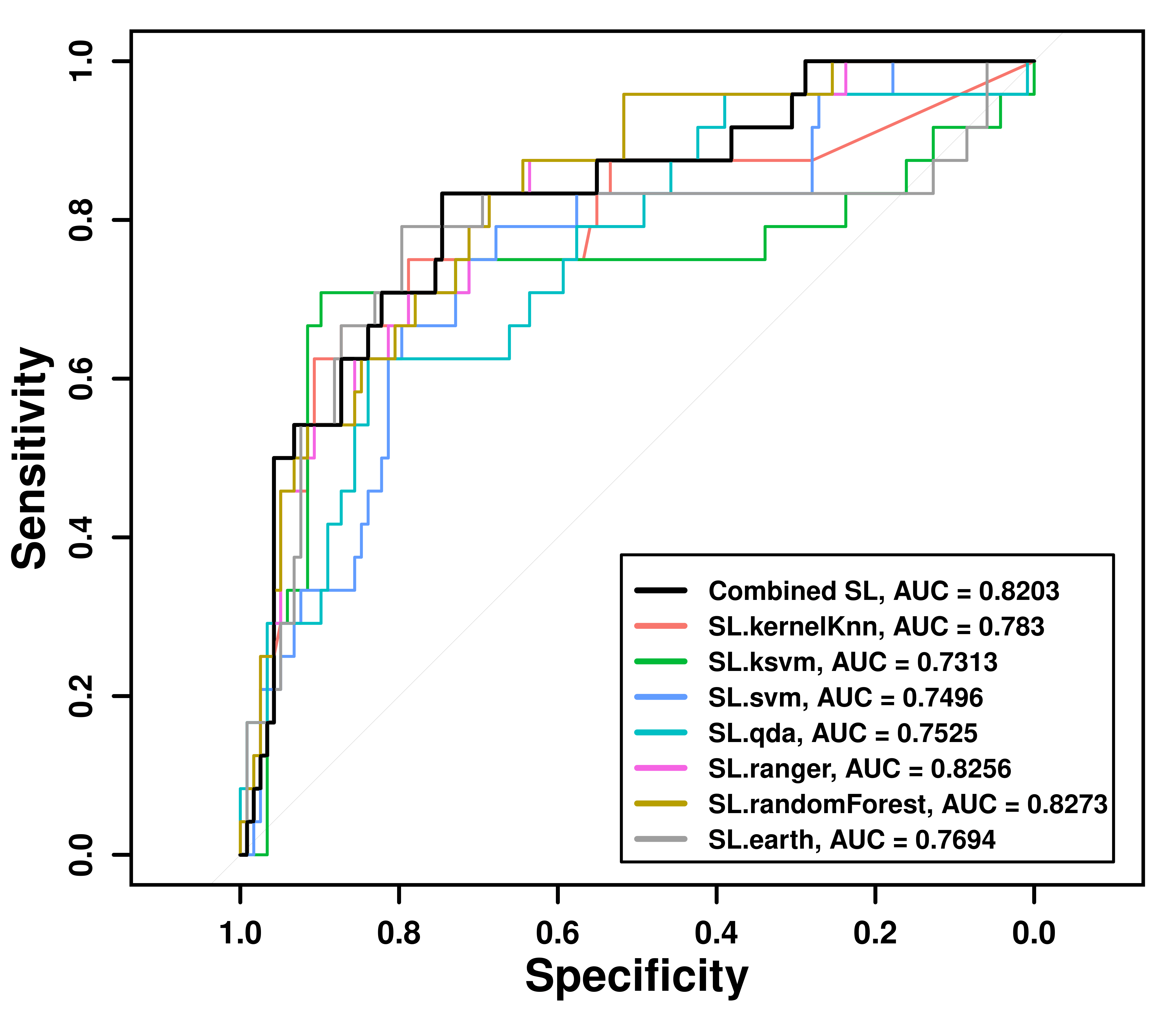}{0.33\textwidth}{}
          \fig{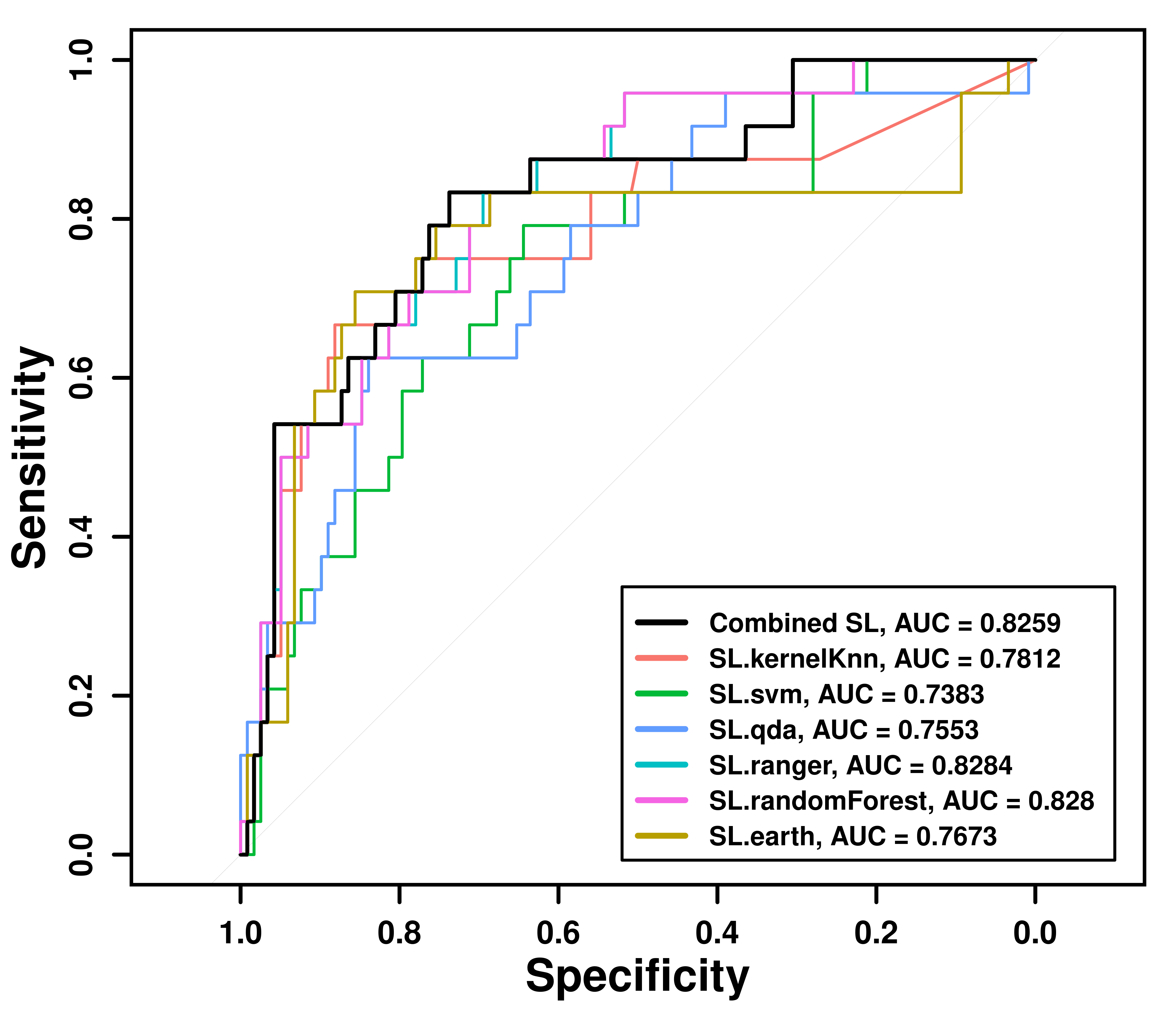}{0.33\textwidth}{}
         }
\vspace{-30pt}
\gridline{\fig{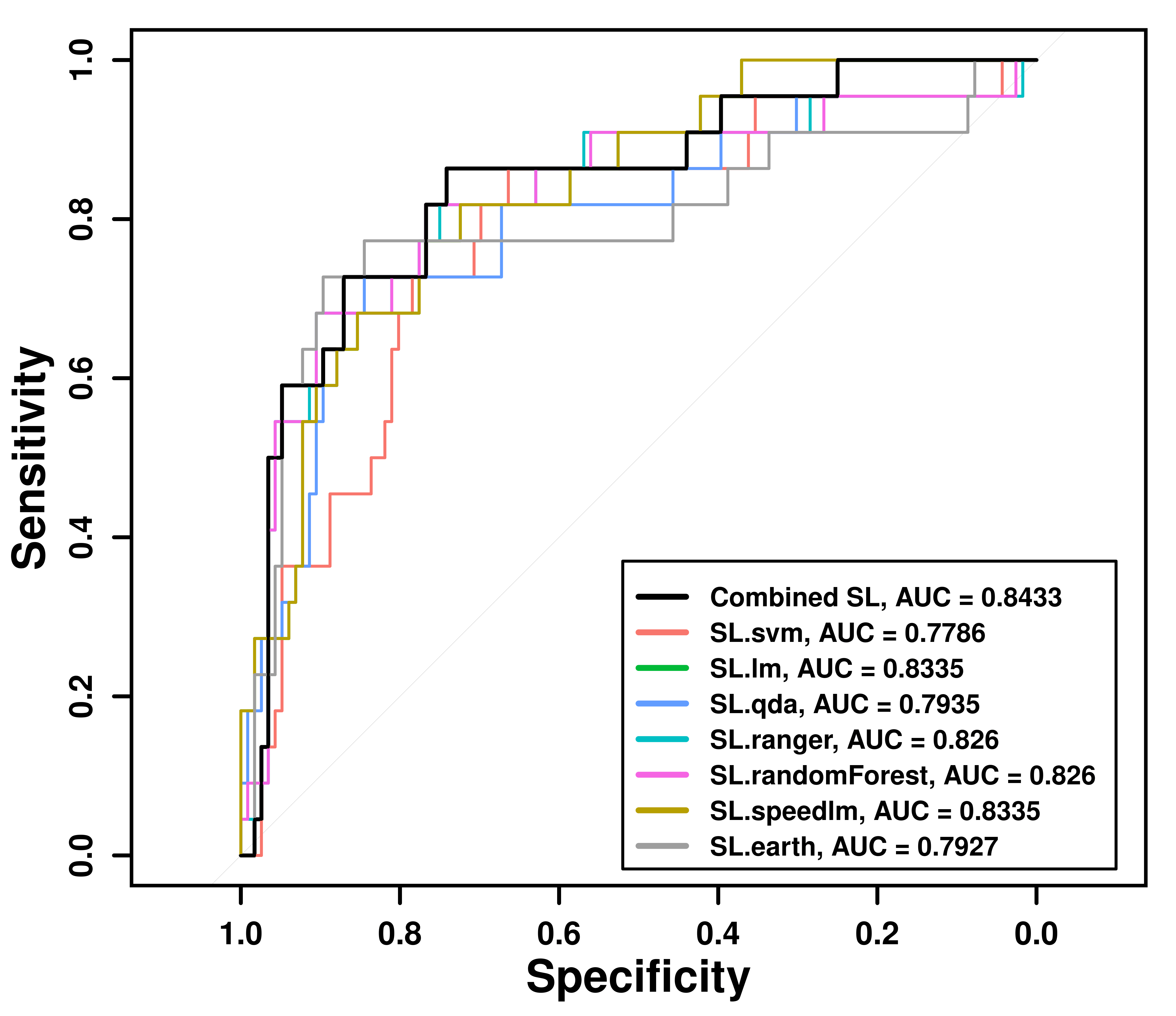}{0.33\textwidth}{}
          \fig{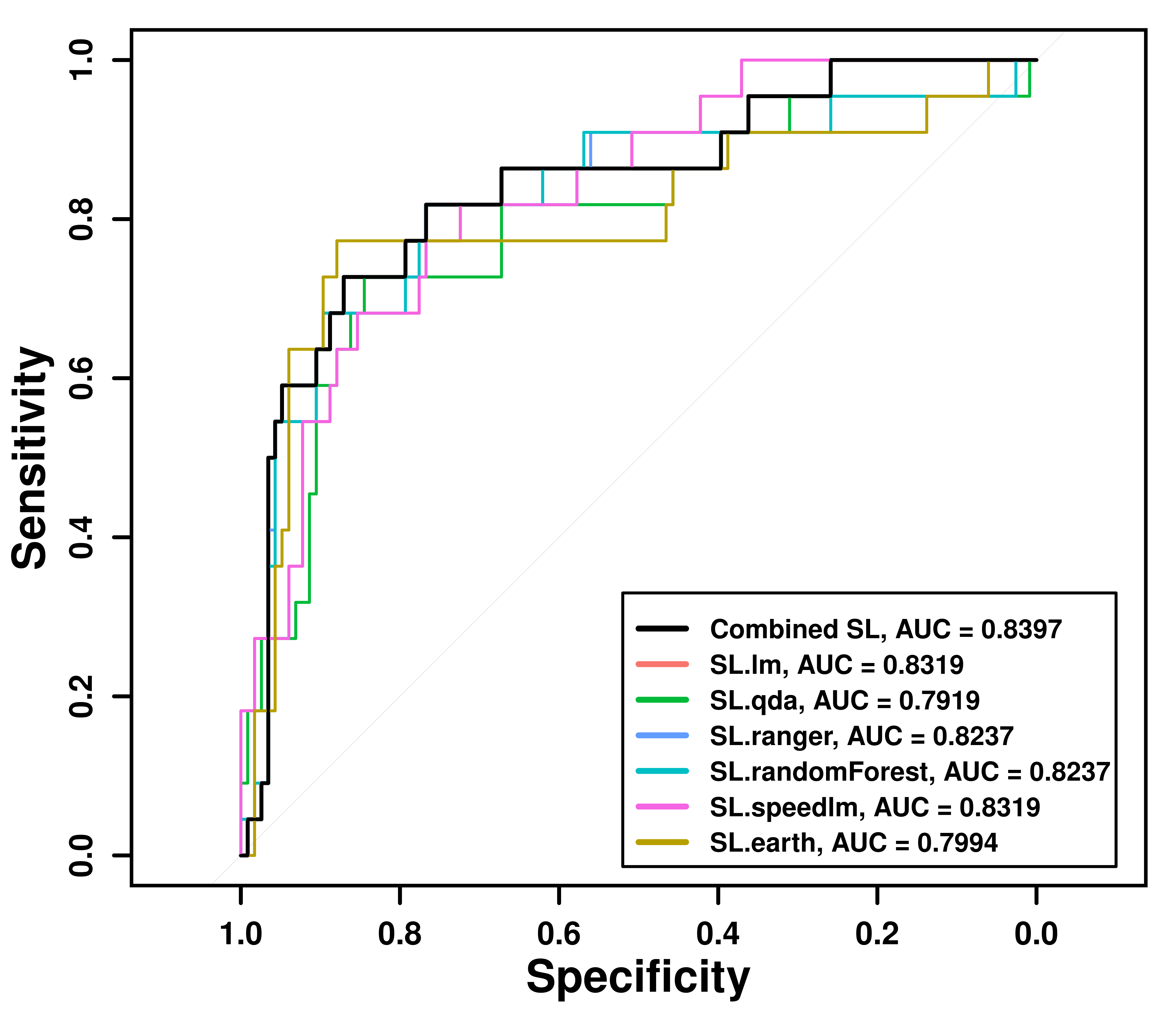}{0.33\textwidth}{}
         }

\vspace{-30pt}
\gridline{\fig{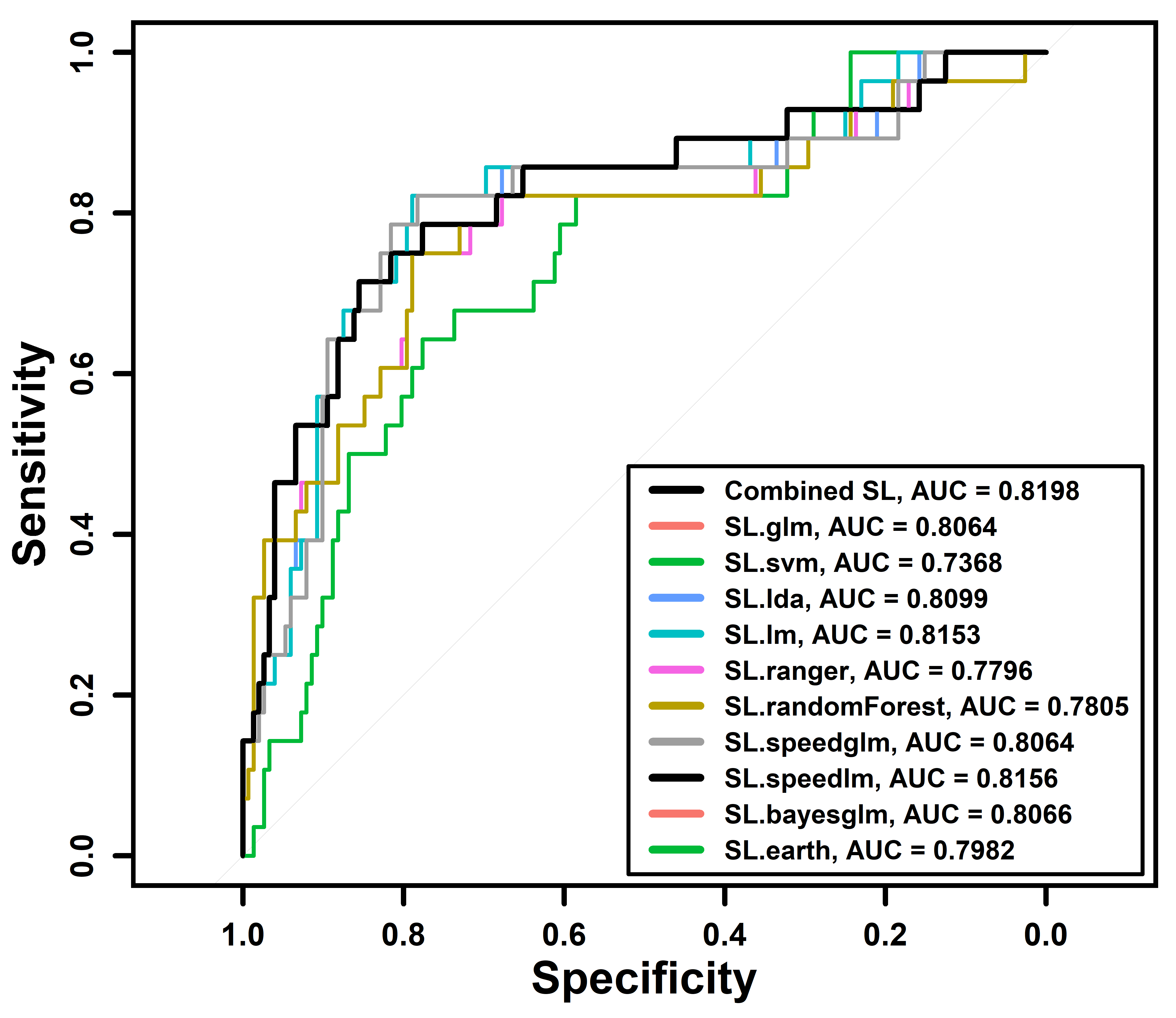}{0.33\textwidth}{}
          \fig{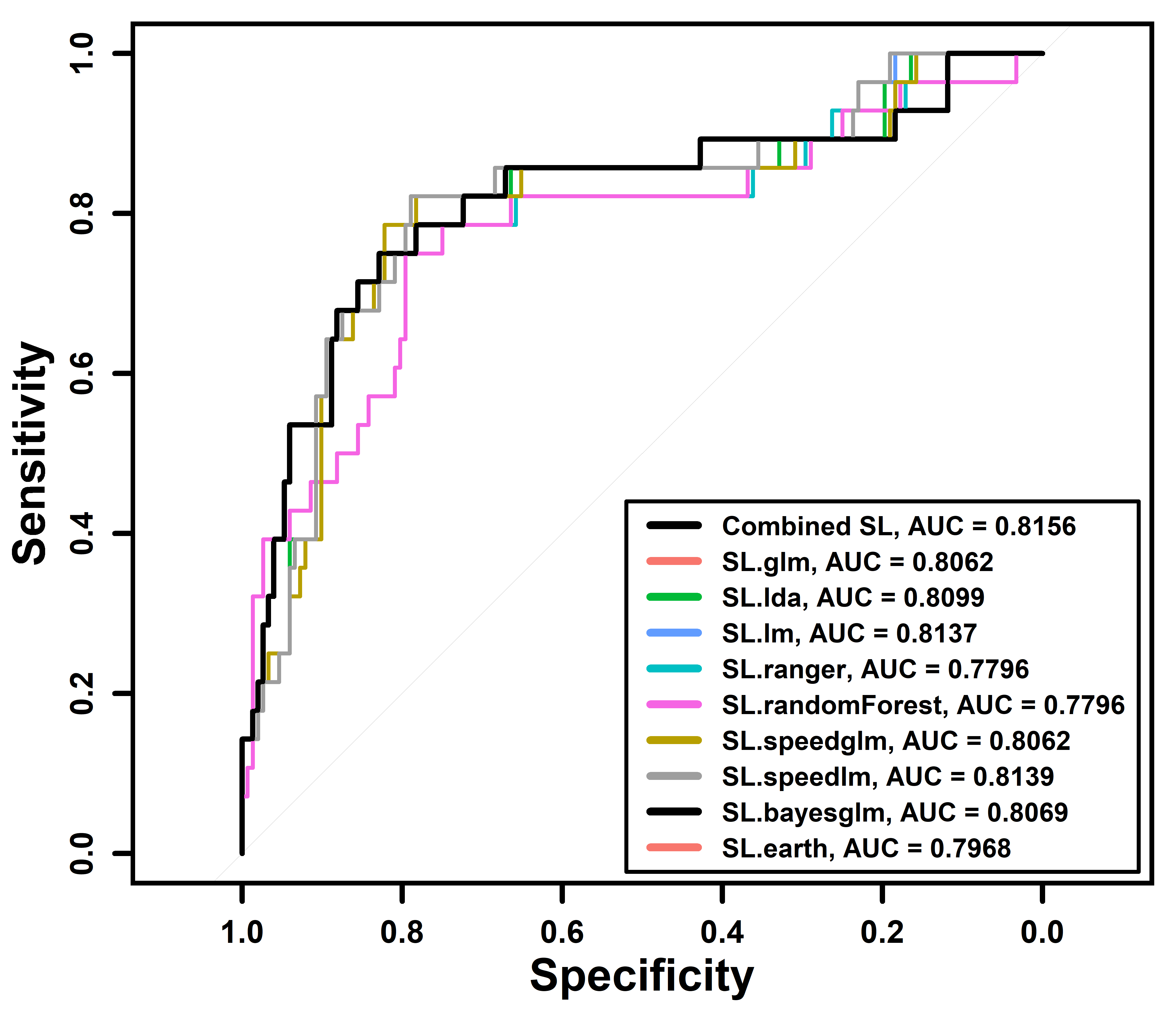}{0.33\textwidth}{}
         }

\vspace{-30pt}
\gridline{\fig{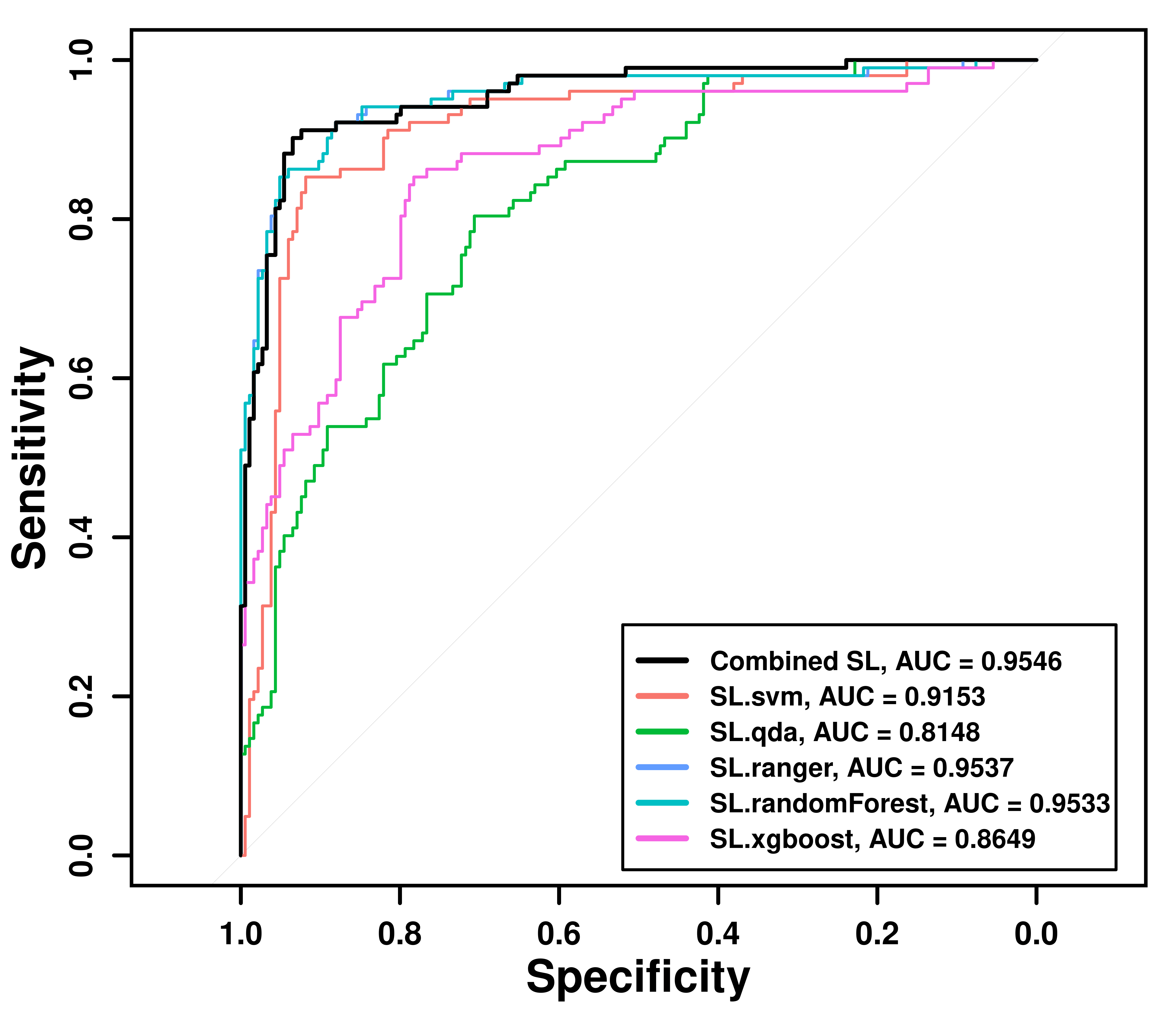}{0.33\textwidth}{}
          \fig{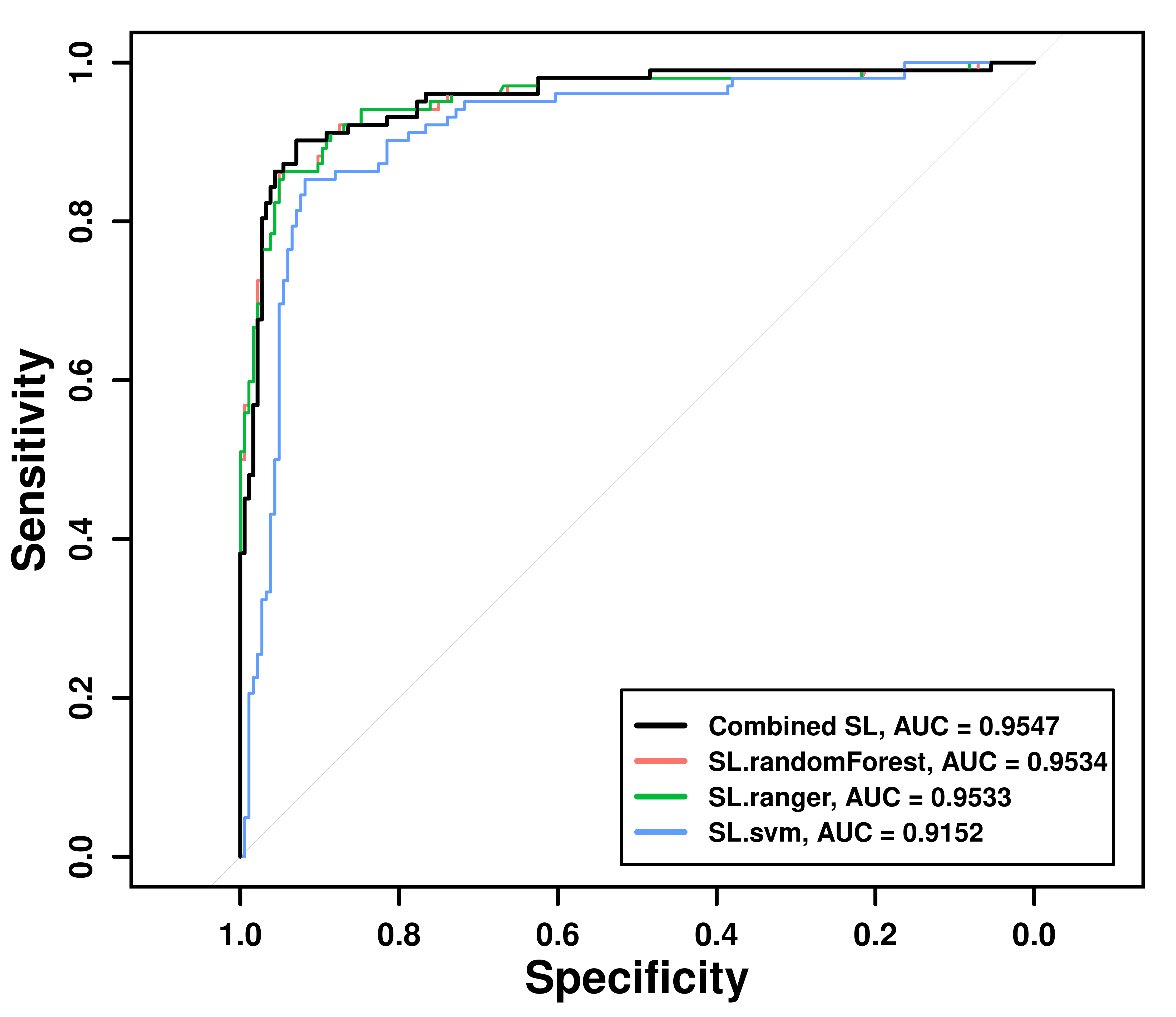}{0.35\textwidth}{}
         }
\vspace{-15pt}
\caption{
For $z_{t}=3.5$. Top row: raw without M-estimator. Second row: raw with M-estimator. Third row: MICE-imputed. Bottom row: SMOTE balanced. First column: ROC-AUC curve with best algorithms chosen by SuperLearner with weights $>$ 0.05 in each dataset. Top right: ROC-AUC curve without `ksvm'. Second right: ROC-AUC curve without `earth'. Third right: ROC-AUC curve without `svm'. Bottom right: ROC-AUC curve without `qda' and `xgboost'.
}
\label{fig:combined_raw_z3.5}
\end{figure*}

Table \ref{tab:LASSO} presents the optimal features selected by LASSO with weights $> 2\%$ for $z_{t}$ of 2.0, 2.5, 3.0, and 3.5 across the four datasets. An example of how LASSO operates on the four datasets is shown for $z_{t}$ = 3.5 in Fig. \ref{Fig:LASSO}. Table \ref{tab:SL-algo} displays the best algorithms selected by the SuperLearner with weights $>$ 0.05 for the same $z_{t}$ and datasets; an example with $z_t=3.5$ is given in Fig. \ref{Fig:SL weights}. In the subsequent figures of this section (Figure \ref{fig:combined_raw_z2.0}, \ref{fig:combined_raw_z2.5}, \ref{fig:combined_raw_z3.0}, and \ref{fig:combined_raw_z3.5}), we show the specificity vs. sensitivity in both the left and right panels: the top row corresponds to the raw dataset without the M-estimator, the second row represents the raw dataset with the M-estimator, the third row shows the MICE-imputed dataset, and the bottom row depicts the SMOTE balanced dataset. The left panels in these figures display the performance of the best models selected by SuperLearner, while the right panels illustrate the results after removing the least-performing algorithms. Since there is inherent randomness in results, we have checked to what extent we need to run multiple loops to evaluate this randomness and possibly reduce it. We conducted additional checks with more than 100 iterations, namely 200, 300, 500, and 1000, using the MICE-imputed data, and this procedure was repeated for each $z_{t}$ to ensure consistency. In the analysis of the multiple runs, we also tested the impact of the large error bars by removing GRBs with $\Delta x/x > 1$ on the variables even before applying the M-estimator, and we obtained that 5 GRBs were removed instead of 6, with one of them (GRB 050826) overlapping. However, the results, as we can see from Table \ref{tab:combined_auc_scores_mice}, are very similar to the 100 loops performed without the $\Delta x/x$ removal before the M-estimator in Table \ref{tab:AUC_for_SuperLearner}.
Again, there is no statistically meaningful difference between removing the outliers before or after the M-estimator since the difference is at most 1\%, and the maximum AUC values remain at 85\% for the imputed data.

The results of the AUC values for the training and the test sets are presented in Table \ref{tab:AUC_for_SuperLearner}.
We clarify that the Superlearner Algorithms (SLA), the least-performing removed algorithms (LPR), and the best-performing models (BPM) are calculated according to the test set in all tables.
We note that the smallest percentage difference among the AUC values obtained with the SLA models (the second column), the AUC without the LPR models (the fourth column), and the AUC with the BPM (the sixth column) is represented by $z_t=3.5$ with the SMOTE data showing $3\%<\Delta<4\%$. We also note that these percentage differences are the highest, with 52\% for $z_t=2.0$ in the raw data without the M-estimator. One can note that this percentage is calculated as the
\begin{equation}
\Delta=\frac{|AUC_{training}-AUC_{test}|}{\frac{(AUC_{training}+AUC_{test})}{2}.}
\label{Eq:percentage-deifference}
\end{equation}
The smallest AUC percentage difference between the training and test set AUCs without balancing the sample ranges between 3\% and 8\%, and it is followed by the raw data with M-estimator with $0.5\%<\Delta<10\%$ for $z_{t}=3.5$. The largest difference is for $z_t=2$.
We can see that for the test set, the smallest percentage difference $\Delta$ is consistently seen with the SMOTE-balanced data, ranging from 1\% in the case of $z_{t}=3$ to 5\% for $z_{t} = 2$. However, since we wish to use this analysis for future observations without balancing the sample, we should look at the MICE-imputed data set, where the smallest percentage difference between the training set and the test set AUC values is for $z_{t} = 3.5$ with $3\%<\Delta<15\%$. However, one must evaluate the entire results also considering the confusion matrix values, tabulated in Table \ref{tab:confusion_matrices_2.0}. When we use the $z_{t}=3.5$, the sample contains fewer GRBs categorized as high$-z$, and thus, the training set and test set AUC values have the smallest percentage deviation ($3\%<\Delta<15\%$), and the highest deviation is seen for the  $z_{t}=2$ with the $4\%<\Delta<53\%$. It is clear from Table \ref{tab:AUC_for_SuperLearner} that the MICE-imputed data is the sample we should use for actual categorization with observed data and for possible follow-up observations. For the MICE-imputed data, $z_{t}=2.0$ yields the best performance when looking contemporaneously at the highest TPR and TNR (see Table \ref{tab:confusion_matrices_2.0}) and if we only look at TPR. However, since we are interested in the sensitivity, we thus reiterate that the sample at $z_t=3.5$ is the best suited for the follow-up observations. If we look at the highest TNR, then $z_{t}=3.5$ is also the best-performing sample. Indeed, both the AUC values and the TNR agree in this sample, posing again that this sample is best suited for our purposes.
In the appendices, we have detailed for each $z_t$ (see Appendix \ref{sec:z2.0}, \ref{sec:z2.5}, \ref{sec:z3.0}, and \ref{sec:z3.5}) which is the most performing and least-performing model in the training set.

\begin{table}
    \centering
    \begin{tabular}{lccc}
    \hline
    \multicolumn{3}{c}{\textbf{Number of GRBs in the different redshift ranges}} \\
       \hline
    \hline
        \bf Dataset & \bf High$-z$ GRBs & \bf Low$-z$ GRBs \\
    \hline \hline
        \multicolumn{3}{c}{\bm{$z_{t}=2.0$}} \\
        \hline
        Raw dataset without M-estimator: & 83 & 95 \\
        Raw dataset with M-estimator: & 79 & 93 \\
        MICE-imputed dataset: & 102 & 123 \\
        SMOTE-balanced dataset: & 235 & 123 \\
        \hline
        \hline
        \multicolumn{3}{c}{\bm{$z_{t}=2.5$}} \\
        \hline
        Raw dataset without M-estimator: & 56 & 122 \\
        Raw dataset with M-estimator: & 52 & 120 \\
        MICE-imputed dataset: & 69 & 156 \\
        SMOTE-balanced dataset: & 202 & 156 \\
        \hline
        \hline
        \multicolumn{3}{c}{\bm{$z_{t}=3.0$}} \\
        \hline
        Raw dataset without M-estimator: & 43 & 135 \\
        Raw dataset with M-estimator: & 39 & 133 \\
        MICE-imputed dataset: & 52 & 173 \\
        SMOTE-balanced dataset: & 156 & 202 \\
        \hline
        \hline
        \multicolumn{3}{c}{\bm{$z_{t}=3.5$}} \\
        \hline
        Raw dataset without M-estimator: & 31 & 147 \\
        Raw dataset with M-estimator: & 27 & 145 \\
        MICE-imputed dataset: & 36 & 189 \\
        SMOTE-balanced dataset: & 115 & 243 \\
        \hline
    \end{tabular}
    \caption{Number of high$-$ and low$-z$ GRBs based on the $z_{t}$ of 2.0, 2.5, 3.0, and 3.5 (from top to bottom of the table) across four different datasets (left column).}
    \label{tab:GRBdata}
\end{table}

\begin{table}
    \begin{center}
    \begin{tabular}{lcc}
    \hline
    \multicolumn{2}{c}{\bf Features selected by LASSO in different redshift ranges with weights \bm{$>$} 2\%} \\
    \hline
    \hline
    \bf Dataset & \bf Features Selected \\ 
    \hline
    \hline
        \multicolumn{2}{c}{\bm{$z_{t}=2.0$}} \\
        \hline
        Raw dataset without M-estimator: & $\log{(T_{90})}$, $\log\rm{(PeakFlux)}$, PhotonIndex, and $\log\rm{(NH)}$ \\
        Raw dataset with M-estimator: & $\log{(T_{90})}$, $\log\rm{(PeakFlux)}$, PhotonIndex, $\log\rm{(NH)}$, $\Gamma$ \\
        MICE-imputed dataset: & $\log{(T_{90})}$, $\log\rm{(PeakFlux)}$, PhotonIndex, $\log\rm{(NH)}$, $\log{(F_{a})}$, $\log{(T_{a})}$, $\alpha$ and $\beta$ \\
        SMOTE-balanced dataset: & $\log{(T_{90})}$, $\log\rm{(PeakFlux)}$, PhotonIndex, $\log\rm{(NH)}$, $\log{(F_{a})}$, $\log{(T_{a})}$, and $\beta$ \\
        \hline
        \hline
        \multicolumn{2}{c}{\bm{$z_{t}=2.5$}} \\
        \hline
        Raw dataset without M-estimator: & $\log{(T_{90})}$, $\log\rm{(PeakFlux)}$, $\log\rm{(NH)}$, and $\log{(F_{a})}$ \\
        Raw dataset with M-estimator: & $\log{(T_{90})}$, $\log\rm{(PeakFlux)}$, $\log\rm{(NH)}$, and $\log{(F_{a})}$ \\
        MICE-imputed dataset: & $\log{(T_{90})}$, $\log\rm{(PeakFlux)}$, PhotonIndex, $\log\rm{(NH)}$, $\log{(F_{a})}$, $\log{(T_{a})}$, and $\beta$ \\
        SMOTE-balanced dataset: & $\log{(T_{90})}$, $\log\rm{(PeakFlux)}$, PhotonIndex, $\log\rm{(NH)}$, $\log{(F_{a})}$, and $\beta$ \\
        \hline
        \hline
        \multicolumn{2}{c}{\bm{$z_{t}=3.0$}} \\
        \hline
        Raw dataset without M-estimator: & $\log{(T_{90})}$, $\log\rm{(PeakFlux)}$, $\log\rm{(NH)}$, and $\log{(F_{a})}$ \\
        Raw dataset with M-estimator: & $\log{(T_{90})}$, $\log\rm{(PeakFlux)}$, PhotonIndex, $\log\rm{(NH)}$, $\log{(F_{a})}$, $\beta$ and $\Gamma$ \\
        MICE-imputed dataset: & $\log{(T_{90})}$, $\log\rm{(PeakFlux)}$, PhotonIndex, $\log\rm{(NH)}$, and $\beta$ \\
        SMOTE-balanced dataset: & $\log{(T_{90})}$, $\log\rm{(PeakFlux)}$, PhotonIndex, $\log\rm{(NH)}$, and $\log{(F_{a})}$ \\
        \hline
        \hline
        \multicolumn{2}{c}{\bm{$z_{t}=3.5$}} \\
        \hline
        Raw dataset without M-estimator: & $\log\rm{(PeakFlux)}$, $\log\rm{(NH)}$, and $\log{(F_{a})}$ \\
        Raw dataset with M-estimator: & $\log\rm{(PeakFlux)}$, $\log\rm{(NH)}$, and $\log{(F_{a})}$ \\
        MICE-imputed dataset: & $\log\rm{(PeakFlux)}$, PhotonIndex, $\log\rm{(NH)}$, $\log{(F_{a})}$, and $\Gamma$ \\
        SMOTE-balanced dataset: & $\log(T_{90})$, $\log\rm{(PeakFlux)}$, PhotonIndex, $\log\rm{(NH)}$, $\log{(F_{a})}$, and $\Gamma$ \\
        \hline
    \end{tabular}
    \end{center}
    \caption{Features selected by LASSO with weights $>$ 2\% for $z_{t}$ of 2.0, 2.5, 3.0, and 3.5 (from top to bottom panel) across four different datasets (left column).}
    \label{tab:LASSO}
\end{table}

\begin{table}
    \centering
    \begin{tabular}{lcc}
    \hline
    \multicolumn{2}{c}{\bf Data sets and  Algorithms selected by the Superlearner in the \bm{$z_{t}$} ranges with weights \bm{$>$} 0.05} \\
    \hline
    \hline
    \bf Dataset & \bf Algorithms Selected \\ 
    \hline
    \hline
        \multicolumn{2}{c}{\bm{$z_{t}=2.0$}} \\
        \hline
        Raw dataset without M-estimator: & glm, glmnet, kernelKnn, lda, lm, qda, ranger, \\
        & speedglm, speedlm, xgboost, biglasso, and bayesglm \\
        Raw dataset with M-estimator: & glm, glmnet, lda, lm, qda, Random Forest, speedglm, \\
        & speedlm, biglasso, bayesglm, and caret.rpart \\
        MICE-imputed dataset: & kernelKnn, ksvm, qda, ranger, Random Forest, speedlm, \\
        & xgboost, and earth \\
        SMOTE-balanced dataset: & kernelKnn, ksvm, svm, qda, ranger, and Random Forest \\
        \hline
        \hline
        \multicolumn{2}{c}{\bm{$z_{t}=2.5$}} \\
        \hline
        Raw dataset without M-estimator: & glm, glmnet, kernelKnn, lm, qda, speedglm, \\
        & speedlm, biglasso, bayesglm, and caret.rpart \\
        Raw dataset with M-estimator: & cforest, glm, kernelKnn, svm, qda, speedlm, \\
        & xgboost, and bayesglm \\
        MICE-imputed dataset: & cforest, glm, glmnet, kernelKnn, ksvm, lda, lm, qda, ranger, \\
        & Random Forest, speedglm, speedlm, xgboost, and bayesglm \\
        SMOTE-balanced dataset: & svm, qda, ranger, Random Forest, xgboost, and earth \\
        \hline
        \hline
        \multicolumn{2}{c}{\bm{$z_{t}=3.0$}} \\
        \hline
        Raw dataset without M-estimator: & kernelKnn, ksvm, svm, qda, ranger, Random Forest, and earth \\
        Raw dataset with M-estimator: & cforest, kernelKnn, qda, ranger, Random Forest, xgboost, \\ 
        & caret.rpart, and earth \\
        MICE-imputed dataset: & cforest, kernelKnn, ksvm, svm, qda, ranger, and Random Forest \\
        SMOTE-balanced dataset: & ksvm, svm, ranger, and Random Forest\\
        \hline
        \hline
        \multicolumn{2}{c}{\bm{$z_{t}=3.5$}} \\
        \hline
        Raw dataset without M-estimator: & kernelKnn, ksvm, svm, qda, ranger, Random Forest, and earth \\
        Raw dataset with M-estimator: & svm, lm, qda, ranger, Random Forest, speedlm, and earth\\
        MICE-imputed dataset: & glm, svm, lda, lm, ranger, Random Forest, speedglm, \\
        & speedlm, bayesglm, and earth \\
        SMOTE-balanced dataset: & svm, qda, ranger, Random Forest, and xgboost \\
        \hline
    \end{tabular}
    \caption{Algorithms selected by SuperLearner with weights $>$ 0.05 for $z_{t}$ of 2.0, 2.5, 3.0, and 3.5 (from top to bottom panel) across four different datasets (left column).}
    \label{tab:SL-algo}
\end{table}

\begin{table}[!ht]
    \centering
    {\resizebox{!}{.15\paperheight}{%
    \begin{tabular}{lcccccc}
    \hline
    \multicolumn{7}{c}{\bf AUC for SuperLearner for training and test sets} \\
       \hline
    \hline
        \bf Dataset & \bf AUC - SLA & \bf $\Delta_{SLA}$ (\%) & \bf AUC - LPR & \bf $\Delta_{LPR}$ (\%) & \bf AUC - BPM & \bf $\Delta_{BPM}$ (\%) \\
        \hline \hline
        & \textbf{training, test} &  & \textbf{training, test} &  & \textbf{training, test} \\
        \hline \hline
        \multicolumn{7}{c}{\bm{$z_{t}=2.0$}} \\
        \hline
        Raw dataset without M-estimator: & 0.757, 0.443 & 52 \% & 0.758, 0.452 & 51 \% & 0.762,  0.499 & 42 \%\\
        Raw dataset with M-estimator: & 0.798, 0.491 & 48 \% & 0.809, 0.470 & 53 \% & 0.803, 0.505 & 46 \%\\
        MICE-imputed dataset: & 0.805, 0.554 & 37 \% & 0.811, 0.544 & 39 \% & 0.799, 0.59 & 30 \%\\
        SMOTE-balanced dataset: & 0.879, 0.918 & 4 \% &  0.882, 0.928 & 5 \% & 0.883, 0.928 & 5 \%\\
        \hline
        \hline
        \multicolumn{7}{c}{\bm{$z_{t}=2.5$}} \\
        \hline
        Raw dataset without M-estimator: & 0.774, 0.581 & 28 \% & 0.772, 0.581 & 28 \% & 0.787, 0.612 & 25 \%\\
        Raw dataset with M-estimator: & 0.802, 0.606 & 28 \% & 0.805, 0.606 & 28 \% & 0.811, 0.639 & 24 \% \\
        MICE-imputed dataset: & 0.792, 0.639 & 21 \% & 0.802, 0.615  & 26 \% & 0.797, 0.651 & 20 \% \\
        SMOTE-balanced dataset: & 0.907, 0.938 & 3 \% & 0.916, 0.957  & 4 \% & 0.916, 0.957 & 4 \%\\
        \hline
        \hline
        \multicolumn{7}{c}{\bm{$z_{t}=3.0$}} \\
        \hline
        Raw dataset without M-estimator: & 0.843, 0.532 & 45 \% & 0.846, 0.522 & 57 \% & 0.832, 0.616 & 30 \% \\
        Raw dataset with M-estimator: & 0.840, 0.635 & 28 \% & 0.844, 0.621  & 30 \% & 0.841, 0.752 & 11 \%\\
        MICE-imputed dataset: & 0.848, 0.577 & 38 \% & 0.849, 0.571 & 39 \% & 0.827, 0.617 & 29 \%\\
        SMOTE-balanced dataset: & 0.950, 0.963 & 1 \% & 0.951, 0.970 & 2 \% &  0.951, 0.970 & 2 \%\\
        \hline
        \hline
        \multicolumn{7}{c}{\bm{$z_{t}=3.5$}} \\
        \hline
        Raw dataset without M-estimator: & 0.820, 0.714 & 14 \% & 0.826, 0.709 & 15 \% & 0.828, 0.724 & 13 \% \\
        Raw dataset with M-estimator: &  0.843, 0.759 & 10 \% & 0.840, 0.779 & 8 \% & 0.832, 0.828 & 0.5 \%\\
        MICE-imputed dataset: & 0.820, 0.791 & 4 \% & 0.816, 0.757 & 8 \% & 0.814, 0.787 & 3 \% \\
        SMOTE-balanced dataset: & 0.955, 0.986 & 3 \% & 0.955, 0.987  & 3 \% & 0.953, 0.992 & 4 \%\\
        \hline
    \end{tabular}}}
    \caption{AUC - SLA is the AUC from the SuperLearner selected algorithms, AUC - LPR is the AUC without the least-performing models and AUC - BPM is the AUC with the best-performing models. The second, fourth, and sixth columns contain the training and test AUC values for each $z_{t}$. The left value in each of these coloumns is the training AUC and the right value is the testing AUC. The third, fifth, and seventh columns show the percentage difference between the training and testing AUC. The percentage difference here is calculated as the absolute deviation of the training vs. the test set divided by their mean. These values are already rounded at the third significant digit from our analysis.}
    
    \label{tab:AUC_for_SuperLearner}
\end{table}

\begin{table}[!ht]
\centering
{\resizebox{!}{.38\paperheight}{%
\begin{tabular}{c | c c | c c | c }
\hline
\multicolumn{6}{c}{\bf Confusion Matrix for the several datasets including TPR and TNR} \\
\hline
\multicolumn{6}{c}{\bm{$z_{t}=2.0$}} \\ 
\hline
\multicolumn{1}{c}{TNR and TPR} & \multicolumn{2}{|c|}{\textbf{Raw without M-estimator}} & \multicolumn{2}{c|}{\textbf{Raw without M-estimator - algos removed}} & \multicolumn{1}{c}{TNR and TPR} \\

\hline
0.740 & 57 & 20 & 58 & 20 & 0.744\\
0.708 & 19 & 46 & 18 & 46 & 0.719 \\
\hline
\multicolumn{1}{c}{TNR and TPR} & \multicolumn{2}{|c|}{\textbf{Raw with M-estimator}}  &  \multicolumn{2}{c|}{\textbf{Raw with M-estimator - algos removed}} & \multicolumn{1}{c}{TNR and TPR} \\
\hline
0.763 & 58 & 18 & 57 & 22 & 0.722\\
0.742 & 16 & 46 & 17 & 42 & 0.712\\
\hline
\multicolumn{1}{c}{TNR and TPR} & \multicolumn{2}{|c|}{\textbf{MICE Imputed}} & \multicolumn{2}{c|}{\textbf{MICE Imputed - algos removed}} & \multicolumn{1}{c}{TNR and TPR} \\
\hline
0.736 & 78 & 28 & 79 & 24 & 0.767\\
0.730 & 20 & 54 & 19 & 58 & 0.753\\
\hline
\multicolumn{1}{c}{TNR and TPR} & \multicolumn{2}{|c|}{\textbf{SMOTE Balanced}} & \multicolumn{2}{c|}{\textbf{SMOTE Balanced - algos removed}} & \multicolumn{1}{c}{TNR and TPR}\\
\hline
0.800 & 40 & 10 & 47 & 12 & 0.797 \\
0.805 & 46 & 190 & 39 & 188 & 0.828\\
\hline
\multicolumn{6}{c}{\bm{$z_{t}=2.5$}} \\
\hline
\multicolumn{1}{c}{TNR and TPR} & \multicolumn{2}{|c|}{\textbf{Raw without M-estimator}} & \multicolumn{2}{c|}{\textbf{Raw without M-estimator - algos removed}} & \multicolumn{1}{c}{TNR and TPR} \\
\hline
0.757 & 84 & 27 & 82 & 24 & 0.774\\
0.613 & 12 & 19 & 14 & 22 & 0.611 \\
\hline
\multicolumn{1}{c}{TNR and TPR} & \multicolumn{2}{|c|}{\textbf{Raw with M-estimator}}  &  \multicolumn{2}{c|}{\textbf{Raw with M-estimator - algos removed}} & \multicolumn{1}{c}{TNR and TPR} \\
\hline
0.775 & 79 & 23 & 80 & 23 & 0.777\\
0.583 & 15 & 21 & 14 & 21 & 0.600\\
\hline
\multicolumn{1}{c}{TNR and TPR} & \multicolumn{2}{|c|}{\textbf{MICE Imputed}} & \multicolumn{2}{c|}{\textbf{MICE Imputed - algos removed}} & \multicolumn{1}{c}{TNR and TPR} \\
\hline
0.778 & 112 & 32 & 112 & 31 & 0.783\\
0.667 & 12 & 24 & 12 & 25 & 0.676\\
\hline
\multicolumn{1}{c}{TNR and TPR} & \multicolumn{2}{|c|}{\textbf{SMOTE Balanced}} & \multicolumn{2}{c|}{\textbf{SMOTE Balanced - algos removed}} & \multicolumn{1}{c}{TNR and TPR}\\
\hline
0.960 & 71 & 3 & 72 & 4 & 0.947\\
0.826 & 37 & 175 & 36 & 174 & 0.829 \\
\hline
\multicolumn{6}{c}{\bm{$z_{t}=3.0$}} \\
\hline
\multicolumn{1}{c}{TNR and TPR} & \multicolumn{2}{|c|}{\textbf{Raw without M-estimator}} & \multicolumn{2}{c|}{\textbf{Raw without M-estimator - algos removed}} & \multicolumn{1}{c}{TNR and TPR} \\
\hline
0.815 & 97 & 22 & 97 & 18 & 0.844\\
0.609 & 9 & 14 & 9 & 18 & 0.667\\
\hline
\multicolumn{1}{c}{TNR and TPR} & \multicolumn{2}{|c|}{\textbf{Raw with M-estimator}}  &  \multicolumn{2}{c|}{\textbf{Raw with M-estimator - algos removed}} & \multicolumn{1}{c}{TNR and TPR} \\
\hline
0.803 & 98 & 24 & 97 & 23 & 0.808\\
0.625 & 6 & 10 & 7 & 11 & 0.611\\
\hline
\multicolumn{1}{c}{TNR and TPR} & \multicolumn{2}{|c|}{\textbf{MICE Imputed}} & \multicolumn{2}{c|}{\textbf{MICE Imputed - algos removed}} & \multicolumn{1}{c}{TNR and TPR} \\
\hline
0.818 & 130 & 29 & 131 & 30 & 0.814 \\
0.667 & 7 & 14 & 6 & 13 & 0.684 \\
\hline
\multicolumn{1}{c}{TNR and TPR} & \multicolumn{2}{|c|}{\textbf{SMOTE Balanced}} & \multicolumn{2}{c|}{\textbf{SMOTE Balanced - algos removed}} & \multicolumn{1}{c}{TNR and TPR}\\
\hline
0.942 & 131 & 8 & 128 & 10 & 0.928\\
0.871 & 19 & 128 & 22 & 126 & 0.851\\
\hline
\multicolumn{6}{c}{\bm{$z_{t}=3.5$}} \\
\hline
\multicolumn{1}{c}{TNR and TPR} & \multicolumn{2}{|c|}{\textbf{Raw without M-estimator}} & \multicolumn{2}{c|}{\textbf{Raw without M-estimator - algos removed}} & \multicolumn{1}{c}{TNR and TPR} \\
\hline
0.856 & 113 & 19 & 113 & 17 & 0.869\\
0.500 & 5 & 5 & 5 & 7 & 0.583\\
\hline
\multicolumn{1}{c}{TNR and TPR} & \multicolumn{2}{|c|}{\textbf{Raw with M-estimator}}  &  \multicolumn{2}{c|}{\textbf{Raw with M-estimator - algos removed}} & \multicolumn{1}{c}{TNR and TPR} \\
\hline
0.875 & 112 & 16 & 112 & 14 & 0.889\\
0.6 & 4 & 6 & 4 & 8 & 0.667\\
\hline
\multicolumn{1}{c}{TNR and TPR} & \multicolumn{2}{|c|}{\textbf{MICE Imputed}} & \multicolumn{2}{c|}{\textbf{MICE Imputed - algos removed}} & \multicolumn{1}{c}{TNR and TPR} \\
\hline
0.862 & 150 & 24 & 150 & 24 & 0.862\\
0.667 & 2 & 4 & 2 & 4 & 0.667\\
\hline
\multicolumn{1}{c}{TNR and TPR} & \multicolumn{2}{|c|}{\textbf{SMOTE Balanced}} & \multicolumn{2}{c|}{\textbf{SMOTE Balanced - algos removed}} & \multicolumn{1}{c}{TNR and TPR}\\
\hline
0.935 & 173 & 12 & 174 & 13 & 0.931\\
0.891 & 11 & 90 & 10 & 89 & 0.900\\
\end{tabular}}}
\caption{
Confusion matrices for the training set of each dataset with their respective TNR (every first row) and TPR (every second row).
}
\label{tab:confusion_matrices_2.0}
\end{table}

\begin{table}
    \centering
    \begin{tabular}{lcccccc}
    \hline
    \multicolumn{6}{c}{\bf Loop for each \bm{$z_{t}$} repeated 5 times for the MICE imputed data} \\
    \hline
    \hline
    \textbf{Run} & \textbf{100 loops} & \textbf{200 loops} & \textbf{300 loops} & \textbf{500 loops} & \textbf{1000 loops} \\
    \hline
    \hline
    & \textbf{AUC} & \textbf{AUC} & \textbf{AUC} & \textbf{AUC} & \textbf{AUC} \\
    & \textbf{training, test} & \textbf{training, test} & \textbf{training, test} & \textbf{training, test} & \textbf{training, test} \\
    \hline
    \hline
    \multicolumn{6}{c}{\bm{$z_{t}=2.0$}} \\
    \hline
    \hline
        1 & 0.802 0.552 & 0.811 0.538 & 0.802 0.548 & 0.803 0.554 & 0.801 0.538 \\
        2 & 0.812 0.544 & 0.808 0.548 & 0.806 0.552 & 0.807 0.548 & 0.802 0.542 \\
        3 & 0.802 0.552 & 0.803 0.540 & 0.801 0.536 & 0.806 0.540 & 0.807 0.552 \\
        4 & 0.807 0.546 & 0.809 0.552 & 0.801 0.528 & 0.810 0.526 & 0.806 0.552 \\
        5 & 0.802 0.520 & 0.801 0.544 & 0.805 0.550 & 0.804 0.570 & 0.803 0.538 \\
    \hline
    \hline
    \multicolumn{6}{c}{\bm{$z_{t}=2.5$}} \\
    \hline
    \hline
        1 & 0.801 0.628 & 0.793 0.649 & 0.792 0.644 & 0.797 0.635 & 0.790 0.637 \\
        2 & 0.804 0.618 & 0.793 0.637 & 0.796 0.639 & 0.795 0.632 & 0.803 0.618\\
        3 & 0.793 0.651 & 0.797 0.639 & 0.794 0.635 & 0.790 0.651 & 0.792 0.639 \\
        4 & 0.794 0.642 & 0.797 0.632 & 0.792 0.642 & 0.793 0.649 & 0.794 0.635 \\
        5 & 0.803 0.613 & 0.792 0.635 & 0.803 0.630 & 0.804 0.632 & 0.807 0.613 \\
    \hline
    \hline
    \multicolumn{6}{c}{\bm{$z_{t}=3.0$}} \\
    \hline
    \hline
        1 & 0.852, 0.557 & 0.844, 0.593 & 0.842, 0.627 & 0.851, 0.568 & 0.843, 0.590 \\
        2 & 0.849, 0.569 & 0.843, 0.590 & 0.846, 0.580 & 0.842, 0.586 & 0.846, 0.580 \\
        3 & 0.842, 0.617 & 0.844, 0.602 & 0.849, 0.571 & 0.844, 0.565 & 0.844, 0.590 \\
        4 & 0.843, 0.593 & 0.844, 0.611 & 0.842, 0.590 & 0.851, 0.568 & 0.842, 0.590 \\
        5 & 0.844, 0.574 & 0.844, 0.586 & 0.847, 0.577 & 0.846, 0.577 & 0.844, 0.590 \\
    \hline
    \hline
    \multicolumn{6}{c}{\bm{$z_{t}=3.5$}} \\
    \hline
    \hline
        1 & 0.819 0.767 & 0.822 0.818 & 0.818 0.797 & 0.819 0.797 & 0.815 0.791 \\
        2 & 0.825 0.821 & 0.825 0.811 & 0.816 0.797 & 0.820 0.811 & 0.828 0.801 \\
        3 & 0.817 0.811 & 0.821 0.801 & 0.818 0.794 & 0.812 0.784 & 0.820, 0.804 \\
        4 & 0.821 0.811 & 0.822 0.801 & 0.813 0.801 & 0.818 0.804 & 0.826 0.784 \\
        5 & 0.821 0.807 & 0.822 0.794 & 0.829 0.814 & 0.813 0.797 & 0.835 0.807 \\
    \hline
    \end{tabular}
    \caption{Combined training and testing data AUC scores with MICE-imputed dataset for $z_{t}$ = 2.0, 2.5, 3.0, and 3.5 (from top to bottom). The left value in each loop column shows the AUC value for the training dataset, while the right value in each loop column shows the AUC value for the testing dataset.}
    \label{tab:combined_auc_scores_mice}
\end{table}

\section{Discussion}
\label{sec:discussion}

\subsection{Relevance and Scope of Methods}
The most important aim of the paper is to craft a reliable and as much as possible complete methodology that allows the GRB classification in accordance to observing strategies and the time our classifier is able to allow follow-up studies (for details see sec. \ref{New Missions}). This strategy entails Superlearner, including 17 methods ranging from parametric, semi-parametric, and fully non-parametric methods, and it assesses which methods are the most predictive and which are the least predictive. Indeed, we have also used Superlearner as a method to discriminate among the most predictive algorithms, which ranges from 4 models for the raw data with M-estimator (right upper panel of \ref{Fig:SL weights}),  5 models for the SMOTE balanced data (lower right panel of \ref{Fig:SL weights} ), 7 models for raw data (upper left panel of \ref{Fig:SL weights}), 10 models for MICE imputed data (lower left of \ref{Fig:SL weights})  in the configuration of $z_{t}$=3.5. In these 4 data sets, the common models used are: Random Forest and ranger. However, the models common in the raw data, raw data with M estimator and the MICE imputed data is only earth, besides Random Forest and ranger. The common algorithms in the SMOTE, and MICE imputed and the raw data set is svm, besides random forest and ranger. 
The common algorithms in the SMOTE balanced data and the raw data are qda and svm, in addition to Random Forest and earth. We have also analyzed the methods and described the variables' relative importance.

\subsection{The predictive power of the algorithms}
We discuss the predictive power of our algorithms in relation to the methodology and the $z_{t}$ used. 
We found that removing outliers using the M-estimator (Section \ref{sec:Mestimator}) enhances the predictive power of our ensemble model. Specifically, increasing the AUC value from $\sim$ 76\% to $\sim$ 80\% for $z_{t}$ = 2.0, from $\sim$ 77\% to $\sim$ 80\% for $z_{t}$ = 2.5, and from $\sim$ 82\% to $\sim$ 84\% for $z_{t}$ = 3.5. Notably, for $z_{t}$ = 3.0, the AUC value remains nearly constant at $\sim$ 84\%. 
In general, $z_t = 3.0$ and 3.5 are the ones for which the SuperLearner consistently picks the same or similar (more complex or less complex) algorithms among the four data sets, with `ranger' and `Random Forest' having similar AUC values. Furthermore, for $z_{t}$ = 3.0, the AUC differences are $0.3\%$ before and after applying an M-estimator to the data, while $z_t=3.5$ shows the smallest $\Delta$ among the training and the test set AUC values. 
The model's performance further improves when we expand our data sample using the MICE imputation technique (Section \ref{sec:mice}). In this case, our ensemble model achieves its best performance at $z_{t}=3.0$, yielding an AUC of $\sim$ 85\%.
While there exists an inherent uncertainty in each SuperLearner model, we show in Table \ref{tab:combined_auc_scores_mice} that the performance of SuperLearner with the MICE-imputed dataset across 100, 200, 300, 500, and 1000 loops remains consistent within a 1\% change in the AUC value considering the training set within the same $z_{t}$, even though the best algorithms selected may vary. Thus, we set the loop count to 100 in our analysis since there is no need for extra computational analysis. This selection is not entirely dissimilar across multiple SuperLearner runs, and similar base learners, such as more advanced versions of the same selected algorithms, are often selected. For example, if in one run, the best-selected algorithm is `lm', in a subsequent run, it is `speedlm', an advanced implementation of `lm'. This demonstrates the robustness and consistency of our model's predictive capabilities regardless of loop number (see Table \ref{tab:combined_auc_scores_mice}).  While two models could have similar AUC values, further performance characteristics such as generalization power may differ. However, the generalization power will be investigated in a forthcoming paper.
The difference in the performance power can be seen in the confusion matrices for each model and each $z_{t}$ in Table \ref{tab:confusion_matrices_2.0}.

Another significant increase is observed when we balance the dataset using the SMOTE balance technique (Section \ref{sec:databalancing}), which increases the AUC to $\sim$ 95\% for both $z_t=3$ and $z_t=3.5$ in the training set.
In the test set, the AUC values for $z_t=3$ change from 58\% to 96\%. We should stress that our test set is small (it is composed of 20\% of the total sample) and contains 45 GRBs in the MICE imputed data and 72 GRBs in the balance sampling.
We also note that, in $z_{t}=3.0$, the smallest variation in the AUC values from the training set to the test set is 28\% for the M-estimator sample. However, the M-estimator sample with $z_{t}=3.5$ carries the least percentage difference between the AUC values of the training and the test set with a percentage difference of 10\%. 
This makes the sample of $z_t=3.5$ the most stable in terms of the smallest percentage difference between the training and test set AUC. 
In this case, the AUC value increases from 82\% with the MICE-imputed data to even 96\% with the SMOTE balanced data, making $z_t=3.5$ the most performant in terms of both the highest of AUC and the smallest $\Delta$. 
Given the smallest number of high$-z$ GRBs, a higher AUC is expected in this range. 
We also note that these percentage differences are the highest for $z_{t}=2.0$, with 52\% for the raw data without the M-estimator. 
If we do not consider the SMOTE balanced data, we also note that for $z_t=3.5$, the highest AUC is the raw with M-estimator instead of the MICE-imputed data. This suggests that the $z_t=3.5$ sample is the most stable in terms of considering observed data without augmentation. In general, when we quote the most performant sample and algorithms for the balance sampling, we need to pose a caveat that said analysis involves augmented data. Then, the difference between the AUC values among the training and test sets is minimal compared to the other samples because the training and testing sets are augmented by synthetic data extrapolating the minority subsample. This caveat does not apply to the raw with M-estimator findings, so the low difference between the training and test set for $z_t=3.5$ is inherent to the procedure. In the future, it is worth estimating the performance of SMOTE on the outlier distribution of the minority events, but this analysis goes beyond the scope of the current paper.
This percentage increase, considering the raw sample with the M-estimator, is also valid for both $z_{t}$ = 2.5. 

For the balanced data set, the most predictive models identified, regardless of the chosen $z_{t}$, are `Random Forest' and `ranger,' similar to what we have for the MICE-imputed data at $z_{t}=3$. These two algorithms also consistently emerged as the most frequently occurring in all datasets for all $z_t$. This demonstrates their overall robustness and effectiveness in GRB classification based on $z$, which is consistent with \cite{ukwatta2016machine}'s finding that `Random Forest' itself achieves an AUC of 89\%. 
Additionally, regarding the least predictive algorithm, we note that the non-parametric algorithm `caret.rpart' either has the lowest predictive power or is not selected as a top-performing algorithm based on the 0.05 weight cutoff.

\begin{figure*}[htbp]
\gridline{\fig{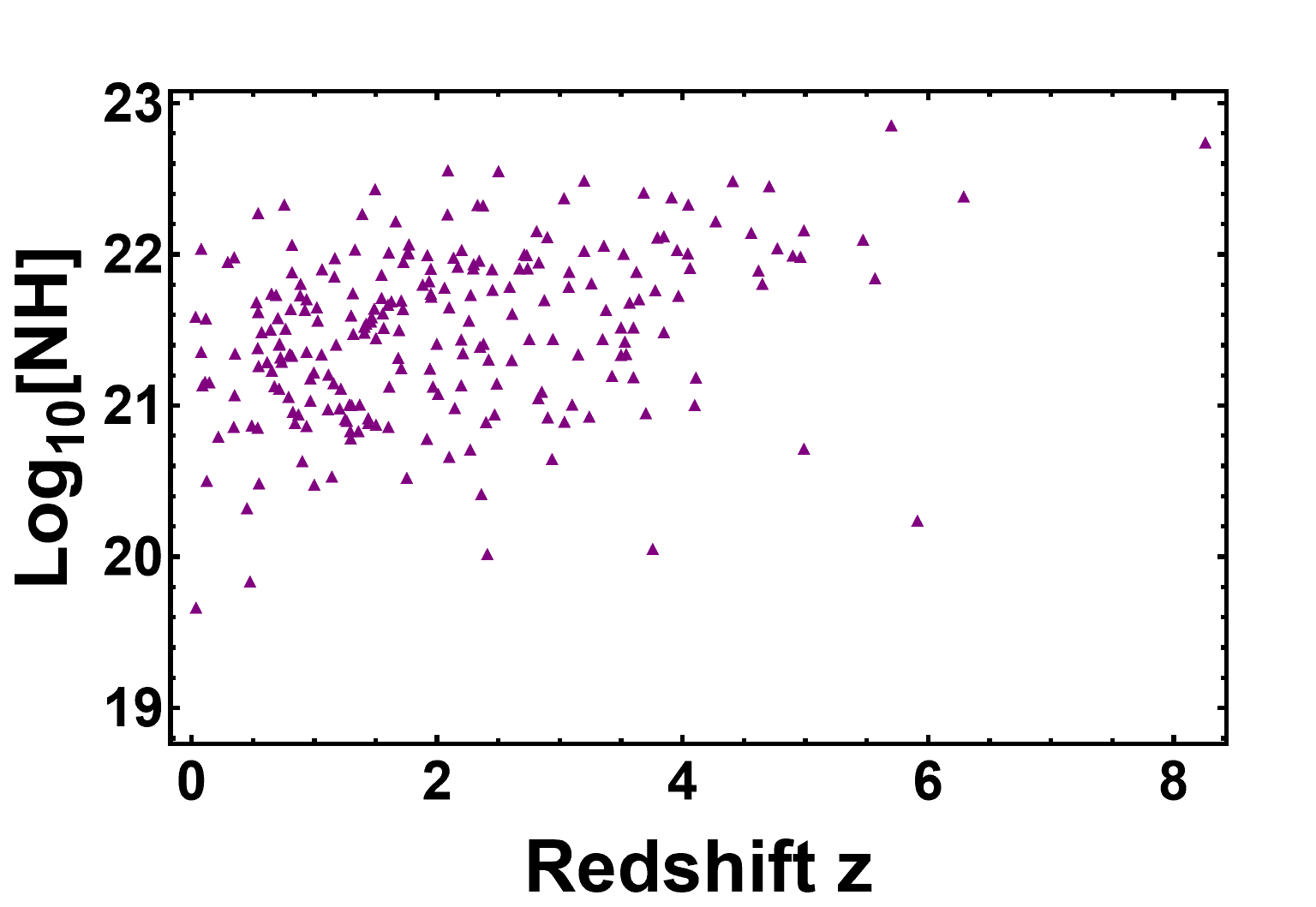}{0.4\textwidth}{}
          \fig{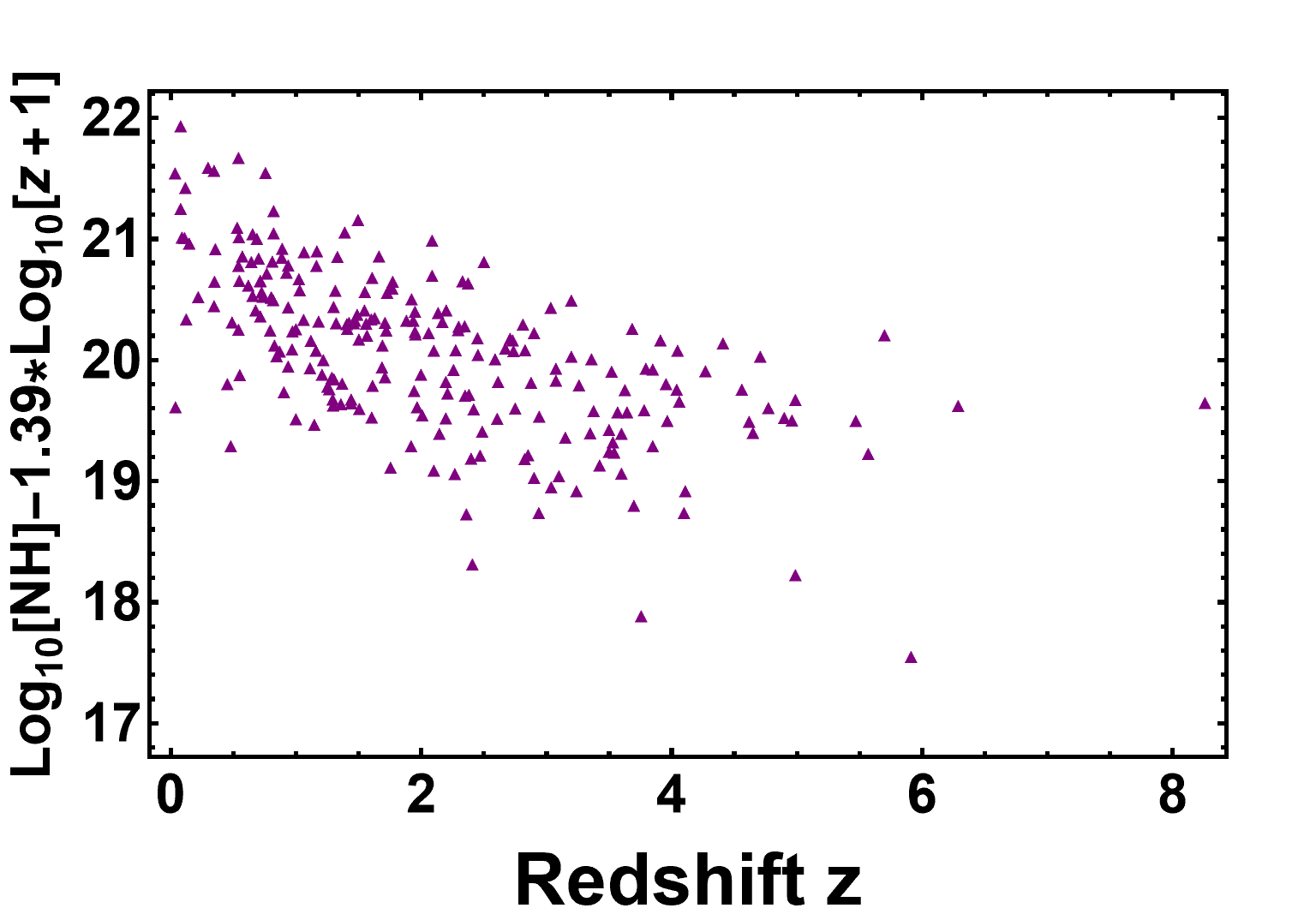}{0.4\textwidth}{}
          }
\gridline{\fig{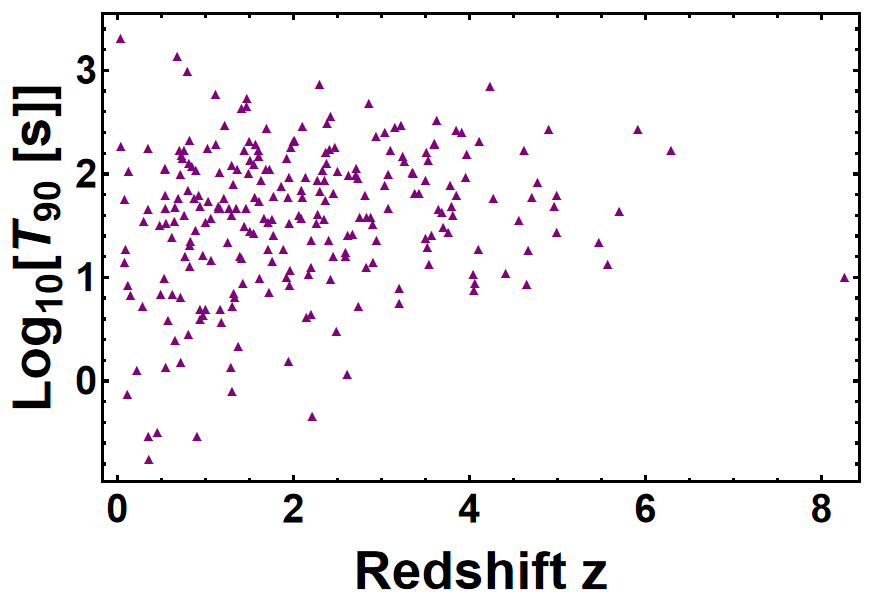}{0.4\textwidth}{}
          \fig{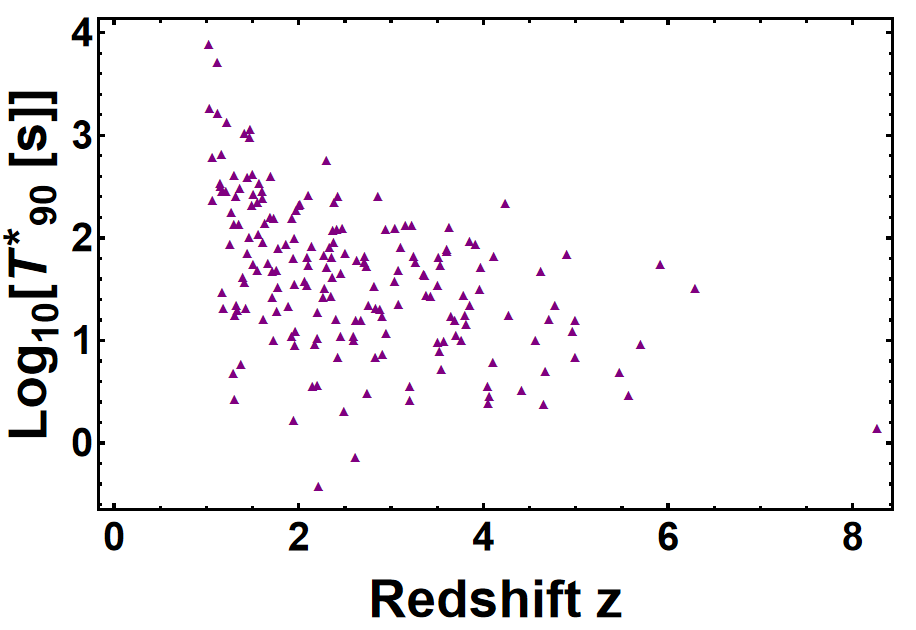}{0.4\textwidth}{}
         }
\gridline{\fig{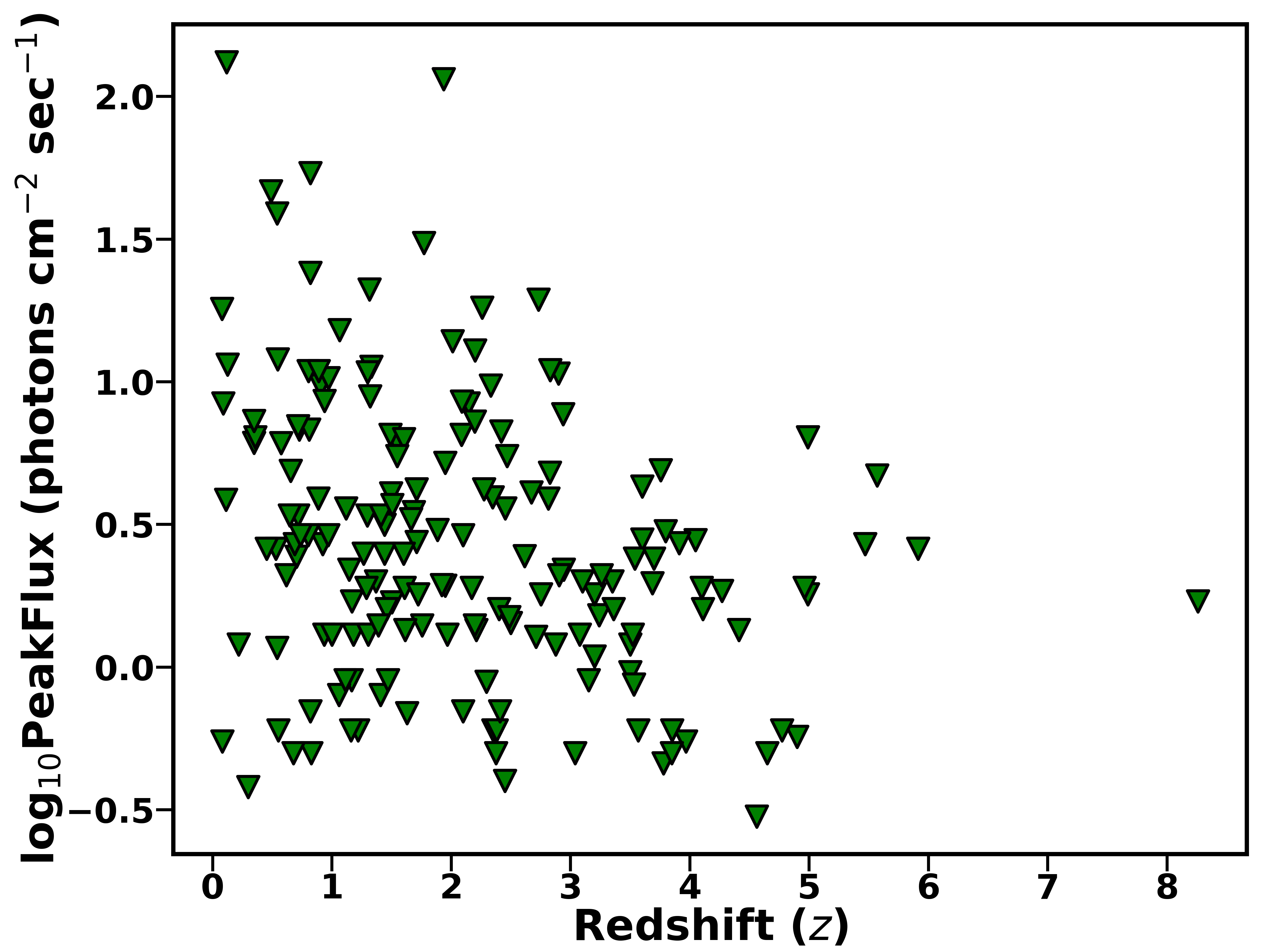}{0.4\textwidth}{}
          \fig{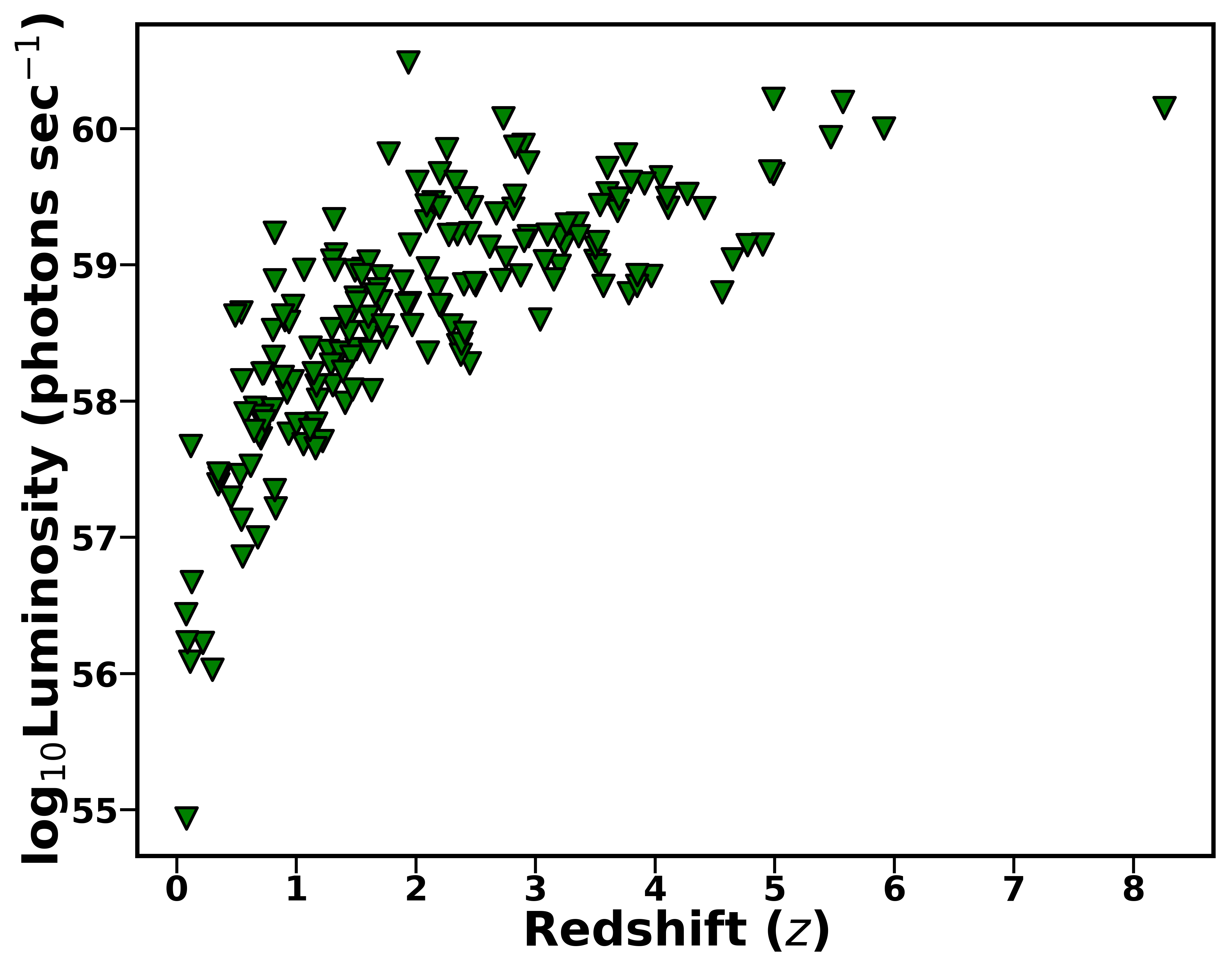}{0.4\textwidth}{}
          }
\caption{
Top Left: The distribution of $\log(\rm{NH})$ vs. $z$, Top Right: $\log(\rm{NH}) - 1.39*\log(z + 1)$ vs. $z$, Middle Left: $\log(T_{90})$ vs. $z$, Middle Right: $\log(T^*_{90})$ vs. $z$, Bottom Left: The distribution of $\log(\rm{PeakFlux})$ vs. $z$, and Bottom Right: $\log(\rm{Luminosity})$ vs. $z$ for the GRBs in our sample.
}
\label{fig:NHz}
\end{figure*}

\subsection{The predictive power of the features}
We discuss the predictive power of the three most important predictors and highlight their connection with GRB physics.

\subsubsection{The dependence of $\log(\rm{NH})$ vs redshift}
This $\log(NH)$ that we have used in our analysis accounts for the total neutral hydrogen column density along our line of sight between the burst and the Milky Way, including both the host galaxy and a fixed Galactic absorption. It was derived by the automated time-averaged spectral fitting of the data from Swift-XRT Catalog \citep{Evans2009} using the HEASoft software. Specifically, the Galactic column density in the direction of the burst has been computed with the NHFtool, and the spectrum is fitted in with the model phabs*phabs*pow. The first absorption component is frozen at the Galactic value and the second is free to vary.
The scatter displayed in the log NH vs. z in the scatter matrix plot in Fig. \ref{Fig:M-estimator Scatter matrix plot-raw} poses a considerable challenge to a simple interpretation. 
Analysis of the $\log(\rm{NH})$ vs. z is not straightforward to handle. Indeed, to determine the truncation in the plane $\log(\rm{NH})$ vs. z, one needs to evaluate both the flux limit of Swift XRT and $\log(\rm{NH})$, whose accuracy is simultaneously affected by the observed flux. In this case, \cite{galaxies12050051} constrained the flux limit with 495,000
spectral simulations, including the response matrix file to generate mock data and then estimate the accuracy of the $\log(\rm{NH})$. 
We here discuss some of the possible interpretation in relation to $\log(\rm{NH})$ shown in \cite{Rahin_2019,2020MNRAS.495.2342D,Campana2010,Starling2013}.
$\log(\rm{NH})$ is affected by multiple factors:
\begin{enumerate}
    \item The first factor is the dependence on the metal content. Higher metallicity signals a greater abundance of heavy elements in the surrounding medium. Heavier elements absorb X-rays more efficiently than lighter ones. Hence, metallicity plays a critical role in determining the X-ray absorption. Star clusters undergo more significant mass loss in higher metallicity than those in lower-metallicity environments \citep{Trani_2014}.  This can contribute to significant extinction in the surrounding Interstellar medium (ISM) during the progenitor’s lifetime, as the expelled material increases the overall density of the surrounding gas and dust. The increased extinction due to higher metallicity-driven mass loss can influence the observed X-ray and optical GRB emission properties. To consider that metallicity has a major role in the dependence on the redshift, we should also consider the metallicity evolution \citep{Niino_2017,Heintz:2022ozz}. As shown by \cite{chincarini2010MNRAS.406.2113C} and \cite{graham_2023}, high-$z$ GRBs tend to occur in low-metallicity regions. If we use a lower metallicity $0.07 \pm 0.05 Z/Z\odot$, the calculated $\log NH$ is reduced only by a tiny percentage (0.7-0.9\%) \citep{2020MNRAS.495.2342D}. Hence, the evolution of $\log(\rm{NH})$ cannot be straightforwardly associated with the metals present in the general interstellar medium of the host galaxy.
    
    \item The second factor is the density of the ISM within the host galaxy. High X-ray column densities in galaxies with intense star formation are often associated with the dense interstellar medium and emissions of stellar processes. So, for GRB hosts, high star formation rates may suggest young, actively forming galaxies with more dense ISM. Additionally, starbursts are associated with a high specific star formation rate, and intense processes may lead to significant ejection of gas/dust into the nearby medium. This can result in higher $\log(\rm{NH})$ values, as the burst’s X-ray emissions interact with the host galaxy’s dense material. Past works showed that GRBs undergo an unusually high formation rate at low$-z$, exceeding the corresponding SFR by more than an order of magnitude \citep{Dainotti2024b, 2024ApJ...963L..12P}. However, almost all works show that between redshifts 1 and 3, the GRB rate follows an almost constant SFR, with a noticeable decline beyond this range \citep{petrosian2015, Dong:2021wdc}. As shown by \citep{Cucchiara_2011,Palmerio2019A&A...623A..26P} and \citep{graham2023}, high$-z$ GRBs tend to occur in low-metallicity regions. \cite{bolmer2018} further found that GRB host galaxies at $z > 4$ have lower dust and extinction values ($A_V < 0.5$ mag) than low$-z$sources, suggesting that the evolution of metallicity at $z > 4$ is not driven by higher dust content. However, several factors come into play, and the host galaxy's local ISM can also influence this trend. Results from \cite{Arumaningtyas2024} showed that 13\% of GRBs remain in high-density ($10^4 \, \mathrm{cm}^{-3}$) low-temperature star-forming regions. Since most GRBs (87\%) occur in low-density ($\sim 10^{-2.5} \, \mathrm{cm}^{-3}$) high-temperature regions \citep{2014ApJ...794...50C}, 
    this is in contradiction with the high $\log(\rm{NH})$ and the evolution with redshift and gives support to the actual trend observed. However, this evidence can be explained by several models of dust and star distribution in a very clumpy ISM \citep{2018A&A...617A.141C}.
 
    \item The third factor responsible for the evolution of $\log(\rm{NH})$ with redshift may be the absorption of the GRB's emission by the intergalactic medium (IGM) between our galaxy and the GRB's host galaxy. This evolution may be due to the ionized and diffuse gas accumulation in the IGM \citep{Rahin:2019dwy}.
        
    \item The stellar initial mass function (IMF) may also play a role in the evolution of $\log(\rm{NH})$. Starting from the assumption that LGRBs originate from massive stars, the IMF in the early universe could have been skewed toward larger stellar masses. \cite{2008MNRAS.385..687W, 2017MNRAS.468.3071N} suggested that the early universe likely favored the formation of more massive stars than we observe today. Population III stars, with masses ranging from tens of $M_{\odot}$ \citep{10.1111/j.1365-2966.2009.16113.x} to values larger than a hundred $M_{\odot}$ \citep{Toma_2011}, can possibly be progenitors for LGRBs at high redshift \citep{Bromm:2005ep}. \cite{2011MNRAS.416.2760C} found that the expected fraction of Pop III GRBs is $\sim 10\%$ of the full GRB population at $z > 6$ and becomes as high as $40\%$ at $z > 10$. 
    \item The fifth factor that can influence the $\log(\rm{NH})$ evolution is the existence of an anti-correlation of the average escape fraction with the column density \citep{Chen_2007}. The total escape fraction inferred from the GRB column density is lower by one order of magnitude than the expected value \citep{Tanvir:2018pbq}. Simulations show that some GRBs with the same massive progenitor can remain in the star-forming region, and others may migrate to a less dense environment \citep{2014ApJ...794...50C}. This migration could occur due to significant relative motions between stars and their birth clouds: a relative motion of 10 km s\(^{-1}\) could result in a displacement of 100 pc over 10 Myr. Runaway OB stars generally have a relative motion of 20 km s\(^{-1}\) \citep{phillips2024runawayobstarssmall}, thus leading a pivotal role in the GRB displacement from their birth clouds. The presence of GRBs in a hot, low-density medium also depends on supernova heating from earlier supernovae within the birth clouds. These two effects have roughly equal weight in locating most GRBs in low-density, high-temperature regions \citep{2014ApJ...794...50C}. As a conclusion, the evolution of $\log(\rm{NH})$ suggests that there is a population of GRBs with low column density (less than log 19) that have been missed in detections; see the plot of $\log(\rm{NH})$ vs $z$  and log \( NH - 1.39 \cdot \log(z + 1) \) vs \( z \).

\end{enumerate}

With all these factors, it is even more clear how challenging is the explanation of this observed trend.

\subsubsection{The dependence of $\log(\rm{PeakFlux})$ vs redshift}
In all the cases with different redshift values ($z = 2.0, 2.5, 3.0$, and 3.5) with different datasets (Raw without M-estimator, Raw with M-estimator, MICE imputed, and SMOTE balanced), we found that another crucial parameter from the prompt phase is the observed peak photon flux, $\log(\rm{PeakFlux})$. We discuss the significance of $\log(\rm{PeakFlux})$ in $z$-classification and its dependence on redshift.

\cite{2002ApJ...574..554L} found the evolution of luminosity with redshift, showing that luminosity scales as $L \propto (1 + z)^{1.4 \pm 0.5}$ by studying the redshifts and luminosities of 220 GRBs derived using the luminosity-variability relation from \cite{2000astro.ph..4176F}.  \cite{2003MNRAS.345..743W} also found a similar luminosity evolution with redshift: $L \propto (1 + z)^{2.5 \pm 0.1}$ when disregarding flux truncation effects, and $L \propto (1 + z)^{1.7 \pm 0.5}$ when accounting for selection biases. In line with these findings, \cite{yonetoku2004} also found a similar trend in the luminosity evolution, with $L \propto (1 + z)^{2.6^{+0.15}_{-0.2}}$, which is consistent with the one obtained by \cite{2002ApJ...574..554L}. 
From the parametric approach, a study done by \cite{salvaterra2012complete} revealed a strong luminosity evolution with redshift ($L \propto (1 + z)^{2.3 \pm 0.6}$) using a sample of 58 LGRBs observed by \textit{Swift}-BAT, with 1-s peak photon flux $P \geq 2.6$ photons $\rm{cm}^{-2}$ $\rm{s}^{-1}$ in the 15--150 KeV BAT energy band, 52 of them with known redshifts. Furthermore, in the non-parametric approach, \cite{petrosian2015} examined a sample of 200 \textit{Swift} LGRBs with known redshift by applying the Efron-Petrosian (EP) method \citep{Efron1992} and revealed an intrinsic luminosity-redshift correlation, quantified as $L \propto (1 + z)^{2.3 \pm 0.8}$, extending up to $z \leqslant 3$. Additionally, \cite{Yu2015ApJS..218...13Y} confirmed similar findings with 127 LGRBs with well-measured spectral parameters from \textit{Fermi}-GBM \citep{meegan2009} and Konus-\textit{Wind} \citep{Konus1995} and observed redshift from \textit{Swift}, identifying a comparable evolution trend of $L \propto (1 + z)^{2.43^{+0.41}_{-0.38}}$. Recent work of \cite{pescalli2016A&A...587A..40P} reinforced these results using an expanded \textit{Swift}-BAT sample of 99 LGRBs (82 GRBs with measured redshift and luminosity) with 1-s peak photon flux $P \geq 2.6$ photons $\rm{cm}^{-2}$ $\rm{s}^{-1}$ in the 15--150 KeV BAT energy band, confirming a luminosity evolution with redshift as $L \propto (1 + z)^{2.3 \pm 0.8}$, using the non-parametric EP method.

From the parametric approach, a study done by \cite{salvaterra2012complete} revealed a strong luminosity evolution with redshift ($L \propto (1 + z)^{2.3 \pm 0.6}$) using a sample of 58 LGRBs observed by \textit{Swift}-BAT, with 1-s peak photon flux $P \geq 2.6$ photons $\rm{cm}^{-2}$ $\rm{s}^{-1}$ in the 15--150 KeV BAT energy band, 52 of them with known redshifts.
Similarly to \cite{salvaterra2012complete}, \cite{2019MNRAS.488.4607L} conducted an extended analysis with 99 LGRBs from \textit{Swift}-BAT having 1-s peak photon flux $P \geq 2.6$ photons $\rm{cm}^{-2}$ $\rm{s}^{-1}$ in the 15--150 KeV BAT energy band, with measured redshift and luminosity of 82 GRBs and confirmed this trend, finding $L \propto (1 + z)^{2.22^{+0.32}_{-0.31}}$. Recently, \cite{2021MNRAS.508...52L} expanded this sample to 733 LGRBs with a peak photon flux threshold of $P \geq 1$ photons $\rm{cm}^{-2}$ $\rm{s}^{-1}$ in the 15--150 KeV BAT energy band, including 302 bursts with known redshift. They also found a strong luminosity evolution with the redshift, with $L \propto (1 + z)^{1.85^{+0.24}_{-0.23}}$ when the GRB luminosity function (LF) is broken power-law (BPL), and $L \propto (1 + z)^{1.92^{+0.25}_{-0.37}}$ when the LF is a PL+BPL law. They found that the evolutionary effect appears to be independent of the specific functional form of the GRB LF. 

These consistent findings across various methods highlight a strong correlation between LGRB luminosity and redshift, supporting intrinsic evolution rather than observational bias alone. This underscores the need for strong luminosity evolution to account for the observations. Since $\log(\rm{PeakFlux})$ is related to the intrinsic luminosity, these relationships imply that $\log(\rm{PeakFlux})$ would tend to increase at higher redshifts, consistent with the observed evolution in luminosity and this is what we observe in our results. In the left and right bottom panels in Figure \ref{fig:NHz}, we have shown the distribution of $\log(\rm{PeakFlux})$ vs $z$ and $\log(\rm{Luminosity})$ vs $z$, respectively. The $\log(\rm{Luminosity})$ vs $z$ plot (right bottom panel in Figure \ref{fig:NHz}) distinctly shows an evolutionary trend, indicating that the luminosity of GRBs evolves as a function of $z$.


\subsubsection{The dependence of $T_{90}$ vs redshift}

In the case of the dataset with MICE imputation, we observe that $\log(T_{90})$ is selected as one of the important variables; see the left panel of Fig. \ref{fig:NHz} where   $\log(T_{90})$ is shown as a function of z, and here we discuss why $\log(T_{90})$ is an important predictor for the $z$ estimate.

Although in our sample, we have almost a negligible correlation with a coefficient of 0.1 (see the left panel of Fig. \ref{fig:NHz}), we need to investigate the more profound meaning of this correlation in the rest frame (see the right panel of Fig. \ref{fig:NHz} ). Thus, the discussion below is in relation to the rest-frame quantities. Another caveat is that we cannot introduce the redshift dependence and the selection biases since the redshift is the unknown parameter.

For our discussion, we here connect the progenitor of LGRBs with $T^{*}_{90}$, with the symbol $*$ we denote the rest-frame quantity.

The progenitor of a GRB can be a red supergiant (RSG), blue supergiant (BSG), or a Wolf-Rayet star, and regardless of the progenitor type, we can explain the $T^{*}_{90}$-$z$ correlation as a function of the metallicity. According to \cite{Lloyd2023ApJ...947...85L}, isolated single stars with higher metallicity have a lower density 
\citep{maeder&meynet2001,Georgy_2013,Sanyal_2017,Woosley2006ARA&A}, larger stellar radius, and smaller iron core mass. The implication for the lifetime of the GRB jet produced by the single star progenitor can be estimated by using the free-fall time,$t_{ff}$, defined by $t_{ff}\propto \sqrt{(R^3/M)}$, where M is the mass of the collapsing star and $R$ is its radius.

This time, $t_{ff}$, is the interval over which the stellar material feeds into the central engine, powering the GRB jet. Thus, we can relate the GRB jet's duration directly with the star's radius. Hence, during the core collapse, stars with the same mass have larger radii at higher metallicities, giving rise to longer-duration GRB jets. The 1D simulations in \cite{Lloyd2023ApJ...947...85L} support this idea, where the difference between the stellar radii at different metallicities possibly can explain the trend of the $T^*_{90}$ vs $z$ after correction for selection biases. 
This hypothesis is further reinforced by the fact that higher metallicity stars also gain a larger accretion disk and a lower accretion rate \citep{Wei_2022,Suwa_2010,Fukushima_2017,Toyouchi_2018}, thus fueling the central engine longer and producing a longer-duration jet.

Recently, as the sample has become larger and larger, more GRBs have been found in high-metallicity galaxies \citep{graham2023}. Because higher metallicity stars are, on average, located at lower redshift, this leads to an anti-correlation between $T^{*}_{90}$ and $z$ \citep{graham2023, Zahid_2013}.

 \subsection{The dependence of $F_a$ at the end of the plateau from the redshift}

Another important variable is the flux at the plateau emission, which is among the third most important predictors for z=3.5. If the plateau emission is produced by a magnetar, a fast-rotating spinning newborn neutron star, its luminosity, which is the flux in the observer frame, is directly linked to the time of the plateau. The observed relation between time and flux carries a correlation of 0.71 when MICE is applied. Then, naturally the magnetar can be explained given that the slope of this correlation is -1.

\section{Comparison with earlier results}
\label{sec:comparison}
Compared to the work of \cite{ukwatta2016machine}, where they achieved a sensitivity of 80\% with $z_{t} = 4.0$, our classification of GRBs based on different $z_{t}$ demonstrates a percentage increase of 9\% and 11\% in the sensitivity with the SMOTE balanced data, achieving values of 87\% and 89\% (see the first column of Figure \ref{tab:confusion_matrices_2.0}) for the training sets with $z_{t}$ = 3.0 and 3.5, respectively. 
In terms of AUC, our analysis achieved a value of 95\% for $z_{t}$ = 3.0 and 3.5 with SMOTE balanced data, showing a 13\% and 7\% percentage increase over prior works where only the Random Forest algorithm was employed, which yielded an AUC of 84\% \citep{morgan2012} and 89\% \citep{ukwatta2016machine}, respectively.
However, if we consider the MICE imputed data with M-estimator, our results are comparable with the one of \cite{morgan2012} since the AUC is 85\% and they are slightly less performant (with 4\% of decrease) than the work presented by \cite{ukwatta2016machine} in terms of AUC. However, these AUC values presented in our work do not yet carry the optimization performed in \cite{ukwatta2016machine}. 
The primary novelty of our study, beyond the enhanced performance and the adoption of a more comprehensive methodology, as detailed in Section \ref{sec:methodology} which removes outliers and impute data, is the inclusion of the plateau properties, which has not been explored in previous studies besides some of us \citep{Dainotti2024b, Dainotti_2024}.

Furthermore, while \cite{morgan2012} utilized a sample of 135 GRBs detected by \textit{Swift} and trained a single ML model on the entire dataset without reserving a subset of the dataset for testing the trained ML model, thus in this case the predictive power has not been checked on a separate test set. 
Similarly, \cite{ukwatta2016machine} also trained a single model on their complete data sample for $z_{t}$ = 4.0 only, leading to a limited number of high$-z$ GRBs, and without setting a subset of the sample for testing the trained ML model. Our approach is more comprehensive compared to these past studies. 
We trained and tested our ensemble model on an 80\% training set and 20\% test set across four different $z_{t}$ (2.0, 2.5, 3.0, and 3.5) with four different datasets (raw data, the data with selection of M-estimator, the imputed data with MICE, and the balanced data with SMOTE), thereby enhancing the stability and robustness of the ensemble model.

\subsection{The prompt emission properties for the classifier}
Another critical point is that we can obtain the classification in some samples (raw data with and without M-estimator for $z_{t}=2$) also without the plateau emission properties, such as $T_{90}$, the log Peak Flux, the Photon Index, and the addition of $\log(\rm{NH})$.

If we compare these results with those in \cite{ukwatta2016machine}, we can conclude that our analysis with the MICE imputed data carries an AUC=0.88, which is 9.5\% higher. However, if we consider the raw data, the prediction is consistent with 80\% similar to  \cite{ukwatta2016machine}

This will allow us to use a much broader sample and to classify GRBs more rapidly in the future with a larger sample with only prompt emission properties. This part of using only prompt emission parameters is similar to the previous study by \cite{morgan2012,ukwatta2016machine}.

\section{The user-friendly web-app}
\label{sec:broader impact}
\begin{figure}[h!]
    \centering
    \fbox{\includegraphics[scale=0.4]{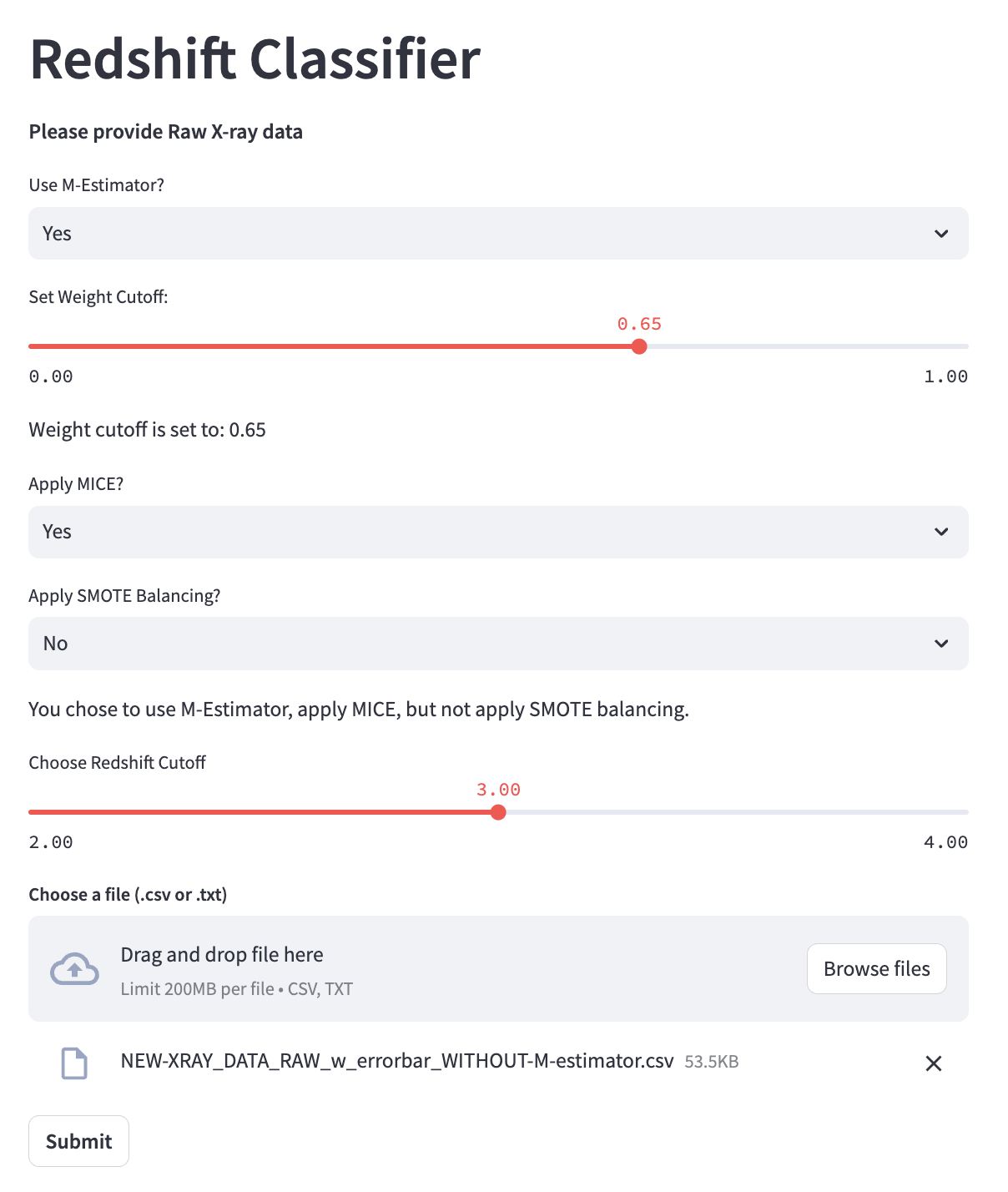}}%
    \hfill
    \fbox{\includegraphics[scale=0.4]{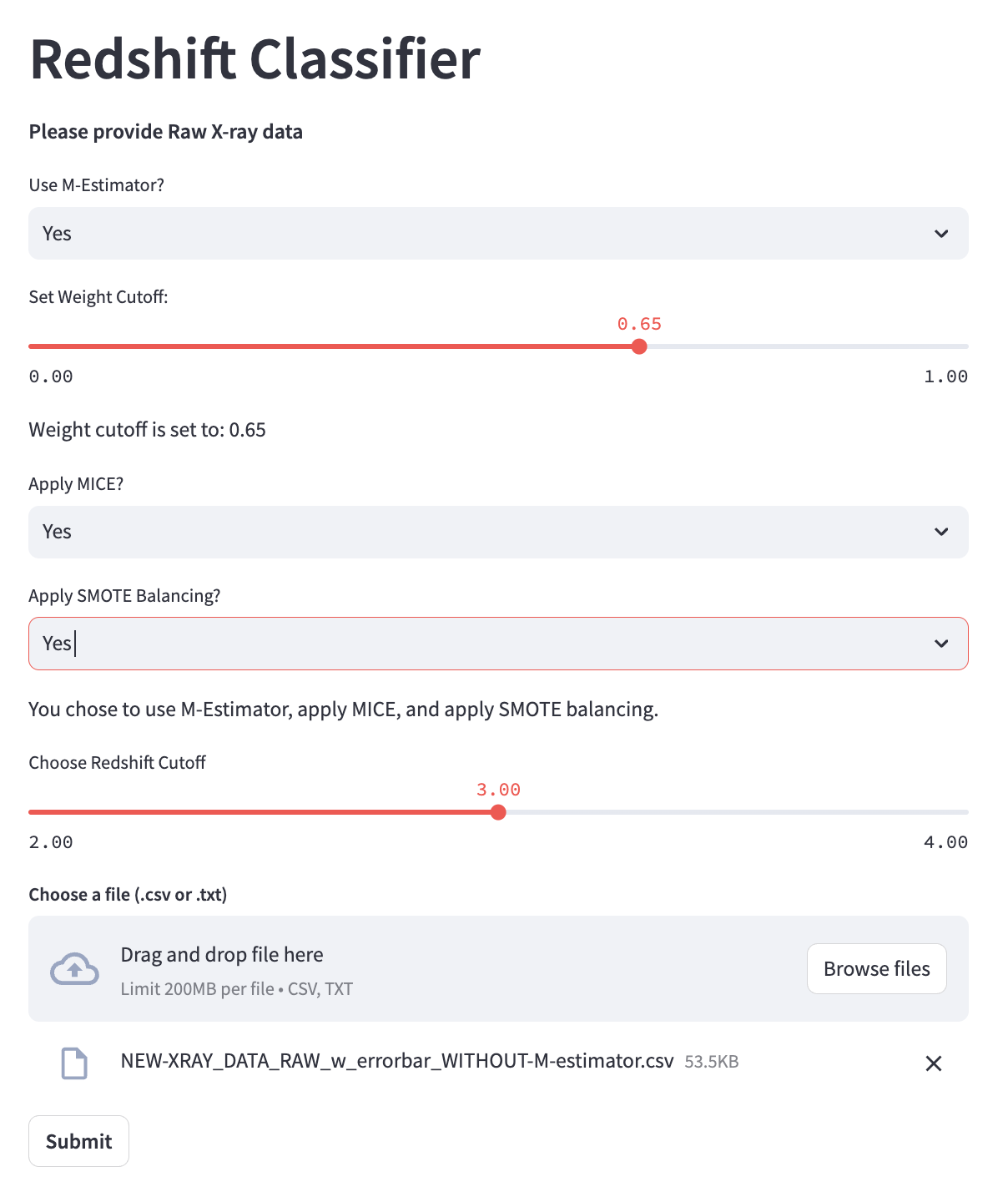}}
    \caption{Left: shows the web-app interface when the user chooses to apply both the M-estimator and MICE-imputation before running the SuperLearner. Right: depicts the web-app interface where the user chooses to apply the M-estimator, MICE-imputation, and SMOTE balancing, followed by the SuperLearner. The user can select the desired value $z_{t}$ within the web app before running the SuperLearner.}
    \label{fig:APP}
\end{figure}

To ensure a broader impact of this work, we have developed a user-friendly publicly available web app on Streamlit, an open-source framework for interactive web applications for data science and ML projects. The app can be run locally by the user on their machine and can be found at the current GitHub: \href{https://github.com/Milind018/Redshift-Classifier}{Redshift-Classifier}. Figure \ref{fig:APP} illustrates some parts of the web app interface for our redshift classifier.
The app allows users to customize the classification process based on their specific requirements. Users can select whether to apply the M-estimator for outlier removal based on their desired threshold, as well as opt for MICE imputation and SMOTE data balancing. Additionally, they can select their preferred $z_{t}$ for a flexible GRB redshift classification.
 
When the user opts not to apply the M-estimator, the app runs the SuperLearner using the raw data provided by the user. The first step within the SuperLearner is data cleaning (removing NA variables and data points with $\Delta x/x > 1$).
After applying the M-estimator and either MICE or SMOTE, data points with $\Delta x/x > 1$ are discarded in each app section.
When the user selects the M-estimator to remove the outliers, the app first runs the M-estimator and then the SuperLearner using the output data from the M-estimator.
When the user chooses to apply the MICE imputation after the M-estimator, the app runs the SuperLearner using the MICE's output data.
If the user opts to apply the SMOTE balancing, the app uses the output data from the SMOTE.

Subsequent steps within the SuperLearner pipeline include performing LASSO feature selection, selecting the best algorithms, and then, finally, performing the 100 nested 10fCV. The final output from the SuperLearner includes all the plots related to the LASSO feature selection, SuperLearner coefficient, and ROC-AUC for training and test datasets.

\section{The observing strategies in relation to our high-z classifier and the future missions and Telescopes}\label{New Missions}
By using GRB observational properties, this machine learning web-app tool will allow the community to classify GRBs as high$-z$ or low$-z$. This will enable rapid and intriguing GRB follow-up programs using both space-based satellites and ground-based telescopes. For instance, when a GRB is triggered by \textit{Swift}, the SVOM \citep{2015arXiv151203323C,bernardini2021svom}, CALET \citep{Sugita2022}, the Einstein Probe \citep{Yuan2022} and the several  Cube Sat satellites now orbiting \citep{Bomani2021}, users can use the web app to quickly discern whether the triggered GRB is of high$-$ or low$-z$. 
In the future, this classification can even be automatized and then promptly communicated through Gamma-ray Coordinate Network (GCN, \href{https://gcn.gsfc.nasa.gov/}{https://gcn.gsfc.nasa.gov/}), facilitating rapid and detailed follow-up observations by other telescopes, such as the BOOTES telescopes \citep{https://doi.org/10.1155/2016/1928465, 10.3389/fspas.2023.952887}, as well as the Subaru telescope \citep{Kodaira1992}, Kiso observatory \citep{1977AnTok..16...74T}, and other future observatories, such as POLAR-2 \citep{2023arXiv230900518P}, THESEUS \citep{2018AdSpR..62..191A} and High$-z$ GUNDAM \citep{GUNDAM}. The Kiso Observatory has already begun conducting optical follow-up observations of GRBs using the wide-field CMOS camera, Tomo-e Gozen \citep{10.1117/12.2310049}, which is mounted on the 1.05-meter Schmidt telescope. This follow-up program, initiated by Dainotti and her collaboration in January 2022, has already contributed six data points using a clear filter \citep{2024MNRAS.tmp.1527D}.

For follow-up programs desiring reliable detection of high$-z$ events, the primary goal is to achieve the highest possible sensitivity (TPR) in classification. Our ensemble method is optimized for this task at $z_{t} = 2.0$, trained on the MICE-imputed dataset (with worst-performing algorithms removed). This configuration of our procedure achieves a TPR of 75\% (see the last column of Table \ref{tab:confusion_matrices_2.0}) in a 100 nested 10fCV, thus ensuring that the vast majority of high$-z$ events will signal the need for follow-up. In this case, if one would like to use only raw data with the M-estimator, the TPR = 71\% after removing the least performing algorithm (see the last column of Table \ref{tab:confusion_matrices_2.0}). 
For follow-up programs interested in low$-z$ detections, such as kilonovae or X-ray flashes or low-luminosity GRBs, our procedure is optimized at $z_{t}=3.5$ for the raw dataset with M-estimator (with least performing algorithms removed), achieving a TNR of 89\% in a 100 nested 10fCV, which ensures that the vast majority of low$-z$ events will trigger a follow-up. The versatility of our ensemble method in the selection of sensitivity and specificity demonstrates its utility across a wide range of follow-up programs, maximizing the scientific return of follow-up observations.

These timely follow-up observations will significantly increase the sample of high$-z$ GRBs, enabling them to serve as standard candles to probe the early universe. The inclusion of redshift information is vital for GRB population studies. For instance, we can better estimate the LF and density rate evolution, providing further constraints on models of star formation and cosmic evolution in the early Universe \citep{Dainotti2024b, Dainotti_2024, 2024ApJ...963L..12P}.
So with this expanded high$-z$ sample, we can enhance the use of GRBs for cosmological purposes as shown in \cite{2022MNRAS.tmp.2639D, Dainotti2022}.

\begin{figure}[hbt!]
    \centering
    \includegraphics[width=0.9\textwidth,height=0.5\textheight]
    {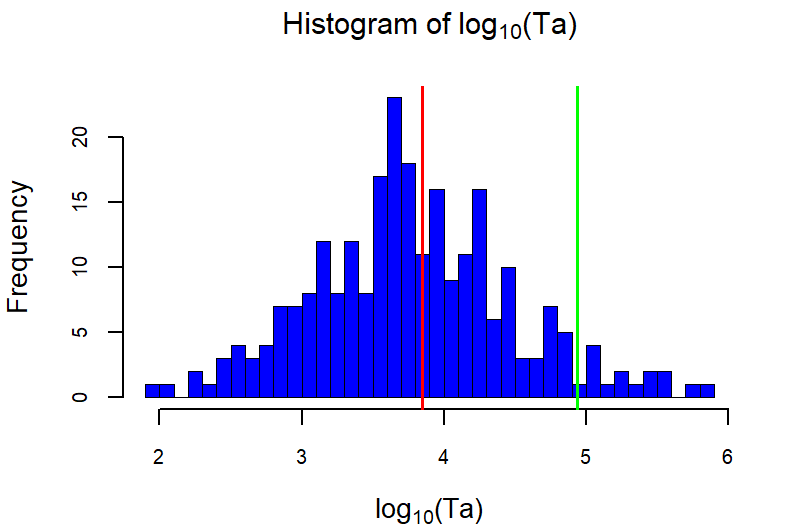}
    \includegraphics[width=0.8\textwidth,height=0.5\textheight]
    {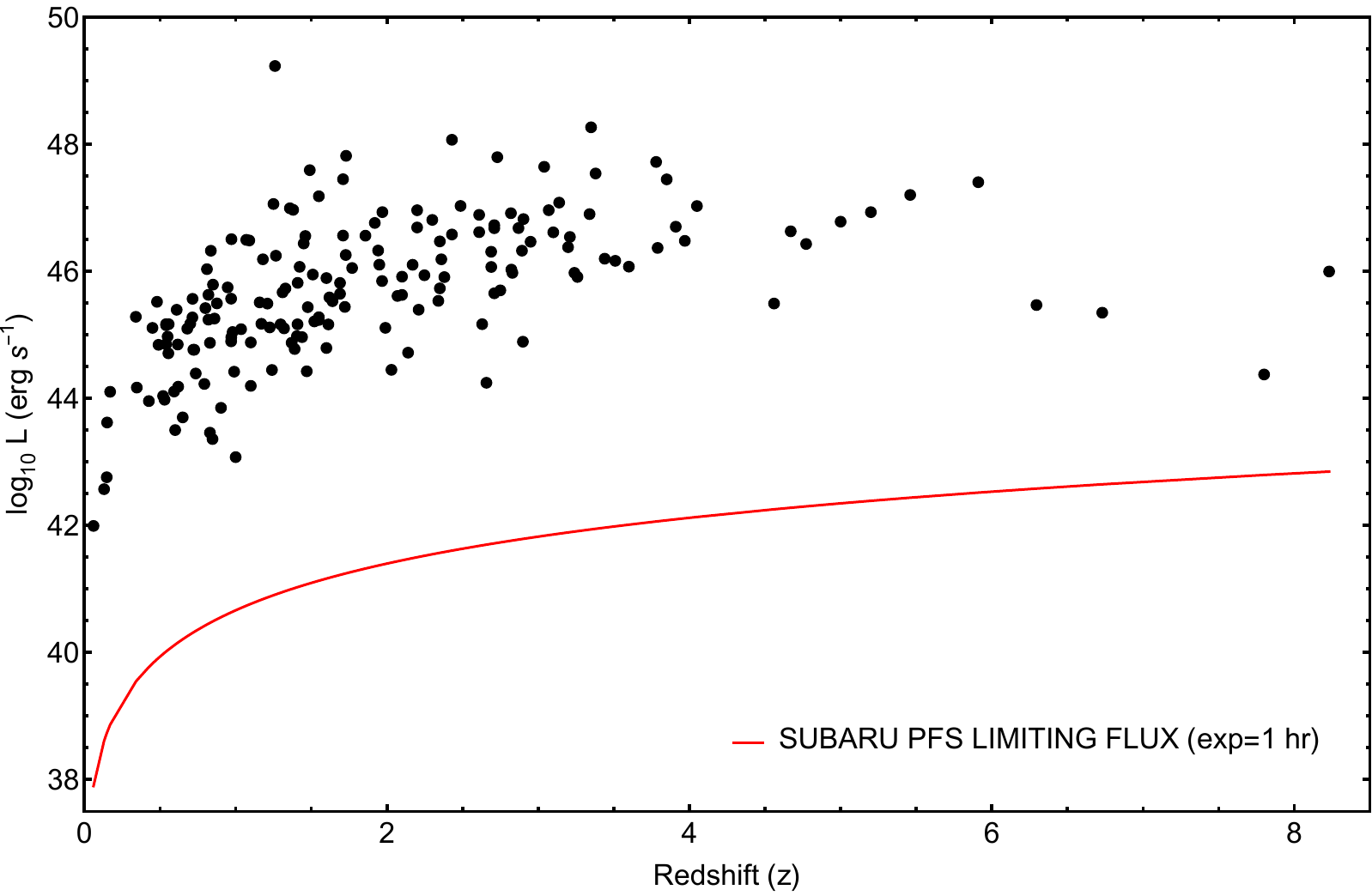}
    \caption{The upper panel shows the differential histogram of $log_{10}Ta$ with the red line corresponding to 3.85 (2 hours) and 4.94 (1 day). The bottom panel is the plot of the luminosities in optical with the current nominal limiting flux of the PFS.}
    \label{Fig: histogram_logTa}
\end{figure}

Our machine learning method is meant to be complementary with other current observing strategies. 
Regarding the follow-up observations that can be tackled within our method, we can pursue them pretty quickly. 
Our method is also complementary and competitive with the VLT-like instruments, since almost all $T_a$ observations are $< 1$ day=$4.94$ seconds on the log scale (see the green vertical line in \ref{Fig: histogram_logTa}. To be precise, only 14 cases out of 251 have $\log T_a> 4.94$. Indeed, this is only $5.5\%$. For clarity, we plot the differential distribution of $T_a$, which ranges from 91 seconds to $6.9 * 10^5$ seconds. In addition, the great advantage of this method is that we can make the prediction regardless of the weather and other conditions, such as GRB being above the horizon, the schedule for other observations, etc. 
Moreover, more than half of GRBs $(58\%)$ have $T_a<2$ hours (see the red line in Fig. \ref{Fig: histogram_logTa}, thus becoming complementary and, in some cases (where the conditions of a given telescope are not available), competitive with smaller aperture telescopes. 
It is important to stress that we do not need to wait several days for having our algorithm working, because $\alpha$, as shown in Fig. \ref{Fig:LASSO} in all $z_{t}$ configurations, is not present as the most influential predictor. 
Most importantly, we also have a configuration in which only prompt emission parameters along with NH are needed; thus, we do not need to wait for the plateau parameters. 
Considering the prompt emission properties, we can be competitive with other diagnostics because we only need variables such as $T_{90}$, the $\log$ Peak Flux, the Photon Index, and $\log$ NH.  Although the large aperture telescopes would wait for several hours, there are still worldwide medium-size and smaller aperture telescopes, privately owned by universities and research centers where they own some time for these follow-up observations, and they can leverage to observe with this opportunity. In addition, if we have bright GRBs at high$-z$, these can be observed with large aperture telescopes even after a few hours of our trigger strategies. Since the Malmquist bias will allow us to see only the brightest GRBs at high$-z$, we will likely observe the most distant ones, which are the most intriguing for the detection of the high$-z$ Universe.

If the wavelength of the $Ly-\alpha$ break at $z > 5$ is more than $730$ nm, observation in the optical band becomes difficult for the follow-up. In these cases, the afterglow is typically detected by telescopes with an aperture of more than 2 meters and a near-infrared detector.

Thus, our tool is important for current and future observations and for helping the community expand the sample at high$-z$, thus working on important questions about the GRB rate in general and population III stars. For example, a possible future mission, THESEUS, which, if approved, will be launched in 2032, will identify more than 20 GRBs at z $>$ 6 in 3.5 years of nominal operation. Hence,  the final sample of high$-z$ GRBs will be much larger than what has been achieved so far \citep{Tanvir_2021}. High$-z$ Gundam mission is comparable to THESEUS (5 events per year at high$-z$, \citep{2023HEAD...2010318Y}. SVOM has already been operational since June 26, 2024, equipped with the 4–150 keV wide-field trigger camera, and Eclair \citep{10.1117/12.2055507} will be able to detect 200 GRB of all types, including also high redshift ones. It is expected to detect around 80 to 100 GRBs annually \citep{Lanza:2024sls, Fortin:2024ofp}, with roughly 4- 5\% being high redshift $(z > 5$) \citep{Wei:2016eox}

The SVOM mission also comprises other ground-based components, such as COLIBRI and CAGIRE \citep{Wei:2016eox}. The latter will begin imaging one minute after the initial trigger from ECLAIRs. If observations are performed in the H band, the detectability of CAGIRE combined with COLIBRI can reach even z=13.3 for $S/N>3$ and they can reach $z=9.6$ if observations are performed in the J band.

Continuing on the ground-based facilities, in February 2025, the Prime Focus Spectrograph (PFS), a cutting-edge ultra-wide-field multi-object spectrograph, will commence open science observations. PFS is designed to simultaneously capture spectra from up to 2,400 objects, significantly enhancing the capabilities of the Subaru Telescope. With this addition, Subaru will achieve a field of view 50 times larger and the ability to observe 20 times more spectroscopic targets at once compared to traditional instruments. These advances will enable the identification of spectroscopic redshifts for a greater number of high-redshift GRBs and facilitate extensive observational campaigns.  We have added to show that all plateaus will be observed with PSF a plot of the maximum sensitivity reached by PSF with one hour of integration, see the lower panel of Fig. \ref{Fig: histogram_logTa}.

To avoid mismatching the observed redshift with another object in the same field of view, increasing both ``The High spatial resolution" and the wide field of view, which are competing elements, is important. To address this challenge, the ``ULTIMATE-Subaru" project is nearing completion, expected at the end of 2020. This instrument will integrate adaptive optics with a field 200 times wider than the conventional systems used by the Subaru Telescope, offering the largest field of view among telescopes of its scale worldwide. With a sensitivity reaching approximately 23 magnitudes in the R band for a 300-second exposure, ULTIMATE-Subaru will also enable the observation of all high-redshift plateau phases in the optical range.

JWST so far has broken the record of the highest galaxy at z=14.32 \citep{highz14galaxy2024Natur.633..318C} even surpassing the previous record at z=13.2 \citep{highzgalaxyRobertson_2023} and in principle can observe galaxies and thus possibly hosting GRBs beyond z=16, this will give the unprecedented opportunity to observe and catch Pop III stars.

Once we have a more extensive training set, we can improve the classification method and refine the most predictive algorithms.

\section{Summary and Conclusion}
\label{sec:conclusion}
In this study, we developed a reliable ensemble ML framework to classify GRBs based on their $z$, distinguishing between high$-$ and low$-z$ GRBs. Our analysis incorporated the prompt, plateau, and afterglow features of 251 GRBs from BAT and XRT instruments aboard the \textit{Swift} satellite. We trained and tested our models on this dataset to develop a procedure to quickly identify newly detected high-$z$ GRBs. This approach facilitates more precise follow-up and extends the available high-$z$ dataset for future research.  In summary, training of the classifier consists of several steps, as shown in the flowchart (Figure \ref{Fig:flowchart}):

\begin{enumerate}
    \item Cleaning the data by removing non-physical quantities and transforming some variables into log base-10.
    \item Removing outliers using the M-estimator method.
    \item Imputing missing values with the MICE technique.
    \item Using the SMOTE technique to balance our dataset.
    \item Applying the LASSO method to steps (1-4) to identify the most relevant predictors from a high-dimensional dataset and reduce the number of predictors because of the small data sample.
    \item Leveraging the SuperLearner to perform 100-fold nested 10fCV on 18 models to select the optimal predictive model based on a desired threshold.
    \item Creating our final ensemble model to perform the 100 nested 10fCV using SuperLearner on the training dataset to maximize the AUC and minimize the difference in AUC between the training and test datasets.
   \item creation of a web-app.
    \item discussion of the use of the algorithm for observing follow-up.
\end{enumerate}

Our study's primary innovation lies in including the plateau phase in the feature set, which has not been included in prior analysis. Our method leveraged an ensemble approach through the SuperLearner algorithm, comprehensively evaluating the various $z_{t}$ and advanced preprocessing techniques. 
Results demonstrate that the $z_{t}=3.5$ offers the best overall performance, achieving an AUC value of 84\% for the raw dataset with M-estimator and a value of 95\% for the balanced dataset, which is similar to previous results \citep{morgan2012,ukwatta2016machine} except for the case of the balance data.
Another advantage of $z_{t} = 3.5$ is the consistently lower percentage difference between the test and training sets compared to other redshifts.
The substantial benefit of this analysis is that our publicly available methodology can be modified and fine-tuned to achieve various levels of sensitivity and specificity depending on the needs of follow-up programs.

We stress that the initial results show a prediction in terms of the metric of the AUC of 76\% when we use raw data with no outlier removal or imputation methods and without balancing the sample. Adding more ML methods and removing outliers lead to an AUC of 80\%, imputing missing variables leads to an AUC of 85\%, and balancing the dataset finally leads to an AUC of 95\% for the classification, an improvement of 19\% compared to our initial analysis, paving the way for new, exciting follow-up observations. 
However, it is essential to note that when we use the balanced data sample, we are not considering the outliers that can occur in the actual observations.

Our findings demonstrate that ensemble machine learning-based methods can significantly advance astrophysics and cosmology. The implications of this work are significant for the broader community, as this robust methodology can pave the way for new and intriguing follow-up observations of high$-z$ GRBs, which will expand the high$-z$ GRB sample with detailed photometric and spectroscopic studies.  Expanding the high$-z$ GRB dataset with known redshifts will enable us to better estimate the luminosity function and density rate evolution, thereby deepening our understanding of the evolution of the high$-z$ universe. With this newly enlarged dataset, we can employ the Dainotti relation \citep{Dainotti2008, dainotti2010, Dainotti11b, Dainotti2013b, Dainotti2015b, 2016ApJ...825L..20D, delvecchio16, Dainotti2017a, 2020ApJ...904...97D, Dainotti2020ApJ, 2022MNRAS.tmp.2639D, Dainotti2022} as a reliable cosmological tool. 
Although we have a precise redshift estimator
developed by \cite{Dainotti2024b, Dainotti_2024, 2024MNRAS.529.2676A}, it is crucial that the classifier has an independent cross-check of our methods and provides the community with more reliable results.

The accuracy and sensitivity of $z-$classification for GRBs are expected to improve as more data becomes available and ML techniques continue to evolve, offering exciting prospects for future cosmological research. This approach can also be applied to other astrophysical objects and datasets, such as AGN, potentially revolutionizing how classification problems within astrophysics are addressed. Our research advances the broader objective of utilizing GRBs as cosmic tools to investigate the early universe. By enabling more precise $z-$classification, we can gain more insights into the high$-z$ universe and its evolution, ultimately contributing to a better understanding of population III stars.

\section{Acknowledgments}
\label{sec:acknowledgments}
This research was supported by Swift GI Cycle XIX, grant No. 22-SWIFT22-0032. This article is based upon work from COST Action CA21136 – ``Addressing observational tensions in cosmology with systematics and fundamental physics (CosmoVerse)", supported by COST (European Cooperation in Science and Technology). NL is grateful for the financial support of the SOKENDAI Graduate Winter School Program training when this paper started. Research at SMU were partially supported by the Dedman College Dean's Research Council via the grant ``Data Science Applications in Theoretical Particle Physics". We are grateful to Mark Ward and the Data Mine at Purdue University for facilitating the collaboration involving SMU participants. AP is grateful to the grant number PLG/2024/017205 and the Polish National Science Centre grant 2023/50/A/ST9/00579. We acknowledge Aditya Narendra for his help with the web app development, Max Ponce-Chavez for mentoring SMU students, and Anekah Kelley and Trevor Knotts for their support in running the very initial analysis. We also acknowledge the supercomputing facility at the O’Donnell Data Science and Research Computing Institute and the Office of Information Technology at Southern Methodist University. SB and MGD are grateful to NAOJ for supporting part of the page charges.

\section{Data Availability}
Data in this paper have been downloaded from the \textit{Swift} BAT+XRT repository. This work made use of data provided by the UK Swift Science Data Centre located at the University of Leicester.

\begin{acknowledgments}
\end{acknowledgments}

%




\appendix
\label{appendix}
\section{Redshift threshold of 2.0}
\label{sec:z2.0}

\subsection{Raw dataset without M-estimator}
\label{sec:raw_z2.0}
The SuperLearner shows a combined AUC value of 0.757, with `lm' and `speedlm' emerging as the individual best-performing algorithm, both with an AUC value of 0.763 and `caret.rpart' as the individual least-performing algorithm, with an AUC value of 0.631 (Figure \ref{fig:combined_raw_z2.0} top left). When we removed the least-performing algorithm (`caret.rpart'), the combined AUC increased marginally from 0.757 to 0.758 (Figure \ref{fig:combined_raw_z2.0} top right).

\subsection{Raw dataset with M-estimator}
\label{sec:raw_Mestimator_z2.0}
We apply the M-estimator to identify and remove the outliers (see Section \ref{sec:Mestimator}). The SuperLearner yields an AUC value of 0.798 on the training dataset. The best-performing individual algorithm is again `lm' and `speedlm', both with an AUC value of 0.799, while `care.rpart' is again the least-performing individual algorithm, with an AUC value of 0.661 (Figure \ref{fig:combined_raw_z2.0} second row left). By removing the least-performing algorithm (`caret.rpart'), the combined AUC value increased from 0.798 to 0.809 (Figure \ref{fig:combined_raw_z2.0} second row right).

\subsection{Imputed dataset with MICE}
\label{sec:MICE_Mestimator_z2.0}
We use MICE imputed data with the M-estimator. 
The combined performance of the SuperLearner on the training dataset resulted in an AUC value of 0.805. The highest-performing individual algorithm is `ranger' with an AUC value of 0.799, while `xgboost' is the least-performing individual algorithm, with an AUC value of 0.726 (Figure \ref{fig:combined_raw_z2.0} third row left). Removing the least-performing algorithm (`xgboost' and `kernelKnn') resulted in an increase in the combined AUC value from 0.805 to 0.811 (Figure \ref{fig:combined_raw_z2.0} third row right). 

\subsection{Balanced dataset with SMOTE}
\label{sec:BALANCED_Mestimator_z2.0}
Given our imbalanced dataset, we applied the SMOTE technique (see Section \ref{sec:databalancing}) to balance the sample for better prediction in terms of AUC. 
The combined performance of the SuperLearner on the training dataset yielded an AUC value of 0.879. The highest-performing individual algorithm is `Random Forest' with an AUC value of 0.882, while `kernelKnn' is the least-performing individual algorithm, with an AUC value of 0.764 (Figure \ref{fig:combined_raw_z2.0} bottom left). We also removed the less-performing algorithms (`kernelKnn', `ksvm', `svm', and `qda'), which resulted in an increased combined AUC value from 0.879 to 0.882 (Figure \ref{fig:combined_raw_z2.0} bottom right).

\section{Redshift threshold of 2.5}
\label{sec:z2.5}
\subsection{Raw dataset without M-estimator}
\label{sec:raw_z2.5}
The SuperLearner's combined performance on the training dataset yielded an AUC of 0.774, with `qda' being the highest-performing individual algorithm (AUC = 0.788), while `caret.rpart' was the lowest-performing individual algorithm (AUC = 0.596) (Figure \ref{fig:combined_raw_z2.5} top left). Upon removing the least performing algorithms (`caret.rpart'), the combined AUC value decreased slightly from 0.774 to 0.772 (Figure \ref{fig:combined_raw_z2.5} top right). This decrease in the combined AUC is not statistically significant (0.002).

\subsection{Raw dataset with M-estimator}
\label{sec:raw_Mestimator_z2.5}
The SuperLearner's combined performance on the training dataset achieved an AUC value of 0.802. Among the individual algorithms, again `qda' performed the best with an AUC value of 0.812, while `svm' showed the lowest performance with an AUC value of 0.675 (Figure \ref{fig:combined_raw_z2.5} second row left). Removing the least performing algorithms (`svm', and `xgboost') led to a slight increase in the combined AUC value from 0.802 to 0.805 (Figure \ref{fig:combined_raw_z2.5} second row right).

\subsection{Imputed dataset with MICE}
\label{sec:MICE_Mestimator_z2.5}
The combined performance of the SuperLearner on the training dataset resulted in an AUC value of 0.792. The top-performing individual algorithms are `lm' and `speedlm', each achieving an AUC value of 0.787, while `kernelKnn' performed the least with an AUC value of 0.738 (Figure \ref{fig:combined_raw_z2.5} third row left). 
We also removed the least performing algorithm (`kernelKnn' and `xgboost'), which resulted in an increased combined AUC value from 0.792 to 0.802 (Figure \ref{fig:combined_raw_z2.5} third row right). 

\subsection{Balanced dataset with SMOTE}
\label{sec:BALANCED_Mestimator_z2.5}
SuperLearner's combined performance on the training dataset achieved an AUC value of 0.907. The best-performing individual algorithm is `Random Forest' with an AUC value of 0.915, while `qda' has the lowest individual AUC value of 0.807 (Figure \ref{fig:combined_raw_z2.5} bottom left). 
Removing the least-performing algorithms, `svm', `qda', `xgboost', and `earth' resulted in a slight increase in the combined AUC from 0.907 to 0.916 (Figure \ref{fig:combined_raw_z2.5} bottom right).

\section{Redshift threshold of 3.0}
\label{sec:z3.0}
\subsection{Raw dataset without M-estimator}
\label{sec:raw_z3.0}
The SuperLearner yielded a combined AUC value of 0.843, with `ranger' emerging as the individual best-performing algorithm with an AUC value of 0.833 and `svm' as the individual least-performing algorithm with AUC = 0.784 (Figure \ref{fig:combined_raw_z3.0} top left). 
When we removed the least-performing algorithms (`svm' and `ksvm'), the combined AUC increased from 0.843 to 0.846 (Figure \ref{fig:combined_raw_z3.0} top right).

\subsection{Raw dataset with M-estimator}
\label{sec:raw_Mestimator_z3.0}
The SuperLearner's combined performance on the training dataset yielded an AUC value of 0.840. The best-performing individual algorithm is again `kernelKnn' with an AUC value of 0.839, while `caret.rpart' is the least-performing individual algorithm, with an AUC value of 0.719 (Figure \ref{fig:combined_raw_z3.0} second row left). 
By removing the least performing algorithms (`caret.rpart'), the combined AUC value increased from 0.840 to 0.844 (Figure \ref{fig:combined_raw_z3.0} second row right).

\subsection{Imputed dataset with MICE}
\label{sec:MICE_Mestimator_z3.0}
The combined performance of the SuperLearner on the training dataset resulted in an AUC value of 0.848. The highest-performing individual algorithm is `cforest' with an AUC value of 0.826, while `ksvm' is the least-performing individual algorithm, with an AUC value of 0.786 (Figure \ref{fig:combined_raw_z3.0} third row left).  
Removing the least-performing algorithm (`ksvm') resulted in an increase in the combined AUC value from 0.848 to 0.850 (Figure \ref{fig:combined_raw_z3.0} third row right).

\subsection{Balanced dataset with SMOTE}
\label{sec:BALANCED_Mestimator_z3.0}
The combined performance of the SuperLearner on the training dataset yielded an AUC value of 0.950. The highest-performing individual algorithm is `Random Forest' with an AUC value of 0.951, while `ksvm' is the least-performing individual algorithm, with an AUC value of 0.863 (Figure \ref{fig:combined_raw_z3.0} bottom left). 
We removed the less-performing algorithms (`ksvm' and `svm'), which resulted in an increased combined AUC value from 0.950 to 0.951 (Figure \ref{fig:combined_raw_z3.0} bottom right).

\section{Redshift threshold of 3.5}
\label{sec:z3.5}
\subsection{Raw dataset without M-estimator}
\label{sec:raw_z3.5}
The SuperLearner's combined performance on the training dataset yielded an AUC of 0.820, with `Random Forest' being the highest-performing individual algorithm with AUC = 0.827, while `ksvm', the lowest-performing individual algorithm (AUC = 0.731) (Figure \ref{fig:combined_raw_z3.5} top left). 
Upon removing the least performing algorithm (`ksvm'), the combined AUC value improved from 0.820 to 0.826 (Figure \ref{fig:combined_raw_z3.5} top right).

\subsection{Raw dataset with M-estimator}
\label{sec:raw_Mestimator_z3.5}
The SuperLearner's combined performance on the training dataset achieved an AUC value of 0.843. Among the individual algorithms, both `lm' and `speedlm' performed the best with an AUC value of 0.834, while `svm' showed the lowest performance with an AUC value of 0.779 (Figure \ref{fig:combined_raw_z3.5} second row left). 
Removing the least performing algorithm (`svm') led to a slight decrease in the combined AUC value from 0.843 to 0.840 (Figure \ref{fig:combined_raw_z3.5} second row right).

\subsection{Imputed dataset with MICE}
\label{sec:MICE_Mestimator_z3.5}
The combined performance of the SuperLearner on the training dataset resulted in an AUC value of 0.820. The top-performing individual algorithm is `speedlm', achieving an AUC value of 0.816, while `svm' remains the least-performing algorithm with an AUC value of 0.737 (Figure \ref{fig:combined_raw_z3.5} third row left). 
We also removed the least performing algorithm (`svm'), which resulted in a slight decrease in the combined AUC value from 0.820 to 0.816 (Figure \ref{fig:combined_raw_z3.5} third row right).

\subsection{Balanced dataset with SMOTE}
\label{sec:BALANCED_Mestimator_z3.5}
SuperLearner's combined performance on the training dataset achieved an AUC value of 0.955. The best-performing individual algorithm is `ranger' with an AUC value of 0.954, while `qda' has the lowest individual AUC value of 0.815 (Figure \ref{fig:combined_raw_z3.5} bottom left). 
Removing the least-performing algorithms, `qda', `svm', and `xgboost', does not change the combined AUC (Figure \ref{fig:combined_raw_z3.5} bottom right).

\bibliography{sample631}{}

\begin{thebibliography}{}
\expandafter\ifx\csname natexlab\endcsname\relax\def\natexlab#1{#1}\fi
\providecommand{\url}[1]{\href{#1}{#1}}
\providecommand{\dodoi}[1]{doi:~\href{http://doi.org/#1}{\nolinkurl{#1}}}
\providecommand{\doeprint}[1]{\href{http://ascl.net/#1}{\nolinkurl{http://ascl.net/#1}}}
\providecommand{\doarXiv}[1]{\href{https://arxiv.org/abs/#1}{\nolinkurl{https://arxiv.org/abs/#1}}}

\bibitem[{{Aldowma} \& {Razzaque}(2024)}]{2024MNRAS.529.2676A}
{Aldowma}, T., \& {Razzaque}, S. 2024, \mnras, 529, 2676, \dodoi{10.1093/mnras/stae535}

\bibitem[{Altman \& Bland(1994)}]{altman1994diagnostic}
Altman, D.~G., \& Bland, J.~M. 1994, BMJ: British Medical Journal, 308, 1552

\bibitem[{{Amati} {et~al.}(2018){Amati}, {O'Brien}, {G{\"o}tz}, {Bozzo}, {Tenzer}, {Frontera}, {Ghirlanda}, {Labanti}, {Osborne}, {Stratta}, {Tanvir}, {Willingale}, {Attina}, {Campana}, {Castro-Tirado}, {Contini}, {Fuschino}, {Gomboc}, {Hudec}, {Orleanski}, {Renotte}, {Rodic}, {Bagoly}, {Blain}, {Callanan}, {Covino}, {Ferrara}, {Le Floch}, {Marisaldi}, {Mereghetti}, {Rosati}, {Vacchi}, {D'Avanzo}, {Giommi}, {Piranomonte}, {Piro}, {Reglero}, {Rossi}, {Santangelo}, {Salvaterra}, {Tagliaferri}, {Vergani}, {Vinciguerra}, {Briggs}, {Campolongo}, {Ciolfi}, {Connaughton}, {Cordier}, {Morelli}, {Orlandini}, {Adami}, {Argan}, {Atteia}, {Auricchio}, {Balazs}, {Baldazzi}, {Basa}, {Basak}, {Bellutti}, {Bernardini}, {Bertuccio}, {Braga}, {Branchesi}, {Brandt}, {Brocato}, {Budtz-Jorgensen}, {Bulgarelli}, {Burderi}, {Camp}, {Capozziello}, {Caruana}, {Casella}, {Cenko}, {Chardonnet}, {Ciardi}, {Colafrancesco}, {Dainotti}, {D'Elia}, {De Martino}, {De Pasquale}, {Del Monte}, {Della Valle}, {Drago}, {Evangelista}, {Feroci},
  {Finelli}, {Fiorini}, {Fynbo}, {Gal-Yam}, {Gendre}, {Ghisellini}, {Grado}, {Guidorzi}, {Hafizi}, {Hanlon}, {Hjorth}, {Izzo}, {Kiss}, {Kumar}, {Kuvvetli}, {Lavagna}, {Li}, {Longo}, {Lyutikov}, {Maio}, {Maiorano}, {Malcovati}, {Malesani}, {Margutti}, {Martin-Carrillo}, {Masetti}, {McBreen}, {Mignani}, {Morgante}, {Mundell}, {Nargaard-Nielsen}, {Nicastro}, {Palazzi}, {Paltani}, {Panessa}, {Pareschi}, {Pe'er}, {Penacchioni}, {Pian}, {Piedipalumbo}, {Piran}, {Rauw}, {Razzano}, {Read}, {Rezzolla}, {Romano}, {Ruffini}, {Savaglio}, {Sguera}, {Schady}, {Skidmore}, {Song}, {Stanway}, {Starling}, {Topinka}, {Troja}, {van Putten}, {Vanzella}, {Vercellone}, {Wilson-Hodge}, {Yonetoku}, {Zampa}, {Zampa}, {Zhang}, {Zhang}, {Zhang}, {Zhang}, {Antonelli}, {Bianco}, {Boci}, {Boer}, {Botticella}, {Boulade}, {Butler}, {Campana}, {Capitanio}, {Celotti}, {Chen}, {Colpi}, {Comastri}, {Cuby}, {Dadina}, {De Luca}, {Dong}, {Ettori}, {Gandhi}, {Geza}, {Greiner}, {Guiriec}, {Harms}, {Hernanz}, {Hornstrup}, {Hutchinson}, {Israel},
  {Jonker}, {Kaneko}, {Kawai}, {Wiersema}, {Korpela}, {Lebrun}, {Lu}, {MacFadyen}, {Malaguti}, {Maraschi}, {Melandri}, {Modjaz}, {Morris}, {Omodei}, {Paizis}, {P{\'a}ta}, {Petrosian}, {Rachevski}, {Rhoads}, {Ryde}, {Sabau-Graziati}, {Shigehiro}, {Sims}, {Soomin}, {Sz{\'e}csi}, {Urata}, {Uslenghi}, {Valenziano}, {Vianello}, {Vojtech}, {Watson}, \& {Zicha}}]{2018AdSpR..62..191A}
{Amati}, L., {O'Brien}, P., {G{\"o}tz}, D., {et~al.} 2018, Advances in Space Research, 62, 191, \dodoi{10.1016/j.asr.2018.03.010}

\bibitem[{Anderson \& Darling(1954)}]{AD-1}
Anderson, T.~W., \& Darling, D.~A. 1954, Journal of the American Statistical Association, 49, 765, \dodoi{10.1080/01621459.1954.10501232}

\bibitem[{Anderson \& Darling(2008)}]{AD}
---. 2008, Anderson--Darling Test (New York, NY: Springer New York), 12--14, \dodoi{10.1007/978-0-387-32833-1_11}

\bibitem[{{Aptekar} {et~al.}(1995){Aptekar}, {Frederiks}, {Golenetskii}, {Ilynskii}, {Mazets}, {Panov}, {Sokolova}, {Terekhov}, {Sheshin}, {Cline}, \& {Stilwell}}]{Konus1995}
{Aptekar}, R.~L., {Frederiks}, D.~D., {Golenetskii}, S.~V., {et~al.} 1995, \ssr, 71, 265, \dodoi{10.1007/BF00751332}

\bibitem[{Arumaningtyas {et~al.}(2024{\natexlab{a}})Arumaningtyas, Al~Rasyid, Dainotti, \& Yonetoku}]{galaxies12050051}
Arumaningtyas, E.~P., Al~Rasyid, H., Dainotti, M.~G., \& Yonetoku, D. 2024{\natexlab{a}}, Galaxies, 12, \dodoi{10.3390/galaxies12050051}

\bibitem[{Arumaningtyas {et~al.}(2024{\natexlab{b}})Arumaningtyas, Al~Rasyid, Dainotti, \& Yonetoku}]{Arumaningtyas2024}
---. 2024{\natexlab{b}}, Galaxies, 12, \dodoi{10.3390/galaxies12050051}

\bibitem[{{Barthelmy} {et~al.}(2005){Barthelmy}, {Chincarini}, {Burrows}, {Gehrels}, {Covino}, {Moretti}, {Romano}, {O'Brien}, {Sarazin}, {Kouveliotou}, {Goad}, {Vaughan}, {Tagliaferri}, {Zhang}, {Antonelli}, {Campana}, {Cummings}, {D'Avanzo}, {Davies}, {Giommi}, {Grupe}, {Kaneko}, {Kennea}, {King}, {Kobayashi}, {Melandri}, {Meszaros}, {Nousek}, {Patel}, {Sakamoto}, \& {Wijers}}]{Barthelmy2005}
{Barthelmy}, S.~D., {Chincarini}, G., {Burrows}, D.~N., {et~al.} 2005, Nature, 438, 994, \dodoi{10.1038/nature04392}

\bibitem[{Bernardini {et~al.}(2021)Bernardini, Cordier, \& Wei}]{bernardini2021svom}
Bernardini, M.~G., Cordier, B., \& Wei, J. 2021, Galaxies, 9, 113

\bibitem[{{Beskin} {et~al.}(2010){Beskin}, {Karpov}, {Bondar}, {Greco}, {Guarnieri}, {Bartolini}, \& {Piccioni}}]{Beskin2010ApJ}
{Beskin}, G., {Karpov}, S., {Bondar}, S., {et~al.} 2010, ApJL, 719, L10, \dodoi{10.1088/2041-8205/719/1/L10}

\bibitem[{Bhardwaj {et~al.}(2023)Bhardwaj, Dainotti, Venkatesh, Narendra, Kalsi, Rinaldi, \& Pollo}]{10.1093/mnras/stad2593}
Bhardwaj, S., Dainotti, M.~G., Venkatesh, S., {et~al.} 2023, Monthly Notices of the Royal Astronomical Society, 525, 5204, \dodoi{10.1093/mnras/stad2593}

\bibitem[{Birnbaum(1962)}]{birnbaum1962foundations}
Birnbaum, A. 1962, JASA, 57, 269

\bibitem[{{Bolmer, J.} {et~al.}(2018){Bolmer, J.}, {Greiner, J.}, {Krühler, T.}, {Schady, P.}, {Ledoux, C.}, {Tanvir, N. R.}, \& {Levan, A. J.}}]{bolmer2018}
{Bolmer, J.}, {Greiner, J.}, {Krühler, T.}, {et~al.} 2018, A\&A, 609, A62, \dodoi{10.1051/0004-6361/201731255}

\bibitem[{Bomani(2021)}]{Bomani2021}
Bomani, B. 2021, NASA

\bibitem[{Breiman(2001)}]{breiman2001randomforest}
Breiman, L. 2001, Machine Learning, 45, 5, \dodoi{10.1023/A:1010933404324}

\bibitem[{{Breiman} {et~al.}(1984){Breiman}, {Friedman}, {Olshen}, \& {Stone}}]{caret-rpart}
{Breiman}, L., {Friedman}, J., {Olshen}, R., \& {Stone}, C. 1984, Classification and Regression Trees (Chapman and Hall/CRC), \dodoi{https://doi.org/10.1201/9781315139470}

\bibitem[{Bromm(2006)}]{Bromm:2005ep}
Bromm, V. 2006, Astrophys. J., 642, 382, \dodoi{10.1086/500799}

\bibitem[{Brown {et~al.}(2009)Brown, Tauler, \& Walczak}]{Brown2009}
Brown, S.~D., Tauler, R., \& Walczak, B., eds. 2009, Comprehensive Chemometrics: Chemical and Biochemical Data Analysis (Elsevier), \dodoi{10.1016/B978-0-444-64163-0.X0001-7}

\bibitem[{Bunke \& Varga(2007)}]{Bunke2007}
Bunke, H., \& Varga, T. 2007, Off-Line Roman Cursive Handwriting Recognition (London: Springer London), 165--183, \dodoi{10.1007/978-1-84628-726-8_8}

\bibitem[{{Burrows} {et~al.}(2005){Burrows}, {Hill}, {Nousek}, {Kennea}, {Wells}, {Osborne}, {Abbey}, {Beardmore}, {Mukerjee}, {Short}, {Chincarini}, {Campana}, {Citterio}, {Moretti}, {Pagani}, {Tagliaferri}, {Giommi}, {Capalbi}, {Tamburelli}, {Angelini}, {Cusumano}, {Br{\"a}uninger}, {Burkert}, \& {Hartner}}]{2005SSRv..120..165B}
{Burrows}, D.~N., {Hill}, J.~E., {Nousek}, J.~A., {et~al.} 2005, SSRv, 120, 165, \dodoi{10.1007/s11214-005-5097-2}

\bibitem[{{Campana} {et~al.}(2007){Campana}, {Tagliaferri}, {Malesani}, {Stella}, {D'Avanzo}, {Chincarini}, \& {Covino}}]{2007A&A...464L..25C}
{Campana}, S., {Tagliaferri}, G., {Malesani}, D., {et~al.} 2007, \aap, 464, L25, \dodoi{10.1051/0004-6361:20066592}

\bibitem[{Campana {et~al.}(2010)Campana, Thöne, de~Ugarte~Postigo, Tagliaferri, Moretti, \& Covino}]{Campana2010}
Campana, S., Thöne, C.~C., de~Ugarte~Postigo, A., {et~al.} 2010, Monthly Notices of the Royal Astronomical Society, 402, 2429, \dodoi{10.1111/j.1365-2966.2009.16006.x}

\bibitem[{{Campisi} {et~al.}(2011){Campisi}, {Maio}, {Salvaterra}, \& {Ciardi}}]{2011MNRAS.416.2760C}
{Campisi}, M.~A., {Maio}, U., {Salvaterra}, R., \& {Ciardi}, B. 2011, \mnras, 416, 2760, \dodoi{10.1111/j.1365-2966.2011.19238.x}

\bibitem[{{Carniani} {et~al.}(2024){Carniani}, {Hainline}, {D'Eugenio}, {Eisenstein}, {Jakobsen}, {Witstok}, {Johnson}, {Chevallard}, {Maiolino}, {Helton}, {Willott}, {Robertson}, {Alberts}, {Arribas}, {Baker}, {Bhatawdekar}, {Boyett}, {Bunker}, {Cameron}, {Cargile}, {Charlot}, {Curti}, {Curtis-Lake}, {Egami}, {Giardino}, {Isaak}, {Ji}, {Jones}, {Kumari}, {Maseda}, {Parlanti}, {P{\'e}rez-Gonz{\'a}lez}, {Rawle}, {Rieke}, {Rieke}, {Del Pino}, {Saxena}, {Scholtz}, {Smit}, {Sun}, {Tacchella}, {{\"U}bler}, {Venturi}, {Williams}, \& {Willmer}}]{highz14galaxy2024Natur.633..318C}
{Carniani}, S., {Hainline}, K., {D'Eugenio}, F., {et~al.} 2024, \nat, 633, 318, \dodoi{10.1038/s41586-024-07860-9}

\bibitem[{{Cen} \& {Kimm}(2014)}]{2014ApJ...794...50C}
{Cen}, R., \& {Kimm}, T. 2014, \apj, 794, 50, \dodoi{10.1088/0004-637X/794/1/50}

\bibitem[{{Chawla} {et~al.}(2011){Chawla}, {Bowyer}, {Hall}, \& {Kegelmeyer}}]{2011arXiv1106.1813C}
{Chawla}, N.~V., {Bowyer}, K.~W., {Hall}, L.~O., \& {Kegelmeyer}, W.~P. 2011, arXiv e-prints, arXiv:1106.1813, \dodoi{10.48550/arXiv.1106.1813}

\bibitem[{Chen {et~al.}(2007)Chen, Prochaska, \& Gnedin}]{Chen_2007}
Chen, H.-W., Prochaska, J.~X., \& Gnedin, N.~Y. 2007, The Astrophysical Journal, 667, L125–L128, \dodoi{10.1086/522306}

\bibitem[{Chen \& Guestrin(2016)}]{chen2016xgboost}
Chen, T., \& Guestrin, C. 2016, in Proceedings of the 22nd acm sigkdd international conference on knowledge discovery and data mining, 785--794

\bibitem[{{Chincarini} {et~al.}(2010){Chincarini}, {Mao}, {Margutti}, {Bernardini}, {Guidorzi}, {Pasotti}, {Giannios}, {Della Valle}, {Moretti}, {Romano}, {D'Avanzo}, {Cusumano}, \& {Giommi}}]{chincarini2010MNRAS.406.2113C}
{Chincarini}, G., {Mao}, J., {Margutti}, R., {et~al.} 2010, MNRAS, 406, 2113, \dodoi{10.1111/j.1365-2966.2010.17037.x}

\bibitem[{{Cordier} {et~al.}(2015){Cordier}, {Wei}, {Atteia}, {Basa}, {Claret}, {Daigne}, {Deng}, {Dong}, {Godet}, {Goldwurm}, {G{\"o}tz}, {Han}, {Klotz}, {Lachaud}, {Osborne}, {Qiu}, {Schanne}, {Wu}, {Wang}, {Wu}, {Xin}, {Zhang}, \& {Zhang}}]{2015arXiv151203323C}
{Cordier}, B., {Wei}, J., {Atteia}, J.~L., {et~al.} 2015, arXiv e-prints, arXiv:1512.03323, \dodoi{10.48550/arXiv.1512.03323}

\bibitem[{{Corre} {et~al.}(2018){Corre}, {Buat}, {Basa}, {Boissier}, {Japelj}, {Palmerio}, {Salvaterra}, {Vergani}, \& {Zafar}}]{2018A&A...617A.141C}
{Corre}, D., {Buat}, V., {Basa}, S., {et~al.} 2018, \aap, 617, A141, \dodoi{10.1051/0004-6361/201832926}

\bibitem[{Cortes \& Vapnik(1995)}]{cortes1995support}
Cortes, C., \& Vapnik, V. 1995, Machine learning, 20, 273, \dodoi{10.1007/BF00994018}

\bibitem[{{Costa} {et~al.}(1997){Costa}, {Frontera}, {Heise}, {Feroci}, {in't Zand}, {Fiore}, {Cinti}, {Dal Fiume}, {Nicastro}, {Orlandini}, {Palazzi}, {Rapisarda\#}, {Zavattini}, {Jager}, {Parmar}, {Owens}, {Molendi}, {Cusumano}, {Maccarone}, {Giarrusso}, {Coletta}, {Antonelli}, {Giommi}, {Muller}, {Piro}, \& {Butler}}]{costa1997}
{Costa}, E., {Frontera}, F., {Heise}, J., {et~al.} 1997, Nature, 387, 783, \dodoi{10.1038/42885}

\bibitem[{{Cucchiara} {et~al.}(2011){Cucchiara}, {Levan}, {Fox}, {Tanvir}, {Ukwatta}, {Berger}, {Kr{\"u}hler}, {K{\"u}pc{\"u} Yolda{\c{s}}}, {Wu}, {Toma}, {Greiner}, {Olivares}, {Rowlinson}, {Amati}, {Sakamoto}, {Roth}, {Stephens}, {Fritz}, {Fynbo}, {Hjorth}, {Malesani}, {Jakobsson}, {Wiersema}, {O'Brien}, {Soderberg}, {Foley}, {Fruchter}, {Rhoads}, {Rutledge}, {Schmidt}, {Dopita}, {Podsiadlowski}, {Willingale}, {Wolf}, {Kulkarni}, \& {D'Avanzo}}]{2011ApJ...736....7C}
{Cucchiara}, A., {Levan}, A.~J., {Fox}, D.~B., {et~al.} 2011, ApJ, 736, 7, \dodoi{10.1088/0004-637X/736/1/7}

\bibitem[{Cucchiara {et~al.}(2011)Cucchiara, Levan, Fox, Tanvir, Ukwatta, Berger, Krühler, Yoldaş, Wu, Toma, Greiner, E.~Olivares, Rowlinson, Amati, Sakamoto, Roth, Stephens, Fritz, Fynbo, Hjorth, Malesani, Jakobsson, Wiersema, O’Brien, Soderberg, Foley, Fruchter, Rhoads, Rutledge, Schmidt, Dopita, Podsiadlowski, Willingale, Wolf, Kulkarni, \& D’Avanzo}]{Cucchiara_2011}
Cucchiara, A., Levan, A.~J., Fox, D.~B., {et~al.} 2011, The Astrophysical Journal, 736, 7, \dodoi{10.1088/0004-637x/736/1/7}

\bibitem[{{Cusumano} {et~al.}(2006){Cusumano}, {Mangano}, {Chincarini}, {Panaitescu}, {Burrows}, {La Parola}, {Sakamoto}, {Campana}, {Mineo}, {Tagliaferri}, {Angelini}, {Barthelemy}, {Beardmore}, {Boyd}, {Cominsky}, {Gronwall}, {Fenimore}, {Gehrels}, {Giommi}, {Goad}, {Hurley}, {Kennea}, {Mason}, {Marshall}, {M{\'e}sz{\'a}ros}, {Nousek}, {Osborne}, {Palmer}, {Roming}, {Wells}, {White}, \& {Zhang}}]{cusumano2006}
{Cusumano}, G., {Mangano}, V., {Chincarini}, G., {et~al.} 2006, Nature, 440, 164, \dodoi{10.1038/440164a}

\bibitem[{{Cusumano} {et~al.}(2007){Cusumano}, {Mangano}, {Chincarini}, {Panaitescu}, {Burrows}, {La Parola}, {Sakamoto}, {Campana}, {Mineo}, {Tagliaferri}, {Angelini}, {Barthelmy}, {Beardmore}, {Boyd}, {Cominsky}, {Gronwall}, {Fenimore}, {Gehrels}, {Giommi}, {Goad}, {Hurley}, {Immler}, {Kennea}, {Mason}, {Marshal}, {M{\'e}sz{\'a}ros}, {Nousek}, {Osborne}, {Palmer}, {Roming}, {Wells}, {White}, \& {Zhang}}]{cusumano2007}
---. 2007, Astronomy \& Astrophysics, 462, 73, \dodoi{10.1051/0004-6361:20065173}

\bibitem[{{Dainotti} {et~al.}(2015){Dainotti}, {Petrosian}, {Willingale}, {O'Brien}, {Ostrowski}, \& {Nagataki}}]{Dainotti2015b}
{Dainotti}, M., {Petrosian}, V., {Willingale}, R., {et~al.} 2015, \mnras, 451, 3898, \dodoi{10.1093/mnras/stv1229}

\bibitem[{{Dainotti} {et~al.}(2008){Dainotti}, {Cardone}, \& {Capozziello}}]{Dainotti2008}
{Dainotti}, M.~G., {Cardone}, V.~F., \& {Capozziello}, S. 2008, MNRAS, 391, L79, \dodoi{10.1111/j.1745-3933.2008.00560.x}

\bibitem[{{Dainotti} {et~al.}(2022{\natexlab{a}}){Dainotti}, {Lenart}, {Chraya}, {Sarracino}, {Nagataki}, {Fraija}, {Capozziello}, \& {Bogdan}}]{2022MNRAS.tmp.2639D}
{Dainotti}, M.~G., {Lenart}, A.~L., {Chraya}, A., {et~al.} 2022{\natexlab{a}}, MNRAS, \dodoi{10.1093/mnras/stac2752}

\bibitem[{{Dainotti} {et~al.}(2020{\natexlab{a}}){Dainotti}, {Lenart}, {Sarracino}, {Nagataki}, {Capozziello}, \& {Fraija}}]{2020ApJ...904...97D}
{Dainotti}, M.~G., {Lenart}, A.~{\L}., {Sarracino}, G., {et~al.} 2020{\natexlab{a}}, \apj, 904, 97, \dodoi{10.3847/1538-4357/abbe8a}

\bibitem[{{Dainotti} {et~al.}(2017){Dainotti}, {Nagataki}, {Maeda}, {Postnikov}, \& {Pian}}]{Dainotti2017a}
{Dainotti}, M.~G., {Nagataki}, S., {Maeda}, K., {Postnikov}, S., \& {Pian}, E. 2017, A\&A, 600, A98, \dodoi{10.1051/0004-6361/201628384}

\bibitem[{{Dainotti} {et~al.}(2011){Dainotti}, {Ostrowski}, \& {Willingale}}]{Dainotti11b}
{Dainotti}, M.~G., {Ostrowski}, M., \& {Willingale}, R. 2011, \mnras, 418, 2202, \dodoi{10.1111/j.1365-2966.2011.19433.x}

\bibitem[{{Dainotti} {et~al.}(2013){Dainotti}, {Petrosian}, {Singal}, \& {Ostrowski}}]{Dainotti2013b}
{Dainotti}, M.~G., {Petrosian}, V., {Singal}, J., \& {Ostrowski}, M. 2013, \apj, 774, 157, \dodoi{10.1088/0004-637X/774/2/157}

\bibitem[{{Dainotti} {et~al.}(2016){Dainotti}, {Postnikov}, {Hernandez}, \& {Ostrowski}}]{2016ApJ...825L..20D}
{Dainotti}, M.~G., {Postnikov}, S., {Hernandez}, X., \& {Ostrowski}, M. 2016, ApJL, 825, L20, \dodoi{10.3847/2041-8205/825/2/L20}

\bibitem[{{Dainotti} {et~al.}(2010){Dainotti}, {Willingale}, {Capozziello}, {Fabrizio Cardone}, \& {Ostrowski}}]{dainotti2010}
{Dainotti}, M.~G., {Willingale}, R., {Capozziello}, S., {Fabrizio Cardone}, V., \& {Ostrowski}, M. 2010, ApJL, 722, L215, \dodoi{10.1088/2041-8205/722/2/L215}

\bibitem[{{Dainotti} {et~al.}(2020{\natexlab{b}}){Dainotti}, {Livermore}, {Kann}, {Li}, {Oates}, {Yi}, {Zhang}, {Gendre}, {Cenko}, \& {Fraija}}]{dainotti2020b}
{Dainotti}, M.~G., {Livermore}, S., {Kann}, D.~A., {et~al.} 2020{\natexlab{b}}, ApJL, 905, L26, \dodoi{10.3847/2041-8213/abcda9}

\bibitem[{{Dainotti} {et~al.}(2020{\natexlab{c}}){Dainotti}, {Livermore}, {Kann}, {Li}, {Oates}, {Yi}, {Zhang}, {Gendre}, {Cenko}, \& {Fraija}}]{Dainotti2020ApJ}
---. 2020{\natexlab{c}}, \apjl, 905, L26, \dodoi{10.3847/2041-8213/abcda9}

\bibitem[{{Dainotti} {et~al.}(2021){Dainotti}, {Omodei}, {Srinivasaragavan}, {Vianello}, {Willingale}, {O'Brien}, {Nagataki}, {Petrosian}, {Nuygen}, {Hernandez}, {Axelsson}, {Bissaldi}, \& {Longo}}]{2021ApJS..255...13D}
{Dainotti}, M.~G., {Omodei}, N., {Srinivasaragavan}, G.~P., {et~al.} 2021, ApJS, 255, 13, \dodoi{10.3847/1538-4365/abfe17}

\bibitem[{Dainotti {et~al.}(2021)Dainotti, Bogdan, Narendra, Gibson, Miasojedow, Liodakis, Pollo, Nelson, Wozniak, Nguyen, \& Larrson}]{dainotti2021predicting}
Dainotti, M.~G., Bogdan, M., Narendra, A., {et~al.} 2021, ApJ, 920, 118, \dodoi{10.3847/1538-4357/ac1748}

\bibitem[{{Dainotti} {et~al.}(2022{\natexlab{b}}){Dainotti}, {Young}, {Li}, {Levine}, {Kalinowski}, {Kann}, {Tran}, {Zambrano-Tapia}, {Zambrano-Tapia}, {Cenko}, {Fuentes}, {S{\'a}nchez-V{\'a}zquez}, {Oates}, {Fraija}, {Becerra}, {Watson}, {Butler}, {Gonz{\'a}lez}, {Kutyrev}, {Lee}, {Prochaska}, {Ramirez-Ruiz}, {Richer}, \& {Zola}}]{Dainotti2022}
{Dainotti}, M.~G., {Young}, S., {Li}, L., {et~al.} 2022{\natexlab{b}}, \apjs, 261, 25, \dodoi{10.3847/1538-4365/ac7c64}

\bibitem[{Dainotti {et~al.}(2024{\natexlab{a}})Dainotti, Narendra, Pollo, Petrosian, Bogdan, Iwasaki, Prochaska, Rinaldi, \& Zhou}]{Dainotti2024b}
Dainotti, M.~G., Narendra, A., Pollo, A., {et~al.} 2024{\natexlab{a}}, The Astrophysical Journal, \dodoi{0.3847/2041-8213/ad4970}

\bibitem[{Dainotti {et~al.}(2024{\natexlab{b}})Dainotti, Taira, Wang, Lehman, Narendra, Pollo, Madejski, Petrosian, Bogdan, Dey, \& Bhardwaj}]{Dainotti_2024}
Dainotti, M.~G., Taira, E., Wang, E., {et~al.} 2024{\natexlab{b}}, The Astrophysical Journal Supplement Series, 271, 22, \dodoi{10.3847/1538-4365/ad1aaf}

\bibitem[{{Dainotti} {et~al.}(2024){Dainotti}, {De Simone}, {Malik}, {Pasumarti}, {Levine}, {Saha}, {Gendre}, {Kido}, {Watson}, {Becerra}, {Belkin}, {Desai}, {do E S Pedreira}, {Das}, {Li}, {Oates}, {Cenko}, {Pozanenko}, {Volnova}, {Hu}, {Castro-Tirado}, {Orange}, {Moriya}, {Fraija}, {Niino}, {Rinaldi}, {Butler}, {Gonz{\'a}lez}, {Kutyrev}, {Lee}, {Prochaska}, {Ramirez-Ruiz}, {Richer}, {Siegel}, {Misra}, {Rossi}, {Lopresti}, {Quadri}, {Strabla}, {Ruocco}, {Leonini}, {Conti}, {Rosi}, {Ramirez}, {Zola}, {Jindal}, {Kumar}, {Chan}, {Fuentes}, {Lambiase}, {Kalinowski}, \& {Jamal}}]{2024MNRAS.tmp.1527D}
{Dainotti}, M.~G., {De Simone}, B., {Malik}, R.~F.~M., {et~al.} 2024, \mnras, \dodoi{10.1093/mnras/stae1484}

\bibitem[{{Dalton} \& {Morris}(2020)}]{2020MNRAS.495.2342D}
{Dalton}, T., \& {Morris}, S.~L. 2020, \mnras, 495, 2342, \dodoi{10.1093/mnras/staa1321}

\bibitem[{{de Menezes} {et~al.}(2021){de Menezes}, Prata, Secchi, \& Pinto}]{DEMENEZES2021107254}
{de Menezes}, D., Prata, D., Secchi, A., \& Pinto, J. 2021, Computers \& Chemical Engineering, 147, 107254, \dodoi{https://doi.org/10.1016/j.compchemeng.2021.107254}

\bibitem[{{Del Vecchio} {et~al.}(2016){Del Vecchio}, {Dainotti}, \& {Ostrowski}}]{delvecchio16}
{Del Vecchio}, R., {Dainotti}, M.~G., \& {Ostrowski}, M. 2016, \apj, 828, 36, \dodoi{10.3847/0004-637X/828/1/36}

\bibitem[{Dong {et~al.}(2022)Dong, Li, Zhang, \& Zhang}]{Dong:2021wdc}
Dong, X.~F., Li, X.~J., Zhang, Z.~B., \& Zhang, X.~L. 2022, Mon. Not. Roy. Astron. Soc., 513, 1078, \dodoi{10.1093/mnras/stac949}

\bibitem[{Efron \& Petrosian(1992)}]{Efron1992}
Efron, B., \& Petrosian, V. 1992, ApJ, 399, 345, \dodoi{10.1086/171931}

\bibitem[{Enea(2009)}]{speedglmspeedlm}
Enea, M. 2009, in Fitting Linear Models and Generalized Linear Models with large data sets in R

\bibitem[{{Evans} {et~al.}(2009){Evans}, {Beardmore}, {Page}, {Osborne}, {O'Brien}, {Willingale}, {Starling}, {Burrows}, {Godet}, {Vetere}, {Racusin}, {Goad}, {Wiersema}, {Angelini}, {Capalbi}, {Chincarini}, {Gehrels}, {Kennea}, {Margutti}, {Morris}, {Mountford}, {Pagani}, {Perri}, {Romano}, \& {Tanvir}}]{Evans2009}
{Evans}, P.~A., {Beardmore}, A.~P., {Page}, K.~L., {et~al.} 2009, MNRAS, 397, 1177, \dodoi{10.1111/j.1365-2966.2009.14913.x}

\bibitem[{{Fenimore} \& {Ramirez-Ruiz}(2000)}]{2000astro.ph..4176F}
{Fenimore}, E.~E., \& {Ramirez-Ruiz}, E. 2000, arXiv e-prints, astro, \dodoi{10.48550/arXiv.astro-ph/0004176}

\bibitem[{Fortin {et~al.}(2024)}]{Fortin:2024ofp}
Fortin, F., {et~al.} 2024.
\newblock \doarXiv{2410.16979}

\bibitem[{Fox(2015)}]{fox_2015}
Fox, J. 2015, Applied Regression Analysis and Generalized Linear Models (SAGE Publications)

\bibitem[{{Frail} {et~al.}(2006){Frail}, {Cameron}, {Kasliwal}, {Nakar}, {Price}, {Berger}, {Gal-Yam}, {Kulkarni}, {Fox}, {Soderberg}, {Schmidt}, {Ofek}, \& {Cenko}}]{frail2006}
{Frail}, D.~A., {Cameron}, P.~B., {Kasliwal}, M., {et~al.} 2006, ApJL, 646, L99, \dodoi{10.1086/506934}

\bibitem[{Friedman {et~al.}(2001)Friedman, Hastie, \& Tibshirani}]{friedman2001elements}
Friedman, J., Hastie, T., \& Tibshirani, R. 2001, The elements of statistical learning, Springer Series in statistics (New York, NY, USA: Springer)

\bibitem[{Friedman {et~al.}(2010)Friedman, Hastie, \& Tibshirani}]{friedman2010regularization}
---. 2010, Journal of statistical software, 33, 1

\bibitem[{Friedman {et~al.}(2000)Friedman, Hastie, Tibshirani, {et~al.}}]{friedman2000additive}
Friedman, J., Hastie, T., Tibshirani, R., {et~al.} 2000, Annals of statistics, 28, 337

\bibitem[{Friedman \& Roosen(1995)}]{FriedmanMARS}
Friedman, J.~H., \& Roosen, C.~B. 1995, Statistical Methods in Medical Research, 4, 197, \dodoi{10.1177/096228029500400303}

\bibitem[{Fukushima {et~al.}(2017)Fukushima, Omukai, \& Hosokawa}]{Fukushima_2017}
Fukushima, H., Omukai, K., \& Hosokawa, T. 2017, Monthly Notices of the Royal Astronomical Society, 473, 4754–4772, \dodoi{10.1093/mnras/stx2620}

\bibitem[{{Gehrels} {et~al.}(2004){Gehrels}, {Chincarini}, {Giommi}, {Mason}, {Nousek}, {Wells}, {White}, {Barthelmy}, {Burrows}, {Cominsky}, {Hurley}, {Marshall}, {M{\'e}sz{\'a}ros}, {Roming}, {Angelini}, {Barbier}, {Belloni}, {Campana}, {Caraveo}, {Chester}, {Citterio}, {Cline}, {Cropper}, {Cummings}, {Dean}, {Feigelson}, {Fenimore}, {Frail}, {Fruchter}, {Garmire}, {Gendreau}, {Ghisellini}, {Greiner}, {Hill}, {Hunsberger}, {Krimm}, {Kulkarni}, {Kumar}, {Lebrun}, {Lloyd-Ronning}, {Markwardt}, {Mattson}, {Mushotzky}, {Norris}, {Osborne}, {Paczynski}, {Palmer}, {Park}, {Parsons}, {Paul}, {Rees}, {Reynolds}, {Rhoads}, {Sasseen}, {Schaefer}, {Short}, {Smale}, {Smith}, {Stella}, {Tagliaferri}, {Takahashi}, {Tashiro}, {Townsley}, {Tueller}, {Turner}, {Vietri}, {Voges}, {Ward}, {Willingale}, {Zerbi}, \& {Zhang}}]{2004ApJ...611.1005G}
{Gehrels}, N., {Chincarini}, G., {Giommi}, P., {et~al.} 2004, ApJ, 611, 1005, \dodoi{10.1086/422091}

\bibitem[{Georgy {et~al.}(2013)Georgy, Ekström, Eggenberger, Meynet, Haemmerlé, Maeder, Granada, Groh, Hirschi, Mowlavi, Yusof, Charbonnel, Decressin, \& Barblan}]{Georgy_2013}
Georgy, C., Ekström, S., Eggenberger, P., {et~al.} 2013, Astronomy \&; Astrophysics, 558, A103, \dodoi{10.1051/0004-6361/201322178}

\bibitem[{{Geurts} {et~al.}(2006){Geurts}, {Ernst}, \& {Wehenkel}}]{geurts06extremetrees}
{Geurts}, P., {Ernst}, D., \& {Wehenkel}, L. 2006, Machine Learning, 63, 42, \dodoi{10.1007/s10994-006-6226-1}

\bibitem[{Gibson {et~al.}(2022)Gibson, Narendra, Dainotti, Bogdan, Pollo, Poliszczuk, Rinaldi, \& Liodakis}]{gibson2022}
Gibson, S.~J., Narendra, A., Dainotti, M.~G., {et~al.} 2022, Frontiers in Astronomy and Space Sciences, 9, \dodoi{10.3389/fspas.2022.836215}

\bibitem[{Godet {et~al.}(2014)Godet, Nasser, Atteia, Cordier, Mandrou, Barret, Triou, Pons, Amoros, Bordon, Gevin, Gonzalez, G{\"o}tz, Gros, Houret, Lachaud, Lacombe, Marty, Mercier, Rambaud, Ramon, Rouaix, Schanne, \& Waegebaert}]{10.1117/12.2055507}
Godet, O., Nasser, G., Atteia, J.-., {et~al.} 2014, in Space Telescopes and Instrumentation 2014: Ultraviolet to Gamma Ray, ed. T.~Takahashi, J.-W.~A. den Herder, \& M.~Bautz, Vol. 9144, International Society for Optics and Photonics (SPIE), 914424, \dodoi{10.1117/12.2055507}

\bibitem[{{Gorbovskoy} {et~al.}(2012){Gorbovskoy}, {Lipunova}, {Lipunov}, {Kornilov}, {Belinski}, {Shatskiy}, {Tyurina}, {Kuvshinov}, {Balanutsa}, {Chazov}, {Kuznetsov}, {Zimnukhov}, {Kornilov}, {Sankovich}, {Krylov}, {Ivanov}, {Chvalaev}, {Poleschuk}, {Konstantinov}, {Gress}, {Yazev}, {Budnev}, {Krushinski}, {Zalozhnich}, {Popov}, {Tlatov}, {Parhomenko}, {Dormidontov}, {Senik}, {Yurkov}, {Sergienko}, {Varda}, {Kudelina}, {Castro-Tirado}, {Gorosabel}, {S{\'a}nchez-Ram{\'\i}rez}, {Jelinek}, \& {Tello}}]{2012MNRAS.421.1874G}
{Gorbovskoy}, E.~S., {Lipunova}, G.~V., {Lipunov}, V.~M., {et~al.} 2012, MNRAS, 421, 1874, \dodoi{10.1111/j.1365-2966.2012.20195.x}

\bibitem[{Graham {et~al.}(2023{\natexlab{a}})Graham, Schady, \& Fruchter}]{graham_2023}
Graham, J.~F., Schady, P., \& Fruchter, A.~S. 2023{\natexlab{a}}, A Surprising Lack of Metallicity Evolution with Redshift in the Long Gamma-Ray Burst Host Galaxy Population.
\newblock \doarXiv{1904.02673}

\bibitem[{Graham {et~al.}(2023{\natexlab{b}})Graham, Schady, \& Fruchter}]{graham2023}
---. 2023{\natexlab{b}}, A Surprising Lack of Metallicity Evolution with Redshift in the Long Gamma-Ray Burst Host Galaxy Population.
\newblock \doarXiv{1904.02673}

\bibitem[{{Grupe} {et~al.}(2007){Grupe}, {Nousek}, {vanden Berk}, {Roming}, {Burrows}, {Godet}, {Osborne}, \& {Gehrels}}]{2007AJ....133.2216G}
{Grupe}, D., {Nousek}, J.~A., {vanden Berk}, D.~E., {et~al.} 2007, \aj, 133, 2216, \dodoi{10.1086/513014}

\bibitem[{Hastie \& Tibshirani(1987)}]{hastie1987generalized}
Hastie, T., \& Tibshirani, R. 1987, Journal of the American Statistical Association, 82, 371, \dodoi{10.2307/2289439}

\bibitem[{Hastie {et~al.}(2009)Hastie, Tibshirani, \& Friedman}]{hastie2009elements}
Hastie, T., Tibshirani, R., \& Friedman, J. 2009, The elements of statistical learning: data mining, inference, and prediction (Springer Science \& Business Media)

\bibitem[{Hastie \& Tibshirani(1990)}]{hastie1990generalized}
Hastie, T.~J., \& Tibshirani, R.~J. 1990, Generalized additive models, Vol.~43 (CRC press), \dodoi{10.1201/9780203753781}

\bibitem[{Heintz {et~al.}(2023)}]{Heintz:2022ozz}
Heintz, K.~E., {et~al.} 2023, Astrophys. J. Lett., 944, L30, \dodoi{10.3847/2041-8213/acb2cf}

\bibitem[{Ho(1995)}]{ho1995random}
Ho, T.~K. 1995, in Proceedings of 3rd international conference on document analysis and recognition, Vol.~1, IEEE, 278--282

\bibitem[{Hu {et~al.}(2023)Hu, Fernández-García, Caballero-García, Pérez-García, Carrasco-García, Castellón, Pérez~del Pulgar, Reina~Terol, \& Castro-Tirado}]{10.3389/fspas.2023.952887}
Hu, Y.-D., Fernández-García, E., Caballero-García, M.~D., {et~al.} 2023, Frontiers in Astronomy and Space Sciences, 10, \dodoi{10.3389/fspas.2023.952887}

\bibitem[{Huber(1964)}]{10.1214/aoms/1177703732}
Huber, P.~J. 1964, The Annals of Mathematical Statistics, 35, 73 , \dodoi{10.1214/aoms/1177703732}

\bibitem[{Huber \& Ronchetti(2009)}]{huber2009robust}
Huber, P.~J., \& Ronchetti, E.~M. 2009, Robust Statistics, 2nd edn. (Wiley), \dodoi{10.1002/9780470434697}

\bibitem[{Jelínek {et~al.}(2016)Jelínek, Castro-Tirado, Cunniffe, Gorosabel, Vítek, Kubánek, de~Ugarte~Postigo, Guziy, Tello, Páta, Sánchez-Ramírez, Oates, Jeong, Štrobl, Castillo-Carrión, Mateo~Sanguino, Rabaza, Pérez-Ramírez, Fernández-Muñoz, de~la Morena~Carretero, Hudec, Reglero, \& Sabau-Graziati}]{https://doi.org/10.1155/2016/1928465}
Jelínek, M., Castro-Tirado, A.~J., Cunniffe, R., {et~al.} 2016, Advances in Astronomy, 2016, 1928465, \dodoi{https://doi.org/10.1155/2016/1928465}

\bibitem[{{Kann} {et~al.}(2006){Kann}, {Klose}, \& {Zeh}}]{Kann2006}
{Kann}, D.~A., {Klose}, S., \& {Zeh}, A. 2006, ApJ, 641, 993, \dodoi{10.1086/500652}

\bibitem[{Karatzoglou {et~al.}(2004)Karatzoglou, Smola, Hornik, \& Zeileis}]{JSSv011i09}
Karatzoglou, A., Smola, A., Hornik, K., \& Zeileis, A. 2004, Journal of Statistical Software, 11, 1–20, \dodoi{10.18637/jss.v011.i09}

\bibitem[{Karson(1968)}]{doi:10.1080/01621459.1968.11009335}
Karson, M. 1968, Journal of the American Statistical Association, 63, 1047, \dodoi{10.1080/01621459.1968.11009335}

\bibitem[{Kodaira(1992)}]{Kodaira1992}
Kodaira, K. 1992, in ESO Conference and Workshop Proceedings, 43

\bibitem[{{Koen}(2009)}]{2009MNRAS.396.1499K}
{Koen}, C. 2009, \mnras, 396, 1499, \dodoi{10.1111/j.1365-2966.2009.14795.x}

\bibitem[{{Koen}(2010)}]{2010MNRAS.401.1369K}
---. 2010, \mnras, 401, 1369, \dodoi{10.1111/j.1365-2966.2009.15737.x}

\bibitem[{{Kouveliotou} {et~al.}(1993){Kouveliotou}, {Meegan}, {Fishman}, {Bhat}, {Briggs}, {Koshut}, {Paciesas}, \& {Pendleton}}]{1993ApJ...413L.101K}
{Kouveliotou}, C., {Meegan}, C.~A., {Fishman}, G.~J., {et~al.} 1993, \apjl, 413, L101, \dodoi{10.1086/186969}

\bibitem[{{Kumar} \& {Zhang}(2015)}]{2015PhR...561....1K}
{Kumar}, P., \& {Zhang}, B. 2015, \physrep, 561, 1, \dodoi{10.1016/j.physrep.2014.09.008}

\bibitem[{{Lamb}(2002)}]{2002luml.conf..157L}
{Lamb}, D.~Q. 2002, in Lighthouses of the Universe: The Most Luminous Celestial Objects and Their Use for Cosmology, ed. M.~{Gilfanov}, R.~{Sunyeav}, \& E.~{Churazov}, 157, \dodoi{10.1007/10856495_20}

\bibitem[{{Lamb} \& {Reichart}(2000)}]{lamb_reichart2000}
{Lamb}, D.~Q., \& {Reichart}, D.~E. 2000, in American Institute of Physics Conference Series, Vol. 526, Gamma-ray Bursts, 5th Huntsville Symposium, ed. R.~M. {Kippen}, R.~S. {Mallozzi}, \& G.~J. {Fishman}, 658--662, \dodoi{10.1063/1.1361618}

\bibitem[{{Lan} {et~al.}(2021){Lan}, {Wei}, {Zeng}, {Li}, \& {Wu}}]{2021MNRAS.508...52L}
{Lan}, G.-X., {Wei}, J.-J., {Zeng}, H.-D., {Li}, Y., \& {Wu}, X.-F. 2021, \mnras, 508, 52, \dodoi{10.1093/mnras/stab2508}

\bibitem[{{Lan} {et~al.}(2019){Lan}, {Zeng}, {Wei}, \& {Wu}}]{2019MNRAS.488.4607L}
{Lan}, G.-X., {Zeng}, H.-D., {Wei}, J.-J., \& {Wu}, X.-F. 2019, \mnras, 488, 4607, \dodoi{10.1093/mnras/stz2011}

\bibitem[{Lanza {et~al.}(2024)Lanza, Godet, Arcier, Yassine, Atteia, \& Bouchet}]{Lanza:2024sls}
Lanza, M.~L., Godet, O., Arcier, B., {et~al.} 2024, Astron. Astrophys., 685, A163, \dodoi{10.1051/0004-6361/202347966}

\bibitem[{{Levine} {et~al.}(2022){Levine}, {Dainotti}, {Zvonarek}, {Fraija}, {Warren}, {Chandra}, \& {Lloyd-Ronning}}]{2022ApJ...925...15L}
{Levine}, D., {Dainotti}, M., {Zvonarek}, K.~J., {et~al.} 2022, ApJ, 925, 15, \dodoi{10.3847/1538-4357/ac4221}

\bibitem[{{Li} {et~al.}(2018){Li}, {Wu}, {Lei}, {Dai}, {Liang}, \& {Ryde}}]{Li2018b}
{Li}, L., {Wu}, X.-F., {Lei}, W.-H., {et~al.} 2018, ApJ Supplement, 236, 26, \dodoi{10.3847/1538-4365/aabaf3}

\bibitem[{{Liang} {et~al.}(2007){Liang}, {Zhang}, \& {Zhang}}]{Liang2007}
{Liang}, E.-W., {Zhang}, B.-B., \& {Zhang}, B. 2007, ApJ, 670, 565, \dodoi{10.1086/521870}

\bibitem[{{Lien} {et~al.}(2016){Lien}, {Sakamoto}, {Barthelmy}, {Baumgartner}, {Cannizzo}, {Chen}, {Collins}, {Cummings}, {Gehrels}, {Krimm}, {Markwardt}, {Palmer}, {Stamatikos}, {Troja}, \& {Ukwatta}}]{2016ApJ...829....7L}
{Lien}, A., {Sakamoto}, T., {Barthelmy}, S.~D., {et~al.} 2016, ApJ, 829, 7, \dodoi{10.3847/0004-637X/829/1/7}

\bibitem[{Little \& Rubin(2019)}]{little2019statistical}
Little, R.~J., \& Rubin, D.~B. 2019, Statistical analysis with missing data, Vol. 793 (John Wiley \& Sons), \dodoi{10.1002/9781119013563}

\bibitem[{{Lloyd-Ronning} {et~al.}(2023){Lloyd-Ronning}, {Johnson}, {Cheng}, {Luu}, {Sanderbeck}, {Kenoly}, \& {Toral}}]{Lloyd2023ApJ...947...85L}
{Lloyd-Ronning}, N., {Johnson}, J., {Cheng}, R.~M., {et~al.} 2023, \apj, 947, 85, \dodoi{10.3847/1538-4357/acc795}

\bibitem[{{Lloyd-Ronning} {et~al.}(2002){Lloyd-Ronning}, {Fryer}, \& {Ramirez-Ruiz}}]{2002ApJ...574..554L}
{Lloyd-Ronning}, N.~M., {Fryer}, C.~L., \& {Ramirez-Ruiz}, E. 2002, \apj, 574, 554, \dodoi{10.1086/341059}

\bibitem[{{Maeder} \& {Meynet}(2001)}]{maeder&meynet2001}
{Maeder}, A., \& {Meynet}, G. 2001, \aap, 373, 555, \dodoi{10.1051/0004-6361:20010596}

\bibitem[{Maronna {et~al.}(2006)Maronna, Martin, \& Yohai}]{maronna2006robust}
Maronna, R.~A., Martin, R.~D., \& Yohai, V.~J. 2006, Robust Statistics: Theory and Methods (Wiley), \dodoi{10.1002/0470010940}

\bibitem[{McLachlan(2004)}]{mclachlan2004}
McLachlan, G.~J. 2004, Discriminant Analysis and Statistical Pattern Recognition (Wiley-Interscience)

\bibitem[{{Meegan} {et~al.}(2009){Meegan}, {Lichti}, {Bhat}, {Bissaldi}, {Briggs}, {Connaughton}, {Diehl}, {Fishman}, {Greiner}, {Hoover}, {van der Horst}, {von Kienlin}, {Kippen}, {Kouveliotou}, {McBreen}, {Paciesas}, {Preece}, {Steinle}, {Wallace}, {Wilson}, \& {Wilson-Hodge}}]{meegan2009}
{Meegan}, C., {Lichti}, G., {Bhat}, P.~N., {et~al.} 2009, \apj, 702, 791, \dodoi{10.1088/0004-637X/702/1/791}

\bibitem[{{Morgan} {et~al.}(2012){Morgan}, {Long}, {Richards}, {Broderick}, {Butler}, \& {Bloom}}]{morgan2012}
{Morgan}, A.~N., {Long}, J., {Richards}, J.~W., {et~al.} 2012, ApJ, 746, 170, \dodoi{10.1088/0004-637X/746/2/170}

\bibitem[{{Nanayakkara} {et~al.}(2017){Nanayakkara}, {Glazebrook}, {Kacprzak}, {Yuan}, {Fisher}, {Tran}, {Kewley}, {Spitler}, {Alcorn}, {Cowley}, {Labbe}, {Straatman}, \& {Tomczak}}]{2017MNRAS.468.3071N}
{Nanayakkara}, T., {Glazebrook}, K., {Kacprzak}, G.~G., {et~al.} 2017, \mnras, 468, 3071, \dodoi{10.1093/mnras/stx605}

\bibitem[{Narendra {et~al.}(2022)Narendra, Gibson, Dainotti, Bogdan, Pollo, Liodakis, Poliszczuk, \& Rinaldi}]{narendra2022predicting}
Narendra, A., Gibson, S.~J., Dainotti, M.~G., {et~al.} 2022, ApJ Supplement Series, 259, 55, \dodoi{10.3847/1538-4365/ac545a}

\bibitem[{Nelder \& Wedderburn(1972{\natexlab{a}})}]{nelder1972generalized}
Nelder, J.~A., \& Wedderburn, R.~W. 1972{\natexlab{a}}, Journal of the Royal Statistical Society: Series A (General), 135, 370

\bibitem[{Nelder \& Wedderburn(1972{\natexlab{b}})}]{68aee965-a8a0-3e72-9f89-8d89ae91a62b}
Nelder, J.~A., \& Wedderburn, R. W.~M. 1972{\natexlab{b}}, Journal of the Royal Statistical Society. Series A (General), 135, 370.
\newblock \url{http://www.jstor.org/stable/2344614}

\bibitem[{Niino {et~al.}(2017)Niino, Aoki, Hashimoto, Hattori, Ishikawa, Kashikawa, Kosugi, Onoue, Toshikawa, \& Yabe}]{Niino_2017}
Niino, Y., Aoki, K., Hashimoto, T., {et~al.} 2017, Publications of the Astronomical Society of Japan, 69, \dodoi{10.1093/pasj/psw133}

\bibitem[{{Nousek} {et~al.}(2006){Nousek}, {Kouveliotou}, {Grupe}, {Page}, {Granot}, {Ramirez-Ruiz}, {Patel}, {Burrows}, {Mangano}, {Barthelmy}, {Beardmore}, {Campana}, {Capalbi}, {Chincarini}, {Cusumano}, {Falcone}, {Gehrels}, {Giommi}, {Goad}, {Godet}, {Hurkett}, {Kennea}, {Moretti}, {O'Brien}, {Osborne}, {Romano}, {Tagliaferri}, \& {Wells}}]{Nousek2006}
{Nousek}, J.~A., {Kouveliotou}, C., {Grupe}, D., {et~al.} 2006, ApJ, 642, 389, \dodoi{10.1086/500724}

\bibitem[{{Oates} {et~al.}(2012){Oates}, {Page}, {De Pasquale}, {Schady}, {Breeveld}, {Holland}, {Kuin}, \& {Marshall}}]{Oates2012}
{Oates}, S.~R., {Page}, M.~J., {De Pasquale}, M., {et~al.} 2012, MNRAS, 426, L86, \dodoi{10.1111/j.1745-3933.2012.01331.x}

\bibitem[{{O'Brien} {et~al.}(2006){O'Brien}, {Willingale}, {Osborne}, {Goad}, {Page}, {Vaughan}, {Rol}, {Beardmore}, {Godet}, {Hurkett}, {Wells}, {Zhang}, {Kobayashi}, {Burrows}, {Nousek}, {Kennea}, {Falcone}, {Grupe}, {Gehrels}, {Barthelmy}, {Cannizzo}, {Cummings}, {Hill}, {Krimm}, {Chincarini}, {Tagliaferri}, {Campana}, {Moretti}, {Giommi}, {Perri}, {Mangano}, \& {LaParola}}]{OBrien2006}
{O'Brien}, P.~T., {Willingale}, R., {Osborne}, J., {et~al.} 2006, ApJ, 647, 1213, \dodoi{10.1086/505457}

\bibitem[{{Paczy{\'n}ski}(1998)}]{paczynski1998ApJ...494L..45P}
{Paczy{\'n}ski}, B. 1998, ApJL, 494, L45, \dodoi{10.1086/311148}

\bibitem[{{Palmerio} {et~al.}(2019){Palmerio}, {Vergani}, {Salvaterra}, {Sanders}, {Japelj}, {Vidal-Garc{\'\i}a}, {D'Avanzo}, {Corre}, {Perley}, {Shapley}, {Boissier}, {Greiner}, {Le Floc'h}, \& {Wiseman}}]{Palmerio2019A&A...623A..26P}
{Palmerio}, J.~T., {Vergani}, S.~D., {Salvaterra}, R., {et~al.} 2019, \aap, 623, A26, \dodoi{10.1051/0004-6361/201834179}

\bibitem[{{Panaitescu} \& {Vestrand}(2008)}]{panaitescu2008taxonomy}
{Panaitescu}, A., \& {Vestrand}, W.~T. 2008, MNRAS, 387, 497, \dodoi{10.1111/j.1365-2966.2008.13231.x}

\bibitem[{{Panaitescu} \& {Vestrand}(2011)}]{panaitescu2011optical}
---. 2011, MNRAS, 414, 3537, \dodoi{10.1111/j.1365-2966.2011.18653.x}

\bibitem[{Parikh {et~al.}(2008)Parikh, Mathai, Parikh, Sekhar, \& Thomas}]{parikh2008understanding}
Parikh, R., Mathai, A., Parikh, S., Sekhar, G.~C., \& Thomas, R. 2008, Indian journal of ophthalmology, 56, 45

\bibitem[{{Pescalli} {et~al.}(2016){Pescalli}, {Ghirlanda}, {Salvaterra}, {Ghisellini}, {Vergani}, {Nappo}, {Salafia}, {Melandri}, {Covino}, \& {G{\"o}tz}}]{pescalli2016A&A...587A..40P}
{Pescalli}, A., {Ghirlanda}, G., {Salvaterra}, R., {et~al.} 2016, A\&A, 587, A40, \dodoi{10.1051/0004-6361/201526760}

\bibitem[{{Petrosian} \& {Dainotti}(2024)}]{2024ApJ...963L..12P}
{Petrosian}, V., \& {Dainotti}, M.~G. 2024, \apjl, 963, L12, \dodoi{10.3847/2041-8213/ad2763}

\bibitem[{{Petrosian} {et~al.}(2015){Petrosian}, {Kitanidis}, \& {Kocevski}}]{petrosian2015}
{Petrosian}, V., {Kitanidis}, E., \& {Kocevski}, D. 2015, ApJ, 806, 44, \dodoi{10.1088/0004-637X/806/1/44}

\bibitem[{Phillips {et~al.}(2024)Phillips, Oey, Cuevas, Castro, \& Kothari}]{phillips2024runawayobstarssmall}
Phillips, G.~D., Oey, M.~S., Cuevas, M., Castro, N., \& Kothari, R. 2024, Runaway OB Stars in the Small Magellanic Cloud III. Updated Kinematics and Insights on Dynamical vs. Supernova Ejections.
\newblock \doarXiv{2403.17198}

\bibitem[{{Piro} {et~al.}(1998){Piro}, {Amati}, {Antonelli}, {Butler}, {Costa}, {Cusumano}, {Feroci}, {Frontera}, {Heise}, {in 't Zand}, {Molendi}, {Muller}, {Nicastro}, {Orlandini}, {Owens}, {Parmar}, {Soffitta}, \& {Tavani}}]{Piro1998}
{Piro}, L., {Amati}, L., {Antonelli}, L.~A., {et~al.} 1998, A\&A, 331, L41.
\newblock \doarXiv{astro-ph/9710355}

\bibitem[{Polley {et~al.}(2011)Polley, Rose, \& van~der Laan}]{polley2010super}
Polley, E.~C., Rose, S., \& van~der Laan, M.~J. 2011, Super Learning (New York, NY: Springer New York), 43--66

\bibitem[{{Produit} {et~al.}(2023){Produit}, {Kole}, {Wu}, {De Angelis}, {Li}, {Rybka}, {Pollo}, {Mianowski}, {Greiner}, {Burgess}, {Sun}, \& {Zhang}}]{2023arXiv230900518P}
{Produit}, N., {Kole}, M., {Wu}, X., {et~al.} 2023, arXiv e-prints, arXiv:2309.00518, \dodoi{10.48550/arXiv.2309.00518}

\bibitem[{{R Core Team}(2022)}]{R}
{R Core Team}. 2022, R: A Language and Environment for Statistical Computing, R Foundation for Statistical Computing, Vienna, Austria.
\newblock \url{https://www.R-project.org/}

\bibitem[{Rahin \& Behar(2019{\natexlab{a}})}]{Rahin_2019}
Rahin, R., \& Behar, E. 2019{\natexlab{a}}, The Astrophysical Journal, 885, 47, \dodoi{10.3847/1538-4357/ab3e34}

\bibitem[{Rahin \& Behar(2019{\natexlab{b}})}]{Rahin:2019dwy}
---. 2019{\natexlab{b}}, \dodoi{10.3847/1538-4357/ab3e34}

\bibitem[{Robertson {et~al.}(2023)Robertson, Tacchella, Johnson, Hainline, Whitler, Eisenstein, Endsley, Rieke, Stark, Alberts, Dressler, Egami, Hausen, Rieke, Shivaei, Williams, Willmer, Arribas, Bonaventura, Bunker, Cameron, Carniani, Charlot, Chevallard, Curti, Curtis-Lake, D’Eugenio, Jakobsen, Looser, Lützgendorf, Maiolino, Maseda, Rawle, Rix, Smit, Übler, Willott, Witstok, Baum, Bhatawdekar, Boyett, Chen, de~Graaff, Florian, Helton, Hviding, Ji, Kumari, Lyu, Nelson, Sandles, Saxena, Suess, Sun, Topping, \& Wallace}]{highzgalaxyRobertson_2023}
Robertson, B.~E., Tacchella, S., Johnson, B.~D., {et~al.} 2023, Nature Astronomy, 7, 611–621, \dodoi{10.1038/s41550-023-01921-1}

\bibitem[{{Roming} {et~al.}(2005){Roming}, {Kennedy}, {Mason}, {Nousek}, {Ahr}, {Bingham}, {Broos}, {Carter}, {Hancock}, {Huckle}, {Hunsberger}, {Kawakami}, {Killough}, {Koch}, {McLelland}, {Smith}, {Smith}, {Soto}, {Boyd}, {Breeveld}, {Holland}, {Ivanushkina}, {Pryzby}, {Still}, \& {Stock}}]{Roming2005}
{Roming}, P. W.~A., {Kennedy}, T.~E., {Mason}, K.~O., {et~al.} 2005, SSRv, 120, 95, \dodoi{10.1007/s11214-005-5095-4}

\bibitem[{{Rowlinson} {et~al.}(2014){Rowlinson}, {Gompertz}, {Dainotti}, {O'Brien}, {Wijers}, \& {van der Horst}}]{Rowlinson2014}
{Rowlinson}, A., {Gompertz}, B.~P., {Dainotti}, M., {et~al.} 2014, MNRAS, 443, 1779, \dodoi{10.1093/mnras/stu1277}

\bibitem[{Rubin(1976)}]{rubin1976inference}
Rubin, D.~B. 1976, Biometrika, 63, 581, \dodoi{10.2307/2335739}

\bibitem[{{Sakamoto} {et~al.}(2007){Sakamoto}, {Hill}, {Yamazaki}, {Angelini}, {Krimm}, {Sato}, {Swindell}, {Takami}, \& {Osborne}}]{Sakamoto2007}
{Sakamoto}, T., {Hill}, J.~E., {Yamazaki}, R., {et~al.} 2007, ApJ, 669, 1115, \dodoi{10.1086/521640}

\bibitem[{Sako {et~al.}(2018)Sako, Ohsawa, Takahashi, Kojima, Doi, Kobayashi, Aoki, Arima, Arimatsu, Ichiki, Ikeda, Inooka, Ita, Kasuga, Kokubo, Konishi, Maehara, Matsunaga, Mitsuda, Miyata, Mori, Morii, Morokuma, Motohara, Nakada, Okumura, Sarugaku, Sato, Shigeyama, Soyano, Tanaka, Tarusawa, Tominaga, Totani, Urakawa, Usui, Watanabe, Yamashita, \& Yoshikawa}]{10.1117/12.2310049}
Sako, S., Ohsawa, R., Takahashi, H., {et~al.} 2018, in Ground-based and Airborne Instrumentation for Astronomy VII, ed. C.~J. Evans, L.~Simard, \& H.~Takami, Vol. 10702, International Society for Optics and Photonics (SPIE), 107020J, \dodoi{10.1117/12.2310049}

\bibitem[{{Salvaterra} {et~al.}(2007){Salvaterra}, {Campana}, {Chincarini}, {Tagliaferri}, \& {Covino}}]{2007MNRAS.380L..45S}
{Salvaterra}, R., {Campana}, S., {Chincarini}, G., {Tagliaferri}, G., \& {Covino}, S. 2007, \mnras, 380, L45, \dodoi{10.1111/j.1745-3933.2007.00345.x}

\bibitem[{Salvaterra {et~al.}(2012)Salvaterra, Campana, Vergani, Covino, D’Avanzo, Fugazza, Ghirlanda, Ghisellini, Melandri, Nava, {et~al.}}]{salvaterra2012complete}
Salvaterra, R., Campana, S., Vergani, S., {et~al.} 2012, ApJ, 749, 68

\bibitem[{Sanyal {et~al.}(2017)Sanyal, Langer, Szécsi, C~Yoon, \& Grassitelli}]{Sanyal_2017}
Sanyal, D., Langer, N., Szécsi, D., C~Yoon, S., \& Grassitelli, L. 2017, Astronomy \&; Astrophysics, 597, A71, \dodoi{10.1051/0004-6361/201629612}

\bibitem[{Schafer \& Graham(2002)}]{schafer2002missing}
Schafer, J.~L., \& Graham, J.~W. 2002, Psychological methods, 7, 147, \dodoi{10.1037/1082-989X.7.2.147}

\bibitem[{Simon {et~al.}(2011)Simon, Friedman, Hastie, \& Tibshirani}]{JSSv039i05}
Simon, N., Friedman, J.~H., Hastie, T., \& Tibshirani, R. 2011, Journal of Statistical Software, 39, 1–13, \dodoi{10.18637/jss.v039.i05}

\bibitem[{{Srinivasaragavan} {et~al.}(2020){Srinivasaragavan}, {Dainotti}, {Fraija}, {Hernandez}, {Nagataki}, {Lenart}, {Bowden}, \& {Wagner}}]{Srinivasaragavan2020}
{Srinivasaragavan}, G.~P., {Dainotti}, M.~G., {Fraija}, N., {et~al.} 2020, ApJ, 903, 18, \dodoi{10.3847/1538-4357/abb702}

\bibitem[{Stacy {et~al.}(2010)Stacy, Greif, \& Bromm}]{10.1111/j.1365-2966.2009.16113.x}
Stacy, A., Greif, T.~H., \& Bromm, V. 2010, Monthly Notices of the Royal Astronomical Society, 403, 45, \dodoi{10.1111/j.1365-2966.2009.16113.x}

\bibitem[{Starling {et~al.}(2013)Starling, Willingale, Tanvir, Scott, Wiersema, O’Brien, Levan, \& Stewart}]{Starling2013}
Starling, R. L.~C., Willingale, R., Tanvir, N.~R., {et~al.} 2013, Monthly Notices of the Royal Astronomical Society, 431, 3159, \dodoi{10.1093/mnras/stt400}

\bibitem[{{Sugita} {et~al.}(2022){Sugita}, {Yoshida}, {Sakamoto}, {Kawakubo}, {Yamaoka}, {Nakahira}, {Asaoka}, {Torii}, {Akaike}, {Kobayashi}, {Shimizu}, {Tamura}, {Cannady}, {Cherry}, {Ricciarini}, {Marrocchesi}, \& {Calet Collaboration}}]{Sugita2022}
{Sugita}, S., {Yoshida}, A., {Sakamoto}, T., {et~al.} 2022, GRB Coordinates Network, 32124, 1

\bibitem[{Suwa \& Ioka(2010)}]{Suwa_2010}
Suwa, Y., \& Ioka, K. 2010, The Astrophysical Journal, 726, 107, \dodoi{10.1088/0004-637x/726/2/107}

\bibitem[{{Takase} {et~al.}(1977){Takase}, {Ishida}, {Shimizu}, {Maehara}, {Hamajima}, {Noguchi}, \& {Ohashi}}]{1977AnTok..16...74T}
{Takase}, B., {Ishida}, K., {Shimizu}, M., {et~al.} 1977, Annals of the Tokyo Astronomical Observatory, 16, 74

\bibitem[{Tanvir {et~al.}(2019)}]{Tanvir:2018pbq}
Tanvir, N.~R., {et~al.} 2019, Mon. Not. Roy. Astron. Soc., 483, 5380, \dodoi{10.1093/mnras/sty3460}

\bibitem[{Tanvir {et~al.}(2021)Tanvir, Le~Floc’h, Christensen, Caruana, Salvaterra, Ghirlanda, Ciardi, Maio, D’Odorico, Piedipalumbo, Campana, Noterdaeme, Graziani, Amati, Bagoly, Balázs, Basa, Behar, De~Cia, Valle, De~Pasquale, Frontera, Gomboc, Götz, Horvath, Hudec, Mereghetti, O’Brien, Osborne, Paltani, Rosati, Sergijenko, Stanway, Szécsi, Tot́h, Urata, Vergani, \& Zane}]{Tanvir_2021}
Tanvir, N.~R., Le~Floc’h, E., Christensen, L., {et~al.} 2021, Experimental Astronomy, 52, 219–244, \dodoi{10.1007/s10686-021-09778-w}

\bibitem[{Tibshirani(1996)}]{TibshiraniLasso}
Tibshirani, R. 1996, Journal of the Royal Statistical Society Series B, 58, 267

\bibitem[{Toma {et~al.}(2011)Toma, Sakamoto, \& Mészáros}]{Toma_2011}
Toma, K., Sakamoto, T., \& Mészáros, P. 2011, The Astrophysical Journal, 731, 127, \dodoi{10.1088/0004-637x/731/2/127}

\bibitem[{Torsten~Hothorn \& Zeileis(2006)}]{doi:10.1198/106186006X133933}
Torsten~Hothorn, K.~H., \& Zeileis, A. 2006, Journal of Computational and Graphical Statistics, 15, 651, \dodoi{10.1198/106186006X133933}

\bibitem[{Toyouchi {et~al.}(2018)Toyouchi, Hosokawa, Sugimura, Nakatani, \& Kuiper}]{Toyouchi_2018}
Toyouchi, D., Hosokawa, T., Sugimura, K., Nakatani, R., \& Kuiper, R. 2018, Monthly Notices of the Royal Astronomical Society, \dodoi{10.1093/mnras/sty3012}

\bibitem[{Trani {et~al.}(2014)Trani, Mapelli, \& Bressan}]{Trani_2014}
Trani, A.~A., Mapelli, M., \& Bressan, A. 2014, Monthly Notices of the Royal Astronomical Society, 445, 1967–1976, \dodoi{10.1093/mnras/stu1898}

\bibitem[{{Ukwatta} {et~al.}(2009){Ukwatta}, {Sakamoto}, {Dhuga}, {Parke}, {Barthelmy}, {Gehrels}, {Stamatikos}, \& {Tueller}}]{2009AIPC.1133..437U}
{Ukwatta}, T.~N., {Sakamoto}, T., {Dhuga}, K.~S., {et~al.} 2009, in American Institute of Physics Conference Series, Vol. 1133, Gamma-ray Burst: Sixth Huntsville Symposium, ed. C.~{Meegan}, C.~{Kouveliotou}, \& N.~{Gehrels} (AIP), 437--439, \dodoi{10.1063/1.3155945}

\bibitem[{{Ukwatta} {et~al.}(2008){Ukwatta}, {Sakamoto}, {Stamatikos}, {Gehrels}, \& {Dhuga}}]{2008AIPC.1000..166U}
{Ukwatta}, T.~N., {Sakamoto}, T., {Stamatikos}, M., {Gehrels}, N., \& {Dhuga}, K.~S. 2008, in American Institute of Physics Conference Series, Vol. 1000, Gamma-ray Bursts 2007, ed. M.~{Galassi}, D.~{Palmer}, \& E.~{Fenimore} (AIP), 166--169, \dodoi{10.1063/1.2943435}

\bibitem[{{Ukwatta} {et~al.}(2016){Ukwatta}, {Wo{\'z}niak}, \& {Gehrels}}]{ukwatta2016machine}
{Ukwatta}, T.~N., {Wo{\'z}niak}, P.~R., \& {Gehrels}, N. 2016, MNRAS, 458, 3821, \dodoi{10.1093/mnras/stw559}

\bibitem[{van Buuren \& Groothuis-Oudshoorn(2011)}]{van2011mice}
van Buuren, S., \& Groothuis-Oudshoorn, K. 2011, JSS, 45, 1–67, \dodoi{10.18637/jss.v045.i03}

\bibitem[{Van~der Laan {et~al.}(2007)Van~der Laan, Polley, \& Hubbard}]{van2007super}
Van~der Laan, M.~J., Polley, E.~C., \& Hubbard, A.~E. 2007, Statistical applications in genetics and molecular biology, 6

\bibitem[{{van Paradijs} {et~al.}(1997){van Paradijs}, {Groot}, {Galama}, {Kouveliotou}, {Strom}, {Telting}, {Rutten}, {Fishman}, {Meegan}, {Pettini}, {Tanvir}, {Bloom}, {Pedersen}, {N{\o}rdgaard-Nielsen}, {Linden-V{\o}rnle}, {Melnick}, {Van der Steene}, {Bremer}, {Naber}, {Heise}, {in't Zand}, {Costa}, {Feroci}, {Piro}, {Frontera}, {Zavattini}, {Nicastro}, {Palazzi}, {Bennett}, {Hanlon}, \& {Parmar}}]{vanParadijs1997}
{van Paradijs}, J., {Groot}, P.~J., {Galama}, T., {et~al.} 1997, Nature, 386, 686, \dodoi{10.1038/386686a0}

\bibitem[{{vanden Berk} {et~al.}(2008){vanden Berk}, {Grupe}, {Racusin}, {Roming}, \& {Koch}}]{2008AIPC.1000...80V}
{vanden Berk}, D.~E., {Grupe}, D., {Racusin}, J., {Roming}, P., \& {Koch}, S. 2008, in American Institute of Physics Conference Series, Vol. 1000, Gamma-ray Bursts 2007, ed. M.~{Galassi}, D.~{Palmer}, \& E.~{Fenimore} (AIP), 80--83, \dodoi{10.1063/1.2943555}

\bibitem[{Vapnik(1995)}]{vapnik1995nature}
Vapnik, V.~N. 1995, The Nature of Statistical Learning Theory (New York, NY: Springer-Verlag New York, Inc.)

\bibitem[{{Vestrand} {et~al.}(2005){Vestrand}, {Wozniak}, {Wren}, {Fenimore}, {Sakamoto}, {White}, {Casperson}, {Davis}, {Evans}, {Galassi}, {McGowan}, {Schier}, {Asa}, {Barthelmy}, {Cummings}, {Gehrels}, {Hullinger}, {Krimm}, {Markwardt}, {McLean}, {Palmer}, {Parsons}, \& {Tueller}}]{Vestrand2005Natur}
{Vestrand}, W.~T., {Wozniak}, P.~R., {Wren}, J.~A., {et~al.} 2005, Nature, 435, 178, \dodoi{10.1038/nature03515}

\bibitem[{{Vestrand} {et~al.}(2014){Vestrand}, {Wren}, {Panaitescu}, {Wozniak}, {Davis}, {Palmer}, {Vianello}, {Omodei}, {Xiong}, {Briggs}, {Elphick}, {Paciesas}, \& {Rosing}}]{2014Sci...343...38V}
{Vestrand}, W.~T., {Wren}, J.~A., {Panaitescu}, A., {et~al.} 2014, Science, 343, 38, \dodoi{10.1126/science.1242316}

\bibitem[{Wager \& Athey(2017)}]{wager2017estimation}
Wager, S., \& Athey, S. 2017, Estimation and Inference of Heterogeneous Treatment Effects using Random Forests.
\newblock \doarXiv{1510.04342}

\bibitem[{{Wei} \& {Gao}(2003)}]{2003MNRAS.345..743W}
{Wei}, D.~M., \& {Gao}, W.~H. 2003, \mnras, 345, 743, \dodoi{10.1046/j.1365-8711.2003.06971.x}

\bibitem[{Wei \& Cordier(2016)}]{Wei:2016eox}
Wei, J., \& Cordier, B. 2016.
\newblock \doarXiv{1610.06892}

\bibitem[{Wei \& Liu(2022)}]{Wei_2022}
Wei, Y.-F., \& Liu, T. 2022, The Astrophysical Journal, 936, 182, \dodoi{10.3847/1538-4357/ac8bd1}

\bibitem[{{Wilkins} {et~al.}(2008){Wilkins}, {Trentham}, \& {Hopkins}}]{2008MNRAS.385..687W}
{Wilkins}, S.~M., {Trentham}, N., \& {Hopkins}, A.~M. 2008, \mnras, 385, 687, \dodoi{10.1111/j.1365-2966.2008.12885.x}

\bibitem[{{Willingale} {et~al.}(2007){Willingale}, {O'Brien}, {Osborne}, {Godet}, {Page}, {Goad}, {Burrows}, {Zhang}, {Rol}, {Gehrels}, \& {Chincarini}}]{Willingale2007}
{Willingale}, R., {O'Brien}, P.~T., {Osborne}, J.~P., {et~al.} 2007, ApJ, 662, 1093, \dodoi{10.1086/517989}

\bibitem[{{Woosley}(1993)}]{woosley1993ApJ...405..273W}
{Woosley}, S.~E. 1993, ApJ, 405, 273, \dodoi{10.1086/172359}

\bibitem[{{Woosley} \& {Bloom}(2006)}]{Woosley2006ARA&A}
{Woosley}, S.~E., \& {Bloom}, J.~S. 2006, ARA\&A, 44, 507, \dodoi{10.1146/annurev.astro.43.072103.150558}

\bibitem[{{Woosley} {et~al.}(1993){Woosley}, {Langer}, \& {Weaver}}]{woosley1993}
{Woosley}, S.~E., {Langer}, N., \& {Weaver}, T.~A. 1993, ApJ, 411, 823, \dodoi{10.1086/172886}

\bibitem[{{Yonetoku} \& {Hiz-Gundam Collaboration}(2023)}]{2023HEAD...2010318Y}
{Yonetoku}, D., \& {Hiz-Gundam Collaboration}. 2023, in AAS/High Energy Astrophysics Division, Vol.~20, AAS/High Energy Astrophysics Division, 103.18

\bibitem[{{Yonetoku} {et~al.}(2004){Yonetoku}, {Murakami}, {Nakamura}, {Yamazaki}, {Inoue}, \& {Ioka}}]{yonetoku2004}
{Yonetoku}, D., {Murakami}, T., {Nakamura}, T., {et~al.} 2004, ApJ, 609, 935, \dodoi{10.1086/421285}

\bibitem[{{Yonetoku} {et~al.}(2014){Yonetoku}, {Mihara}, {Sawano}, {Ikeda}, {Harayama}, {Takata}, {Yoshida}, {Seta}, {Toyanago}, {Kagawa}, {Kawai}, {Kawai}, {Sakamoto}, {Serino}, {Kurosawa}, {Gunji}, {Tanimori}, {Murakami}, {Yatsu}, {Yamaoka}, {Yoshida}, {Kawabata}, {Matsumoto}, {Tsumura}, {Matsuura}, {Shirahata}, {Okita}, {Yanagisawa}, {Yoshida}, \& {Motohara}}]{GUNDAM}
{Yonetoku}, D., {Mihara}, T., {Sawano}, T., {et~al.} 2014, in Society of Photo-Optical Instrumentation Engineers (SPIE) Conference Series, Vol. 9144, Space Telescopes and Instrumentation 2014: Ultraviolet to Gamma Ray, ed. T.~{Takahashi}, J.-W.~A. {den Herder}, \& M.~{Bautz}, 91442S, \dodoi{10.1117/12.2055041}

\bibitem[{{Yu} {et~al.}(2015){Yu}, {Wang}, {Dai}, \& {Cheng}}]{Yu2015ApJS..218...13Y}
{Yu}, H., {Wang}, F.~Y., {Dai}, Z.~G., \& {Cheng}, K.~S. 2015, ApJ Supplement, 218, 13, \dodoi{10.1088/0067-0049/218/1/13}

\bibitem[{Yu {et~al.}(2002)Yu, Ji, \& Zhang}]{kernelkNN}
Yu, K., Ji, L., \& Zhang, X. 2002, Neural Processing Letters, 15, 147, \dodoi{10.1023/A:1015244902967}

\bibitem[{Yuan {et~al.}(2022)Yuan, Zhang, Chen, \& Ling}]{Yuan2022}
Yuan, W., Zhang, C., Chen, Y., \& Ling, Z. 2022, The Einstein Probe Mission (Singapore: Springer Nature Singapore), 1--30, \dodoi{10.1007/978-981-16-4544-0_151-1}

\bibitem[{Zahid {et~al.}(2013)Zahid, Geller, Kewley, Hwang, Fabricant, \& Kurtz}]{Zahid_2013}
Zahid, H.~J., Geller, M.~J., Kewley, L.~J., {et~al.} 2013, The Astrophysical Journal, 771, L19, \dodoi{10.1088/2041-8205/771/2/l19}

\bibitem[{{Zaninoni} {et~al.}(2013){Zaninoni}, {Bernardini}, {Margutti}, {Oates}, \& {Chincarini}}]{Zaninoni2013}
{Zaninoni}, E., {Bernardini}, M.~G., {Margutti}, R., {Oates}, S., \& {Chincarini}, G. 2013, A\&A, 557, A12, \dodoi{10.1051/0004-6361/201321221}

\bibitem[{{Zeh} {et~al.}(2006){Zeh}, {Klose}, \& {Kann}}]{Zeh2006}
{Zeh}, A., {Klose}, S., \& {Kann}, D.~A. 2006, ApJ, 637, 889, \dodoi{10.1086/498442}

\bibitem[{Zeng \& Breheny(2017)}]{zeng2017biglasso}
Zeng, Y., \& Breheny, P. 2017, arXiv preprint arXiv:1701.05936

\bibitem[{Zhang(2014)}]{2014IJMPD..2330002Z}
Zhang, B. 2014, International Journal of Modern Physics D, 23, 1430002, \dodoi{10.1142/S021827181430002X}

\bibitem[{Zhang(2018)}]{zhang_2018}
---. 2018, The Physics of Gamma-Ray Bursts (Cambridge University Press), \dodoi{10.1017/9781139226530}

\bibitem[{{Zhang} {et~al.}(2006){Zhang}, {Fan}, {Dyks}, {Kobayashi}, {M{\'e}sz{\'a}ros}, {Burrows}, {Nousek}, \& {Gehrels}}]{Zhang2006}
{Zhang}, B., {Fan}, Y.~Z., {Dyks}, J., {et~al.} 2006, ApJ, 642, 354, \dodoi{10.1086/500723}

\bibitem[{{Zhang} \& {M{\'e}sz{\'a}ros}(2004)}]{Zhang2004IJMPA..19.2385Z}
{Zhang}, B., \& {M{\'e}sz{\'a}ros}, P. 2004, Int J Mod Phys A, 19, 2385, \dodoi{10.1142/S0217751X0401746X}

\bibitem[{{Zhang} {et~al.}(2009){Zhang}, {Zhang}, {Virgili}, {Liang}, {Kann}, {Wu}, {Proga}, {Lv}, {Toma}, {M{\'e}sz{\'a}ros}, {Burrows}, {Roming}, \& {Gehrels}}]{Zhang2009}
{Zhang}, B., {Zhang}, B.-B., {Virgili}, F.~J., {et~al.} 2009, ApJ, 703, 1696, \dodoi{10.1088/0004-637X/703/2/1696}

\bibitem[{Zhang {et~al.}(2014)Zhang, Zhang, Murase, Connaughton, \& Briggs}]{zhang2014long}
Zhang, B.-B., Zhang, B., Murase, K., Connaughton, V., \& Briggs, M.~S. 2014, ApJ, 787, 66, \dodoi{10.1088/0004-637X/787/1/66}

\bibitem[{Zou \& Hastie(2005)}]{10.1111/j.1467-9868.2005.00503.x}
Zou, H., \& Hastie, T. 2005, Journal of the Royal Statistical Society Series B: Statistical Methodology, 67, 301, \dodoi{10.1111/j.1467-9868.2005.00503.x}

\end{thebibliography}
\bibliographystyle{aasjournal}



\end{document}